**The Role of Social Movements, Coalitions, and Workers in Resisting Harmful Artificial Intelligence and Contributing to the Development of Responsible AI**

**Working Paper**
**Susan von Struensee, JD, MPH**
**Global Research Initiative**
**June 2021**

**Introduction**

The rapid spread of artificial intelligence (AI) systems has precipitated a rise in ethical and rights-based frameworks intended to guide the development and use of these technologies. Despite the proliferation of these "AI principles, " there is mounting public concern over the influence that the AI systems have in our society, and coalitions in all sectors are organizing to resist harmful applications of AI worldwide.

The globe has witnessed an exponential growth in the use of AI and other automated decision-making systems. Government institutions increasingly rely on automated decision-making technologies in many areas, such as managing traffic,[1] conducting risk assessments,[2] screening immigrants,[3] allocating social services,[4] and more.[5] Private companies have integrated AI into their hiring processes,[6] lending and loan management,[7] and other functions.[8] Many decisions that were once done by humans are gradually being performed by automated AI systems.[9]

The growing pervasiveness of AI-based systems that govern human behavior presents numerous advantages, but also heightens the need to ensure sufficient oversight of such automated decision-making processes.[10] A major concern is that AI systems exhibit and intensify human biases and unfair, discriminatory, and derogatory practices.[11]

---

[1] Miguel Carrasco , Steven Mills , Adam Whybrew, and Adam Jura, The Citizen's Perspective on the Use of AI in Government, BCG DIGITAL GOVERNMENT BENCHMARKING (March 1, 2019) https://www.bcg.com/en-il/publications/2019/citizen-perspective-use-artificial-intelligence-government-digital-benchmarking.aspx;

[2] State v. Loomis, 881 N.W.2d 749 (Wis. 2016). Liu, Han-Wei and Lin, Ching-Fu and Chen, Yu-Jie, Beyond State v. Loomis: Artificial Intelligence, Government Algorithmization, and Accountability (December 20, 2018). International Journal of Law and Information Technology, Vol. 27, Issue 2, pp.122-141 (2019). , Available at SSRN: https://ssrn.com/abstract=3313916  See also Karl Manheim & Lyric Kaplan, Artificial Intelligence: Risks to Privacy and Democracy, 21 YALE J. L. & TECH. 106, 188 (2019).

[3] See e.g., Margaret Hu, Algorithmic Jim Crow, 86 FORDHAM L. REV. 633, 641 (2017).

[4] See e.g., Aaron Rieke, Miranda Bogen and David G. Robinson, Public Scrutiny of Automated Decisions: Early Lessons and Emerging Methods An Upturn and Omidyar Network Report (February 2018) https://www.omidyar.com/insights/public-scrutiny-automated-decisions-early-lessons-and-emerging-methods.

[5] The Computational Journalism Lab at the University of Maryland curated set of algorithms being used in the US Federal government. See, http://algorithmtips.org/

[6] Hilke Schellmann & Jason Bellini, Artificial Intelligence: The Robots Are Hiring, WALL ST. J. (Sept. 20, 2018), https://www.wsj.com/articles/artificial-intelligence-the-robots-arenow-hiring-moving-upstream-1537435820

[7] Jon Walker, Artificial Intelligence Applications for Lending and Loan Management, EMERJ (May 19, 2019) https://emerj.com/ai-sector-overviews/artificial-intelligence-applications-lending-loan-management/.

[8] For instance, Uber is using AI to identify and circumvent officials in cities all over the world. See, Mike Isaac, How Uber Deceives the Authorities Worldwide, THE NEW YORK TIMES (March 3, 2017) https://www.nytimes.com/2017/03/03/technology/uber-greyballprogram-evade-authorities.html.

[9] Shlomit Yanisky-Ravid, Sean K. Hallisey, Equality and Privacy by Design": A New Model of Artificial Intelligence Data Transparency Via Auditing, Certification, and Safe Harbor Regimes, 46 FORDHAM URB. L.J. 428, 431 (2019).

[10]  Sonia K. Katyal, Private Accountability in the Age of Artificial Intelligence, 66 UCLA L. REV. 54, 111–13 (2019); John O. McGinnis, Accelerating AI, 104 NW. U. L. REV. 1253 (2010). The legal scholarship can be roughly divided into two opposing views: those who acknowledge the threat AI poses and those who dismiss

**AI and Machine Learning**

With the recent rise of and attention given to deep learning technologies, the terms artificial intelligence, machine learning, and deep learning have been used somewhat interchangeably by the general public to reflect the concept of replicating "intelligent" behavior in machines.

Generally, AI refers to a subfield of computer science focused on building machines and software that can mimic such behavior. Machine learning is the subfield of artificial intelligence that focuses on giving computer systems the ability to learn from data. Deep learning is a subcategory of machine learning that uses neural networks to learn to represent and extrapolate from a dataset. Machine learning and deep learning processes impact children's lives and ultimately, their human rights and how artificial intelligence technologies are being used in ways that positively or negatively impact children at home, at school, and at play.[12]

**Workers in AI**

Through a combination of surveillance, predictive analytics, and integration into workplace systems such as interviewing, human resources, and watching, employers are implementing algorithmic systems to rank and assess workers, automatically set wages and performance targets, and even fire workers. [13]

The introduction of automation and artificial intelligence-enabled labor management systems raises significant questions about workers' rights and safety, according to the "AI Now 2019 Report,"[14] which explores the social implications of AI technologies.

In almost every case, these systems are optimized from the perspective of business owners and rarely involve or include worker perspectives, needs, or considerations. Most algorithmic management tools, like most algorithmic decision systems, lack meaningful opportunities for workers to understand how the systems work or to contest or change determinations about their livelihood.[15]

---

it. See generally, Brian S. Haney, The Perils and Promises of Artificial General Intelligence, 45 J. LEGIS. 151 (2018); Matthew U. Scherer, Regulating Artificial Intelligence Systems, 29 HARV. J.L. & TECH. 353 (2016)

[11] See e.g. Solon Barocas & Andrew D. Selbst, Big Data's Disparate Impact, 104 CAL. L. REV. 671 (2016); FRANK PASQUALE, THE BLACK BOX SOCIETY: THE SECRET ALGORITHMS THAT CONTROL MONEY AND INFORMATION 1-18 (2015); Joshua New and Daniel Castro, How Policymakers Can Foster Algorithmic Accountability, CENTER FOR DATA INNOVATION, May 21, 2018 https://www.datainnovation.org/2018/05/how-policymakers-can-foster-algorithmic-accountability/ at p. 3. Karen Yeung, Andrew Howes, and Ganna Pogrebna, AI Governance by Human Rights-Centred Design, Deliberation and Oversight: An End to Ethics Washing (June 21, 2019)Available at SSRN: https://ssrn.com/abstract=3435011 (suggesting a "human-rights-centered design, deliberation and oversight" approach to the governance of AI). Muhammad Ali et al., Discrimination through Optimization: How Facebook's Ad Delivery Can Lead to Skewed Outcomes, ARXIV (Apr. 19, 2019), https://arxiv.org/pdf/1904.02095.pdf..

[12] "Office of Innovation, UNICEF Office of Innovation,"UNICEF Innovation Home Page, available at https://www.unicef.org/innovation/.

[13] See, for example, Alex Rosenblat, "When Your Boss Is an Algorithm," New York Times, October 12, 2018, https://www.nytimes.com/2018/10/12/opinion/sunday/uber-driver-life.html; and Jeremias Adams-Prassl, "The Algorithmic Boss", NYU Law, October 28, 2019, https://its.law.nyu.edu/eventcalendar/index.cfm?fuseaction=main.detail&id=73302.

[14] Crawford, Kate, Roel Dobbe, Theodora Dryer, Genevieve Fried, Ben Green, Elizabeth Kaziunas,Amba Kak, Varoon Mathur, Erin McElroy, Andrea Nill Sánchez, Deborah Raji, Joy Lisi Rankin, Rashida Richardson, Jason Schultz, Sarah Myers West, and Meredith Whittaker. AI Now 2019 Report. New York: AI Now Institute, 2019, https://ainowinstitute.org/AI_Now_2019_Report.html.

[15] Crawford, Kate, Roel Dobbe, Theodora Dryer, Genevieve Fried, Ben Green, Elizabeth Kaziunas,Amba Kak, Varoon Mathur, Erin McElroy, Andrea Nill Sánchez, Deborah Raji, Joy Lisi Rankin, Rashida Richardson, Jason

A growing number of employers rely on AI systems to manage workers and set workflows. For example, Amazon uses an AI system that sets performance targets for workers, a so-called "rate."[16] The "rate" is calculated automatically, and changes from day to day. If a worker falls behind, they are subject to disciplinary action. In many warehouses, termination is an automated process.[17] According to Abdirahman Muse, executive director of the Awood Center,[18] an organizer with Amazon warehouse workers in Minneapolis, if workers fall behind the algorithmically set productivity rate three times in one day, they are fired, however long they may have worked for the company, and irrespective of the personal circumstances that led to their "mistakes."[19] Muse recounts workers deciding between going to the bathroom and maintaining their rate. Many workers in the Amazon warehouse where he organizes are Somali immigrants. A report from the Economic Roundtable[20] found that in California's Inland Empire, the home to a major Amazon warehouse hub, "86 percent [of Amazon's logistics employees] earn less than the basic living wage... The typical worker had total annual earnings in 2017 of $20,585, which is slightly over half of the living wage."[21]

Amazon is not alone in using AI to enforce worker productivity. The integration of tracking and productivity technologies in the agriculture sector in Canada, finding that "surveillance technologies are utilized to regiment workers to determine their pace at work and their production levels, much like what we see in warehouses."[22]

When the Philadelphia Marriott Downtown began using an app to give its housekeepers room assignments, workers found the new system sent them zigzagging across a hotel the size of a city block. It reduced their ability to organize their day, making their work more physically demanding.[23] When the Children's Hospital of Philadelphia (CHOP) hired outside contractors to assemble and distribute supplies and essential equipment, they used an opaque algorithmic "rate" that set the amount of work. If anything is off, it's "nearly impossible to meet [the rate]," say workers, "if they're understaffed or overstaffed, if it's a holiday, if there's a person who's new and just getting up to speed." If workers don't meet their rate, they'll be written up.[24] CHOP links the practice of hiring contract workers (whose labor is leased by one firm to another) to algorithmically set productivity rates. This mandates a rate of productivity as part of the contractual agreement and enforces that rate through an algorithm, instead of through on-site supervisors.

---

Schultz, Sarah Myers West, and Meredith Whittaker. AI Now 2019 Report. New York: AI Now Institute, 2019, [hereinafter AI Now Report] https://ainowinstitute.org/AI_Now_2019_Report.pdf

[16] Joshua Brustein, "Warehouses Are Tracking Workers' Every Muscle Movement," Bloomberg, November 5, 2019, https://www.bloomberg.com/news/articles/2019-11-05/am-i-being-tracked-at-work-plenty-of-warehouse-workers-are .

[17] Colin Lecher, "How Amazon Automatically Tracks and Fires Warehouse Workers for 'Productivity,'" The Verge, April 25, 2019, https://www.theverge.com/2019/4/25/18516004/amazon-warehouse-fulfillment-centers-productivity-firing-terminations .

[18] Karen Weise, Somali Workers in Minnesota Force Amazon to Negotiate, NY Times, (Nov. 20, 2018) https://www.nytimes.com/2018/11/20/technology/amazon-somali-workers-minnesota.html

[19] Abdi Muse, Bhairavi Desai, Veena Dubal, and Meredith Whittaker, "Organizing Tech" panel at AI Now Symposium, October 2, 2019, https://ainowinstitute.org/symposia/videos/organizing-tech.html .

[20] https://economicrt.org/about/

[21] Daniel Flaming and Patrick Burns, Economic Roundtable, "Too Big to Govern: Public Balance Sheet for the World's Largest Store," November 2019, https://economicrt.org/wp-content/uploads/2019/11/Too-Big-to-Govern.pdf .

[22] Chris Ramsaroop, "Reality Check 101: Rethinking the Impact of Automation and Surveillance on Farm Workers," Data & Society: Points , September 6, 2019, https://points.datasociety.net/reality-check-101-c6e501c3b9a3 .

[23] Juliana Feliciano Reyes, "Hotel Housekeeping on Demand: Marriott Cleaners Say This App Makes Their Job Harder," Philadelphia Inquirer , July 2, 2018, https://www.inquirer.com/philly/news/hotel-housekeepers-schedules-app-marriott-union-hotsos-20180702.html .

[24] Juliana Feliciano Reyes, "In the Basement of CHOP, Warehouse Workers Say They're Held to Impossible Quotas," Philadelphia Inquirer, April 22, 2019, https://www.inquirer.com/news/warehouse-workers-quotas-rate-childrens-hospital-of-philadelphia-canon-20190422.html .

Such rate-setting systems rely on pervasive worker surveillance to measure how much they are doing. Systems to enable such invasive worker monitoring are becoming more common, including in traditionally "white-collar" working environments. For example, the start-up Humanyze[25] incorporates sensors into employee badges to monitor employee activities, telling employers where workers go, whom they interact with, and how long they stay in a given place. Another company called Workplace Advisor uses heat sensors to achieve a similar aim. And though the usefulness of these products is disputed,[26] they reflect an increasing willingness to engage in invasive surveillance of workers in the name of workplace control and eking out incremental gains in productivity.

**Algorithmic Dynamic Wage Control**

Algorithmic worker management and control systems have also had a severe negative impact on wages across the so-called "gig economy." These platforms treat workers as subjects of constant experimentation, often in ways that destabilize their economic and even psychological security.[27]

In many instances, industries that adopt discourses of technological advancement are driven by precarious worker labor—what Mary Gray and Siddarth Suri describe as ghost work. [28] Such AI systems are correlated with low wages and "flexible" work policies that, in practice, often make it hard for workers to plan their income, schedule, or whether they will even be able to work that day. Similar to other algorithmic management systems, these function by pooling information and power together for the benefit of owners, managers, and a handful of developers, allowing companies to optimize such systems in ways that maximize revenue without regard to the need for stable and livable wages or predictable incomes, schedules, and availability of work. Indeed, many workers have reported being abruptly "kicked off" a gig work platform, and finding themselves unable to work without warning. The process to reinstate an account can be obscure and onerous.[29]

These platforms are continually optimized by companies and owners. Abrupt changes intended to increase revenue for the company can result in significant losses for workers. In one example, Instacart made changes to its interface that misled customers into thinking they were leaving a tip for workers, when in fact they were paying a service fee to the company. [30] This practice is something that DoorDash[31] also engaged in until July of this year. [32]

These examples demonstrate the significant power asymmetry between workers and customers on one hand, and the companies who control worker management platforms on the other. How, and where, companies may be "optimizing" their platforms at the expense of workers remains largely opaque to anyone outside of companies' corporate offices, and what is known comes largely from worker whistleblowers, who place themselves at great risk and lose much. [33]

---

[25] https://humanyze.com/
[26] Rose Eveleth, "Your Employer May Be Spying on You—and Wasting Its Time," Scientific American, August 16, 2019, https://www.scientificamerican.com/article/your-employer-may-be-spying-on-you-and-wasting-its-time/ .
[27] A substantial number of firms adopting this strategy are funded by the same investor: the Japanese firm Softbank. See Nathaniel Popper, Vindu Goel, and Arjun Harindranath, "The SoftBank Effect: How $100 Billion Left Workers in a Hole," New York Times , November 12, 2019, https://www.nytimes.com/2019/11/12/technology/softbank-startups.html .
[28] Mary L. Gray and Siddarth Suri, Ghost Work: How to Stop Silicon Valley from Building a New Global Underclass (Boston: Houghton Mifflin Harcourt, 2019).
[29] See Jaden Urbi, "Some Transgender Drivers Are Being Kicked Off Uber's App," CNBC, August 8, 2018, https://www.cnbc.com/2018/08/08/transgender-uber-driver-suspended-tech-oversight-facial-recognition.html ; and Rob Hayes, "Uber, Lyft Drivers Rally in Downtown Los Angeles to Demand Better Wages, Employment Rights," Eyewitness News / ABC7, https://abc7.com/business/uber-lyft-drivers-rally-in-la-to-demand-better-wages-employment-rights/5353986/ .
[30] Vanessa Bain, "Dear Instacart Customers," October 9, 2019, https://medium.com/@vanessabain/dear-instacart-customers-664dbb59016e ; Megan Rose Dickey, "Instacart Is under Fire for How It Compensates Shoppers," TechCrunch, November 12, 2019, https://techcrunch.com/2019/11/12/instacart-is-under-fire-for-how-it-compensates-shoppers/ .
[31] https://www.doordash.com/
[32] April Glaser, "How DoorDash, Postmates, and Other Delivery Services Tip Workers," Slate, July 23, 2019, https://slate.com/technology/2019/07/doordash-postmates-grubhub-instacart-tip-policies.html .

The ability of automated management platforms to manipulate and arbitrarily cut wages has been at the heart of worker grievances.[34] Instacart workers report that their earnings decreased precipitously over the last year.[35] Uber and Lyft workers report similar drops.[36] Many identify this as part of a tactic to make workers dependent on the platform for wages, drawing them in with promises of a living wage and flexible working conditions, then severely cutting wages once workers have structured their lives around working for the platform.

Legal scholar Veena Dubal, who has worked with Uber and taxi drivers, argues that these practices are not new, but "[reproduce] risky, early 20th century working conditions," enabled by large-scale AI platforms and deregulation.[37] Scholar Jim Stanford states "The only thing truly new about gig employment is its use of digital and on-line techniques to assign work, discipline workers . . . and control the money. That's more effective than the bulletin boards and classified ads of yesteryear—but it hardly negates the inherent power imbalance between an individual worker and the multibillion-dollar company they work for."[38]

Researchers examined platform workers in Bengaluru, India, showing that they exist in a context in which many workers are already stitching together "flexible" work options, without the expectation of social safety nets. Given these existing practices and expectations, AI-enabled platform work provides a comparatively more lucrative source of employment. [39] Similarly, Indian Turk workers[40] (who until recently made up 40 percent of the total platform workforce) benefited from the global pay rates set by the platform, and the cost of living difference between India and the US.[41] This meant that tasks that US workers would not take due to low pay were more attractive for their global South counterparts. In a study of beauty and wellness platforms in India, researchers discovered that women found these platforms attractive because they allowed them to pursue economic activity within the constraints of gendered, religious, and family norms.[42] In these ways, AI-enabled platform work has challenged labor markets globally.[43]

---

[33] Pacella, Jennifer M., Making Whistleblowers Whole (2021). Forthcoming, 2022, UC Irvine Law Review , Available at SSRN: https://ssrn.com/abstract=3788634 or http://dx.doi.org/10.2139/ssrn.3788634; and DeMott, Deborah, Whistleblowers: Implications for Corporate Governance (June 11, 2021). Washington University Law Review, Vol. 98, 2021, Available at SSRN: https://ssrn.com/abstract=3864994;

[34] While the use of AI systems puts more power and control in the hands of the company, it also harms mainly low-wage workers, who are disproportionately people of color, according to the report. These systems don't work for employees when they set unrealistic productivity goals that can lead to injury or psychological stress and when they impose "unpredictable algorithmic wage cuts" on gig workers that undermine their financial stability, for example. Lower-wage workers stand to lose the most with the rise of automation while white-collar workers are generally unaffected, the report noted. It cited a McKinsey & Co. study that concluded "labor automation will further exacerbate the racial wealth gap in the U.S. absent any interventions." https://www.workforce.com/news/artificial-intelligence-ethics-for-managing-low-wage-workers

[35] Sean Captain, "Instacart Delivery Drivers Say Tips Are Mysteriously Decreasing," Fast Company, October 9, 2019, https://www.fastcompany.com/90413156/tips-for-instacart-delivery-drivers-are-mysteriously-decreasing .

[36] Faiz Siddiqui, "Uber and Lyft Slashed Wages. Now California Drivers Are Protesting Their IPOs," Washington Post , March 26, 2019, https://www.washingtonpost.com/technology/2019/03/26/uber-lyft-slashed-wages-now-california-driversare-protesting-their-ipos/ .

[37] Veena Dubal, "The Drive to Precarity: A Political History of Work, Regulation, & Labor Advocacy in San Francisco's Taxi & Uber Economies," Berkeley Journal of Employment and Labor Law 38, no. 1, February 21,2017; UC Hastings Research Paper no. 236. Available at SSRN: https://ssrn.com/abstract=2921486 .

[38] Jim Stanford, "Bring Your Own Equipment and Wait for Work: Working for Uber Is a Lot Like Being a Dock Worker a Century Ago," Star, November 17, 2019, https://www.thestar.com/business/opinion/2019/11/17/bring-your-own-equipment-and-wait-for-work-working-for-uber-is-a-lot-like-being-a-dock-worker-a-century-ago.html .

[39] Aditi Surie and Jyothi Koduganti, "The Emerging Nature of Work in Platform Economy Companies in Bengaluru, India: The Case of Uber and Ola Cab Drivers," E-Journal of International and Comparative Labour Studies 5, no. 3 (September–October 2016), http://ejcls.adapt.it/index.php/ejcls_adapt/article/view/224/ .

[40] Amazon Mechanical Turk ("mTurk") is an online platform where employers, called requesters, post piecework, and workers, many of whom call themselves "Turkers,"complete that work for pay. Turkers collect in online communities, such as forums and Facebook groups, to share information about Turking—information that helps some of them earn a living wage. Kristy Milland, The Unsupported Crowd: Exclusion

**AI in Hiring**

AI systems to manage and control workers are also being applied in hiring, rapidly and actively shaping the labor market and helping determine who is fit for work, and who isn't. Most hiring tech operates in the absence of any specific rules or requirements to disclose their use for candidate selection, ranking, and hiring to the job seekers whose lives these AI systems affect.[44]

Commercial firms across industries, including major employers like Unilever,[45] Goldman Sachs,[46] and Target,[47] are integrating predictive technologies into the process of selecting whom they hire. AI systems also actively shape employment advertising, résumé ranking, and assessment of both active and passive recruitment.[48] Because AI systems often encode and reproduce patterns of bias within categories such as "competence," "success," and "cultural fit," the rapid deployment of such systems in hiring has significantly raised the stakes of their use.[49]

---

of Indian Workers in Amazon Mechanical Turk Communities, GLRC Graduate Student Symposium 2017, https://osf.io/vqfke/download and Silva, Yasmeen, Using Amazon's Mechnical Turk (MTurk) For Modern Survey Experiments: A Review With A Particular Focus on Racial Inequality (June 16, 2021). Available at SSRN: https://ssrn.com/abstract=3868474 or http://dx.doi.org/10.2139/ssrn.3868474 stating MTurk is a multiple sources platform that provides people and companies with the ability to outsource procedures and jobs to distributed workers from across the globe.

41 Kristy "spamgirl" Milland, "The Unsupported Crowd: Exclusion of Indian Workers in Amazon Mechanical Turk Communities," 2017, http://kristymilland.com/papers/Milland.2017.The.Unsupported.Crowd.pdf .

42 Noopur Raval and Joyojeet Pal, "Making a 'Pro': 'Professionalism' after Platforms in Beauty-work," Journal Proceedings of the ACM on Human-Computer Interaction 3, no. CSCW (November 2019), https://dl.acm.org/citation.cfm?id=3359277 .

43 Mark Graham and Mohammed Amir Anwar, "The Global Gig Economy: Towards a Planetary Labour Market?," First Monday 24, no. 4 (April 1, 2019), https://firstmonday.org/ojs/index.php/fm/article/view/9913/7748#p2 .

44 Miranda Bogen and Aaron Rieke, "Help Wanted: An Examination of Hiring Algorithms, Equity, and Bias," Upturn , December 2018, https://www.upturn.org/static/reports/2018/hiring-algorithms/files/Upturn%20--%20Help%20Wanted%20-%20An%20Exploration%20of%20Hiring%20Algorithms,%20Equity%20and%20Bias.pdf .

45 Robert Booth, "Unilever Saves on Recruiters by Using AI to Assess Job Interviews," Guardian , October 25, 2019, https://www.theguardian.com/technology/2019/oct/25/unilever-saves-on-recruiters-by-using-ai-to-assessjob-interviews

46 Rosalind S. Helderman, "HireVue's AI Face-Scanning Algorithm Increasingly Decides Whether You Deserve the Job," Washington Post , October 22, 2019, https://www.washingtonpost.com/technology/2019/10/22/ai-hiring-face-scanning-algorithm-increasinglydecides-whether-you-deserve-job/ .

47 Daniel Greene and Ifeoma Ajunwa, "Automated Hiring Platforms as Technological Intermediaries and Brokers," Dan Greene, 2017, http://dmgreene.net/wp-content/uploads/2014/11/GreeneAjunwaAutomated-Hiring-Plaforms-as-Technological-Intermediaries-and-Brokers.pdf .

48 Bogen and Rieke, "Help Wanted: An Examination of Hiring Algorithms, Equity, and Bias," Upturn, December 2018, https://www.upturn.org/static/reports/2018/hiring-algorithms/files/Upturn%20--%20Help%20Wanted%20-%20An%20Exploration%20of%20Hiring%20Algorithms,%20Equity%20and%20Bias.pdf .

49 See Jim Fruchterman and Joan Mellea, "Expanding Employment Success for People with Disabilities," Benetech, November 2018, https://benetech.org/about/resources/expanding-employment-success-for-people-with-disabilities/ See also Sánchez-Monedero, Javier and Dencik, Lina and Edwards, Lilian, What Does It Mean to 'Solve' the Problem of Discrimination in Hiring? (October 2, 2019). Available at SSRN: https://ssrn.com/abstract=3463141 or http://dx.doi.org/10.2139/ssrn.3463141 ("Discriminatory practices in recruitment and hiring are an ongoing issue that is of concern not just for workplace relations, but also for Indeed, many researchers suspect that these tools most likely exacerbate inequity and reinforce discrimination, creating what legal scholar Pauline Kim terms "classification bias." [50] But without meaningful access to these systems and their processes, workers lack the evidence necessary to challenge their use.[51]

Effective January 1, 2020, the Artificial Intelligence Video Interview Act ("AIVIA")[52] is the governing law in Illinois for any employer who chooses to "use artificial intelligence (AI) to analyze video interview by job candidates."[53] Under AIVIA, employers are required to provide advance notice to the applicant of the use of the video interview technology, and further to "explain to the applicant 'how the [AI] works' and what general characteristics the technology uses to evaluate applicants."[54]

This call for transparency is important. Beyond transparency, the law requires that employers "obtain, in advance, the applicant's consent to use the technology."[55] The law also features provisions for data protection. It imposes limits on "the distribution and sharing of the video," granting access "only to those persons 'whose expertise or technology' is necessary to evaluate the applicant."[56] Further, candidates are given some control over what happens to the video after their assessment. Employers are required to "destroy the video (and all backup copies) within 30 days" of the applicant requesting its destruction.[57]

The law firm, Davis Wright Tremaine, LLP ("DWT") identifies some key issues with the law. The law fails to define "artificial intelligence" and "artificial intelligence analysis" along with other key terms.[58] This ambiguity may

---

wider understandings of economic justice and inequality. Yet the way decisions are made on who is eligible for jobs, and why, are rapidly changing with the advent and growth in uptake of automated hiring systems (AHSs) powered by data-driven tools .A recent report estimated that 98% of Fortune 500 companies use Applicant Tracking Systems of some kind in their hiring process. Several of these AHSs claim to detect and mitigate discriminatory practices against protected groups. Yet whilst these tools have a growing user-base around the world, such claims of 'bias mitigation' are rarely scrutinised and evaluated, and when done so, have almost exclusively been from a US social and legal perspective. In this paper, we introduce a perspective from outside the US by critically examining how three prominent automated hiring systems (AHSs) in regular use in the UK, HireVue, Pymetrics and Applied, understand and attempt to mitigate bias and discrimination. These systems have been chosen as they explicitly claim to address issues of discrimination in hiring and provide some information about how their systems work to do this. Using publicly available documents, we describe how their tools are designed, validated and audited for bias, highlighting assumptions and limitations, before situating these in the social and legal context of the UK. The UK has a very different legal background to the US in terms not only of hiring and equality law, but also in terms of data protection (DP) law. We argue that this might be an important challenge to building bias mitigation into AHSs definitively capable of meeting EU legal standards. Furthermore attempts at bias mitigation intended to meet US law may not map to UK or EU law. AHSs may thus obscure rather than improve systemic discrimination in the workplace.")

50   Pauline Kim, "Data-Driven Discrimination at Work," William & Mary Law Review 48 (2017): 857–936, https://ssrn.com/abstract=2801251
51   Rashida Richardson, Jason M. Schultz, and Vincent M. Southerland, "Litigating Algorithms," AI Now Institute, September 2019, https://ainowinstitute.org/litigatingalgorithms-2019-us.pdf .
52   820 ILL. COMP. STAT. 42 (2020)
53   . Nicole Mormilo, Matthew Jedreski, K.C. Halm & Jeffrey S. Bosley, Employers Using AI in Hiring Take Note: Illinois' Artificial Intelligence Video Interview Act Is Now in Effect, DAVIS WRIGHT TREMAINE LLP (Feb. 10, 2020), https://www.dwt.com/blogs/artificialintelligence-law-advisor/2020/02/illinois-aivia-compliance [https://perma.cc/JL6E-RUQZ]
54   Matthew Jedreski, Jeffrey S. Bosley & K.C. Halm, Illinois Becomes First State to Regulate Employers' Use of Artificial Intelligence to Evaluate Video Interviews, DAVIS WRIGHT TREMAINE LLP (Sept. 3, 2019), https://www.dwt.com/blogs/artificial-intelligencelaw-advisor/2019/09/illinois-becomes-first-state-to-regulate-employers[https://perma.cc/46JD-2T32].
55   Id.
56   Id.
57   Id.
58   Id.

mean that certain employer AI use cases, such as "to track data about its candidates," may not be covered. Further, ambiguity in the transparency mandate of the law could pose serious problems for its effective use. DWT notes that the law does not go in depth to specify "how much detail an employer must provide when 'explaining how artificial intelligence works' to an applicant" or what "'characteristics' of the AI employers must disclose. "Therefore, employers may be permitted to use broad, cursory statements such as "AI will assess a candidate's performance" to satisfy this requirement, statements which do not serve the true spirit of transparency. DWT finds the law to be unclear in several other aspects as well. It notes that there is no requirement that candidates provide express written consent. Further, the law "does not include a private right of action or any explicit penalties," which could raise serious issues in enforcing its provisions.

As for data destruction, DWT points out that it is not clear if "data that an employer extracts or derives from the video interview . . . is subject to the destruction duty under the law." If such data is not protected by the AIVIA, then the extent to which the act allows candidates control over their interview data is potentially limited. Lastly, DWT points out that "there is no guidance on what it means for a job to be 'based in' Illinois, and the statute is silent as to whether employees may refuse to consider applicants who refuse to consent."[59]

Ultimately, AIVIA is a step in the right direction, as it touches on the serious concerns of transparency and data rights. However, in addition to the gaps noted above, while some employers may surely make a good faith effort to comply, many employers themselves are not privy to how the AI they use truly works. Companies such as HireVue[60] keep a close guard over their algorithms and technologies to protect their market share, to the detriment of clients and candidates alike.[61] In order to push AI video interview companies to be more transparent, the law must put in place effective penalties such that employers would not choose to use technology unless AI companies provided enough information. Effective legislation must hold enough weight to impact all stakeholders in the AI video interview universe. AIVIA is commendable as first-of-its-kind legislation attempting to counterbalance the immense power which the AI sphere currently holds often against the public interest.

Tech vendors are also attempting to make the case that their systems help fight against historical and human biases, claiming they have been designed to reduce discrimination and increase diversity. Yet at this point, such claims amount to marketing statements and are unsupported by peer-reviewed research. Instead, studies show that there just isn't enough transparency to assess whether and how these models actually work or to determine whether they are free from bias.[62]

---

[59] Id.
[60] Janine Woodworth & Jake Bauer, Digital Interviewing: The Voice of the Candidate, HIREVUE 7 (2014), http://www.thetalentboard.org/wp-content/uploads/2014/06/DigitalInterviewing-The-Voice-of-the-Candidate.pdf [https://perma.cc/LY2K-PJAG].
[61] See e.g., DAN LYONS, LAB RATS: HOW SILICON VALLEY MADE WORK MISERABLE FOR THE REST OF US 159 (2019). See also See FRANK PASQUALE, THE BLACK BOX SOCIETY: THE SECRET ALGORITHMS THAT CONTROL MONEY AND INFORMATION 16 (2015) (arguing that unregulated and opaque data collection is contributing to social inequality). See, e.g., Rebecca Greenfield, The Rise of the (Truly Awful) Webcam Job Interview, BLOOMBERG (Oct. 12, 2016, 7:00 AM), https://www.bloomberg.com/news/articles/2016-10-12/the-rise-of-the-truly-awful-webcam-job-interview [https://perma.cc/M93J-QTY8].
[62] Manish Raghavan, Solon Barocas, Jon Kleinberg, and Karen Levy, "Mitigating Bias in Algorithmic Hiring:Evaluating Claims and Practices," June 21, 2019, https://papers.ssrn.com/sol3/papers.cfm?abstract_id=3408010 .

This led the Electronic Privacy Information Center[63] to file a complaint with the Federal Trade Commission alleging that one AI hiring company, HireVue, is engaging in "unfair and deceptive" business practices by failing to ensure the accuracy, reliability, or validity of its algorithmically driven results.[64]

Employers, not workers, are the "customers" whom AI hiring companies seek to court with promises of efficiency and fewer worries about accountability and liability. In fact, several prominent companies, such as Pymetrics, offer to cover their customers' legal fees or liabilities that might arise from the use of their products or services.[65] AI-driven hiring systems are only the starting point of a push to use AI to monitor and control workers and workplaces as these platforms "create a managerial framework for workers as fungible human capital, available on demand and easily ported between job tasks and organizations."[66]

It is critical that researchers and advocates not only examine the application of artificial intelligence in the hiring process in isolation, but also consider how AI is being implicated in broader shifts in labor practices, and how it might be serving to define and redefine notions of competence and ability.[67]

**Labor Automation's Disparate Impacts**

In recent years, two predominant narratives have emerged around the future of work and labor automation. One insists that labor automation will yield a net gain for society, increasing productivity, growing the economy, and creating more jobs and demand for workers that will offset any technological displacement that happens along the way.[68] The other predicts a labor apocalypse, where automation will ultimately take over the workforce, create massive unemployment, and serve only the financial interests of those who own them and the engines of our economy.[69] Both narratives are predicated on the assumption that automation in the workplace is inevitable and that automated systems are capable of performing tasks that had previously been the work of humans.

---

[63] EPIC is a public interest research center in Washington, DC. EPIC was established in 1994 to focus public attention on emerging privacy and civil liberties issues and to protect privacy, freedom of expression, and democratic values in the information age. EPIC pursues a wide range of program activities including policy research, public education, conferences, litigation, publications, and advocacy. EPIC routinely files amicus briefs in federal courts, pursues open government cases, defends consumer privacy, organizes conferences for NGOs, and speaks before Congress and judicial organizations about emerging privacy and civil liberties issues. EPIC works closely with a distinguished advisory board, with expertise in law, technology and public policy. EPIC maintains one of the most popular privacy web sites in the world. https://epic.org/

[64] Drew Harwell, "Rights Group Files Federal Complaint against AI-hiring Firm Citing Unfair, Deceptive Practices," Washington Post, November 6, 2019, https://www.washingtonpost.com/technology/2019/11/06/prominent-rights-group-files-federal-complaintagainst-ai-hiring-firm-hirevue-citing-unfair-deceptive-practices/ .

[65] Pymetrics, "Pymetrics End User Agreement," https://www.pymetrics.ai/terms-of-service/ .

[66] Ifeoma Ajunwa and Daniel Greene, "Platforms at Work: Automated Hiring Platforms and Other New Intermediaries in the Organization of Work," in Work and Labor in the Digital Age, https://papers.ssrn.com/sol3/papers.cfm?abstract_id=3248675 .

[67] See, for example, Ifeoma Ajunwa, Kate Crawford, and Jason Schultz, " Limitless Worker Surveillance," 105 Cal. Rev. 735, March 10, 2016, https://papers.ssrn.com/sol3/papers.cfm?abstract_id=2746211 ; Ifeoma Ajunwa, "Algorithms at Work: Productivity Monitoring Applications and Wearable Technology as the New Data-Centric Research Agenda for Employment and Labor Law," 63 St. Louis U.L.J, September 10, 2018, https://papers.ssrn.com/sol3/papers.cfm?abstract_id=3247286 ; and Meredith Whittaker et al., "Disability, Bias, and AI," AI Now Institute, November 2019, https://ainowinstitute.org/disabilitybiasai-2019.pdf .

[68] Kweilin Ellingrud, "The Upside of Automation: New Jobs, Increased Productivity and Changing Roles for Workers," Forbes, October 23, 2019, https://www.forbes.com/sites/kweilinellingrud/2018/10/23/the-upside-of-automation-new-jobs-increasedproductivity-and-changing-roles-for-workers/#9bae2fb7df04 .

[69] See Kwan, Martin, Automation and the International Human Right to Work (May 13, 2021). Emory International Law Review Recent Developments, Vol. 35, 2021, Available at SSRN: https://ssrn.com/abstract=3845262Carl Benedikt Frey and Michael A. Osborne, "The Future of Employment: How Susceptible Are Jobs to Computerisation?" Oxford Martin, September 17, 2013, https://www.oxfordmartin.ox.ac.uk/downloads/academic/The_Future_of_Employment.pdf .

A study from the Brookings Institute predicts that certain demographic groups and places will likely bear more of the burden of adjusting to labor automation than others, and advocates for a universal basic income.[70] First, the study found that lower-wage workers stand to lose the most due to automation while white-collar workers will likely remain largely unaffected. Using a model that views a job as a bundle of tasks (some of which can be automated and others not), Brookings concluded that the average "automation potential" for US occupations requiring less than a bachelor's degree is 55 percent—more than double the 24 percent susceptibility among occupations requiring a bachelor's degree or more.[71] That means US workers in occupations that pay the least, like food preparation and serving, production jobs in factories, and administrative support—which pay wages of only 50 to 75 percent of the national average—could experience 60 to 80 percent task-level disruption.[72] Meanwhile, higher-paying jobs in business and financial operations or engineering, where US workers earn 150 percent of the average wage, will likely experience as little as 14 percent of their current tasks being displaced by automation. This has serious implications in terms of the risk exposure faced by certain communities. Black, Native American, and Latinx workers who make up a larger proportion of the workforce in occupations like construction, agriculture, and transportation[73] face average task-automation potentials of 44 to 47 percent. That's anywhere from five to eight percent more than their white counterparts.[74]

The disparate effects of task automation will also likely entail disproportionate job losses. Even McKinsey & Company, which believes AI could lift productivity and economic growth, concluded that labor automation will further exacerbate the racial wealth gap in the US absent any interventions.[75] One study from July 2019 found that more than a quarter of Latinx workers—as many as seven million people—are in jobs that could be automated by 2030.[76] That translates to a potential displacement rate of 25.5 percent for Latinx workers, three percentage points higher than the national average.[77] McKinsey calculated that 4.6 million Black workers will be displaced by 2030 due to automation, with a potential displacement rate of 23.1 percent. [78]

---

[70] See Mark Muro, Robert Maxim, and Jacob Whiton, "Automation and Artificial Intelligence: How Machines Are Affecting People and Places," Brookings Institution (January 2019), https://www.brookings.edu/wp-content/uploads/2019/01/2019.01_BrookingsMetro_Automation-AI_Report_Muro-Maxim-Whiton-FINAL-version.pdf ; and "Artificial Intelligence, Automation, and the Economy," Executive Office of the President (December 2016): 14,https://obamawhitehouse.archives.gov/sites/whitehouse.gov/files/documents/Artificial-Intelligence-Automation-Economy.PDF .

[71] Brookings employed a backward- and forward-looking analysis of the impacts of automation from 1980 to 2016 and 2016 to 2030 across approximately 800 occupations. See Muro et al., "Automation and Artificial Intelligence," 33, https://www.brookings.edu/wp-content/uploads/2019/01/2019.01_BrookingsMetro_Automation-AI_Report_Muro-Maxim-Whiton-FINAL-version.pdf

[72] Muro et al., "Automation and Artificial Intelligence," 33–34, https://www.brookings.edu/wp-content/uploads/2019/01/2019.01_BrookingsMetro_Automation-AI_Report_Muro-Maxim-Whiton-FINAL-version.pdf .

[73] Muro et al., "Automation and Artificial Intelligence," 45–46,https://www.brookings.edu/wp-content/uploads/2019/01/2019.01_BrookingsMetro_Automation-AI_Report_Muro-Maxim-Whiton-FINAL-version.pdf .

[74] Muro et al., "Automation and Artificial Intelligence," 7, https://www.brookings.edu/wp-content/uploads/2019/01/2019.01_BrookingsMetro_Automation-AI_Report_Muro-Maxim-Whiton-FINAL-version.pdf .

[75] Kelemwork Cook, Duwain Pinder, Shelley Stewart III, Amaka Uchegbu, and Jason Wright, "The Future of Work in Black America," McKinsey & Company, October 2019, https://www.mckinsey.com/featured-insights/future-of-work/the-future-of-work-in-black-america .

[76] Susan Lund, James Manyika, et al., "The Future of Work in America: People and Places, Today and Tomorrow," McKinsey Global Institute (July 2019): 61,https://www.mckinsey.com/~/media/McKinsey/Featured%20Insights/Future%20of%20Organizations/The%20future%20of%20work%20in%20America%20People%20and%20places%20today%20and%20tomorrow/MGI-The-Future-of-Work-in-America-Report-July-2019.ashx .

The exact job-loss figures caused by automation are ultimately hotly contested. After MIT Technology Review synthesized 18 different reports on the effects of automation on labor with predictions ranging from a gain of nearly 1 billion jobs globally by 2030 to a loss of 2 billion, it aptly noted that "prognostications are all over the map."[79] With all of these projections, the devil is in the details. We may "have no idea how many jobs will actually be lost to the march of technological progress,"[80] but we can begin to answer who will lose their jobs based on the power dynamics and economic disparities that already exist today.

**The Limitations of Corporate AI Ethics**

Many companies, governments, NGOs, and academic institutions follow the path of generating AI ethics principles and statements. These ethics statements are necessary but insufficient in of themselves. These ethics principles are presented as the product of a growing "global consensus" on AI ethics. This promotes a majoritarian view of ethics, which is especially concerning given the widespread evidence showing that AI bias and misuse harms many people whose voices are largely missing from these ethics principles and in official ethics debates.[81]

There are now so many ethics policy statements that some groups began to aggregate them into standalone AI ethics surveys, which attempted to summarize and consolidate a representative sample of AI principle statements in order to identify themes and make normative assertions about the state of AI ethics.[82] These surveys tend to aggregate AI ethics content from a very wide variety of contexts, blending corporate statements released on corporate blogs,[83] publicly informed governing declarations,[84] government policy guidelines from national and coalition strategies,[85] and nonprofit mission statements and charters.[86] However, they usually lack a comprehensive account of the methods used and sometimes equate internal and often secret corporate decision-making processes with grassroots-driven statements and governmental policy recommendations. The vast majority of these documents were generated from countries and organizations in the global North.[87] Principle statements and the ethical priorities of the global South with regard to artificial intelligence are often absent from these surveys. Scholars and advocates have increasingly called attention to the gap between high-level statements and meaningful accountability.[88]

---

[77] Lund, Manyika, et al, "The Future of Work in America," https://www.mckinsey.com/~/media/McKinsey/Featured%20Insights/Future%20of%20Organizations/The%20future%20of%20work%20in%20America%20People%20and%20places%20today%20and%20tomorrow/MGI-The-Future-of-Work-in-America-Report-July-2019.ashx .

[78] Lund, Manyika, et al, "The Future of Work in America," 13, https://www.mckinsey.com/~/media/McKinsey/Featured%20Insights/Future%20of%20Organizations/The%20future%20of%20work%20in%20America%20People%20and%20places%20today%20and%20tomorrow/MGI-The-Future-of-Work-in-America-Report-July-2019.ashx .

[79] Erin Winick, "Every Study We Could Find on What Automation Will Do To Jobs in One Chart," MIT Technology Review, January 25, 2018, https://www.technologyreview.com/s/610005/every-study-we-could-find-on-what-automation-will-do-to-jobs-in-one-chart/ .

[80] Winick, "Every Study We Could Find on What Automation Will Do To Jobs in One Chart," https://www.technologyreview.com/s/610005/every-study-we-could-find-on-what-automation-will-do-to-jobs-in-one-chart/ .

[81] Sarah Myers West, Meredith Whittaker, and Kate Crawford, "Discriminating Systems: Gender, Race, and Power in AI," AI Now Institute, April 2019, https://ainowinstitute.org/discriminatingsystems.pdf .

[82] Şerife Wong, "Fluxus Landscape: An Expansive View of AI Ethics and Governance," Kumu , August 20, 2019, https://icarus..kumu.io/fluxus-landscape ; Jessica Fjeld et al., "Principled Artificial Intelligence: A Map of Ethical and Rights-Based Approaches," Berkman Klein Center for Internet & Society at Harvard University, July 4, 2019, https://ai-hr.cyber.harvard.edu/primp-viz.html ; Luciano Floridi and Josh Cowls, "A Unified Framework of Five Principles for AI in Society," Harvard Data Science Review , June 22, 2019, https://doi.org/10.1162/99608f92.8cd550d1 ; Thilo Hagendorff, "The Ethics of AI Ethics—An Evaluation of Guidelines," arXiv:1903.03425 [Cs, Stat] , October 11, 2019, http://arxiv.org/abs/1903.03425 ; Yi Zeng, Enmeng Lu, and Cunqing Huangfu, "Linking Artificial Intelligence Principles," arXiv:1812.04814 [Cs] ,December 12, 2018, http://arxiv.org/abs/1812.04814 ; Anna Jobin, Marcello Ienca, and Effy Vayena, "The Global Landscape of AI Ethics Guidelines," Nature Machine Intelligence 1, no. 9 (September 2019): 389–99,https://doi.org/10.1038/s42256-019-0088-2 .

Critics have identified conflicting ideals and vague definitions as barriers that are preventing the operationalization of ethics principles in AI product development, deployment, and auditing frameworks. One example is Microsoft's former funding of an Israeli facial-recognition surveillance company AnyVision.[89] AnyVision facilitates surveillance in the West Bank, allowing Israeli authorities to identify Palestinian individuals and track their movements in public space. Given the documented human-rights abuses happening on the West Bank,[90] together with the civil-liberties implications associated with facial recognition in policing contexts,[91] this use case directly contradicted Microsoft's own declared principles of "lawful surveillance" and "non-discrimination," along with the company's promise not to "deploy facial recognition technology in scenarios that we believe will put freedoms at risk."[92] More perplexing was that AnyVision confirmed to reporters that their technology had been vetted against Microsoft's ethical commitments. After public outcry, Microsoft acknowledged that there could be an ethical problem, and hired former Attorney General Eric Holder to investigate the alignment between AnyVision's actions and Microsoft's ethical principles.[93]

---

[83] "Our Approach: Microsoft AI Principles," Microsoft, https://www.microsoft.com/en-us/ai/our-approach-to-ai ; "IBM'S Principles for Data Trust and Transparency," THINKPolicy, May 30, 2018, https://www.ibm.com/blogs/policy/trust-principles/ ; "Our Principles," Google AI, https://ai.google/principles/ .

[84] "Official Launch of the Montréal Declaration for Responsible Development of Artificial Intelligence," Mila, December 4, 2018, https://mila.quebec/en/2018/12/official-launch-of-the-montreal-declaration-for-responsible-development-of-artificial-intelligence/ ; Access Now Policy Team, "The Toronto Declaration: Protecting the Rights to Equality and Non-Discrimination in Machine Learning Systems," Access Now (blog), May 16, 2018, https://www.accessnow.org/the-toronto-declaration-protecting-the-rights-to-equality-and-non-discrimination-in-machine-learning-systems/ .

[85] "OECD Principles on Artificial Intelligence," Organisation for Economic Co-Operation and Development, https://www.oecd.org/going-digital/ai/principles/ .

[86] "OpenAI Charter," OpenAI, , https://openai.com/charter/ ; "Tenets,"Partnership on AI, https://www.partnershiponai.org/tenets/ .

[87] See Vidushi Marda, "Introduction" in APC, Article 19, and SIDA, "Artificial Intelligence: Human Rights,Social Justice and Development," Global Information Watch 2019, November 2019,https://giswatch.org/sites/default/files/gisw2019_artificial_intelligence.pdf .

[88] Daniel Greene, Anna Lauren Hoffmann, and Luke Stark, "Better, Nicer, Clearer, Fairer: A Critical Assessment of the Movement for Ethical Artificial Intelligence and Machine Learning," January 8, 2019, https://doi.org/10.24251/HICSS.2019.258 ; Daniel Greene et al., "A Critical Assessment of the Movement for Ethical Artificial Intelligence and Machine Learning," http://dmgreene.net/wp-content/uploads/2018/12/Greene-Hoffmann-Stark-Better-Nicer-Clearer-Fairer-HICSS-Final-Submission-Revised.pdf ; Jess Whittlestone et al., "The Role and Limits of Principles in AI Ethics:Towards a Focus on Tensions," n.d., 7; Roel Dobbe and Morgan Ames, "Up Next For FAT*: From Ethical Values To Ethical Practices," Medium, February 9, 2019,https://medium.com/@roeldobbe/up-next-for-fat-from-ethical-values-to-ethical-practices-ebbed9f6adee .

[89] Olivia Solon, "Microsoft Funded Firm Doing Secret Israeli Surveillance on West Bank," NBC News,October 28, 2019, https://www.nbcnews.com/news/all/why-did-microsoft-fund-israeli-firm-surveils-west-bank-palestinians-1072116 .

[90] Human Rights Watch, "Israel and Palestine: Events of 2018," https://www.hrw.org/world-report/2019/country-chapters/israel/palestine#1b36d4 .

[91] Evan Selinger and Woodrow Hartzog, "What Happens When Employers Can Read Your Facial Expressions?," New York Times , October 17, 2019, https://www.nytimes.com/2019/10/17/opinion/facial-recognition-ban.html .

[92] Rich Sauer, "Six Principles to Guide Microsoft's Facial Recognition Work," Microsoft on the Issues,December 17, 2018, https://blogs.microsoft.com/on-the-issues/2018/12/17/six-principles-to-guide-microsofts-facial-recognition-work/

In another of many such examples of corporations openly defying their own ethics principles, and despite declaring as one of its AI principles to "avoid creating or reinforcing unfair bias,"[94] Google set up the Advanced Technology External Advisory Council (ATEAC), an ethics board that included Kay Coles James, the president of the Heritage Foundation. Workers and the public objected. A petition signed by over 2,500 Google workers argued: "In selecting James, Google is making clear that its version of 'ethics' values proximity to power over the wellbeing of trans people, other LGBTQ people, and immigrants. . . . Not only are James' views counter to Google's stated values, but they are directly counter to the project of ensuring that the development and application of AI prioritizes justice over profit."[95] Following the backlash, Google dissolved ATEAC after a little over a week.[96]

Yet corporate AI ethics are surely valuable and do somewhat help guide better practices and decisions, it is clear that change in the design, development, and implementation of AI systems largely occurs when there is pressure on companies from workers, grassroots coalitions, movements, the press, and policymakers. For example, the various controversies Facebook has publicly faced demonstrate that public pressure and organized workers appear to be far better at ensuring ethical AI than principles. Facebook advertises its own internal ethics process. However, investigative reports from ProPublica on Facebook's discriminatory online advertising filtering mechanisms together with published studies about Facebook's online ad ecosystem bolstered lawsuits brought by the Department of Urban Housing and Defense, civil rights groups, and labor organizations against the company in 2019.[97]

Given the concerns that ethical promises are inadequate in the face of notable accountability gaps, many have argued that human rights principles, which are based on more established legal interpretations and practice, should replace "ethics" as the dominant framework for conversations about AI governance and oversight. Advocates for this approach describe human rights as ethics "with teeth," or an alternative to the challenge of operationalizing ethics.[98]

---

[93] Olivia Solon, "MSFT Hires Eric Holder to Audit AnyVision's Facial Recognition Tech," CNBC, November 15, 2019, https://www.cnbc.com/2019/11/15/msft-hires-eric-holder-to-audit-anyvisions-facial-recognition-tech.html .

[94] Sundar Pichai, "AI at Google: Our Principles," Google, June 7, 2018, https://blog.google/technology/ai/ai-principles/ .

[95] Googlers Against Transphobia, "Googlers Against Transphobia and Hate," Medium, April 1, 2019, https://medium.com/@against.transphobia/googlers-against-transphobia-and-hate-b1b0a5dbf76 .

[96] Nick Statt, "Google Dissolves AI Ethics Board Just One Week after Forming It," The Verge , April 4, 2019, https://www.theverge.com/2019/4/4/18296113/google-ai-ethics-board-ends-controversy-kay-coles-jamesheritage-foundation .

[97] See Tracy Jan and Elizabeth Dwoskin, "HUD Is Reviewing Twitter's and Google's Ad Practices as Part of Housing Discrimination Probe," Washington Post , March 28, 2019, https://www.washingtonpost.com/business/2019/03/28/hud-charges-facebook-with-housing-discrimination/ ; Julia Angwin, Ariana Tobin and Madeleine Varner, "Facebook (Still) Letting Housing Advertisers Exclude Users by Race," ProPublica , November 21, 2017, https://www.propublica.org/article/facebook-advertising-discrimination-housing-race-sex-national-origin ; and Muhammad Ali et al., "Discrimination through Optimization: How Facebook's Ad Delivery Can Lead to Skewed Outcomes," arXiv:1904.02095v5 [cs.CY] , September 12, 2019, https://arxiv.org/pdf/1904.02095.pdf .

[98] Article 19, "Governance with Teeth: How Human Rights Can Strengthen FAT and Ethics Initiatives on Artificial Intelligence, April 2019, https://www.article19.org/wp-content/uploads/2019/04/Governance-with-teeth_A19_April_2019.pdf .

The human rights legal framework has its own potential shortcomings, especially as it relates to AI technology.[99] One of these limitations is the challenges of enforcement of international human rights law when it pertains to powerful nations. Given that the US and China are considered global AI leaders that have both engaged in varying degrees of documented human rights abuses without facing meaningful consequences under international human rights law,[100] expecting human rights frameworks to constrain governmental and corporate actors within the countries currently dominating AI development may be impractical. Indeed, human rights law is mainly focused on government actors, so beyond the current lack of enforcement, the question of how it might serve to curb corporate malfeasance remains unanswered.

By claiming a commitment to ethics, companies implicitly claim the right to decide what it means to "responsibly" deploy these technologies, and thus the right to decide what "ethical AI" means for the rest of the world.[101] As technology companies rarely suffer meaningful consequences when their ethical principles are violated, true accountability will depend on workers, journalists, researchers, coalitions, policymakers, and the public continuing to be at the forefront of the fight against the harmful uses of AI technology.

Some advocates are also pushing to ensure that engineers and developers are trained in ethics, and thus, the thinking goes, better capable of making more ethical decisions that can ensure more ethical tech. Barbara Grosz, a professor of natural sciences, imagines a world in which "every time a computer scientist logs on to write an algorithm or build a system, a message will flash across the screen that asks, 'Have you thought about the ethical implications of what you're doing?'"[102] The Design Justice Network[103] takes this further, centering justice, not ethics, and calling on developers and designers to center affected communities in the process of creating technology together.[104]

AI developers and researchers make important determinations that can affect billions of people, and helping them consider whom the technology benefits and harms is important. The case of Uber's self-driving car makes clear what could have been had engineers, designers, and executives put more care into ethics and safety (although whether or not these were decisions engineers had the power to make is not something we know). In 2018, an autonomous Uber in Arizona killed Elaine Herzberg, a pedestrian. A recent National Transportation Safety Board investigation found significant problems with Uber's autonomous system, including a shocking disclosure that Uber's self-driving software "wasn't designed to expect that pedestrians outside crosswalks may be crossing the street."[105] Similar engineering failures led to over 37 accidents involving autonomous Uber vehicles.[106]

---

[99] See Filippo Raso et al., " Artificial Intelligence & Human Rights: Opportunities & Risks," Berkman Klein Center for Internet & Society, September 25, 2018, http://nrs.harvard.edu/urn-3:HUL.InstRepos:38021439 ;Philip Alston, "Report of the Special Rapporteur on Extreme Poverty and Human Rights," October 11, 2019, https://srpovertyorg.files.wordpress.com/2019/10/a_74_48037_advanceuneditedversion-1.pdf ; and Jason Pielemeier, " The Advantages and Limitations of Applying the International Human Rights Framework to Artificial Intelligence," Data & Society: Points , June 6, 2018, https://points.datasociety.net/the-advantages-and-limitations-of-applying-the-international-human-rights-framework-to-artificial-291a2dfe1d8a .

[100] For China rights violations, see Marco Rubio, "We Must Stand Up to China's Abuse of Its Muslim Minorities," Guardian , October 31, 2019, https://www.theguardian.com/commentisfree/2019/oct/31/china-uighurs-muslims-religious-minorities-marco-rubio ; for US rights violations, see "UN Rights Chief 'Appalled' by US Border Detention Conditions, Says Holding Migrant Children May Violate International Law," UN News , July 8, 2019, https://news.un.org/en/story/2019/07/1041991 .

[101] Jacob Metcalf, Emanuel Moss, and danah boyd, "Owning Ethics: Corporate Logics, Silicon Valley, and the Institutionalization of Ethics," Data & Society, September 10, 2019, https://datasociety.net/output/owning-ethics-corporate-logics-silicon-valley-and-the-institutionalization-of-ethics/ .

[102] Paul Karoff, "Embedding Ethics in Computer Science Curriculum," Harvard Gazette , January, 25, 2019, https://news.harvard.edu/gazette/story/2019/01/harvard-works-to-embed-ethics-in-computer-science-curriculum/ . See also Greg Epstein, "Teaching Ethics in Computer Science the Right Way with Georgia Tech's Charles Isbell," TechCrunch , September 5, 2019, https://techcrunch.com/2019/09/05/teaching-ethics-in-computer-science-the-right-way-with-georgia-techs-charles-isbell/ ; Zeninjor Enwemeka, "Solving the Tech Industry's Ethics Problem Could Start In The Classroom," National Public Radio, May 31, 2019, https://www.npr.org/2019/05/31/727945689/solving-the-tech-industrys-ethics-problem-could-start-in-theclassroom; and Jenny Anderson, "MIT Developed a Course to Teach Tweens about the Ethics of AI," Quartz, September 4, 2019, https://qz.com/1700325/mit-developed-a-course-to-teach-tweens-about-the-ethics-of-ai/ .

[103] https://designjustice.org/

[104] Design Justice Network Principles, https://designjustice.org/read-the-principles .

It is clear that better testing and engineering practices, grounded in concern for the implications of AI, are urgently needed. However, focusing on engineers without accounting for the broader political economy within which AI is produced and deployed runs the risk of placing responsibility on individual actors within a much larger system, erasing very real power asymmetries. Those at the top of corporate hierarchies have much more power to set direction and shape ethical decision-making than do individual researchers and developers. Such an emphasis on "ethical education" recalls the push for "unconscious bias" training as a way to "improve diversity." Racism and misogyny are treated as "invisible" symptoms latent in individuals, not as structural problems that manifest in material inequities. These formulations ignore the fact that engineers are often not at the center of the decisions that lead to harm, and may not even know about them. For example, the engineers working on Google's Project Maven were not aware that they were building a military drone surveillance system, until the news broke the story.[107]

Indeed, such obscurity is often by design, with sensitive projects being split into sections, making it impossible for any one developer or team to understand the ultimate shape of what they are building, and where it might be applied.[108]

**AI Companies and Geographic Displacement**

Just as the development environments of artificial intelligence and machine learning are filled with disparities, so too are the broader geography in which their development occurs. Whether within large suburban tech campuses or smaller urban tech start-ups, AI and machine learning environments are never contained within company walls. Rather, the racial, gendered, and class-based biases well proven to exist within AI labs are porous, spilling into external spaces. Often this results in processes popularly described as tech-driven gentrification, or the replacement of poor, working-class, and/or racialized residents with wealthier tech employees.[109]

While numerous cities have experienced AI displacement, San Francisco has been especially impacted. With the IPO releases of a number of tech companies this past year, the real estate industry has predicted a new surge of tech wealth. As during the dot-com boom, speculators disproportionately evict Black and Latinx working-class tenants in order to create new housing for wealthier and whiter tech employees.[110]

Northern California's Alameda and Contra Costa counties, which were devastated by the 2008 foreclosure crisis, continue to see the loss of Black and Latinx home ownership and housing. In fact, the subprime crisis and the fintech derivatives market it relied upon can also be understood as a technology of AI displacement.[111] The very algorithms used by lenders and banks relied upon codifying Black homeowners as exploitable.[112]

---

[105] Matt McFarland, "Feds Blame Distracted Test Driver in Uber Self-Driving Car Death," CNN Business, November 19, 2019, https://www.cnn.com/2019/11/19/tech/uber-crash-ntsb/index.html .

[106] Kristen Lee, "Uber's Self-Driving Cars Made It Through 37 Crashes Before Killing Someone," Jalopnik ,November 6, 2019, https://jalopnik.com/ubers-self-driving-cars-made-it-through-37-crashes-befo-1839660767 .

[107] Kate Conger and Cade Metz, "Tech Workers Now Want to Know: What Are We Building This For?," New York Times, October 7, 2018, https://www.nytimes.com/2018/10/07/technology/tech-workers-ask-censorship-surveillance.html .

[108] Ryan Gallagher, "Google Shut Out Privacy and Security Teams from Secret China Project," The Intercept ,November 29, 2018, https://theintercept.com/2018/11/29/google-china-censored-search/ .

[109] AI Now Report 2019

[110] Erin McElroy, "Data, Dispossession, and Facebook: Toponymy and Techno-Imperialism in Gentrifying San Francisco," Urban Geography 40 no. 6 (2019): 826–845,https://doi.org/10.1080/02723638.2019.1591143 .

In the post-2008 era, Wall Street investment firms such as Blackstone/Invitation Homes[113] use machine learning systems to calculate rental acquisitions, buying up properties foreclosed during the subprime crisis and renting them out as single-family homes. [114]They rent such homes out today using proptech AI management systems[115] and property databases known to engage in tenant profiling that disfavors people of color.[116]

This era has also been marked by the 2008 launch of Airbnb, the San Francisco start-up linked to ongoing gentrification of cities worldwide as long-term tenants are replaced with tourists.[117] Even single room occupancy hotels (SROs), which have historically housed precarious residents, have been converted into "tech dorms" and tourist accommodations in cities such as San Francisco and Oakland.[118]

Also characteristic are the private tech luxury buses that facilitate reverse commuting of tech workers to Silicon Valley from urban centers. Landlords have found property adjacent to "Google bus stops" lucrative, leading to increased rental prices and evictions along with new luxury and market-rate development projects. [119] As during the dot-com boom and foreclosure crisis, numerous organizations and collectives formed to organize for housing justice. Rent-control protection groups and tenant unions have been forming monthly, and statewide tenant organizing has been on the rise.[120]

In California, there have been new forms of international solidarity against AI displacement. For instance, current organizing against Google's proposed new campus in San Jose is being led by groups such as Serve the People San Jose. They argue Google's new campus will lead to mass displacement and unaffordability. [121] Thus they have been organizing marches, Google bus blockades, and City Council demonstrations.[122]Much of this has taken place in solidarity with organizers and groups in Berlin such as Google Is Not a Good Neighbor (Google ist kein guter Nachbar), which in 2018 collectively blocked Google from launching a new tech campus in the neighborhood of Kreuzberg.[123]Solidarity has also been found among New York City organizers who successfully

---

[111] William Magnuson, "Why We Should be Worried about Artificial Intelligence on Wall Street," Los Angeles Times , November 1, 2019,https://www.latimes.com/opinion/story/2019-11-01/artificial-intelligence-ai-wall-street .

[112] Paula Chakravartty and Denise Ferreira da Silva, "Accumulation, Dispossession, and Debt: The Racial Logic of Global Capitalism—An Introduction," American Quarterly 64, no. 3 (September 2012): 361–385.

[113] https://www.blackstone.com/

[114] Desiree Fields, "Automated Landlord: Digital Technologies and Post-crisis Financial Accumulation," Environment and Planning A: Economy and Space , May 2019, https://doi.org/10.1177/0308518X19846514 .

[115] Elik Jaeger, Five Things To Consider When Choosing AI-Driven Proptech For Your Multifamily Operation, Forbes, (Apr 2, 2021) https://www.forbes.com/sites/forbesrealestatecouncil/2021/04/02/five-things-to-consider-when-choosing-ai-driven-proptech-for-your-multifamily-operation/?sh=7eac0b757607

[116] Erin McElroy, "Disruption at the Doorstep," Urban Omnibus, November 2019, https://urbanomnibus.net/2019/11/disruption-at-the-doorstep/ .

[117] Agustín Cócola Gant, "Holiday Rentals: The New Gentrification Battlefront," Sociological Research Online 21, no. 3 (2016): 1–9.

[118] Anti-Eviction Mapping Project, "Precarious Housing: Loss of SRO Hotels in Oakland," 2017, http://arcg.is/nymnW .

[119] Manissa M. Maharawal and Erin McElroy. "The Anti-Eviction Mapping Project: Counter Mapping and Oral History Toward Bay Area Housing Justice." Annals of the American Association of Geographers 108, no. 2 (2018): 380–389. and "State of Emergency: Special Report on California's Criminalization of Growing Homeless Encampments," Democracy Now , October 25, 2019, https://www.democracynow.org/2019/10/25/state_of_emergency_special_report_on .

[120] Anti-Eviction Mapping Project, "Rent Control for All," 2018, https://arcg.is/15X5bP .

[121] AI Now Report 2019.

[122] Ramona Giwargis, "Who is Behind the Anti-Google Protests?" San Jose Spotlight , January 9, 2019, https://sanjosespotlight.com/who-is-behind-the-anti-google-protests .

[123] Victoria Turk, "How a Berlin Neighbourhood Took On Google and Won," Wired , October 26, 2018, https://www.wired.co.uk/article/google-campus-berlin-protests .

fought the development of a new Amazon headquarters 2 in 2019, and with activists in Toronto committed to thwarting gentrification induced by Sidewalk Labs.[124]

During demonstrations, banners, light projections, video clips, and statements of support have expressed international solidarity, revealing a new trend toward urban justice.[125] Much work remains to link struggles against forms of tech-sector displacement worldwide.[126]

**Organizing Against and Resisting AI's Harms**

Resistance against AI isn't new, nor is it confined to tech companies and elite universities. Often, it is not even identified as related to biased and harmful AI. This is in part because AI systems are often integrated "in the backend," as part of operationalizing larger policies which are the stated focus of the protest. Because AI technologies are often applied in ways that amplify and exacerbate historical patterns of inequality and discrimination, it is these historical practices, embedded in AI, which organizers and communities seeking justice are reacting.[127] It is important that the bias does not get locked in to AI systems.

**Community Organizing for Responsible AI**

Community organizers have been an important force in the push back against harmful AI. [128]This is visible in the wave of community organizing protesting the use of facial recognition in cities around the world: San Francisco,[129] Oakland,[130] Somerville,[131] Montreal,[132] and Detroit,[133] among others. Community-driven organizing led directly to bans on facial recognition in many of these localities. As we highlight elsewhere in this report, in Brooklyn, tenants of Atlantic Plaza Towers organized and successfully challenged the incorporation of a facial recognition system into their building.[134]

Community organizers played a critical role in mapping the connections between incarceration, the surveillance of communities of color, and the push to adopt predictive policing tools. In Los Angeles, community organizers successfully advocated for a temporary suspension of the Los Angeles Police Department's use of the predictive policing program LASER, which purported to identify individuals likely to predict violent crimes. The Stop LAPD

---

[124] Lara Zarum, "#BlockSidewalk's War Against Google in Canada," The Nation, October 26, 2019, https://www.thenation.com/article/google-toronto-sidewalk-gentrification/ .

[125] Josh O'Kane, "Opponents of Sidewalk Labs Get Advice from German Tech Protestors," Globe and Mail, November 24, 2019, https://www.theglobeandmail.com/business/article-opponents-of-sidewalk-labs-get-advice-from-german-tech-protesters .

[126] AI Now Report 2019.

[127] AI Now Report 2019.

[128] Id.

[129] Jon Fingas, "San Francisco Bans City Use of Facial Recognition," Engadget, May 14, 2019, https://www.engadget.com/2019/05/14/san-francisco-bans-city-use-of-facial-recognition/ .

[130] Christine Fisher, "Oakland Bans City Use of Facial Recognition Software," Engadget, July 17, 2019, https://www.engadget.com/2019/07/17/oakland-california-facial-recognition-ban/ .

[131] Caroline Haskins, "A Second U.S. City Has Banned Facial Recognition," Motherboard, June 27, 2019, https://www.vice.com/en_us/article/paj4ek/somerville-becomes-the-second-us-city-to-ban-facial-recognition .

[132] Colin Harris, "Montreal Grapples with Privacy Concerns as More Canadian Police Forces Use Facial Recognition," CBC, August 8, 2019, https://www.cbc.ca/news/canada/montreal/facial-recognition-artificial-intelligence-montreal-privacy-police-law-enforcement-1.5239892 .

[133] Allie Gross, "Detroiters Concerned over Facial Recognition Technology as Police Commissioners Table Vote," Detroit Free Press, June 27, 2019, https://www.freep.com/story/news/local/michigan/detroit/2019/06/27/detroiters-concerned-over-facial-recognition-technology/1567113001/ .

[134] See Ginia Bellafante, "The Landlord Wants Facial Recognition in Its Rent-Stabilized Buildings. Why?," New York Times, March 28, 2019, https://www.nytimes.com/2019/03/28/nyregion/rent-stabilized-buildings-facial-recognition.html ; and Erin Durkin (@erinmdurkin), "The landlord of Brooklyn's Atlantic Plaza Towers has dropped his application to install facial recognition technology," Twitter, November 21, 2019, 12:45 p.m., https://twitter.com/erinmdurkin/status/1197571728173604864 .

Spying Coalition[135] argued that the department used proxy data to discriminate against Latinx and Black community members.[136]

In this effort, they were joined by UCLA students who signed a public-facing letter denouncing UCLA research and development of the predictive policing tool PredPol. Citing evidence of the role of such tools in perpetuating the overpolicing of communities of color, they requested UCLA researchers abstain from further development and commercialization of the tool.[137] Students and communities acted to operationalize a critique of predictive policing.[138] In St. Louis, Missouri, residents also demonstrated against policing tech, protesting a proposed agreement between St. Louis police and a company called Predictive Surveillance Systems to deploy surveillance planes to collect images of citizens on the ground. They asserted that the "suspicionless tracking" would be an invasion of citizens' privacy.[139]

In Kansas, New York, Pennsylvania, and Connecticut, parents opposed the use of a web-based educational platform called Summit Learning[140] in their schools. High schoolers staged sit-ins, and parents protested at school board meetings, emphasizing that the work of teachers could not be outsourced to technology-based platforms. In Pennsylvania and Connecticut, they were successful in getting the Summit programs cut.[141]

The community group Mijente[142], which describes itself as a political home for multiracial Latinx and Chicanx people, has been at the forefront of mapping the connections between AI and immigration, and building broad coalitions. In July 2019, Mijente joined Media Justice (an organization at the helm of San Francisco's facial-recognition ban)[143] and Tech Workers Coalition[144] to host Take Back Tech. The event convened community organizers alongside tech workers and students, aiming to share strategies and knowledge, and to build coalitions between those harmed by oppressive technologies and those close to the research and development of such tech.[145]

In August 2019, Mijente released a detailed report based on FOIA record requests illuminating the central role certain technologies have played in detaining Black and Brown people, and the use of these technologies in immigration enforcement.[146] The organization also spearheaded the #NoTechforICE campaign in opposition raids and mass deportations of migrants along the southern border of the US. Protesters catalyzed by the campaign have held regular demonstrations at Palantir's headquarters in Palo Alto and at its New York City offices.[147]

---

[135] Stop LAPD Spying Coalition, https://stoplapdspying.org/ .
[136] See City News Service, "LAPD Chief to Outline New Data Policies," NBC Los Angeles, April 9, 2019, https://www.nbclosangeles.com/news/local/LAPD-Chief-to-Outline-New-Data-Policies-508308931.html ; and Stop LAPD Spying Coalition, "The People's Response to OIG Audit of Data-Driven Policing," Stop LAPD Spying, March 2019, https://stoplapdspying.org/wp-content/uploads/2019/03/Peoples_Response_with-hyper-links.pdf .
[137] Stop LAPD Spying, Medium, April 4, 2019, https://medium.com/@stoplapdspying/on-tuesday-april-2nd-2019-twenty-eight-professors-and-forty-gradu\ate-students-of-university-of-8ed7da1a8655 .
[138] Kristian Lum and William Isaac, "To predict and serve?" Significance 13, no. 5 (October 2016): 14–19, https://doi.org/10.1111/j.1740-9713.2016.00960.x ,
[139] Aila Slisco, "Protesters Denounce 'Spy Plane' Plan to Monitor St. Louis," Newsweek, October 10, 2019, https://www.newsweek.com/protesters-denounce-spy-plane-plan-monitor-st-louis-1464535 .
[140] https://www.summitlearning.org/
[141] Nellie Bowles, "Silicon Valley Came to Kansas Schools. That Started a Rebellion," New York Times, April 21, 2019, https://www.nytimes.com/2019/04/21/technology/silicon-valley-kansas-schools.html .
[142] https://mijente.net/
[143] Media Justice, https://mediajustice.org/
[144] Tech Workers Coalition, https://techworkerscoalition.org/ .
[145] https://notechforice.com/convening/
[146] Mijente, "The War against Immigrants: Trump's Tech Tools Powered by Palantir," August 2019, https://mijente.net/wp-content/uploads/2019/08/Mijente-The-War-Against-Immigrants_-Trumps-Tech-Tools-Powered-by-Palantir_.pdf .
[147] Rosalie Chan, "Protesters Blocked Palantir's Cafeteria to Pressure the $20 Billion Big Data Company to Drop Its Contracts with ICE," Business Insider, August 16, 2019, https://www.businessinsider.com/palantir-protest-palo-alto-activists-ice-contracts-2019-8 .

Organizations such as Never Again Action, [148] and Jews for Racial and Economic Justice(JFREJ)[149] also led highly visible actions against Amazon, organizing street protests and sit-ins in Amazon bookstores to protest against the company's ongoing work providing cloud computing services to ICE.[150] Immigrant rights groups such as Make the Road New York,[151] along with Mijente, JFREJ, and other advocates, reached out to academics and computer science and technology professionals through petitions, demanding that prominent conferences drop Palantir as a sponsor, given the company's role in empowering ICE.[152] Community-organized opposition to Palantir's role in ICE's detention of immigrants resulted in UC Berkeley's Privacy Law Scholars Conference,[153] Lesbians Who Tech,[154] and the Grace Hopper Celebration all pulling Palantir as a sponsor.[155]

Athena, a recently launched coalition, includes ALIGN, New York Communities for Change, Make The Road New York, Desis Rising Up and Moving, and many other groups that successfully campaigned to challenge Amazon's plans to build its second headquarters, HQ2[156] in Queens, New York.[157] The campaign against Amazon's HQ2 was notable for its broad multi-issue approach, and for its somewhat unexpected success. Advocates and community organizers criticized the company's tax avoidance, the displacement that would follow in the wake of such a massive tech company headquarters, and the lavish corporate subsidies that New York offered the company. They also organized around Amazon's treatment of warehouse workers and its sale of surveillance tech.[158] Athena expands on this multi-issue approach, recognizing that Amazon is at the heart of a set of interlocking issues, including worker rights at warehouses, climate justice, and mass surveillance. The coalition includes organizations with experience across these domains, and is working to unify the growing opposition to the company and develop strategies capable of tackling AI companies whose reach extends into so many sensitive domains.[159] Amazon's HQ2 was ultimately built in Arlington, Virginia.[160]

AI and Tech ghost workers demand to be seen and heard.[161] Alongside the scientists, there are thousands of low-paid, remotely hired workers whose job it is to classify and label data, the lifeblood of AI machine learning

---

[148] Never Again Action, https://www.neveragainaction.com/ .

[149] Jews for Racial and Economic Justice, https://jfrej.org/ .

[150] Michael Grothaus, "Dozens of People Have Been Arrested at a #JewsAgainstICE Protest at NYC Amazon Books Store," Fast Company , August 12, 2019, https://www.fastcompany.com/90388865/dozens-of-people-have-been-arrested-at-a-jewsagainstice-protest-at-nyc-amazon-books-store .

[151] Make the Road New York, https://maketheroadny.org/ .

[152] Rachel Frazin, "Advocates Start Petition Asking Tech Conference to Drop Palantir as Sponsor over ICE Contracts," The Hill , October 23, 2019, https://thehill.com/policy/technology/467204-advocates-start-petition-asking-tech-conference-to-drop-palantir-as-sponsor .

[153] Lizette Chapman, "Palantir Dropped by Berkeley Privacy Conference After Complaints," Bloomberg, June 5, 2019, https://www.bloomberg.com/news/articles/2019-06-05/palantir-dropped-by-berkeley-privacy-conference-after-complaints .

[154] Rob Price and Rosalie Chan, "LGBTQ Tech Group Lesbians Who Tech Ditches Palantir as a Conference Sponsor over Human-Rights Concerns," Business Insider, August 26, 2019, https://www.businessinsider.com/lesbians-who-tech-ends-sponsorship-deal-palantir-human-rights-2019-8 .

[155] Shirin Gaffary, "The World's Biggest Women's Tech Conference Just Dropped Palantir as a Sponsor," Recode, August 28, 2019, https://www.vox.com/recode/2019/8/28/20837365/anita-b-grace-hopper-palantir-sponsor-worlds-biggestwomens-tech-conference-dropped .

[156] https://www.aboutamazon.com/news/amazon-offices/the-next-chapter-for-hq2-sustainable-buildings-surrounded-by-nature

[157] Athena, https://athenaforall.org/ .

[158] Jimmy Tobias, "The Amazon Deal Was Not Brought Down by a Handful of Politicians: It Was Felled by a Robust Grassroots Coalition," The Nation, February 19, 2019, https://www.thenation.com/article/the-amazon-deal-was-not-brought-down-by-a-handful-of-politicians/ .

[159] David Streitfeld, "Activists Build a Grass-Roots Alliance Against Amazon," New York Times, November 26, 2019, https://www.nytimes.com/2019/11/26/technology/amazon-grass-roots-activists.html .

[160] https://www.aboutamazon.com/news/amazon-offices/the-next-chapter-for-hq2-sustainable-buildings-surrounded-by-nature

[161] Jane Wakefield, AI: Ghost workers demand to be seen and heard, BBC, March 28, 2021 https://www.bbc.com/news/technology-56414491.

systems. Increasingly, there are questions about the exploitation of these workers. The most well-established of these crowdsourcing platforms is Amazon Mechanical Turk[162], owned by the online retail giant and run by its Amazon Web Services[163] division. Other companies using these workers include Sama,[164] CrowdFlower[165] and Microworkers.[166] They all allow businesses to remotely hire workers from anywhere in the world to do tasks that AI cannot yet perform. These tasks could be anything from labelling images to helping computer vision algorithms improve, providing help for natural language processing, or acting as content moderators for YouTube or Twitter.[167]

These are only a handful of instances where community organizers are pushing back against AI and oppressive tech. Collectively, they highlight that the resistance to AI systems is not necessarily just about AI, but about policies and practices that exacerbate inequality and cause harm to our communities. They also demonstrate that AI systems' biases do not exist in isolation, but embed historical surveillance and policing practices that predominantly impact Black communities, communities of color, and the poor.[168]

Acknowledging and making these processes visible is important to demystify AI systems, resist discourses that privilege technology, and listen closely to the communities leading the pushback to harmful AI efforts.[169]

**Workers in AI Organizing**

Although organizing among tech workers has been underway for many years (spurred initially by contract workers), worker organizing around the harms of AI is recent. [170] Such organizing is situated within a broader effort to address overall worker issues, ranging from wages and working conditions to concerns about respect, diversity, and inclusion, that seek to directly confront hostile workplace cultures. This broad organizing platform has resulted from the insight that workers share common concerns when it comes to AI and large-scale technical systems.

For example, tech workers have joined with community organizers in pushing back against tech's role in perpetuating human rights abuses and maltreatment of migrants and Latinx residents at the southern US

---

[162] https://www.mturk.com/
[163] https://aws.amazon.com/about-aws/
[164] Sama formerly known as Samasource is a training-data company, focusing on annotating data for artificial intelligence algorithms. https://www.sama.com/
[165] https://visit.figure-eight.com/People-Powered-Data-Enrichment_T
[166] https://www.microworkers.com/
[167] See also Alexandrine Royer, The urgent need for regulating global ghost work, Brookings, (Feb. 9, 2021) https://www.brookings.edu/techstream/the-urgent-need-for-regulating-global-ghost-work/
[168] See Os Keyes, "The Bones We Leave Behind," Real Life Magazine, October 7, 2019, https://reallifemag.com/the-bones-we-leave-behind/ ; Georgetown Law Center on Privacy and Technology, "The Color of Surveillance," 2019, https://www.law.georgetown.edu/privacy-technology-center/events/color-of-surveillance-2019/ ; and Ed Pilkington, "Digital Dystopia: How Algorithms Punish the Poor," Guardian, October 14, 2019, https://www.theguardian.com/technology/2019/oct/14/automating-poverty-algorithms-punish-poor .
[169] Sasha Costanza Chock, "Design Justice: Towards an Intersectional Feminist Framework for Design Theory and Practice," Proceedings of the Design Research Society 2018 , June 3, 2018, https://papers.ssrn.com/sol3/papers.cfm?abstract_id=3189696 .
[170] AI Now Report 2019.

border. Since the fall of 2018, workers at Salesforce,[171] Microsoft,[172] Accenture,[173] Google,[174] Tableau,[175] and GitHub[176] all signed petitions and open letters protesting their companies' contracts with ICE. Developer Seth Vago pulled his open-source code out of the codebase used by the company Chef[177] after learning of the company's contract with ICE.[178] This led Chef to commit to cancel their contract,[179] and spurred a larger discussion about the ethical responsibility of developers. Seth

Vargo's protest actions caused service outages for Chef customers, though it quickly brought things back to normal. The outage he triggered exposed a new risk for companies using open source. Even workers at Palantir, the tech company at the center of ICE's detention and tracking operations, circulated two open letters, and have expressed mistrust of and frustration with the company's leadership for its decision to keep its contract with ICE.[180] Palantir CEO Alex Karp has publicly defended this work,[181] and in August 2019 the company renewed a contract worth $49 million over three years.[182]

Other workers protested the development of AI systems for military purposes they were uncomfortable with: Microsoft employees signed an open letter to the company asking it not to bid on JEDI, a major Department of Defense cloud-computing contract, which the company ultimately won.[183] In February 2019, employees at the

---

[171] Fight for the Future, "An Open Letter to Salesforce: Drop Your Contract with CBP," Medium, July 17, 2018, https://medium.com/@fightfortheftr/an-open-letter-to-salesforce-drop-your-contract-with-cbp-a8260841b627 .

[172] Sheera Frenkel, "Microsoft Employees Question C.E.O. Over Company's Contract With ICE," New York Times, July 26, 2018, https://www.nytimes.com/2018/07/26/technology/microsoft-ice-immigration.html .

[173] Bryan Menegus, "Accenture Employees Demand Their Company Break Ties With U.S. Border Patrol," Gizmodo, November 15, 2018, https://gizmodo.com/accenture-employees-demand-their-company-break-ties-wit-1830474961 .

[174] Colin Lecher, "Google Employees 'Refuse to Be Complicit' in Border Agency Cloud Contract," The Verge, August 14, 2019, https://www.theverge.com/2019/8/14/20805432/google-employees-petition-protest-customs-border-cloudcomputing-contract .

[175] Lauren Kaori Gurley, "Tech Workers Walked Off the Job after Software They Made Was Sold to ICE," Motherboard, October 31, 2019, https://www.vice.com/en_us/article/43k8mp/tech-workers-walked-off-the-job-after-software-they-made-was-sold-to-ice .

[176] Chris Merriman, "GitHub Devs Warn Microsoft 'Ditch That Contract with ICE or Lose Us,'" Inquirer, June 22, 2018, https://www.theinquirer.net/inquirer/news/3034641/github-devs-warn-microsoft-get-that-contract-on-ice-or-lose-us .

[177] https://www.chef.io/

[178] Ron Miller, "Programmer Who Took Down Open-Source Pieces over Chef ICE Contract Responds," TechCrunch , September 23, 2019, https://techcrunch.com/2019/09/23/programmer-who-took-down-open-source-pieces-over-chef-ice-contract-responds/ .

[179] Rosalie Chan, IT automation startup Chef says it will not renew its contract with ICE, days after an open source programmer brought the service to a temporary halt in protest, BUSINESS INSIDER (Sep 23, 2019, 1:10 PM) https://www.businessinsider.com/chef-ice-contract-expires-next-year-2019-9

[180] See Douglas MacMillan and Elizabeth Dwoskin, "The War inside Palantir: Data-Mining Firm's Ties to ICE under Attack by Employees," Washington Post, August 22, 2019, https://www.washingtonpost.com/business/2019/08/22/war-inside-palantir-data-mining-firms-ties-ice-under-attack-by-employees/ ; and Rosalie Chan, "Palantir Workers Are Split over the Company's Work with ICE, but CEO Alex Karp Won't Budge despite Concerned Employees' Petitions," Business Insider, August 22,2019, https://www.businessinsider.com/palantir-employees-ice-petition-alex-karp-2019-8 .

[181] Alex Karp, "I'm a Tech CEO, and I Don't Think Tech CEOs Should Be Making Policy," Washington Post, September 5, 2019, https://www.washingtonpost.com/opinions/policy-decisions-should-be-made-by-elected-representatives-not-silicon-valley/2019/09/05/e02a38dc-cf61-11e9-87fa-8501a456c003_story.html .

[182] Alie Breland, "ICE Accidentally Just Revealed How Much Its New Contract With Peter Thiel's Palantir Is Worth," Mother Jones, August 20, 2019, https://www.motherjones.com/politics/2019/08/ice-palantir-contract-amount-revealed/ .

[183] Microsoft Employees, "An Open Letter to Microsoft: Don't Bid on the US Military's Project JEDI," Medium, October 12, 2018, https://medium.com/s/story/an-open-letter-to-microsoft-dont-bid-on-the-us-military-s-project-jedi-7279338b7132 .

company followed this with a n unsuccessful demand to cancel a $480 million contract to provide augmented reality headsets to the US military, saying they did "not want to become war profiteers."[184]

Thus organizing around AI is also part of a broader tech-worker movement focused on a broad range of social justice issues, including displacement,[185] two-tiered workforces exploiting contract workers,[186] and climate change. In April 2019, 8,695 Amazon workers publicly signed a letter calling on the company to address its contributions to climate change through a shareholder resolution,[187] and staged a walkout in September 2019 in the face of inaction by the company.[188] The September climate walkout was the first labor action coordinated across multiple tech companies, and provides an indication of the growth of tech-worker organizing.

The 2018 Google Walkout at Google offices around the world protesting the company's treatment of women was effective in many ways. The employees demanded several key changes in how sexual misconduct allegations are dealt with, including a call to end forced arbitration (which was ended).[189] Google CEO Sundar Pichai told staff he supported their right to take the action. "I understand the anger and disappointment that many of you feel," he said in an all-staff email. "I feel it as well, and I am fully committed to making progress on an issue that has persisted for far too long in our society. and, yes, here at Google, too." [190] it was only the first of many employee protests. Employees at Riot Games[191] walked out in protest of the company's stance on forced arbitration, following allegations by multiple employees that the company violated California's Equal Pay Act and claims of gender-based discrimination and harassment.[192]

In China, developers protested what they described as the 996 schedule—9 a.m. to 9 p.m., six days a week—through a GitHub[193] repository of companies and projects asking for excessive hours.[194] And in November, Google workers again walked out, hosting a rally of hundreds of workers in San Francisco protesting retaliation against two organizers.[195] Following this rally, Google fired four organizers, signaling both the growing power of such efforts to impact Google and the company's intolerance of them.[196]

---

[184] Microsoft Workers 4 Good (@MSWorkers4), "On behalf of workers at Microsoft, we're releasing an open letter to Brad Smith and Satya Nadella, demanding for the cancelation of the IVAS contract with a call for stricter ethical guidelines," Twitter, February 22, 2019, https://twitter.com/MsWorkers4/status/1099066343523930112 .

[185] Silicon Valley Rising, "Google Shareholder Meeting," June 19, 2019, https://act.siliconvalleyrising.org/google_shareholder_meeting .

[186] Google Walkout for Real Change, "Not OK, Google," Medium, April 2, 2019, https://medium.com/@GoogleWalkout/not-ok-google-79cc63342c05 .

[187] Amazon Employees for Climate Justice, "Open letter to Jeff Bezos and the Amazon Board of Directors," Medium, April 10, 2019, https://medium.com/@amazonemployeesclimatejustice/public-letter-to-jeff-bezos-and-the-amazon-boardof-directors-82a8405f5e38 .

[188] Louise Matsakis, "Amazon Employees Will Walk Out over the Company's Climate Change Inaction," Wired, September 9, 2019, https://www.wired.com/story/amazon-walkout-climate-change/ .

[189] Fan, Jennifer S., Employees as Regulators: The New Private Ordering in High Technology Companies (December 1, 2019). Utah Law Review, Vol. 2019, No. 5, Pp. 973-1076, 2019, Available at SSRN: https://ssrn.com/abstract=3520230

[190] Dave Lee, "Google Staff Walk Out over Women's Treatment," BBC, November 1, 2018, https://www.bbc.com/news/technology-46054202 .

[191] https://www.riotgames.com/en

[192] Nathan Grayson and Cecilia D'Anastasio, "Over 150 Riot Employees Walk Out to Protest Forced Arbitration and Sexist Culture," Kotaku, May 6, 2019, https://kotaku.com/over-150-riot-employees-walk-out-to-protest-forced-arbi-1834566198 .

[193] https://github.com/

[194] Tracy Qu, "How GitHub Became a Bulletin Board for Chinese Tech Worker Complaints," Quartz, April 9, 2019, https://qz.com/1589309/996-icu-github-hosts-chinese-tech-worker-complaints/

[195] Johana Bhuiyan, "Google Workers Protest Suspensions of Activist Employees," Los Angeles Times ,November 22, 2019, https://www.latimes.com/business/technology/story/2019-11-22/google-workers-rally-activists-protests .

[196] Kate Conger and Daisuke Wakabayashi, "Google Fires 4 Workers Active in Labor Organizing," New York Times, November 25, 2019, https://www.nytimes.com/2019/11/25/technology/google-fires-workers.html .

Contract workers leading the recent wave of tech-worker organizing, [197] protesting the risks they experience as a result of the two-tier labor systems in which they work. In particular, contract workers lack the benefits, stability, and pay of their employee colleagues. A 2016 report from Working Partnerships USA found that 58 percent of blue-collar contract workers in tech are Black and Latinx, and make an average of $19,900 annually.[198] The report found that only 10 percent of "employee" tech workers are Black or Latinx, and that these workers make over $100 thousand annually.[199] Tech workers have called for an end to such discrimination, noting the racial divide and its implications for the perpetuation of structural inequality.[200]

In spite of the uncertainty and disadvantages that come with being classified as a contract worker, these workers continued to organize, from temp workers at Foxconn[201] factories protesting unpaid wages and bonuses promised to them by recruitment agencies,[202] to workers at Amazon warehouses walking out on Prime Day[203] and successfully winning compromises to improve conditions.[204] Beyond protesting workplace conditions, contract workers have been leaders in pushing for ethical company practices, with Amazon-owned Whole Foods workers publishing a letter demanding Amazon end its involvement with ICE [205] and sharing a video revealing the company's union-busting tactics.[206]

Such organizing has led to a wave of unions forming among workers on the corporate campuses of tech firms in recent years, a trend that started in 2014, well before white-collar workers began visibly organizing. These included food-service workers at Airbnb,[207] Facebook,[208] and Yahoo,[209] and shuttle drivers and security guards at a host of Silicon Valley firms.[210]

---

[197] See Steven Greenhouse, "Facebook's Shuttle Bus Drivers Seek to Unionize," New York Times, October 5, 2014, https://www.nytimes.com/2014/10/06/business/facebooks-bus-drivers-seek-union.html ; Mark Harris, "Amazon's Mechanical Turk Workers Protest: 'I Am a Human Being, Not an Algorithm,'" Guardian, December 3, 2014, https://www.theguardian.com/technology/2014/dec/03/amazon-mechanical-turk-workers-protest-jeff-bezos ; Josh Eidelson, "Microsoft's Unionized Contract Workers Get Aggressive," April 30, 2015, https://www.bloomberg.com/news/articles/2015-04-30/microsoft-contract-workers-are-organizing ; and Kia Kokalitcheva, "These Google Workers Have Voted to Join the Teamsters Union," Fortune , August 21, 2015, https://fortune.com/2015/08/21/these-google-workers-have-voted-to-join-the-teamsters-union/ .

[198] AI Now Report 2019.

[199] Working Partnerships USA, "Tech's Invisible Workforce," March, 2016, https://www.wpusa.org/files/reports/TechsInvisibleWorkforce.pdf .

[200] Alexia Fernández Campbell, "Google's Contractors Accuse CEO of Creating Unequal Workforce," Vox , December 7, 2018, https://www.vox.com/2018/12/7/18128922/google-contract-workers-ceo-sundar-pichai .

[201] Hon Hai Precision Industry Co., Ltd., trading as Foxconn Technology Group and better known as Foxconn, is a Taiwanese multinational electronics contract manufacturer with its headquarters in Tucheng, New Taipei City, Taiwan. In 2010, it was the world's largest provider of electronics manufacturing services and the third-largest technology company by revenue. The company is the largest private employer in Taiwan and one of the largest employers worldwide. Terry Gou is the company's founder and former chairman. https://en.wikipedia.org/wiki/Foxconn

[202] Alexia Fernández Campbell, "Google's Contractors Accuse CEO of Creating Unequal Workforce," Vox ,December 7, 2018, https://www.vox.com/2018/12/7/18128922/google-contract-workers-ceo-sundar-pichai .

[203] Prime Day is an annual deal event at Amazon,  exclusively for Prime members, delivering two days of deals on products .https://www.amazon.com/primeday

[204] Josh Dzieza, "'Beat the Machine': Amazon Warehouse Workers Strike to Protest Inhumane Conditions," The Verge , July 16, 2019, https://www.theverge.com/2019/7/16/20696154/amazon-prime-day-2019-strike-warehouse-workers-inhumane-conditions-the-rate-productivity .

[205] Nick Statt, "Whole Foods Employees Demand Amazon Break All Ties with ICE and Palantir," The Verge, August 12, 2019, https://www.theverge.com/2019/8/12/20802893/whole-foods-employees-amazon-ice-protest-palantir-facial-recognition .

[206] Bryan Menegus, "Amazon's Aggressive Anti-Union Tactics Revealed in Leaked 45-Minute Video," Gizmodo, September 26, 2018, https://gizmodo.com/amazons-aggressive-anti-union-tactics-revealed-in-leake-1829305201 .

In Poland, Spain, and Germany, unionized Amazon warehouse workers held strikes to demand higher pay and better working conditions.[211] But Amazon and other tech companies are using tactics to prevent unions from forming: for example, 14 software engineers at the start-up Lanetix were fired shortly after unionizing. The workers filed charges with the National Labor Relations Board and ultimately won their case.[212] Google also hired a consulting firm known for its anti-union work amid employee unrest, a fact disclosed by whistleblowers.[213]

Globally, strikes by transport workers grew in response to ride-sharing apps that are decreasing wages and thus living standards. Uber drivers staged major strikes in cities around the globe[214] including drivers' occupying Uber's offices in France,[215] while Ola[216] drivers in India protested decreasing driver incentives amid increasing fuel prices.[217] China Labor Bulletin recorded nearly 1,400 transport worker protests over a five-year period in cities across the country.[218]

---

[207] Kate Conger, "Food Service Workers at Airbnb Have Unionized," Gizmodo, February 15, 2018, https://gizmodo.com/food-service-workers-at-airbnb-have-unionized-1823049379 .

[208] Alex Heath, "Facebook Cafeteria Workers Vote to Unionize, Demand Higher Wages," Business Insider , July 24, 2017, https://www.businessinsider.com/facebook-cafeteria-workers-unionize-demand-higher-wages-2017-7

[209] Unite Here, "Cafeteria Workers at Yahoo Unionize, Join Workers' Movement for Equality in the Tech Industry," Unite Here, December 13, 2017, http://unitehere.org/cafeteria-workers-at-yahoo-unionize/ .

[210] Josh Eidelson, "Union Power Is Putting Pressure on Silicon Valley's Tech Giants," Bloomberg Businessweek, September 14, 2017, https://www.bloomberg.com/news/articles/2017-09-14/union-power-is-putting-pressure-on-silicon-valley-s-tech-giants See also John Ribeiro, Tech firms are under pressure to improve working conditions for workers like drivers and security guards, Computerworld, March 3, 2015, https://www.computerworld.com/article/2891709/apple-like-google-to-hire-full-time-security-guards-in-silicon-valley.html

[211] Reuters, "Amazon Holds Talks with Workers in Poland as Strike Threatened," Reuters, May 10, 2019, https://www.reuters.com/article/us-amazon-poland-wages/amazon-holds-talks-with-workers-in-poland-asstrike-threatened-idUSKCN1SG1I5 .

[212] Tekla Perry, "Startup Lanetix Pays US $775,000 to Software Engineers Fired for Union Organizing," IEEE Spectrum, November 12, 2018, https://spectrum.ieee.org/view-from-the-valley/at-work/tech-careers/startup-lanetix-pays-775000-to-software-engineers-fired-for-union-organizing .

[213] Noam Scheiber and Daisuke Wakabayashi, "Google Hires Firm Known for Anti-Union Efforts," November 20, 2019, https://www.nytimes.com/2019/11/20/technology/Google-union-consultant.html .

[214] April S. Glaser, "The Ride-Hail Strike Got Just Enough Attention to Terrify Uber," Slate, May 9, 2019, https://slate.com/technology/2019/05/uber-strike-impact-gig-worker-protest.html .

[215] Edward Ongweso Jr., We Spoke To Uber Drivers Who Have Taken Over The Company's Offices In France, VICE, (Nov. 26, 2019) https://www.vice.com/en/article/zmjadx/we-spoke-to-uber-drivers-who-have-taken-over-the-companys-offices-in-france (Uber has run into a multitude of problems that have frustrated the company's perpetual expansion. Things came to a head in 2015, when taxi drivers in France organized massive protests and strikes in response to UberPop, an illegal service that relied on non-professional drivers who didn't need a medallion, license, or insurance. After attempting to make their own fairer version of Uber, VTCs (a French acronym for non-taxi private drivers) banded together to launch a series of protests against Uber's perpetual wage cuts and declining working conditions.)

[216] https://drive.olacabs.com/

[217] People's Dispatch, "Uber, Ola Drivers to Go on Strike in India Seeking Safety Measures and City Taxi Permit," People's Dispatch, July 4, 2019, https://peoplesdispatch.org/2019/07/04/uber-ola-drivers-to-go-on-strike-in-india-seeking-safety-measuresand-city-taxi-permit/ .

[218] China Labour Bulletin, "The Shifting Patterns of Transport Worker Protests in China Present a Major Challenge to the Trade Union," China Labour Bulletin, November 18, 2019, https://clb.org.hk/content/shifting-patterns-transport-worker-protests-china-present-major-challenge-tradeunion.

In the state of California, driver protests resulted in significant and tangible gains—though not from the companies themselves. Instead, California's State Assembly passed Assembly Bill 5 (AB5), which makes it much harder for companies such as Uber to label workers as independent contractors, granting them basic worker protections.[219] In arguing against the change, Uber claimed that drivers weren't core to Uber's business, and thus the company should not have to reclassify them as employees.[220] Based on this argument, Uber and Lyft appear likely to take their case to court.[221] AB-5 in California was followed swiftly by a ruling in New Jersey that argued Uber had misclassified drivers as independent contractors, and demanded the company pay $649 million in unpaid employment taxes.[222]

At the close of 2019, responding to worker protests, a group of US senators wrote Google CEO Sundar Pichai expressing objection to the company's heavy reliance on temporary workers (over half its workforce)[223] and urging the company to end its abuse of worker classifications.[224] Such reclassification of workers would result in thousands of people gaining access to essential benefits, workplace protections, and stability, which are denied contract workers. A move to reclassify all workers as employees would also have significant implications for the production and maintenance of AI systems, since low-paid contract workers are an essential labor force labeling AI training data, and moderating content on large algorithmically driven platforms.[225]

**Students Organize for Responsible AI and Protest Against Tech Giants' Relationships with Sex Trafficker and Tech Financier Jeffrey Epstein**

Engineering students in particular have significant leverage, given that tech companies compete to recruit top talent and view them as "future workers."[226] Facebook has seen its offer acceptance rate dwindle from 85 percent to 35–55 percent at top computer science schools.[227]

In the fall of 2018, students at Stanford first circulated a pledge not to accept interviews from Google until the company canceled its work on Project Maven, a US military effort to build AI-enabled drone surveillance, and committed to no further military involvement.[228] This movement grew significantly during 2019, spearheaded by Mijente's #NoTechForICE campaign. Students around the US demonstrated against recruiting events on campus by technology companies known to be supporting border control or policing activities, such as Amazon, Salesforce, and Palantir.[229] Over 1,200 students representing 17 campuses signed a pledge asserting they would not work at Palantir because of its ties to ICE.[230] In February 2019, students from Central Michigan University fought against the creation of a university Army AI Task Force that was poised to endorse the military use of AI.[231]

---

[219] Kari Paul, "California Uber and Lyft Drivers Rally for Bill Granting Rights to Contract Workers," Guardian, August 27, 2019, https://www.theguardian.com/us-news/2019/aug/27/california-uber-and-lyft-drivers-rally-for-bill-granting-rights-to-contract-workers .

[220] Andrew J. Hawkins, "Uber Argues Its Drivers Aren't Core to Its Business, Won't Reclassify Them as Employees," The Verge, September 11, 2019, https://www.theverge.com/2019/9/11/20861362/uber-ab5-tony-west-drivers-core-ride-share-business-california .

[221] Carolyn Said, "Uber: We'll Fight in Court to Keep Drivers as Independent Contractors," San Francisco Chronicle, September 11, 2019, https://www.sfchronicle.com/business/article/Uber-We-ll-fight-in-court-to-keep-drivers-as-14432241.php .

[222] Matthew Haag and Patrick McGeehan, "Uber Fined $649 Million for Saying Drivers Aren't Employees," New York Times, November 14, 2019, https://www.nytimes.com/2019/11/14/nyregion/uber-new-jersey-drivers.html .

[223] Daisuke Wakabayashi, "Google's Shadow Work Force: Temps Who Outnumber Full-Time Employees," New York Times, May 28, 2019, https://www.nytimes.com/2019/05/28/technology/google-temp-workers.html .

[224] United States Senate, Letter to Sundar Pichai, July 25, 2019, https://int.nyt.com/data/documenthelper/1547-senate-democrats-letter-google-temporary-workers/1ad40d0ad9ac2286b911/optimized/full.pdf#page=1 .

[225] AI Now Report 2019.

[226] Varoon Mathur and Meredith Whittaker (AI Now Institute), "How To Interview a Tech Company: A Guide for Students," Medium, September 17, 2019, https://medium.com/@AINowInstitute/how-to-interview-a-tech-company-d4cc74b436e9 .

[227] Salvador Rodriguez, "Facebook Has Struggled to Hire Talent since the Cambridge Analytica Scandal, according to Recruiters Who Worked There," CNBC, May 16, 2019, https://www.cnbc.com/2019/05/16/facebook-has-struggled-to-recruit-since-cambridge-analytica-scandal.html .

[228] MoveOn.org, "Sign the Petition: Students Pledge to Refrain from Interviewing with Google until Commitment Not to Pursue Future Tech Military Contracts (e.g. Project Maven)," https://petitions.moveon.org/sign/students-pledge-to-refrain .

Student protests also focused on racist, misogynist, and inequitable cultures within universities, tying these to unethical funding practices, and close relationships to surveillance interests.[232] At MIT, graduate student Arwa Mboya was one of the first to call for accountability after revelations surfaced showing the MIT Media Lab's close funding relationship with billionaire donor, sex trafficker, and convicted pedophile Jeffery Epstein.[233] Mboya called on the MIT Media Lab Director Joi Ito to resign from his position. After investigative journalist Ronan Farrow reported on the ties between Epstein and Ito's Media Lab, Ito stepped down.[234] Responding to these disclosures, MIT Students Against War organized protests and town halls, demanding that MIT President L. Rafael Reif and "all senior leadership that was aware of this issue" resign. They also demanded a board made up of students, faculty, and staff to review and approve donations.[235]

Jeffrey Epstein was associated with many other tech moguls.[236] Media reports say that the 2021 divorce of billionaire philanthropists Bill and Melinda Gates has been in the works for years, and that it is tied to concerns about Bill Gates' dealings with Jeffrey Epstein.[237] Andy Rubin, Google developer of the Android, was accused of sexual harassment by employees, and his wife's divorce complaint alleged that the former Google executive had "affairs with multiple women." Some of those affairs, the suit states, included "'ownership' relationships with other women, whereby Rubin would pay for their expenses in exchange for offering them to other men." The complaint includes two messages from Andy Rubin's email account, which his wife claims to have viewed, detailing those relationships. One of these women … was complicit with Rubin in running what appeared to be a sex ring," the complaint reads, alleging that he spent hundreds of thousands of dollars for sexual favors and relationships with other women. [238]

---

[229] See April Glaser, "The Techlash Has Come to Stanford," Slate, August 8, 2019, https://slate.com/technology/2019/08/stanford-tech-students-backlash-google-facebook-palantir.html ; Shirin Ghaffary, "At UC Berkeley, Brown, and Yale, Students Are Fighting to Keep Palantir off Campus over Its ICE Contracts," Recode, September 26, 2019, https://www.vox.com/recode/2019/9/26/20884182/palantir-ice-protests-campus-family-separation-berkeley-yale-brown ; Sebastian Cahill and Olivia Buccieri, "UC Berkeley Students Protest Amazon's Ties with ICE,"Daily Californian , September 27, 2019,https://www.dailycal.org/2019/09/27/uc-berkeley-students-protest-amazons-ties-with-ice/; and Caroline O'Donovan, "Student Groups Don't Want Salesforce and Palantir on Campus," Buzzfeed , February 28, 2019, https://www.buzzfeednews.com/article/carolineodonovan/student-groups-protest-salesforce-palantir-ice-campus .

[230] Mijente, "1,200+ Students at 17 Universities Launch Campaign Targeting Palantir," #NoTechForICE, September 16, 2019, https://notechforice.com/20190916-2/ .

[231] Courtney Linder, "Some Students, Faculty Remain Uneasy about CMU's Army AI Task Force," Pittsburgh Post-Gazette, February 18, 2019, https://www.post-gazette.com/business/tech-news/2019/02/17/army-ai-task-force-pittsburgh-cmu-farnam-jahanian-military-google-project-maven/stories/201902150015 .

[232] AI Now Report 2019 supra

[233] Arwa Mboya, "Why Joi Ito Needs to Resign," The Tech, August 29, 2019, https://thetech.com/2019/08/29/joi-ito-needs-to-resign .

[234] Ronan Farrow, "How an Élite University Research Center Concealed Its Relationship with Jeffrey Epstein," The New Yorker, September 6, 2019, https://www.newyorker.com/news/news-desk/how-an-elite-university-research-center-concealed-its-relationship-with-jeffrey-epstein .

[235] Nicolas Stolte, "After Epstein Protest, MIT Students Host Community Forum," Huntington News , September 26, 2019, https://huntnewsnu.com/59840/city-pulse/after-epstein-protest-mit-students-host-community-forum/ .

[236] Rebecca Aydin, All the Tech Moguls Who Have Been Linked to Jeffrey Epstein After He Became a Convicted Sex Offender, Business Insider, (Sep 9, 2019, 9:31 AM) https://www.businessinsider.com/tech-moguls-jeffrey-epstein-connected-bill-gates-elon-musk-2019-8

[237] Bill And Melinda Gates Divorce Reportedly Linked To Jeffrey Epstein Connection, TODAY, (May 10, 2021), https://www.youtube.com/watch?v=ghDSjoy5JyU

The MIT students' protests in light of the Jeffrey Epstein revelations were one of many examples of a concern with MIT's misogynistic culture, pointing to the university's continued employment of undergraduate professor Seth Lloyd, who visited pedophile and sex trafficker Jeffery Epstein in prison and continues to defend the relationship.[239]

The diverse concerns expressed by the student protests evidence the breadth of focus of the tech-worker movement, and a growing recognition that hostile tech cultures are reflected in the technology produced within such cultures.

**Law and Policy Responses to Harmful AI**

Interest in regulating AI systems is growing, with a focus on data protection, algorithmic accountability, and biometric/facial-recognition safeguards. Building on the emergence of globally oriented data protection approaches such as the European Union's General Data Protection Regulation (GDPR),[240] policymakers are moving quickly, driven both by the current sense of urgency to regulate the mass deployment of AI technologies lacking discernible safeguards and by the failure of ethical frameworks to adequately answer the call for accountability and justice.

**AI Regulatory Frameworks**

Initial attempts to develop new regulatory frameworks to promote public scrutiny have recently emerged. The two most high-profile recent examples of this trend are the proposed U.S. Algorithmic Accountability Act[241] and

---

the E.U.'s broad-reaching General Data Protection Regulation (GDPR).[242] Common to these frameworks is the demand that entities deploying AI-based judgments conduct an "Impact Assessment."[243]

Generally, an impact assessment can be defined as "the process of identifying the future consequences of current or proposed action."[244] A key advantage of impact assessments of AI-driven systems is their ability to influence entities' internal organizational conduct. By requiring an entity to conduct an internal inspection, impact assessments urge coders and designers to conduct a deeper form of analysis, carefully investigating plausible areas of bias error and uncertainty as well as implementing the necessary steps to correct them.[245] The internal and flexible nature of impact assessments shifts the regulated entity's focus away from mere compliance and towards problem solving and improvement.

---

[242] The objective of the EU's General Data Protection Regulation (GDPR) is to give individuals more control over their personal data, and it has come to be regarded as a global gold standard for privacy regulation. Commission Regulation 2016/679 The regulation also has several important provisions pertaining to automated decision-making. In particular, the GDPR states that as a rule there is a prohibition on fully automated individual decision-making, including profiling that has a legal, or similar, effect on th individual. Article 22, Commission Regulation 2016/679. (""The data subject shall have the right not to be subject to a decision based solely on automated processing, including profiling, which produces legal effects concerning him or her or similarly significantly affects him or her.") See also, Bryan Casey, Ashkon Farhangi & Roland Vogl, Rethinking Explainable Machines: The GDPR's 'Right to Explanation' Debate and the Rise of Algorithmic Audits in Enterprise, 34 BERKELEY TECH. L.J. 145, 180 (2019). ("[T]the GDPR's "right to explanation" is no mere remedial mechanism to be invoked by data subjects on an individual basis, but it implies a more general form of oversight with broad implications for the design, prototyping, field testing, and deployment of data processing systems."); Nick Wallace and Daniel Castro, "The Impact of the EU's New General Data Protection Regulation on AI" (Center for Data Innovation, March 2018), http://www2.datainnovation.org/2018-impact-gdpr-ai.pdf;
Nevertheless, there are several exceptions to this rule. See, Article 22(2). See also, THE EUROPEAN COMMISSION'S GUIDELINES ON AUTOMATED INDIVIDUAL DECISION-MAKING AND PROFILING FOR THE PURPOSES OF REGULATION 2016/679, Adopted on 3 October 2017, As last Revised and Adopted on 6 February 2018, https://ec.europa.eu/newsroom/article29/item-detail.cfm?item_id=612053. When one of those exceptions applies, the data controller must implement suitable measures with which to safeguard the individual's (data subject) rights, freedoms, and legitimate interests. These measures should include "at least the right to obtain human intervention on the part of the controller, to express his or her point of view and to contest the decision."See, Article 22(3). Recital 71 adds to this, stating that "In any case, such processing should be subject to suitable safeguards, which should include specific information to the data subject and the right to obtain human intervention, to express his or her point of view, to obtain an explanation of the decision reached after such assessment and to challenge the decision." See Recital 71 Commission Regulation 2016/679 So, the GDPR provides individuals with a "right to explanation." See e.g., Andrew D Selbst & Julia Powles, Meaningful information and the right to explanation, 7 INT'L DATA PRIVACY L. 233 (2017); Bryce Goodman and Seth Flaxman, EU Regulations on Algorithmic Decision-Making and a "Right to Explanation", 38 AI MAGAZINE 2017https://arxiv.org/abs/1606.08813; Francesca Rossi, Artificial Intelligence: Potential Benefits and Ethical Considerations, European Parliament: Policy Department C: Citizens' Rights and Constitutional Affairs (2016), BRIEFING PE 571.380,http://www.europarl.europa.eu/RegData/etudes/BRIE/2016/571380/IPOL_BRI(2016)571380_EN.pdf;INFORMATION COMMISSIONER'S OFFICE, 'OVERVIEW OF THE GENERAL DATA PROTECTION REGULATION (GDPR) (2016) https://ico.org.uk/for-organisations/guide-to-data-protection/guide-to-the-general-data-protection-regulation-gdpr/individual-rights/rights-relatedto-automated-decision-making-including-profiling/; EUROPEAN PARLIAMENT COMMITTEE ON LEGAL AFFAIRS, REPORT WITH RECOMMENDATIONS TO THE COMMISSION ON CIVIL LAW RULES ON ROBOTICS (2017) 2015/2103(INL) http://www.europarl.europa.eu/doceo/document/A-8-2017-0005_EN.html. But see, Aziz Z. Huq, A right to a Human Decision, 105 VIRG. L. REV. (forthcoming 2020); Sandra Wachter, Brent Mittelstadt and Luciano Floridi, Why a Right to Explanation of Automated Decision-Making Does Not Exist in the General Data Protection Regulation (December 28, 2016). Available at SSRN: https://ssrn.com/abstract=2903469 (claiming that the GDPR's right of access allows for a limited right to explanation of the functionality of The idea of a rigorous, standardized process in the form of an impact assessment as a tool to facilitate public accountability and oversight is not new.[246] For instance, many jurisdictions require an environmental impact assessment (EIA) to evaluate the effects of a proposed project and its alternatives on the environment.[247] EIAs are considered a powerful tool for assessing a project's environmental impact.[248] Consequently, several scholars and policymakers have suggested adopting the impact assessment model in other contexts.[249] The concept of impact assessments drew the attention of interest groups, scholars, and policymakers with regard to the use of AI in automated decision-making systems.[250] The following discussion introduces these two novel legislative initiatives, along with the central aspects of each initiative, laying the foundations for the consideration of impact assessments as a tool to provide oversight and accountability of AI use.

---

automated decision-making systems – what they refer to as the 'right to be informed'.") However, the right to explanation is limited to circumstances where a decision is based solely on automated processing.The European Commission's Guidelines on automated individual decisions stated that Article 22 applies only where there is "no human involvement in the decision process." See, THE EUROPEAN COMMISSION'S GUIDELINES ON AUTOMATED INDIVIDUAL DECISION-MAKING AND PROFILING FOR THE PURPOSES OF REGULATION 2016/679. An additional clause of the right to explanation provides data subjects with the right to receive notice of solely automated decision-making processes and to request access to meaningful information.See Articles 13-14, Recital 60 Commission Regulation 2016/679. See also, THE EUROPEAN COMMISSION'S GUIDELINES ON AUTOMATED INDIVIDUAL DECISION-MAKING AND PROFILING FOR THE PURPOSES OF REGULATION 2016/679. For further discussion of the potential effects of the GDPR on AI based decision-making systems, see Finale Doshi-Velez et al., Accountability of AI Under the Law: The Role of Explanation, BERKMAN CENTER RESEARCH PUBLICATION FORTHCOMING; HARVARD PUBLIC LAW WORKING PAPER NO. 18-07(November 3, 2017).. Available at SSRN: https://ssrn.com/abstract=3064761; However, even when the individual fails to invoke any of these rights, the GDPR will still establish and enforce accountability through an array of tools, including mandatory DPIAs. Margot E Kaminski,. and Gianclaudio Malgieri, , Algorithmic Impact Assessments under the GDPR: Producing Multi-layered Explanations, U OF COLORADO LAW LEGAL STUDIES RESEARCH PAPER NO. 19-28, 3,7 (September 18, 2019).. Available at SSRN: https://ssrn.com/abstract=3456224. (arguing that the DPIA is best understood as a nexus between the GDPR's two approaches to algorithmic accountability. Meaning, individual rights and collaborative governance.) The GDPR requires data controllers to carry out a DPIA on any type of processing that is likely to result in "high risk" to an individual's rights and freedoms prior to adoption.9 See, Art. 35 Commission Regulation 2016/679; DATA PROTECTION IMPACT ASSESSMENTS, ICO, https://ico.org.uk/for-organisations/guide-to-the-general-data-protectionregulation-gdpr/accountability-and-governance/data-protection-impact-assessments

[243] The notion of impact assessment have been promulgated in a variety of areas. DILLON REISMAN ET AL., AI NOW INST., ALGORITHMIC IMPACT ASSESSMENTS: A PRACTICAL FRAMEWORK FOR PUBLIC AGENCY ACCOUNTABILITY 5 (2018), https://ainowinstitute.org/aiareport2018.pdf; Andrew D. Selbst, Disparate Impact in Big Data Policing, 52 GA. L. REV. 109 (2017). See also, Organization for Economic Cooperation and Development (OECD), What Is Impact Assessment? (OECD), https://www.oecd.org/sti/inno/What-is-impact-assessmentOECDImpact.pdf.

[244] See, www.iaia.org.

[245] Nicholas Diakopoulos et al., Principles for Accountable Algorithms and a Social Impact Statement for Algorithms, FAT/ML, https://www.fatml.org/resources/principles-for-accountable-algorithms.

[246] See Sibout Nooteboom & Greet Teisman, Sustainable Development: Impact Assessment in the Age of Networking, 5 J. OF ENV. POL'Y AND PLANNING 285, 289 (2004). This is particularly true in the areas of human rights, environmental as well as privacy and data protection

[247] See e.g., Leonard Ortolano & Anne Shepherd, Environmental Impact Assessment: Challenges And Opportunities, 13 IMPACT ASSESSMENT 3 (1995) (published online: 6 February 2012) https://doi.org/10.1080/07349165.1995.9726076; Erika L. Preiss, The International Obligation to Conduct an Environmental Impact Assessment: The Icj Case Concerning the Gabcikovo-Nagymaros Project, 7 N.Y.U. ENVTL. L.J. 307 (1999); Matthew Cashmore, et al., The Interminable Issue of Effectiveness: Substantive Purposes, Outcomes and Research Challenges in the Advancement of Environmental Impact Assessment Theory, 22 IMPACT ASSESSMENT AND PROJECT APPRAISAL 295, 295-96 (2004); see also Jie Zhang, et al.,

Data-protection laws to confront and contain harmful behaviors by technology companies provides a natural foundation for approaches to new forms of algorithmic activity.[251] In particular, the right to access one's personal data,[252] to access information about automated decision-making,[253] requirements such as data protection impact assessments (DPIAs), and privacy by design align well with most AI accountability frameworks.[254] DPIAs are a bridge between "the two faces of the GDPR's approach to algorithmic accountability: individual rights and systemic collaborative governance."[255]

Impact assessments have some important advantages. They improve organizational behavior, promote information sharing, and incentivize private entities to consider the effect of their AI systems on individuals as well as on the public. Yet, they fall short of being a key oversight mechanism for all AI systems.[256] The way that current initiatives structure impact assessments fall short of facilitating sufficient accountability. In particular, impact assessments provide only limited transparency, insufficiently secure due process, and provide only limited room for public review. The initiatives do not require regulated entities to disclose any part of the self-assessment to the public, nor do they provide other means for the public to know that specific conduct took place.[257]

---

Critical Factors for EIA Implementation: Literature Review and Research Options, 114 J. OF ENVTL. MGMT. 148, 151 (2013); Douglas C. Baker & James N. McLelland, Evaluating the effectiveness of British Columbia's environmental assessment process for first nations' participation in mining development, 23 ENVTL.IMPACT ASSESSMENT REV. 581, 582-83 (2003); Matthew J. Rowe et. al., Accountability or Merely "Good Words"? An Analysis of Tribal Consultation Under the National Environmental Policy Act and the National Historic Preservation Act, 8 ARIZ. J. ENVTL. L. & POL'Y 1, 47 (2018).

[248] These assessments present information to the public and decision makers about potential negative environmental impacts . See Jameson Tweedie, Transboundary Environmental Impact Assessment Under the North American Free Trade Agreement, 63 WASH & LEE L. REV. 849, 860-1 (2006).

[249] See e.g., Michael Froomkin, Regulating Mass Surveillance as Privacy Pollution: Learning from Environmental Impact Statements, U. ILL. L. REV. 1713 (2015); Alessandro Mantelero, AI and Big Data: A blueprint for a human rights, social and ethical impact assessment, 34 COMP. L. & SEC. REV. 754 (2018).

[250] Nesta, 10 principles for public sector use of algorithmic decision making, (February 20, 2018) https://www.nesta.org.uk/blog/10-principles-for-public-sector-use-of-algorithmic-decision-making/

[251] See Frank Paquale, "The Second Wave of Algorithmic Accountability," Law and Political Economy, November 25, 2019, https://lpeblog.org/2019/11/25/the-second-wave-of-algorithmic-accountability/ .

[252] This provision in the GDPR was used by journalists to study profiling by dating and social media apps. See Judith Duportail, "I Asked Tinder for My Data. It Sent Me 800 Pages of My Deepest, Darkest Secrets," Guardian , September 26 2017, https://www.theguardian.com/technology/2017/sep/26/tinder-personal-data-dating-app-messages-hacked-sold .

[253] Andrew Selbst and Julia Powles, "Meaningful Information and the Right to Explanation," International Data Privacy Law 7, no. 4 (November 27, 2017): 233–242, https://papers.ssrn.com/sol3/papers.cfm?abstract_id=3039125 .

[254] See Lilian Edwards and Michael Veale, "Slave to the Algorithm? Why a 'Right to an Explanation' Is Probably Not the Remedy You Are Looking For" Duke Law & Technology Review 16, no.18, May 24, 2017, https://papers.ssrn.com/sol3/papers.cfm?abstract_id=2972855 ; and Margot Kaminski, "The GDPR's Version of Algorithmic Accountability," JOTWELL , August 16, 2018, https://cyber.jotwell.com/the-gdprs-version-of-algorithmic-accountability/ .

[255] Margot E. Kaminski and Gianclaudio Malgieri, "Algorithmic Impact Assessments under the GDPR: Producing Multi-layered Explanations," U of Colorado Law Legal Studies Research Paper No. 19-28, https://papers.ssrn.com/sol3/papers.cfm?abstract_id=3456224 .

[256] Nahmias, Yifat and Perel (Filmar), Maayan, The Oversight of Content Moderation by AI: Impact Assessments and Their Limitations (February 13, 2020). Harvard Journal on Legislation, Forthcoming, Available at SSRN: https://ssrn.com/abstract=3565025

[257] Id.

Further, under the US proposed Algorithmic Accountability Act, (the bill has not yet been enacted and faces an uncertain future)[258] individuals are not entitled to notice or the right to be heard.[259] Consequently, the strategy of impact assessment fails to facilitate sufficient public oversight and proper opportunities for correcting erroneous decisions. To improve their oversight potential, commentators thus recommend some improvements to the existing impact assessment schemes, including periodical impact assessments, mandatory notice-and-comment procedure, and mandatory publication.[260]

Present impact assessments models might not fit the oversight challenges raised by different forms of AI-based systems. Focusing on the case of AI-based online content moderation by online platforms, commentators argue that an oversight mechanism of self-assessment is insufficient to oversee private moderation of speech that directly and substantially affects shared public interests.[261]

The application of AI-based content-moderation systems by prominent online platforms directly affects people's ability to engage in certain forms of expression, communication, and sharing of thoughts and critical information. Consequently, it shapes our online public sphere and impacts the free flow of information.[262] Since platforms are private actors, at first glance, mechanisms of self-assessment seem to be the most suitable way to hold them accountable. Indeed, despite concerns that "the real threat to free speech today comes from private entities such as Internet service providers, not from the Government,"[263] interfering with the editorial discretion of platforms is seen as a violation of platforms' First Amendment rights under the United States Constitution.[264] As commercial speakers, platforms might be entitled to constitutional protection of free speech.[265] Impact assessments fit well within this deeply rooted scheme, because they are non-coercive and collaborative and therefore can be generally regarded as a form of self-regulation.[266] They do not force platforms to speak by demanding them to host content against their will,[267] but instead require them to be more transparent about their goals and evaluate the possible implications of their systems.

---

[258] H. Mark Lyon, Cassandra L. Gaedt-Sheckter, and Frances A. Waldmann, Gibson, Dunn "United States: Artificial Intelligence,"published in the Global Data Review Insight Handbook 2021 https://www.gibsondunn.com/wp-content/uploads/2021/01/Lyon-Gaedt-Sheckter-Waldmann-United-States-Artificial-Intelligence-GDR-Insight-Handbook-2021-12-2020.pdf

[259] Gibson, Dunn, 2019 Artificial Intelligence and Automated Systems Annual Legal Review, (Feb. 11, 2020)"The bill casts a wide net, such that many technology companies would find common practices to fall within the purview of the Act. The Act would not only regulate AI systems but also any "automated decision system," which is broadly defined as any "computational process, including one derived from machine learning, statistics, or other data processing or artificial intelligence techniques, that makes a decision or facilitates human decision making, that impacts consumers." For processes within the definition, companies would be required to audit for bias and discrimination and take corrective action to resolve these issues, when identified. The bill would allow regulators to take a closer look at any "[h]igh-risk automated decision system"—those that involve "privacy or security of personal information of consumers[,]" "sensitives aspects of [consumers'] lives, such as their work performance, economic situation, health, personal preferences, interests, behavior, location, or movements[,]" "a significant number of consumers regarding race [and several other sensitive topics]," or "systematically monitors a large, publicly accessible physical place[.]" For these "high-risk" topics, regulators would be permitted to conduct an "impact assessment" and examine a host of proprietary aspects relating to the system. Additional regulations will be needed to give these key terms meaning but, for now, the bill is a harbinger for AI regulation that identifies key areas of concern for lawmakers. Although the bill still faces an uncertain future, if it is enacted, businesses would face a number of challenges, not least significant uncertainty in defining and, ultimately, seeking to comply with the proposed requirements for implementing "high risk" AI systems and utilizing consumer data, as well as the challenges of sufficiently explaining to the FTC the operation of their AI systems. Moreover, the bill expressly states that it does not preempt state law—and states that have already been developing their own consumer privacy protection laws would likely object to any attempts at federal preemption—potentially creating a complex patchwork of federal and state rules. At a minimum, companies operating in this space should certainly anticipate further congressional action on this subject in the near future, and proactively consider how their own "high-risk" systems may raise concerns related to bias." https://www.gibsondunn.com/2019-artificial-intelligence-and-automated-systems-annual-legal-review/#_IVA_ALGORITHMIC_ACCOUNTABILITY

[260] Nahmias, Yifat and Perel (Filmar), Maayan, The Oversight of Content Moderation by AI: Impact Assessments and Their Limitations (February 13, 2020). Harvard Journal on Legislation, Forthcoming, Available at SSRN: https://ssrn.com/abstract=3565025 (" While the impact assessments enunciated under the Accountability Act and the GDPR are tailored to mitigate concerns about the ways general AI-based decision-making systems affect individuals, the most worrying consequences of poorly performed AI driven content-moderation concern our online public sphere. Although the removal of legitimate content affects the speaker's freedom of expression, it also affects the interest of the public in freely consuming and accessing information. Hence, the use of AI for content moderation can impose costs, not only upon the individual speaker, but especially upon society. Nonetheless, in contrast to evaluating the impact of AI-based systems on individuals (such as assessing the impact of an incorrect credit score), it is extremely difficult to evaluate the public impact of AI-based content moderation.")

[261] Id.

Even if platforms disclose how they minimize the spread of harmful content, they do not apply a common threshold of content legitimacy for all users. Instead, each individual views a different curated segment of the online content that meets a personal profile. AI-based content moderation systems create personally tailored, but fragmented "publics" of information.[268] As a result, it is difficult to detect illegitimate deprivations of information. If a user does not see a specific piece of information, it is not necessarily because this piece of content was removed, but possibly because it did not match her personal interests.[269] Thus it is extremely challenging to determine if a platform's AI-based system of content moderation complies with what is disclosed in its impact assessment.

**Public Oversight of AI Developments**

Openness, transparency, and disclosure are the keys for good governance.[270] These features are meant to ensure that decision-makers do not abuse their power, but rather exert it in a fair and effective manner for the benefit of the public. Indeed, "persons with public responsibilities should be answerable to 'the people' for the performance of their duties."[271] Such persons are expected to justify their choices to those affected by these choices and be held responsible for their failures and wrongdoings.[272] Numerous doctrines, procedures, laws, and regulations exist in order to hold government and public officials accountable for their decision-making processes.[273] In addition, freedom of information laws[274] and sunshine laws[275] assure that governmental decision-making processes are open to review, either by requiring governmental bodies to make their records available for public scrutiny,[276] or by giving the public access to observe agency meetings.[277]

---

[262] Kate Klonick, The New Governors: The People, Rules, and Processes Governing Online Speech, 131 HARV. L. REV. 1598, 1622-4 (2018).

[263] United States Telecom Association v. FCC ("USTA"), 855 F.3d 381, 433–34 (D.C. Cir. 2017) (per curiam) (Kavanaugh, J., dissenting).

[264] See Daphne Keller, Who Do You Sue? State and Platform Hybrid Power Over Online Speech, HOOVER INSTITUTION, AEGIS SERIES PAPER NO. 1902 (2019), https://www.hoover.org/sites/default/files/research/docs/who-do-you-sue-state-and-platform-hybrid-powerover-online-speech_0.pdf

[265] Jack Balkin, Free Speech in the Algorithmic Society: Big Data, Private Governance, and New School Speech Regulation,51 UC DAVIS L. REV. 1149 (2018).

[266] Michael Guihot et al., Nudging Robots: Innovative Solutions to Regulate Artificial Intelligence, 20 VAND. J. ENT. & TECH. L. 385, 427 (2017) and Sonia K. Katyal, Private Accountability in the Age of Artificial Intelligence, 66 UCLA L. REV. 54, 111–13 (2019).

[267] La'Tiejira v. Facebook, Inc., 272 F. Supp. 3d 981, 991–922 (S.D. Tex.2017); Zhang D.Y., Q, Li, H. Tong, J. Badilla, Y. Zhang, D. Wang, Crowdsourcing-based copyright infringement detection in live video streams, Proceedings of the 2018 IEEE/ACM International Conference on Advances in Social Networks Analysis and Mining (2018).; Search King, Inc. v. Google Tech., Inc., No. CIV-02-1457-M, 2003 WL 21464568 (W.D. Okla. 2003).

[268] ANAT BEN DAVID, DATA IN DOUBT: CONTEXTUALISING FACEBOOK PUBLICS IN THE AGE OF POLITICAL ASTROTURFING, INFORMATION COMMUNICATIONS & SOCIETY (FORTHCOMING 2019).

[269] Maayan Perel and Niva Elkin-Koren, Separation of Functions for AI: Restraining Speech Regulation by Online Platforms, Available at SSRN: https://ssrn.com/abstract=3439261.

[270] Erin Daly, Let The Sun Shine In: The First Amendment and the War on Terrorism, 21 DELAWARE LAWYER 14, 14 ( 2003).

[271] MICHAEL W. DOWDLE, PUBLIC ACCOUNTABILITY: CONCEPTUAL, HISTORICAL, AND EPISTEMIC MAPPINGS, IN PUBLIC ACCOUNTABILITY: DESIGN, DILEMMAS AND EXPERIENCES 1, 3 (Michael W. Dowdle ed., 2006).

Accountability can be enforced on private entities through legal rules and regulations, but also through informal means, such as market forces that check decision-makers' discretion and promote voluntary disclosure in relation to their choices and related outcomes. Using market forces, members of the public can penalize private entities for unacceptable behavior,[278] and sometimes force organizations to change their practices and alter their behavior.[279]

---

[272] Maayan Perel and Niva Elkin-Koren, Accountability in Algorithmic Copyright Enforcement, 19 STAN. TECH. L. REV. 473, 481 (2016).

[273] Jim Rossie, Participation Run Amok: The Costs of Mass Participation for Deliberative Agency Decisionmaking, 92 NW. U. L. REV, 173, 175 (1997). Kenneth A. Bamberger, Regulation As Delegation: Private Firms, Decisionmaking, and Accountability in the Administrative State, 56 DUKE L.J. 377, 399–400 (2006); Michele Estrin Gilman, Charitable Choice" and the Accountability Challenge: Reconciling the Need for Regulation with the First Amendment Religion Clauses, 55 VAND. L. REV. 799, 803 (2002); Bovens M, Analyzing and Assessing Public Accountability. A Conceptual Framework, EUROGOV EUROPEAN GOVERNANCE PAPERS, (C-06-01), 2006. https://www.ihs.ac.at/publications/lib/ep7.pdf at p. 7-8 (noting that in contemporary scholarly discourse the term accountability is often used to detonate various distinct concepts including transparency, equity, democracy, efficiency, responsiveness, responsibility, and integrity. However, arguing its narrower definition can be understood as the obligation to explain and justify conduct.)

[274] See e.g., The federal Freedom of Information Act ("FOIA"), 5 U.S.C. § 552 (2006), amended by OPEN Government Act of 2007, 5 U.S.C. § 552 (Supp. II 2008); District of Columbia Freedom of Information Act, D.C. Code §§ 2-531 through 2-540 (2006); Arizona Public Records Law, Ariz. Rev. Stat. §§ 39-121 to 39-126 (2009); Kentucky Open Records Act, Ky. Rev. Stat. Ann. §§ 61.870-61.884 (2009). For further discussion See, Justin Cox, Maximizing Information's Freedom: The Nuts, Bolts, and Levers of FOIA, 13 N.Y. CITY L. REV. 387, footnotes 117-121 (2010).

[275] See the Government in the Sunshine Act, 5 U.S.C. § 552b (1994)

[276] U.S. Dep't of Justice v. Reporters Comm. for Freedom of the Press, 489 U.S. 749, 773 (1989) (stating the one key aim of FOIA is informing citizens with regards to "what their government is up to").See 5 U.S.C. § 552 (2006). It is important to note that there are nine statutory exemptions (Id. § 553(b)(1)-(9) and three exclusions (Id. §§ 552(c)(1)-(3) from to the open records requirement. Moreover, FOIA " does not obligate agencies to create or retain documents; it only obligates them to provide access to those which it in fact has created and retained." See, Kissinger v. Reporters Comm. for Freedom of the Press, 445 U.S. 136, 152 (1980).

[277] The Government in the Sunshine Act ("Sunshine Act") governs the right of access to federal agency meetings. Congress passed the Sunshine Act motivated by the idea that citizens have a right to know how the government makes decisions that affect the public interest. The Sunshine Act allows you to attend meetings in which federal agency heads deliberate on agency business. It requires all federal agencies governed by "collegial bodies" to hold open meetings and to provide sufficient notice to allow the public to attend those meetings. The term "collegial bodies" refers to groups of two or more decision-makers that act jointly, such as boards of directors or multiple commissioners. Agencies covered by the Sunshine Act include powerful and important bodies such as the Securities and Exchange Commission (SEC), the Federal Communications Commission (FCC), and the Federal Trade Commission (FTC), as well as lesser known agencies, such as the Marine Mammal Commission, the Railroad Retirement Board, and the Advisory Board for Cuba Broadcasting. If a covered agency has improperly prevented you from attending a meeting, you may sue the agency in federal court. Among other things, you can obtain a court order prohibiting future violations of the Sunshine Act and a transcript of an improperly closed meeting. https://www.dmlp.org/legal-guide/access-federal-agency-meetings

[278] Andreas Schedler. Conceptualizing Accountability, in THE SELF-RESTRAINING STATE: POWER AND ACCOUNTABILITY IN NEW DEMOCRACIES 14-16 (eds. Andreas Schedler, Larry Diamond and Marc F. Plattner, 1999). Available at: http://works.bepress.com/andreas_schedler/22/

**The Challenges to Public Oversight in AI-based Governance**

Governance by artificial intelligence [280] challenges existing notions of accountability.[281] First, AI-systems operate behind closed doors and are therefore considered a "black box"[282] in the sense that the public has only limited access to, and very little understanding, of how they work in practice.[283] Most members of the public have no way of knowing how the AI decision-making process works, what the goals are that the system was designed to carry out, or how a specific recommendation or decision was derived.[284] Moreover, when faced with a black box, the public has little chance of pressuring private entities into modifying their behavior.[285]

Even when the system is not completely closed and the public is privy to some information, AI-based decision-making systems are highly complex and constantly changing.[286] Thus, making any attempt to review these decision-making processes and their results difficult.[287] To illustrate, consider the Cambridge Analytica scandal and Facebook CEO Mark Zuckerberg's subsequent testimony before a US congressional hearing.[288] While the congressional hearing clearly helped raise public awareness in regards to privacy, information misuse, and data security,[289] it also revealed the limited understanding of lawmakers in regard to the ways tech companies work.[290]

---

[279] Thomas N. Hale, Transparency, Accountability, and Global Governance, 14 GLOBAL GOVERNANCE 73, 77-85 (2008) (discussing market pressure and its limitations). See also, Orna Rabinovitch-Einy, Technology's Impact: The Quest for a New Paradigm for Accountability in Mediation, 11 HARV. NEGOT. L. REV. 253, 260-1 (2006).

[280] The term "Artificial Intelligence" is an umbrella term for a wide area of technologies, and generally it can be understood as denoting various technologies that employ some combination of algorithms, machine learning, feedback systems, and automation. See generally, Scherer, supra note 10, at359-62; Tasioulas, First Steps Towards an Ethics of Robots and Artificial Intelligence, 7 JPE 49, 51 (2019); Adam Thierer, Andrea Castillo O'Sullivan, and Raymond Russell, Artificial Intelligence and Public Policy, MERCATUS RESEARCH (2017) https://www.mercatus.org/publications/artificial-intelligencepublic-policy, at p. 5-6; International Electro technical Commission (2018). WHITE PAPER: ARTIFICIAL INTELLIGENCE ACROSS INDUSTRIES, Geneva: IEC, p. 16.

[281] Maayan Perel and Niva Elkin-Koren, Accountability in Algorithmic Copyright Enforcement, 19 STAN. TECH. L. REV. 473, 481 (2016)

[282] FRANK PASQUALE, THE BLACK BOX SOCIETY: THE SECRET ALGORITHMS THAT CONTROL MONEY AND INFORMATION 1-18 (2015), Nicholas Diakopoulos, Algorithmic Accountability Reporting: On The Investigation Of Black Boxes 3 (2013), http://www.nickdiakopoulos.com/wp-content/uploads/2011/07/Algorithmic-Accountability-Reporting_final.pdf.

[283] Transparency "is a particularly pronounced problem in the case of machine learning, as its value lies largely in finding patterns that go well beyond human intuition." Andrew D. Selbst and Solon Barocas, The Intuitive Appeal of Explainable Machines, 87 FORDHAM L REV 1085, 1129 (2018).

[284] Jenna Burrell, How the machine 'thinks': Understanding opacity in machine learning algorithms, 1 BIG DATA & SOCIETY 3, no. 1 (2016); https://www.europarl.europa.eu/RegData/etudes/STUD/2019/624262/EPRS_STU(2019)624262_EN.pdf.

[285] Roger Bickerstaff, Does your machine mind? Ethics and potential bias in the law of algorithms, DIGITALBUSINESS.LAW (June 19, 2017) https://digitalbusiness.law/2017/06/doesyour-machine-mind-ethics-and-potential-bias-in-the-law-of-algorithms/#page=1 ("Greater transparency of the principles, parameters and logic underpiining AI and algorithms in particular may lead to public review and scrutiny. This is likely to a lot more effective in putting pressure on digital players to conform with good principles."). https://www.europarl.europa.eu/RegData/etudes/STUD/2019/624262/EPRS_STU(2019)624262_EN.pdf at p. 6

[286] Berkman Center, What is Artificial Intelligence, THE BERKMAN KLEIN CENTER FOR INTERNET & SOCIETY RESEARCH PUBLICATION SERIES 10 https://cyber.harvard.edu/publication/2018/artificial-intelligence-human-rights.

[287] Jef Ausloos et al., Algorithmic Transparency and Accountability in Practice, Montreal(2018) https://uploads-ssl.webflow.com/5a2007a24a11ce000164d272/5ac883392c10d1baaa4358f2_Algorithmic_Transparency_and_Accountability_in_Practice_CameraReady.pdf; https://www.europarl.europa.eu/RegData/etudes/STUD/2019/624262/EPRS_STU(2019)624262_EN.pdf at p. 6.

[288] See e.g., Chloe Watson, The key moments from Mark Zuckerberg's testimony to Congress, THE GUARDIAN (April 11, 2018) https://www.theguardian.com/technology/2018/apr/11/mark-zuckerbergs-testimony-to-

In addition, AI-driven systems not only implement specific rules and policies—whether originating from a private entity or the legislature, but also constantly re-shape rules and policies in order to accommodate changes and new information. AI systems continuously improve their decision-making processes based on their accumulated information (e.g., via machine learning and deep learning),[291] thus rendering decision-making a continuous process.[292] The dynamic nature of AI-driven systems makes them unpredictable and difficult to monitor. In fact, even successful attempts to perform retrospective and independent oversight essentially providing only partial insights into how the system works.[293]

If the public cannot understand, or is unaware of a decision or an act taken by an actor deploying AI in its decision-making process, they are unable to identify unfair, discriminatory, and derogatory practices that may be the result of tainted training data or biased algorithms. The ability of the public to utilize market forces to penalize or otherwise affect private entities' behavior is hence limited. Furthermore, unless the public has access to significant monetary and/or legal means to dispute erroneous or unfair decisions and cause their correction, oversight cannot be meaningful.

---

congress-the-key-moments

[289] See, Jennifer Grygiel & Nina Brown, Are Social Media Companies Motivated to be Good Corporatecitizens? Examination of the Connection Between Corporate Social Responsibility and Social Media Safety, 43 TELECOMMUNICATIONS POL'Y 445, 450 (2019); Lee Raine, Americans' complicated feelings about social media in an era of privacy concerns, PEW RESEARCH CENTER, March 27, 2018 https://www.pewresearch.org/facttank/2018/03/27/americans-complicated-feelings-about-social-media-in-an-era-of-privacyconcerns/.

[290] Ian Sherr, Rep. Schiff says lawmakers need to learn technology to legislate, CNET (April 12, 2018) https://www.cnet.com/news/schiff-says-lawmakers-need-to-learn-technology-tolegislate-facebook-cambridge-analytica/; Shara Tibken, Questions to Mark Zuckerberg show many senators don't get Facebook, CNET (April 11, 2018) https://www.cnet.com/news/some-senators-in-congress-capitol-hill-just-dont-get-facebookand-mark-zuckerberg/; See Karen Kornbluh & Ellen P. Goodman, Bringing Truth to the Internet, 53 DEMOCRACY A JOURNAL OF IDEAS (2019) https://democracyjournal.org/magazine/53/bringing-truth-to-the-internet/.

[291] Harry Surden, Machine Learning and Law, 89 WASH. L. REV. 87, 90 (2014). (claiming that AI based systems "learn from experience and thus improve their performance over time."); Joshua A. Kroll, Joanna Huey, Solon Barocas, Edward W. Felten, Joel R. & Reidenberg, David G. Robinson, Harlan Yu, Accountable Algorithms, 165 U. PA. L. REV. 633, 680 (2017) ("A significant concern about automated decision-making is that it may simultaneously systematize and conceal discrimination. Because it can be difficult to predict the effects of a rule in advance (especially for large, complicated rules or rules that are machine-derived from data)."

[292] AI-based systems do not instinctively know whether a specific content is offensive or otherwise unwanted. For it to make such a distinction the system requires large amount of training data. Gradually, based on this training data the system learns to distinguish between suitable and offensive content. See, Facebook's Newsroom, Hard Questions: How We Counter Terrorism https://newsroom.fb.com/news/2017/06/how-we-counter-terrorism/

[293] Amanda Levendowski, How Copyright Law Can Fix Artificial Intelligence's Implicit Bias Problem, 93 WASH. L. REV. 579, 599 (2018). See also, Frank Pasquale, Restoring Transparency to Automated Authority, 9 J. ON TELECOMM. & HIGH TECH. L. 235, 246 (2011). Barb Darrow, How Hackers Broke into U.S. Voting Machines in Less Than 2 Hours, FORTUNE (July 31, 2017), http://fortune.com/2017/07/31/defcon-hackers-us-voting-machines/

As governments now move to regulate algorithmic systems, they are not doing so in a policy vacuum. More than 130 countries[294] have now passed comprehensive data protection laws, with Kenya[295] and Brazil[296] being the latest to have modeled their laws largely on the GDPR. While the US still lacks a general data protection law, momentum appears to be growing to address this gap, with a dramatic increase in activity at both the federal and state levels.[297]

However, there is still an ongoing debate about whether GDPR-style frameworks can or should offer a "right to explanation" about specific automated decisions. Some scholars argue that no such right presently exists in the GDPR,[298] while others argue that multiple provisions of the GDPR can be pieced together to obtain meaningful information about the logic involved in automated decisions.[299] It remains to be seen if this is an effective tool for accountability, as there continues to be a debate over the ways in which transparency[300] and other ways of "seeing through data protection laws" can work with the goals of algorithmic accountability frameworks.

These factors, in conjunction with the fact that automated decisionmaking systems may produce discriminatory or biased outcomes,[301] could undermine public trust and confidence in AI,[302] thereby threatening all of its potential benefits. Nevertheless, a carefully constructed accountability mechanism of public scrutiny should be able to mitigate these risks. The use of impact assessments as a means to achieve accountability is a growing trend in AI oversight policy, and, with improvements, their effectiveness can be increased. Subjecting AI based decision-making systems to public scrutiny remains an important goal and tool for fostering trust.[303]

---

[294] Graham Greenleaf, "Global Data Privacy Laws 2019: 132 National Laws & Many Bills," Privacy Laws & Business International Report 157 (May 29, 2019): 14–18, 2019, https://papers.ssrn.com/sol3/papers.cfm?abstract_id=3381593 .

[295] See Yomi Kazeem, "Kenya Is Stepping Up Its Citizens' Digital Security with a New EU-Inspired Data Protection Law," Quartz Africa , November 12, 2019; and Alice Munyua, "Kenya Considers Protection of Privacy and Personal Data," Mozilla Blog, January 2, 2019, https://blog.mozilla.org/netpolicy/2019/01/02/kenya-considers-protection-of-privacy-and-personal-data/ .

[296] Anna Carolina Cagnoni, "Brazilian Data Protection Law: A Complex Patchwork," IAPP Privacy Tracker, April 10, 2019, https://iapp.org/news/a/brazilian-data-protection-law-a-complex-patchwork/ .

[297] See "S.2577 - Data Broker Accountability and Transparency Act of 2019," https://www.congress.gov/bill/116th-congress/senate-bill/2577/text?q=%7B%22search%22%3A%5B%22data%22%5D%7D&r=20&s=7 ; "Following Equifax Settlement, Senators Markey, Blumenthal and Smith Reintroduce Legislation to Hold Data Broker Industry Accountable," Ed Markey, September 26, 2019, https://www.markey.senate.gov/news/press-releases/following-equifax-settlement-senators-markey-blumenthal-and-smith-reintroduce-legislation-to-hold-data-broker-industry-accountable ; "S.1951 – Designing Accounting Safeguards To Help Broaden Oversight and Regulations on Data," https://www.congress.gov/bill/116th-congress/senate-bill/1951/related-bills?q=%7B%22search%22%3A%5B%22%5C%22Designing+Accounting+Safeguards+to+Help+Broaden+Oversight+And+Regulations+on+Data%5C%22%22%5D%7D&r=1&s=9 ; "S.2658 - Augmenting Compatibility and Competition by Enabling Service Switching Act of 2019," https://www.congress.gov/bill/116th-congress/senate-bill/2658/text ; "H.R.2013 - Information Transparency & Personal Data Control Act," https://www.congress.gov/bill/116th-congress/house-bill/2013?q=%7B%22search%22%3A%5B%22data%22%5D%7D&s=7&r=3 ; "H.R.4978 - Online Privacy Act of 2019," https://www.congress.gov/bill/116th-congress/house-bill/4978/text?q=%7B%22search%22%3A%5B%22algorithm%22%5D%7D&r=29&s=1 ; and "California Consumer Privacy Act (CCPA)," https://oag.ca.gov/privacy/ccpa .

[298] Sandra Wachter, Brent Mittelstadt, and Luciano Floridi, "Why a Right to Explanation of Automated Decision-Making Does Not Exist in the General Data Protection Regulation," International Data Privacy Law, December 28, 2016, https://papers.ssrn.com/sol3/papers.cfm?abstract_id=2903469 .

[299] Andrew Selbst and Julia Powles, "Meaningful Information and the Right to Explanation," International Data Privacy Law 7, no. 4 (November 27, 2017): 233–242, https://papers.ssrn.com/sol3/papers.cfm?abstract_id=3039125 .

[300] See Mike Ananny and Kate Crawford, "Seeing without Knowing: Limitations of the Transparency Ideal and Its Application to Algorithmic Accountability," New Media & Society , December 13, 2016; Jenna Burrell, "How the Machine 'Thinks': Understanding Opacity in Machine Learning Algorithms," Big Data & Society , January 6, 2016; Christopher Kuner, Dan Jerker B. Svantesson, Fred H. Cate, Orla Lynskey, Christopher Millard, "Machine Learning with Personal Data: Is Data Protection Law Smart Enough to Meet the Challenge?," International Data Privacy Law 7, no. 1 (February 2017): 1–2, https://doi.org/10.1093/idpl/ipx003 .

**Biometric Recognition Regulation**

Numerous regulatory attempts have emerged to address the privacy, discrimination, and surveillance concerns associated with biometrics—the measurement of unique biological characteristics, including data used in facial and affect recognition. These regulatory attempts range from bans or moratoriums to laws that would allow the technology on a case-by-case basis with specific forms of oversight.

In Europe, the Swedish government fined a high school for its facial-recognition attendance registry as a violation of GDPR.[304] France's data protection authority, CNIL, declared it illegal to use facial recognition in schools based on privacy concerns.[305]

The Australian Parliament took a more aggressive approach, ordering a complete pause on the use of a national face database. The moratorium will not be lifted until legislation emerges that will allow the government to manage and build the system while acknowledging citizen digital rights and develop a proposal that prioritizes "privacy, transparency and . . . robust safeguards."[306]

American cities such as San Francisco, Oakland, Seattle, and Somerville similarly have voted to ban all forms of government use of the technology.[307]

In 2019, members of the United States Congress proposed several biometric bills, including the Commercial Facial Recognition Privacy Act of 2019,[308] the Facial Recognition Technology Warrant Act,[309] and the No Biometric Barriers to Housing Act of 2019.[310] The latter seeks to prohibit biometric recognition in public housing, highlighting many of the same concerns as the tenant organizing at Atlantic Plaza Towers in Brownsville,

---

[301] Danielle Keats Citron, Technological Due Process, 85 WASH. U. L. REV. 1249, 1271-72 (2008); Charles Lane, Will Using Artificial Intelligence to Make Loans Trade One Kind of Bias for Another?, NPR: MORNING EDITION (Mar. 31, 2017, 5:06 AM), http://www.npr.org/sections/alltechconsidered/2017/03/31/521946210/will-using-artificial-intelligence-to-make-loans-trade-onekindof-bias-for-anot; Jeff Larson et al., Breaking the Black Box: How Machines Learn to Be Racist, PROPUBLICA (Oct. 19, 2016), https://www.propublica.org/article/breaking-theblackbox-how-machines-learn-tobe-racist?word=blackness; Ted Greenwald, How AI Is Transforming the Workplace, WALL ST. J. (Mar. 10, 2017), https://www.wsj.com/articles/how-ai-is-transformingtheworkplace-1489371060; Julia Angwin et al., supra note 13; Cathy O'Neil, I'll Stop Calling Algorithms Racist when You Stop Anthropomorphizing AI, MATHBABE (Apr. 7, 2016), https://mathbabe.org/2016/04/07/ill-stopcalling-algorithmsracist-when-you-stop-anthropomorphizing-ai/

[302] RUSSELL T. VOUGHTM MEMORANDUM FOR THE HEADS OF EXECUTIVE DEPARTMENTS AND AGENCIES, GUIDANCE FOR REGULATION OF ARTIFICIAL INTELLIGENCE APPLICATIONS, January 2020 https://www.whitehouse.gov/wp-content/uploads/2020/01/Draft-OMBMemo-on-Regulation-of-AI-1-7-19.pdf.

[303] Janelle Berscheld & Francois Rower-Despres, Beyond Transparency: A Proposed Framework for Accountability in Decision-Making AI Systems, 5 AI MATTERS 13, 15 (2019).

[304] Sofia Edvardsen, "How to Interpret Sweden's First GDPR Fine on Facial Recognition in School," IAPP, August 27, 2019, https://iapp.org/news/a/how-to-interpret-swedens-first-gdpr-fine-on-facial-recognition-in-school/ .

[305] "CNIL Bans High Schools' Facial-Recognition Programs," IAPP, October 29, 2019, https://iapp.org/news/a/cnil-bans-high-school-facial-recognition-programs/ .

[306] Sarah Martin, "Committee Led by Coalition Rejects Facial Recognition Database in Surprise Move," Guardian , October 23, 2019, https://www.theguardian.com/australia-news/2019/oct/24/committee-led-by-coalition-rejects-facial-recognition-database-in-surprise-move .

[307] Rachel Metz, "Beyond San Francisco, More Cities Are Saying No to Facial Recognition," CNN Business, July 17, 2019, https://www.cnn.com/2019/07/17/tech/cities-ban-facial-recognition/index.html .

[308] S. 847 Commercial Facial Recognition Privacy Act of 2019, https://www.govinfo.gov/content/pkg/BILLS-116s847is/pdf/BILLS-116s847is.pdf .

[309] Alfred Ng, "Facial Recognition Surveillance Would Require Warrant under Bipartisan Bill," CNET, November 14, 2019, https://www.cnet.com/news/facial-recognition-surveillance-would-require-warrant-under-bipartisan-bill/ .

Brooklyn, where residents sought to keep their landlord from installing an invasive facial-recognition system in their rent-stabilized apartment complex.[311]

**Facial Recognition Software**

Biometric surveillance, or "facial recognition technology," has emerged as a lightning rod for public debate regarding the risk of improper algorithmic bias and data privacy concerns.[312] Until recently, there were few if any laws or guidelines governing the use of facial recognition technology. Amid widespread fears that the current state of the technology is not sufficiently accurate or reliable to avoid discrimination, regulators have seized the opportunity to act in the AI space—proposing and passing outright bans on the use of facial recognition technology with no margin for discretion or use case testing while a broader regulatory approach develops and the technology evolves. This tentative consensus stands in stark contrast to the generally permissive approach to the development of AI systems in the private sector to date. While much of the regulatory activity to date has been at the local level, momentum is also building for additional regulatory actions at both the state and federal levels.

Several states in the US: Washington,[313] Texas,[314] California,[315] Arkansas,[316] New York,[317] and Illinois,[318] have begun actively restricting and regulating in these areas, including limits on some forms of biometric collection and recognition. In addition, Washington,[319] Michigan,[320] California,[321] Massachusetts,[322] Arizona,[323] and Florida[324] have introduced efforts seeking to do the same.[325]

---

[310] See H.R. 4008, 116th Congress (House of Representatives), currently pending before the House Committee on Financial Services. If passed, the bill would prohibit facial recognition in public housing units that receive Department of Housing and Urban Development ("HUD") funding. It would also require HUD to submit a report on facial recognition and its impacts on public housing units and tenants.

[311] See Caroline Spivack, "New Bill Would Ban Facial Recognition Technology from Public Housing," Curbed, July 29, 2019, https://ny.curbed.com/2019/7/29/8934279/bill-ban-facial-recognition-public-housing-brooklyn-nyc ; and Elizabeth Kim, "Hell's Kitchen Landlord Sued for Keyless Entry System Agrees to Provide Keys," Gothamist ,

[312] https://www.gibsondunn.com/2019-artificial-intelligence-and-automated-systems-annual-legal-review/#_IVB_FACIAL_RECOGNITION May 8, 2019, https://gothamist.com/news/hells-kitchen-landlord-sued-for-keyless-entry-system-agrees-to-provide-keys .

[313] S. 5528, http://lawfilesext.leg.wa.gov/biennium/2019-20/Pdf/Bills/Senate%20Bills/5528.pdf .

[314] Texas 503 Business and Commerce Code, https://statutes.capitol.texas.gov/Docs/BC/htm/BC.503.htm .

[315] See A.B. 1281, "Privacy: Facial Recognition Technology: Disclosure," https://leginfo.legislature.ca.gov/faces/billTextClient.xhtml?bill_id=201920200AB1281 ; and A.B. 1215, "Law Enforcement: Facial Recognition and Other Biometric Surveillance," https://leginfo.legislature.ca.gov/faces/billTextClient.xhtml?bill_id=201920200AB1215 .

[316] Arkansas Act 1030, https://legiscan.com/AR/bill/HB1943/2019 .

[317] NY S.B. 1203, https://assembly.state.ny.us/leg/?default_fld=&leg_video=&bn=S01203&term=2019&Summary=Y&Text=Y ; and A.B. A6787B, https://www.nysenate.gov/legislation/bills/2019/a6787 .

[318] Keep Internet Devices Safe Act, http://www.ilga.gov/legislation/BillStatus.asp?DocNum=1719&GAID=15&DocTypeID=SB&SessionID=108&GA=101 ; Biometric Information Privacy Act (BIPA) from 2008, http://www.ilga.gov/legislation/ilcs/ilcs3.asp?ActID=3004&ChapterID=57 .

[319] Second Substitute Senate Bill 5376, http://lawfilesext.leg.wa.gov/biennium/2019-20/Pdf/Bills/Senate%20Bills/5376-S2.pdf .

[320] Michigan Senate Bill 342, https://legiscan.com/MI/bill/SB0342/2019 .

[321] AB-1215 Law Enforcement: Facial Recognition and Other Biometric Surveillance, https://leginfo.legislature.ca.gov/faces/billCompareClient.xhtml?bill_id=201920200AB1215 .

[322] Bill S. 1385: An Act Establishing a Moratorium on Face Recognition and Other Remote Biometric Surveillance Systems, https://malegislature.gov/Bills/191/S1385/Bills/Joint .

[323] Arizona House Bill 2478, https://legiscan.com/AZ/text/HB2478/id/1857901 .

[324] Florida House of Representatives, H.B. 1153, https://www.flsenate.gov/Session/Bill/2019/1153/BillText/Filed/PDF .

[325] AI Now Report 2019

Several proposals, such as the Florida Biometric Privacy Act, the California Consumer Privacy Act, Bill S. 1385 in Massachusetts, NY SB 1203 in New York, and HB1493 in Washington, are explicitly modeled after Biometric Information Privacy Act, a 2008 Illinois privacy act that serves as a gold standard.

In 2008, the Illinois legislature enacted the Biometric Information Privacy Act (BIPA).[326] BIPA defines a "biometric identifer" to include a "scan of hand or face geometry,"[327] and the law "imposes various obligations regarding the collection, retention, disclosure, and destruction of biometric identifers."[328] Some of these requirements include "establishing a retention schedule and guidelines for permanently destroying biometric identifiers" within three years of an individual's last interaction with the company,[329] and the statute also requires the company to notify the individual in writing and secure a written release before obtaining a biometric identifier.[330] The statute includes a private right of action. It provides that "[a]ny person aggrieved" by a violation of its provisions "shall have a right of action" against an "offending party."[331]

Since 2010, Facebook has employed the use of facial recognition software as part of a photo-tagging suggestion feature.[332] Facebook users living in Illinois brought a class-action suit against Facebook, alleging that Facebook's facial-recognition technology violates BIPA because the company collects, uses, and stores biometric identifiers without obtaining a written release and without a compliant retention schedule.[333] Facebook sought to dismiss the suit, arguing that the Facebook users have not suffered a concrete injury in fact under *Spokeo*.[334] The Supreme Court's 2016 decision in *Spokeo, Inc. v. Robins*[335] requires federal courts to investigate the "concreteness" of a plaintiff's injury.[336] The Ninth Circuit ruled against the company, holding that the users had suffered a concrete injury under the Ninth Circuit's interpretation of *Spokeo*.[337] The court laid out a two-step inquiry for concreteness questions: "We ask (1) whether the statutory provisions at issue were established to protect [the plaintiff's] concrete interests (as opposed to purely procedural rights), and if so, (2) whether the specific procedural violations alleged in this case actually harm, or present a material risk of harm, to such interests."[338]

---

[326] 740 Ill. Comp. Stat. 14/1 et seq.
[327] 740 Ill. Comp. Stat. 14/10.
[328] Rosenbach v. Six Fags Entm't Corp., ___ N.E.3d ___, 2019 IL 126186, at *6 (Ill. 2019).
[329] 740 Ill. Comp. Stat. 14/15(a).
[330] 740 Ill. Comp. Stat. 14/15(b).
[331] 740 Ill. Comp. Stat. 14/20.
[332] Patel v. Facebook, Inc., 932 F.3d 1264, 1268 (9th Cir. 2019).
[333] Patel, 932 F.3d at 1268.
[334] Patel, 932 F.3d at 1269–70.
[335] Spokeo, Inc. v. Robins, 578 U.S. ___, 136 S. Ct. 1540 (2016).
[336] The U.S. Supreme Court in Spokeo, Inc. v. Robins, a 6–2 decision by Justice Samuel Alito, recognized the significant authority of Congress to define intangible injuries such as a statutory right to information for the purposes of Article III standing to sue in federal courts. However, the Court emphasized that a plaintiff must prove a concrete injury and, therefore, a procedural statutory violation may not automatically establish standing.  The Court remanded the case back to the Ninth Circuit because the court of appeals had found that the plaintiff had suffered an individualized injury from defendant Spokeo, Inc.'s alleged misreporting of his personal information, but failed to address whether the alleged injury was sufficiently concrete to confer informational standing. While the Constitution does not explicitly mandate that plaintiffs have standing to file suit in federal courts, the Supreme Court has inferred limitations on justiciability from the Constitution's Article III restriction of judicial decisions to "Cases" and "Controversies" to ensure that the plaintiff has a genuine interest and stake in a case. The Court's three-part standing test requires a plaintiff to show that: (1) she has "suffered an injury-in-fact," which is (a) "concrete and particularized" and (b) "actual or imminent, not conjectural or hypothetical"; (2) "there must be a causal connection between the injury and the conduct complained of—the injury has to be fairly . . .trace[able] to the challenged action of the defendant, and not . . . th[e] result [of] the independent action of some third party not before the court"; and (3) "it must be likely, as opposed to merely speculative, that the injury will be redressed by a favorable decision." A plaintiff has the burden of establishing all three prongs of the standing test for each form of relief sought. A federal court must dismiss a case without deciding the merits if the plaintiff fails to meet the constitutional standing test.Mank, Bradford C., The Supreme Court Acknowledges Congress' Authority to Confer Informational Standing in Spokeo, Inc. v. Robins (July 21, 2017). Washington University Law Review, Vol. 94, p. 1387, 2017, U of Cincinnati Public Law Research Paper No. 16-12, Available at SSRN: https://ssrn.com/abstract=2833032
[337] See Patel, 932 F.3d at 1275.

On the first step, the court held that "the statutory provisions at issue in BIPA were established to protect an individual's concrete interests in privacy, not merely procedural rights."[339] To arrive at that conclusion, the court relied on the Supreme Court's recent Fourth Amendment cases: "In light of this historical background and the Supreme Court's views regarding enhanced technological intrusions on the right to privacy, we conclude that an invasion of an individual's biometric privacy rights 'has a close relationship to a harm that has traditionally been regarded as providing a basis for a lawsuit in English or American courts.'"[340] On the second step, the court concluded that Facebook's practice of "creat[ing] and us[ing] a face template and . . . retain[ing] this template for all time" constituted a violation of the plaintiffs' substantive privacy interests "[b]ecause the privacy right protected by BIPA is the right not to be subject to the collection and use of such biometric data."[341] Accordingly, the court held, "the plaintiffs have alleged a concrete injury-in-fact sufficient to confer Article III standing."[342]
BIPA represents a potent example of a statutory information use restriction. Facebook legally acquired the photos it uses for facial recognition purposes, but the statute seeks to restrain the company from that specific purpose without first satisfying its requirements: "The judgment of the Illinois General Assembly . . . supports the conclusion that the capture and use of a person's biometric information invades concrete interests."[343]

Despite the fact the legislature made the law privately enforceable, Facebook nonetheless argued that its violations of the statute did not implicate any concrete privacy rights. The US Supreme Court denied Facebook's cert. petition on January 21, 2020.[344] Facebook announced on January 29, 2020, that it had settled the case for $550 million. [345]

Key corporate developers of the technology—including Microsoft[346] and Amazon[347]—have come out in support of various forms of regulation on biometric identifier use but have generally resisted calls for bans or moratoriums. This strategy mirrors the historic approaches tech companies have taken to data protection and other regulatory frameworks that emphasize production pathways and compliance over regulatory approach, oversight, and intervention.[348]

Amazon attempted to block a shareholder vote on pausing the company's sale of facial-recognition technology until a third-party confirmation that "it does not cause or contribute to actual or potential violations of human rights." Even after the vote was allowed to go forward by the Securities and Exchange Commission (SEC),[349] Amazon aggressively campaigned for shareholders to vote against the ban.[350]

---

[338] Patel, 932 F.3d at 1270–71, citing Robins v. Spokeo, 867 F.3d 1108, 1113 (9th Cir. 2017).

[339] Patel, 932 F.3d at 1274, quoting Spokeo II, 867 F.3d at 1113.

[340] Patel, 932 F.3d at 1274, quoting Spokeo, 136 S. Ct. at 1549. Tis is an excellent example of how elastic the Court's command to consider historical practice can be.

[341] Patel, 932 F.3d at 1274.

[342] Patel, 932 F.3d at 1274. The Supreme Court of the United States has interpreted the Case or Controversy Clause of Article III of the United States Constitution (found in Art. III, Section 2, Clause 1)"at an irreducible minimum," the constitutional requisites under Article III for the existence of standing are that the plaintiff must personally have: 1) suffered some actual or threatened injury; 2) that injury can fairly be traced to the challenged action of the defendant; and 3) that the injury is likely to be redressed by a favorable decision. https://www.law.cornell.edu/constitution-conan/article-3/section-2/clause-1/constitutional-standards-injury-in-fact-causation-and-redressability

[343] Patel, 932 F.3d at 1273

[344] The Supreme Court denied Facebook's cert. petition on January 21, 2020. Order List, 589 U.S.___, 2020 WL 283288 (Jan. 21, 2020), https://www.supremecourt.gov/orders/courtorders/012120zor_7k47.pdf.

[345] Daniel Stoller, Facebook to Pay $550 Million in Biometric Privacy Accord, BLOOMBERG, Jan. 29, 2020, https://www.bloomberg.com/news/articles/2020-01-29/facebook-to-pay-550-million-to-setlebiometric-privacy-suit.

[346] Brad Smith, "Facial Recognition: It's Time for Action," Microsoft on the Issues , December 6, 2018, https://blogs.microsoft.com/on-the-issues/2018/12/06/facial-recognition-its-time-for-action/ .

[347] Michael Punke, "Some Thoughts on Facial Recognition Legislation," AWS Machine Learning Blog , February 7, 2019, https://aws.amazon.com/blogs/machine-learning/some-thoughts-on-facial-recognition-legislation/ .

[348] AI Now Report 2019

Body camera manufacturer Axon,[351] in contrast, has adopted an internal ban policy.[352] It remains to be seen what concrete impact these efforts will have. As the biometric-recognition industry moves full speed ahead with massive investments in production and deployment, governments are adopting the technology at a faster rate than they are regulating it. France has announced plans to establish a national facial-recognition database.[353]

In the UK, police in Cardiff and London both began trial use of facial-recognition technology, leading to legal challenges and objections by civil society groups, academics, and at least one department's ethics committee.[354] In 2019, news of China's use of biometric recognition as weapons of state power to target a Muslim minority[355] and Hong Kong protestors[356] made international headlines.

**Algorithmic Accountability and Algorithmic Impact Assessments**

As AI systems are increasingly shaping our environment, as well as our access to and exclusion from opportunities and resources, it is essential to ensure AI governance and oversight. Such oversight will help to maintain the rule of law, to protect individual rights, and to ensure the protection of core democratic values. Nevertheless, achieving AI oversight is challenging due to the dynamic and opaque nature of AI systems. In an attempt to increase oversight and accountability for AI systems, the US proposed the Algorithmic Accountability Act and introduced a mandatory impact assessment for private entities that deploy automated decision-making systems. Impact assessment as a means to enhance oversight was likewise recently adopted under the EU's General Data Protection Regulation.[357]

Algorithmic accountability bills and initiatives proliferated in 2019, especially in the United States. As noted, US lawmakers introduced the "Algorithmic Accountability Act," which would authorize the Federal Trade Commission (FTC) to assess whether corporate automated decision systems (ADS) products are biased, discriminatory, or pose a privacy risk to consumers. It also requires ADS vendors to submit impact assessments to the FTC for evaluation.[358] There have also been efforts within the U.S. to require additional disclosures surrounding AI systems. A number of bills imposing disclosure requirements on data driven systems have been proposed within the United States Congress, perhaps the most comprehensive of which is the Algorithmic Accountability Act of 2019.[359] While this bill, and its House counterpart, would use impact assessments to require significant disclosures on the logic used in an AI-based algorithm, along with the rationale behind any actions or decisions, the bill has been held up in committee and seems unlikely to proceed to a floor vote during the current Congress. AI and data-focused legislation has been more successful in the state legislatures, notably California, which passed both the California Consumer Privacy Act (CCPA),[360] and several other measures directed to more specific uses of automated systems.[361] All of these statutes require specific disclosures and notices with regard to automated processing.

---

[349] Steve Dent, "Amazon Shareholders Will Vote to Ban Facial Recognition Tech," Engadget, April 15, 2019, https://www.engadget.com/2019/04/15/amazon-shareholder-vote-facial-recognition/ .

[350] Zack Whittaker, "Amazon Defeated Shareholder's Vote on Facial Recognition by a Wide Margin," TechCrunch, May 28, 2019, https://techcrunch.com/2019/05/28/amazon-facial-recognition-vote/ .

[351] https://www.axon.com/

[352] Charlie Warzel, "A Major Police Body Cam Company Just Banned Facial Recognition," New York Times, June 27, 2019, https://www.nytimes.com/2019/06/27/opinion/police-cam-facial-recognition.html .

[353] Helene Fouquet, "France Set to Roll Out Nationwide Facial Recognition ID Program," Bloomberg, October 3, 2019, https://www.bloomberg.com/news/articles/2019-10-03/french-liberte-tested-by-nationwide-facial-recognition-id-plan .

[354] See, for example, Liberty, "Resist Facial Recognition," https://www.libertyhumanrights.org.uk/resist-facial-recognition ; Big Brother Watch, "Face Off," May 2019, https://bigbrotherwatch.org.uk/all-campaigns/face-off-campaign/ ; Pete Fussey and Daragh Murray, "Independent Report on the London Metropolitan Police Service's Trial of Live Facial Recognition Technology," Human Rights, Big Data and Technology Project, 2019, https://www.hrbdt.ac.uk/download/independent-report-on-the-london-metropolitan-police-services-trial-of-live-facial-recognition-technology/ ; Owen Bowcott, "Police Face Legal Action over Use of Facial Recognition Cameras," Guardian, June 14, 2018, https://www.theguardian.com/technology/2018/jun/14/police-face-legal-action-over-use-of-facial-recognition-cameras ; and Sarah Marsh, "Ethics Committee Raises Alarm over 'Predictive Policing' Tool," Guardian, April 20, 2019, https://www.theguardian.com/uk-news/2019/apr/20/predictive-policing-tool-could-entrench-bias-ethics-committee-warns .

[355] Paul Mozur, "One Month, 500,000 Face Scans: How China Is Using A.I. to Profile a Minority," New York Times, April 14, 2019, https://www.nytimes.com/2019/04/14/technology/china-surveillance-artificial-intelligence-racial-profiling.html .

[356] Paul Mozur, "In Hong Kong Protests, Faces Become Weapons," New York Times, July 26, 2019, https://www.nytimes.com/2019/07/26/technology/hong-kong-protests-facial-recognition-surveillance.html

[357] Nahmias, Yifat and Perel (Filmar), Maayan, The Oversight of Content Moderation by AI: Impact Assessments and Their Limitations (February 13, 2020). Harvard Journal on Legislation, Forthcoming, Available at SSRN: https://ssrn.com/abstract=3565025

Thus, in all of the above contexts—whether through the existing GDPR or CCPA, or as a result of future legislation from the EU or US focused on AI-specific contexts—compliance with the law is going to require developers to provide some form of disclosure as to the operation of the underlying algorithm, and (at least for high-risk applications that can deprive individuals of important rights and expectations) likely also as to the reasoning behind individual decisions or recommendations. It seems clear that demands for transparency imposed by laws in the future are simply going to increase, and thus developers of AI systems need to proactively start thinking about these problems now.[362]

As AI Now's 2018 report highlighted,[363] the use of algorithmic impact assessments (AIAs) has been gaining traction in both policy circles and various countries, states, and cities.[364] Built on the success of data-protection, environmental, human-rights, and privacy-impact assessments, AIAs require AI vendors and their customers to understand and assess the social implications of their technologies before they are used to impact people's lives. As we outline in our AIA framework,[365] these assessments would be made publicly available for comment by interested individuals and communities as well as researchers, policymakers, and advocates to ensure they are safe for deployment and that those who make and use them are acting responsibly.

---

[358] Gibson Dunn ,2019 Artificial Intelligence and Automated Systems Annual Legal Review (Feb. 11, 2020) https://www.gibsondunn.com/2019-artificial-intelligence-and-automated-systems-annual-legal-review/ #_IVA_ALGORITHMIC_ACCOUNTABILITY

[359] S.1108, 116th Congress, at https://www.congress.gov/bill/116th-congress/senatebill/1108/text.

[360] See, e.g., Joshua A. Jessen et al., California Consumer Privacy Act of 2018, available at https://www.gibsondunn.com/california-consumer-privacy-act-of-2018/, and Alex Southwell et al., California Consumer Privacy Act Update: Regulatory Update, available at https://www.gibsondunn.com/california-consumer-privacy-act-update-regulatory-update/

[361] See, e.g., SB-327 Security of Connected Devices (2018) (imposes obligations on providing increased cybersecurity for Internet of Things (IoT) devices), SB-1001 Bot Disclosures (2018) (requiring bots and virtual assistants to notify individuals that they are machines/software and not humans), and AB-1215 Law Enforcement Use of Facial Recognition and Other Biometric Surveillance (2019) (prohibits the use of facial recognition or other biometric surveillance technologies in police body cameras).

[362] H. Mark Lyon, Designing for Why: The Case for Increasing Transparency in AI Systems, n PLI Current: The Journal of PLI Press, Vol. 4, No. 3 (2020) https://www.gibsondunn.com/wp-content/uploads/2020/06/Lyon-Designing-for-Why-The-Case-for-Increasing-Transparency-in-AI-Systems-PLI-Current-The-Journal-of-PLI-Press-06-10-2020.pdf

[363] Meredith Whittaker, Kate Crawford, Roel Dobbe, Genevieve Fried, Elizabeth Kazunias, Varoon Mathur, Sarah Myers West, Rashida Richardson, Jason Schultz, and Oscar Schwartz, "AI Now Report 2018," AI Now Institute, https://ainowinstitute.org/AI_Now_2018_Report.pdf

[364] See also Andrew Selbst, "Accountable Algorithmic Futures: Building Empirical Research into the Future of the Algorithmic Accountability Act," Data & Society: Points , April 19, 2019, https://points.datasociety.net/building-empirical-research-into-the-future-of-algorithmic-accountability-actd230183bb826; Alessandro Mantelero, "AI and Big Data: A Blueprint for a Human Rights, Social and Ethical Impact Assessment," Computer Law & Security Review 34 no. 4 (August 2018): 754–772, https://www.sciencedirect.com/science/article/pii/S0267364918302012 ; and "AI Now Report 2018," https://ainowinstitute.org/AI_Now_2018_Report.pdf .

[365] Reisman et al., "Algorithmic Impact Assessments: A Practical Framework for Public Agency Accountability," https://ainowinstitute.org/aiareport2018.pdf .

For example, Canada's implementation of AIAs appears under its Directive on Automated Decision-Making, as part of the Pan-Canadian AI Strategy, [366] where the Department of Treasury embeds the tool into their government procurement process. Australia's AI Ethics Framework also contemplates the use of AIAs.[367] Washington State became the first state to propose AIAs for government ADS with its House and Senate bills HB 165[368] and SB 5527.[369]

In addition, some scholars have also advocated for a model AIA to complement DPIAs under the GDPR.[370] Another dimension to this year's algorithmic accountability legislation was algorithmic transparency. As law enforcement agencies increasingly turn to proprietary technology in criminal proceedings, the intellectual-property rights of private companies are being pitted against defendants' right to access information about that technology in order to challenge it in court.

In the United States, lawmakers responded to increasing public concern over the perceived dangers of unfettered AI development by proposing a number of high profile draft bills addressing the role of AI and how it should be governed, the real impact of which is yet to be felt across the private sector. U.S. state and local governments pressed forward with concrete legislative proposals regulating the use of AI. There is a growing international consensus that AI technology should be subject to certain technical standards and potentially even certification procedures in the same way other technical systems require certification before deployment. [371] There have also been numerous US federal legislative proposals for AI governance in the areas of Deepfake

---

[366] "Directive on Automated Decision-Making," Government of Canada, February 5, 2019, https://www.tbs-sct.gc.ca/pol/doc-eng.aspx?id=32592 .

[367] D. Dawson, E. Schleiger, et al., "Artificial Intelligence: Australia's Ethics Framework," Commonwealth Scientific and Industrial Research Organisation, April 2019, https://consult.industry.gov.au/strategic-policy/artificial-intelligence-ethics-framework/supporting_documents/ArtificialIntelligenceethicsframeworkdiscussionpaper.pdf .

[368] Washington House Bill 1655, https://legiscan.com/WA/bill/HB1655/2019 .

[369] Washington SB 5527 - 2019-20, https://app.leg.wa.gov/billsummary?BillNumber=5527&Year=2019 .

[370] TAP Staff, "How the GDPR Approaches Algorithmic Accountability," Technology | Academics | Policy, November 8, 2019, http://www.techpolicy.com/Blog/Featured-Blog-Post/How-the-GDPR-Approaches-Algorithmic-Accountability.aspx .

[371] Gibson, Dunn & Crutcher LLP, 2019 Artificial Intelligence and Automated Systems Annual Legal Review, (Feb. 11, 2020) https://www.gibsondunn.com/2019-artificial-intelligence-and-automated-systems-annual-legal-review/#_IVA_ALGORITHMIC_ACCOUNTABILITY; See also H. Mark Lyon, Cassandra L. Gaedt-Sheckter, and Frances A. Waldmann, "United States: Artificial Intelligence," published in the Global Data Review Insight Handbook 2021 https://www.gibsondunn.com/wp-content/uploads/2021/01/Lyon-Gaedt-Sheckter-Waldmann-United-States-Artificial-Intelligence-GDR-Insight-Handbook-2021-12-2020.pdf

Technology,[372] Autonomous Vehicles,[373] Law Enforcement,[374] Health Care,[375] Financial Services,[376] and Labor and Hiring.[377]

**Task Forces and Commissions on AI**

Predictive analytics and Automated Decision Systems ("ADS") present a number of risks and concerns, especially when used by government agencies to make critical determinations around who receives benefits, which school a child attends, and who is released from jail. Recognizing these risks, governments at all levels have begun working to address these concerns, and developing governance and accountability mechanisms.[378]

Of the current approaches, the most common has been the creation of commissions or task forces which include both external experts and government workers. These bodies are tasked with examining emerging technologies

---

[372]On July 9, 2019, Sen. Rob Portman (R-OH) introduced the "Deepfake Report Act" (S. 2065), which would require the Department of Homeland Security to submit five annual reports to Congress on the state of the "digital content forgery" technology and evaluate available methods of detecting and mitigating threats. The reports will include assessments of how the technology can be used to harm national security as well as potential counter measures. The bill defines digital content forgery as "the use of emerging technologies, including artificial intelligence and machine learning techniques, to fabricate or manipulate audio, visual, or text content with the intent to mislead." The bipartisan bill was passed in the Senate by unanimous consent on October 25 and is currently before the House Committee on Energy and Commerce, which is reviewing the same-named companion bill, H.R. 3600.
In the House, H.R. 3230 ("Defending Each and Every Person from False Appearances by Keeping Exploitation Subject to Accountability Act" or the "DEEPFAKES Act") was introduced by Rep. Clarke (D-NY-9) on June 12, 2019. It would require any "advanced technological false personation record" to be digitally watermarked. The watermark would be required to "clearly identifying such record as containing altered audio or visual elements." The bill has been referred to the Subcommittee on Crime, Terrorism, and Homeland Security.
On September 17, 2019, Rep. Anthony Gonzalez (R-OH) introduced the "Identifying Outputs of Generative Adversarial Networks Act" (H.R. 4355), which would direct both the National Science Foundation and NIST to support research on deepfakes to accelerate the development of technologies that could help improve their detection, to issue a joint report on research opportunities with the private sector, and to consider the feasibility of ongoing public and private sector engagement to develop voluntary standards for the outputs of GANs or comparable technologies.
[373]The U.S. House of Representatives passed the Safely Ensuring Lives Future Deployment and Research In Vehicle Evolution (SELF DRIVE) Act (H.R. 3388)[139] by voice vote in September 2017, but its companion bill (the American Vision for Safer Transportation through Advancement of Revolutionary Technologies (AV START) Act (S. 1885)), stalled in the Senate as a result of holds from Democratic senators who expressed concerns that the proposed legislation remains immature and underdeveloped in that it "indefinitely" preempts state and local safety regulations even in the absence of federal standards. Federal regulation of autonomous vehicles ("AVs") has so far faltered in the new Congress, as SELF DRIVE Act and the AV START Act have not been re-introduced since expiring with the close of the 115th Congress.
In the meantime, AVs continue to operate under a complex patchwork of state and local rules, with federal oversight limited to the U.S. Department of Transportation's ("DoT") informal guidance. In January 2020, the DoT published updated guidance for the regulation of the autonomous vehicle industry, "Ensuring American Leadership in Automated Vehicle Technologies" or "AV 4.0." The guidance builds on the AV 3.0 guidance released in October 2018, which introduced guiding principles for AV innovation for all surface transportation modes, and described the DoT's strategy to address existing barriers to potential safety benefits and progress.AV 4.0 includes 10 principles to protect consumers, promote markets and ensure a standardized federal approach to AVs.  In line with previous guidance, the report promises to address legitimate public concerns about safety, security, and privacy without hampering innovation, relying strongly on the industry self-regulating. However, the report also reiterates traditional disclosure and compliance standards that companies leveraging emerging technology should continue to follow. During 2019, several federal agencies announced proposed rule-making to facilitate the integration of autonomous vehicles onto public roads. In May 2019, in the wake of a petition filed by General Motors requesting temporary exemption from Federal Motor Vehicle Safety Standards (FMVSSs) which require manual controls or have requirements that are specific to a human driver, NHTSA announced that it was seeking comments about the possibility of removing 'regulatory barriers' relating to the introduction of automated vehicles in the United States.It is likely that regulatory changes to testing procedures (including preprogrammed execution, simulation, use of external controls, use of a surrogate vehicle with human controls and technical documentation) and modifications to current FMVSSs (such as crashworthiness, crash avoidance and indicator standards) will be finalized in 2021.
[374]Increasingly, algorithms are also being used at every stage of criminal proceedings, from gathering evidence to making sentencing and parole recommendations. H.R. 4368, the "Justice in Forensic Algorithms Act of 2019," was introduced in the House on September 17, 2019, would prohibit the use of trade secrets privileges to prevent defense access to the source

and publishing their findings, along with recommendations for how ADS systems should be held accountable.[379] Task forces and commissions convened on ADS to develop new strategies, policies, standards, or guidance that can inform future legislation or regulation.[380]

**New York City's Automated Decision Systems Task Force**

In May 2018, New York City created the Automated Decision Systems Task Force[381] (the "ADS Task Force") pursuant to Local Law 49 of 2018 to produce a report with recommendations for automated decision making systems. The ADS Task Force released its report in November 2019.[382] Not everyone was pleased with the report.[383] In the report, the ADS Task Force made high-level, principles-based recommendations that aligned with its three core themes of automated decision system management and that provide some guidance for operationalizing them. The ADS Task Force's recommendations are as follows:

---

code of proprietary algorithms used as evidence in criminal proceedings, and require that the Director of the National Institute of Standards and Technology establish a program to provide for the creation and maintenance of standards for the development and use of computational forensic software ("Computational Forensic Algorithm Standards") to protect due process rights. H.R. 4368, 116th Congress (U.S. House of Representatives). The standards would address underlying scientific principles and methods, an assessment of disparate impact on the basis of demographic features such as race or gender, requirements for testing and validating the software and for publicly available documentation, and requirements for reports that are provided to defendants by the prosecution documenting the use and results of computational forensic software in individual cases (e.g., source code). Press Release, Rep. Takano Introduces the Justice in Forensic Algorithms Act to Protect Defendants' Due Process Rights in the Criminal Justice System (Sept. 17, 2019), available at https://takano.house.gov/newsroom/press-releases/rep-takano-introduces-the-justice-in-forensic-algorithms-act-to-protect-defendants-due-process-rights-in-the-criminal-justice-system. Police departments often use predictive algorithms for various functions, such as to help identify suspects. While such technologies can be useful, there is increasing awareness building with regard to the risk of biases and inaccuracies. Karen Hao, AI Is Sending People To Jail – And Getting It Wrong, MIT Technology Review (Jan. 21, 2019), available at https://www.technologyreview.com/s/612775/algorithms-criminal-justice-ai/. See also Rod McCullom, Facial Recognition Technology is Both Biased and Understudied, UnDark (May 17, 2017), available at https://undark.org/article/facial-recognition-technology-biased-understudied/. Private groups, localities, states, and Congress have reacted to concerns fomented by AI applied to policing. In a paper released on February 13, 2019, researchers at the AI Now Institute, a research center that studies the social impact of artificial intelligence, found that police across the United States may be training crime-predicting AIs on falsified "dirty" data, calling into question the validity of predictive policing systems and other criminal risk-assessment tools that use training sets consisting of historical data. Meredith Whittaker, et al, AI Now Report 2018, AI Now Institute, 2.2.1 (December 2018), available at https://ainowinstitute.org/AI_Now_2018_Report; see also Rashida Richardson Schultz, Jason Schultz, and Kate Crawford, Dirty Data, Bad Predictions: How Civil Rights Violations Impact Police Data, Predictive Policing Systems, and Justice (Feb. 13, 2019). New York University Law Review Online, Forthcoming, available at SSRN: https://ssrn.com/abstract=3333423. In some cases, police departments had a culture of purposely manipulating or falsifying data under intense political pressure to bring down official crime rates. In New York, for example, in order to artificially deflate crime statistics, precinct commanders regularly asked victims at crime scenes not to file complaints. In predictive policing systems that rely on machine learning to forecast crime, those corrupted data points become legitimate predictors, creating "a type of tech-washing where people who use these systems assume that they are somehow more neutral or objective, but in actual fact they have ingrained a form of unconstitutionality or illegality." Id.

[375] Unsurprisingly, the use of AI in healthcare draws some of the most exciting prospects and deepest worry, given potential risks. For example, AI has been used in robot-assisted surgery in select fields for years, and studies have shown that AI-assisted procedures can result in far fewer complications. Brian Kalis, Matt Collier and Richard Fu, '10 Promising AI Applications in Health Care', Harvard Business Review (10 May 2018), available at https://hbr.org/2018/05/10-promising-ai-applications-in-health-care. Yet, The New York Times published an article in March 2019 warning of healthcare AI's potential failures, including small changes in vernacular leading to vastly disparate results (eg, 'alcohol abuse' leading to a different diagnosis than 'alcohol dependence'); see Cade Metz and Craig S Smith, 'Warning of a Dark Side to A.I. in Health Care', The New York Times (21 March 2019), available at nytimes.com/2019/03/21/science/health-medicine-artificial-intelligence.html. And these issues are backed by studies, including one released by Science – one of the highest acclaimed journals – just prior to the article, which discusses how 'vulnerabilities allow a small, carefully designed change in how inputs are presented to a system to completely alter its outputs, causing it to confidently arrive at manifestly wrong conclusions.' Samuel G Finlayson, et al, 'Adversarial attacks on medical machine learning', SCIENCE 363:6433, pp. 1287–1289 (22 March 2019) See https://science.sciencemag.org/content/363/6433/1287. As of yet, there are few regulations directed at AI in healthcare specifically, but regulators have recently acknowledged that existing frameworks for medical device approval are not well-suited to AI-related technologies. The US Food and Drug Administration ("FDA") has

**Management capacity**. (1) Develop and centralize resources within the city government that can guide policy and assist agencies in the development, implementation, and use of automated decision systems; (2) adopt a phased approach to developing and institutionalizing agency and citywide automated decision system management practices; and (3) strengthen the capacity of city agencies to develop and use automated decision systems.

**Public involvement.** (1) Facilitate public education about automated decision systems; and (2) engage the public in ongoing work around automated decision systems.

**Operations management.** (1) Establish a framework for agency reporting and publishing of information related to automated decision systems; (2) incorporate information about automated decision systems into processes for public inquiry about or challenge to city agency decisions; and (3) create an internal city process

---

proposed a specific review framework for AI-related medical devices, intended to encourage a pathway for innovative and life-changing AI technologies, while maintaining the FDA's patient safety standards. In April 2019, the FDA recently published a discussion paper – 'Proposed Regulatory Framework for Modifications to Artificial Intelligence/Machine Learning (AI/ML)-Based Software as a Medical Device (SaMD)'—offering that new framework for regulating health products using AI/machine learning ("AI/ML") software as a medical device ("SaMD"), and seeking comment. U.S. Food & Drug Administration, Proposed Regulatory Framework for Modifications to Artificial Intelligence/Machine Learning (AI/ML)-Based Software as a Medical Device (SaMD), at 2 (2 April 2019), available at https://www.fda.gov/media/122535/download. The paper introduces that one of the primary benefits of using AI in an SaMD product is the ability of the product to continuously update in light of an infinite feed of real-world data. But the current review system for medical devices requires a pre-market review, and pre-market review of any modifications, depending on the significance of the modification. The paper mentions that AI-based SaMDs have been approved by the FDA, but they are generally 'locked' algorithms, and any changes would be expected to go through pre-market review. This proposal attempts to anticipate continuously-adapting AI-based SaMD products. If AI-based SaMDs are intended to constantly adjust, the FDA posits that many of these modifications will require pre-market review – a potentially unsustainable framework in its current form. The paper instead proposes an initial pre-market review for AI-related SaMDs that anticipates the expected changes, describes the methodology, and requires manufacturers to provide certain transparency and monitoring, as well as updates to the FDA about the changes that in fact resulted in accordance with the information provided in the initial review. Additional discussion and guidance is expected following the FDA's review of the comments.

[376] On May 9, 2019, Representative Maxine Waters (D-CA) announced that the House Committee on Financial Services would launch two task forces focused on financial technology ("fintech") and AI: a task force on financial intelligence that will focus on the topics of regulating the fintech sector, and an AI task force that will focus on machine learning in financial services and regulation, emerging risks in algorithms and big data, combatting fraud and digital identification technologies, and the impact of automation on jobs in financial services. Katie Grzechnik Neill, Rep. Waters Announces Task Forces on Fintech and Artificial Intelligence (May 13, 2019), available at https://www.insidearm.com/news/00045030-rep-waters-announces-all-democrat-task-fo and See Scott Likens, How Artificial Intelligence Is Already Disrupting Financial Services, Barrons (May 16, 2019), available at https://www.barrons.com/articles/how-artificial-intelligence-is-already-disrupting-financial-services-51558008001. On September 24, 2019, H.R. 4476, the Financial Transparency Act of 2019, was reintroduced into Congress. H.R. 4476, 116th Congress (U.S. House of Representatives).The bipartisan bill, which calls for the Treasury secretary to create uniform, machine-readable data standards for information reported to financial regulatory agencies, has been referred to the Subcommittee on Commodity Exchanges, Energy, and Credit. Id. (including the Securities and Exchange Commission, Commodity Futures Trading Commission, Federal Deposit Insurance Corp., Federal Reserve, Office of the Comptroller of the Currency, the Consumer Financial Protection Bureau, the National Credit Union Association and the Federal Housing Finance Agency). By seeking to make information that is reported to financial regulatory agencies electronically searchable, the bill's supporters aim to "further enable the development of RegTech and Artificial Intelligence applications," "put the United States on a path towards building a comprehensive Standard Business Reporting program," and "harmonize and reduce the private sector's regulatory compliance burden, while enhancing transparency and accountability." Id.

[377] Amid the acceleration in the spread of AI and automated decision-making in the public and private sector, many U.S. and multinational companies have begun to use AI to streamline and introduce objectivity into their employment process. Robert Booth, Unilever saves on recruiters by using AI to assess job interviews, The Guardian (Oct. 25, 2019), available at https://www.theguardian.com/technology/2019/oct/25/unilever-saves-on-recruiters-by-using-ai-to-assess-job-interviews; Lloyd Chinn & Thomas Fiascone, AI In Hiring: Legislative Responses And Litigation Potential, Law360 (Nov. 25, 2019), available at https://www.law360.com/illinois/articles/1220318/ai-in-hiring-legislative-responses-and-litigation-potential. While AI presents an opportunity to eliminate bias from the hiring process, it has also been seen to introduce bias because of

for assessing specific automated decision systems for any risk of disproportionate impact to any individual or group on the basis of protected characteristics.

**Vermont Artificial Intelligence Task Force**

In 2018, Vermont launched the Vermont Artificial Intelligence Task Force (the "AI Task Force") charged with assessing the development and use of AI. The AI Task Force released its report in January 2020, with the following recommendations:
Establish a permanent commission on AI to propose policy initiatives and support its responsible development. Adopt a code of ethics to set standards for responsible AI. Create incentives for the further development of the AI industry in Vermont. Support the responsible use of AI by agencies of state and local government. Enhance education and workforce development programs targeted to AI, with the recommended involvement of Vermont's higher education community for workforce training in the development and use of AI. Expand education of the public on the power and opportunity of AI and the risks created by it so Vermont has an informed citizenry on these issues. The report establishes the following baseline definition for artificial intelligence:

*"Artificial intelligence (A.I.) systems are systems (usually software) capable of perceiving an environment through data acquisition and then processing and interpreting the derived information to take action(s) or imitate intelligent behavior given a specified goal. AI systems can also learn/adapt their behavior by analyzing how the environment is affected by prior actions.*
*"As a scientific discipline, AI includes several approaches and techniques, such as machine learning (of which deep learning and reinforcement learning are specific examples), machine reasoning (which includes planning, scheduling, knowledge representation and reasoning, search, and optimization), and robotics (which includes control, perception, sensors and actuators, as well as the integration of all other techniques into cyber-physical systems)."*

This definition is adapted from the European Union's glossary definition. The European Commission is considering the definition of AI in the context of the white paper mentioned above.[384]

---

inadequate data underlying and powering its algorithms. Legislators are taking action to recognize the potentially vast implications of AI technology on employment and employees' rights. As a result, 2019 saw tentative legislation at federal and state level take on an increased focus upon AI in employment and hiring. On 28 January 2019, the proposed AI JOBS Act of 2019 was introduced and, if enacted, would authorize the Department of Labor to work with businesses and education institutions in creating a report that analyses the future of AI and its impact on the American labor landscape. H.R. 827 – AI JOBS Act of 2019, 116th Cong (2019), available at
https://www.congress.gov/bill/116thcongress/house-bill/827/text. Similar to H.R. 153, this bill indicates federal recognition of the threat the introduction of AI technology poses; however, there is no indication as to what actions the federal government might take in order to offer labor protection, and the bill has not progressed to date.
On September 11, 2019, Sen. Brown (D-OH) introduced S. 2468, the "Workers' Right to Training Act," which would require employers to provide notice and training to employees whose jobs are in danger of being changed or replaced due to technology, and for other purposes. S. 2468, 116th Congress (U.S. Senate). "Technology" is defined in the bill as including "automation, artificial intelligence, robotics, personal computing, information technology, and e-commerce." Id.

[378] AI Now Report 2019.
[379] AI Now Report 2019.
[380] Lee Tiedrich, Nooree Lee & T'Shae Sherman, AI Update: U.S. State and Local Government Task Forces Continue to Examine Artificial Intelligence Trustworthiness, COVINGTON, (May 15, 2020) https://www.insidetechmedia.com/2020/05/15/ai-update-u-s-state-and-local-government-task-forces-continue-to-examine-artificial-intelligence-trustworthiness/
[381] https://www1.nyc.gov/site/adstaskforce/index.page
[382] https://www1.nyc.gov/assets/adstaskforce/downloads/pdf/ADS-Report-11192019.pdf
[383] Colin Lecher, NYC's algorithm task force was 'a waste,' member says, (Nov. 20, 2019) https://www.theverge.com/2019/11/20/20974379/nyc-algorithm-task-force-report-de-blasio some members questioned the city's commitment to transparency. Members of the task force said they had to push to be given examples of automated tools used by the city, among other issues. Id.
[384] Lee Tiedrich, Nooree Lee & T'Shae Sherman, AI Update: U.S. State and Local Government Task Forces Continue to Examine Artificial Intelligence Trustworthiness, COVINGTON, (May 15, 2020)

The AI Task Force did not recommend the promulgation of any new regulations of AI at this time. However, the other recommendations in the report suggest that AI will continue to garner attention in Vermont, both in terms of regulation and government use and procurement.

Vermont currently prohibits the use of biometric identifiers in certain agency processes, including identifying applicants for non-commercial driver licenses. Additionally, in March 2020, Vermont enacted legislation that expanded the categories of personally identifiable information that may trigger notification obligations to individuals and regulators in the event of a breach to include biometric and genetic data.[385]

In May 2020, California appointed its first director of the newly established California Office of Digital Innovation, which will focus on developing applications for use by members of the public when interacting with the state. The California Office of Enterprise Technology Solutions will continue to focus on implementing technology solutions across state governments. The activities of these two state offices potentially may be relevant to AI, in addition to the activities undertaken by the California state legislature.[386]

In May 2019, Alabama established the Commission on Artificial Intelligence and Associated Technologies to review "all aspects of the growth of artificial intelligence and associated technology in the state and the use of artificial intelligence in governance, healthcare, education, environment, transportation, and industries of the future…"[387] Unlike the New York City and Vermont initiatives which are focused primarily on the use of AI in government systems, the Alabama Commission will evaluate AI issues broadly to look beyond government to industries such as "autonomous cars, industrial robots, [and] algorithms for disease diagnosis."[388]

On August 30, 2018, the California legislature endorsed the Asilomar AI Principles[389]. Like the Alabama Commission, the California legislature was not focused specifically on use of AI by the government but rather more broadly on AI development in general.

These preliminary efforts by state and local governments evidence the substantial interest that public procurements of AI and AI development in general have generated at all levels of government. Broadly speaking, public procurement regimes at the federal, state, and local levels follow similar guiding principles, but it is not uncommon for state and local governments to depart from their federal counterparts on specific issues,

---

https://www.insidetechmedia.com/2020/05/15/ai-update-u-s-state-and-local-government-task-forces-continue-to-examine-artificial-intelligence-trustworthiness/

[385] Lee Tiedrich, Nooree Lee & T'Shae Sherman, AI Update: U.S. State and Local Government Task Forces Continue to Examine Artificial Intelligence Trustworthiness, COVINGTON, (May 15, 2020) https://www.insidetechmedia.com/2020/05/15/ai-update-u-s-state-and-local-government-task-forces-continue-to-examine-artificial-intelligence-trustworthiness/

[386] Lee Tiedrich, Nooree Lee & T'Shae Sherman, AI Update: U.S. State and Local Government Task Forces Continue to Examine Artificial Intelligence Trustworthiness, COVINGTON, (May 15, 2020) https://www.insidetechmedia.com/2020/05/15/ai-update-u-s-state-and-local-government-task-forces-continue-to-examine-artificial-intelligence-trustworthiness/

[387] https://legiscan.com/AL/text/SJR71/id/2024111

[388] Lee Tiedrich & Nooree Lee, AI Update: New York City, Vermont, and Other State and Local Governments Evaluating AI Trustworthiness, COVINGTON, July 16, 2019, https://www.insidetechmedia.com/2019/07/16/ai-update-new-york-city-vermont-and-other-state-and-local-governments-evaluating-ai-trustworthiness/

[389] The Asilomar AI Principles are a set of 23 principles intended to promote the safe and beneficial development of artificial intelligence. The principles – which include research issues, ethics and values, and longer-term issues – emerged from a collaboration between AI researchers, economists, legal scholars, ethicists, and philosophers in Asilomar, California in January of 2017. The Principles are the most widely adopted effort of their kind. They have been endorsed by AI research leaders at Google DeepMind, GoogleBrain, Facebook, Apple, and OpenAI. Signatories include Demis Hassabis, Yoshua Bengio, Elon Musk, Ray Kurzweil, the late Stephen Hawking, Tasha McCauley, Joseph Gordon-Levitt, Jeff Dean, Tom Gruber, Anthony Romero, Stuart Russell, and more than 3,800 other AI researchers and experts. With ACR 215 passing the State Senate with unanimous support, the California Legislature has now been added to that list. https://futureoflife.org/2018/08/31/state-of-california-endorses-asilomar-ai-principles/?cn-reloaded=1

as evidenced by several state governments (such as Vermont and Oregon) mandating net neutrality[390] from their internet service providers. Consequently, the actions taken by state and local governments in the area of AI and automated decision systems bear monitoring.

**Litigation Against Harmful AI**

The so-called "FAANG" big-tech firms—Facebook, Apple, Amazon, Netflix and Google—have received considerable scrutiny from competition law enforcers on both sides of the Atlantic.[391] Google, has been the subject of very large fines in EU competition law proceedings[392] and now is defending an antitrust lawsuit in the US.

A group of 36 states and Washington, D.C., sued Google on July 7, 2021 in an antitrust case challenging the company's control over its Android app store — opening a new front in regulators' attempts to rein in the search giant. The suit, filed in California federal court and led by Utah, North Carolina, Tennessee, New York, Arizona, Colorado, Iowa and Nebraska, is the latest in a series of major antitrust cases filed against the tech industry's biggest forces, after years of brewing unhappiness with the growing wealth and power of Silicon Valley. In addition to Wednesday's suit, Google also faces a suit that the Justice Department and 14 states filed in October, focused on Google's efforts to dominate the mobile search market; one from 38 states and territories filed in December, also focused on search; and a third suit by 15 states and territories related to Google's power over the advertising technology.[393]

On October 6, 2020, the Subcommittee on Antitrust, Commercial and Administrative Law of the House Committee on the Judiciary issued a 450-page report on the four dominant online platforms, Amazon, Apple, Facebook, and Google. [394]

---

[390] NET NEUTRALITY IS the idea that internet service providers like Comcast and Verizon should treat all content flowing through their cables and cell towers equally. That means they shouldn't be able to slide some data into "fast lanes" while blocking or otherwise discriminating against other material. In other words, these companies shouldn't be able to block you from accessing a service like Skype, or slow down Netflix or Hulu, in order to encourage you to keep your cable package or buy a different video-streaming service. KLINT FINLEY, The WIRED Guide to Net Neutrality, WIRED, 5/5/2020 https://www.wired.com/story/guide-net-neutrality/

[391] Brennan, Tim, Constructing a Conventional Antitrust Case Against Google (March 11, 2020). Available at SSRN: https://ssrn.com/abstract=3552794 or http://dx.doi.org/10.2139/ssrn.3552794

[392] European Commission, Antitrust: Commission fines Google €2.42 billion for abusing dominance as search engine by giving illegal advantage to own comparison shopping service, press release, June 27, 2017; European Commission, Antitrust: Commission fines Google €4.34 billion for illegal practices regarding Android mobile devices to strengthen dominance of Google's search engine, press release, July 18, 2018; European Commission, Antitrust: Commission fines Google €1.49 billion for abusive practices in online advertising, press release, March 20, 2019; U.S. Federal Trade Commission, FTC to Examine Past Acquisitions by Large Technology Companies, press release, February 11, 2020.

[393] Leah Nylen, 36 states, D.C. sue Google for alleged antitrust violations in its Android app store, POLITICO, (July 7, 2021) https://www.politico.com/news/2021/07/07/36-states-dc-sue-google-for-alleged-antitrust-violations-in-its-android-app-store-498622

[394] The report provides numerous important details that suggest or indicate antitrust activity by the four major technology platforms. Google section at pages 174-247. The Subcommittee concluded: "To put it simply, companies that were once scrappy, underdog start-ups that challenged the status quo have become the kinds of monopolies we last saw in the era of oil barons and railroad tycoons." The report went on to state: "Nearly a century ago, Supreme Court Justice Louis Brandeis wrote, 'We must make our choice. We may have Democracy, or we may have wealth concentrated in the hands of a few, but we cannot have both.' Those words speak to us with great urgency today." Subcommittee on Antitrust, Commercial and Administrative Law of the Committee on the Judiciary. 2020. Majority Staff Report and Recommendations. Investigation of Competition in Digital Markets.
https://judiciary.house.gov/uploadedfiles/competition_in_digital_markets.pdf

In October 2020, the U.S. Department of Justice (DOJ) filed an antitrust suit against Google. It is an alleged monopolization case[395]. The DOJ's suit was joined by 11 states. More states subsequently filed two separate antitrust lawsuits against Google in December 2020. [396] The case is about exclusivity and exclusion in the distribution of search engine services. It alleges that Google paid substantial sums to Apple and to the manufacturers of Android-based mobile phones and tablets and also to wireless carriers and to web browser proprietors – in essence, to distributors – to install the Google search engine as the exclusive pre-set (installed), default search program. The suit alleges that Google thereby made it more difficult for other search engine providers (e.g., Bing; DuckDuckGo) to obtain distribution for their search engine services and thus to attract search-engine users and to sell the online advertising that is associated with search-engine use and that provides the revenue to support the search "platform" in this "two sided market" context.[397]

Exclusion can be seen as a form of "raising rivals' costs".[398] Equivalently, exclusion can be seen as a form of non-price predation. Under either interpretation, the exclusionary action impedes competition. It is important to note that these allegations are different from those Googled settled in 2013[399] with the Federal Trade Commission.[400] That case focused on alleged self-preferencing: that Google was unduly favoring its own products and services (e.g., travel services) in its delivery of search results to users of its search engine. In those cases, the alleged impairment of competition occurred with respect to those competing products and services – not with respect to the search itself.

Google's internet search engine is used for a remarkable 1 billion health related searches per day, constituting 7% of its total search traffic. Search results for health information are typically accompanied by advertisements that appear at the top of the page.

---

[395] The vehicle for the enforcement action is Section 2 of the Sherman Act, which proscribes certain forms of unilateral conduct (i.e., the "bad" or "exclusionary" conduct element) that have a causal connection to the acquisition or maintenance of monopoly power.15 U.S.C. § 2. A Section 2 plaintiff must show that the challenged conduct had some causal connection to the acquisition or maintenance of market power, that is, that the conduct resulted in a less-competitive future. US antitrust law prohibits certain forms of conduct (the "bad" conduct element) that, at some degree of certainty, decreases social welfare, however measured. Joseph Farrell & Michael L. Katz, The Economics of Welfare Standards in Antitrust, COMPETITION POLICY INT'L, Autumn 2006, at 1.

[396] United States of America v. Google LLC. In the United States District Court for the District of Columbia. Case 1:cv-20-03010. Filed 10-20-20.

[397] There is also a related argument: That Google thereby gained greater volume, which allowed it to learn more about its search users and their behavior, and which thereby allowed it to provide better answers to users (and thus a higher-quality offering to its users) and better-targeted (higher-value) advertising to its advertisers. Conversely, Google's search-engine rivals were deprived of that volume, with the mirror-image negative consequences for the rivals. This is just another version of the standard "learning-by-doing" and the related "learning curve" (or "experience curve") concepts that have been well understood in economics for decades. White, Lawrence J., U.S. v. Google: A Tough Slog; But Maybe an Intriguing Possibility? (January 29, 2021). NYU Stern School of Business Forthcoming, Available at SSRN: https://ssrn.com/abstract=3775954 or http://dx.doi.org/10.2139/ssrn.3775954

[398] See, for example, Steven C. Salop and David T. Scheffman, "Raising Rivals' Costs: Recent Advances in the Theory of Industrial Structure," American Economic Review, Vol. 73, No. 2 (May 1983), pp. 267-271; and Thomas G. Krattenmaker and Steven C. Salop, "Anticompetitive Exclusion: Raising Rivals' Costs To Achieve Power Over Price," Yale Law Journal, Vol. 96, No. 2 (December 1986), pp. 209-293.

[399] FTC Press Release, Landmark Agreements Will Give Competitors Access to Standard-Essential Patents;Advertisers Will Get More Flexibility to Use Rival Search Engines, (Jan. 13, 2013) (Google Inc. has agreed to change some of its business practices to resolve Federal Trade Commission concerns that those practices could stifle competition in the markets for popular devices such as smart phones, tablets and gaming consoles, as well as the market for online search advertising. https://www.ftc.gov/news-events/press-releases/2013/01/google-agrees-change-its-business-practices-resolve-ftc

[400] Richard J. Gilbert, "The U.S. Federal Trade Commission Investigation of Google Search," in John E. Kwoka, Jr., and Lawrence J. White, eds. The Antitrust Revolution: Economics, Competition, and Policy, 7th edn. Oxford University Press, 2019, pp. 489-513.

Gregory Curfman, MD, Deputy Editor of JAMA[401] presents arguments, in support of the DOJ antitrust suit, that Google maintains monopolies on online search for health information and advertising for health products. A single, dominant provider of online health information harms consumer welfare, since it discourages innovation in internet search and may result in a biased spectrum of health information.[402]

As many critics of the DOJ's case have pointed out, it is easy for users to switch their default search engine. If internet search were a normal good or service, this ease of switching would leave little room for the exercise of market power. But if this is so: Why is Google willing to pay $8-$12 billion annually for the exclusive default setting on Apple devices and large sums to the manufacturers of Android-based devices (and to wireless carriers and browser proprietors)? Why doesn't Google instead run ads in prominent places that remind users how superior Google's search results are and how easy it is for users (if they haven't already done so) to switch to the Google search engine and make Google the user's default choice?

But suppose that user inertia is important; and suppose that users generally have difficulty in making comparisons with respect to the quality of delivered search results. If this is true, then being the default search engine on Apple and Android-based devices and on other distribution vehicles would be valuable. In this context, the inertia of their customers is a valuable "asset" of the distributors that the distributors may not be able to take advantage of –but that Google can (by providing search services and selling advertising). If Google's taking advantage of this user inertia means that Google exercises market power, a challenge of the case will be defining the relevant market. How to delineate the relevant market will be central to the case.

Even if the DOJ is successful in showing that Google violated Section 2 of the Sherman Act in monopolizing search and/or search-linked advertising, an effective remedy seems problematic. But there also remains the intriguing question: Why was Google willing to pay such large sums for those exclusive default installation rights?[403]

Former President Donald Trump sued Facebook, Twitter and Google's YouTube over their suspensions of his accounts after a mob of his supporters attacked the U.S. Capitol in January 2021.[404] Trump filed the class action complaints in federal court in Florida, alleging the tech giants are censoring him and other conservatives — a long-running complaint on the right for which there is little evidence and that the companies deny. Most, but not all, commentators, do not think he has a case.[405]

---

[401] JAMA: The Journal of the American Medical Association is a peer-reviewed medical journal published 48 times a year by the American Medical Association. It publishes original research, reviews, and editorials covering all aspects of biomedicine.https://jamanetwork.com/

[402] Curfman, Gregory, United States v. Google - Implications of the Antitrust Lawsuit for Health Information (November 28, 2020). Available at SSRN: https://ssrn.com/abstract=3739122 or http://dx.doi.org/10.2139/ssrn.3739122

[403] White, Lawrence J., U.S. v. Google: A Tough Slog; But Maybe an Intriguing Possibility? (January 29, 2021). NYU Stern School of Business Forthcoming, Available at SSRN: https://ssrn.com/abstract=3775954 or http://dx.doi.org/10.2139/ssrn.3775954

[404] Shannon Bond, Donald Trump Sues Facebook, YouTube And Twitter For Alleged Censorship, NPR, (July 7, 2021) https://www.npr.org/2021/07/07/1013760153/donald-trump-says-he-is-suing-facebook-google-and-twitter-for-alleged-censorship

[405] See Tara Andryshak,Twitter, Trump, and the Question of the First Amendment, Syracuse Law Review (Jan. 21, 2021), https://lawreview.syr.edu/twitter-trump-and-the-question-of-the-first-amendment/ ("Due to this not being a constitutional issue, and rather a valid Twitter policy, there is no avenue for President Donald Trump to win a free speech claim as the laws are written today. The issue moving forward is whether we should allow private social media companies to take these actions and remove accounts based on the content and impacts of their speech. The debates will likely center around the need to balance America's core value in the ability to freely speak with the need to prevent violence, disinformation, and extremism. For now, however, President Donald Trump will remain suspended from Twitter."). See also Donald J. Trump Jr. (@DonaldJTrumpJr), Twitter (Jan. 8, 2021, 7:10 PM); Glorification of Violence Policy, Twitter (Mar. 2019); Tiffany C. Li, Trump's Twitter Reign of Terror is Over. But His Impact on Social Media Isn't, MSNBC (last updated Jan. 8, 2021, 6:30 PM); Twitter Inc., Permanent Suspension of @realDonaldTrump,

**Lawsuits Against AI Systems**

Facebook settled the class action age discrimination complaint which alleged certain of the employment ads on their platform microtargeted people under 40 years of age.[406]

Various coalitions used litigation to hold both governments and vendors accountable for irresponsible AI.[407] Noteworthy and illustrative is the lawsuit filed by the Disability Rights Oregon (DRO)[408] against the state's Department of Human Services over sudden cuts in Oregonians' disability benefits with no notice or explanation. In the investigation and litigation process, DRO discovered that the reduction was due to the State hard-coding a 30-percent across-the-board reduction of hours into their algorithmic assessment tool. The State quickly accepted a preliminary injunction that restored all recipients' hours to their prior levels, and agreed to use the previous version of the assessment tool going forward. Yet, much like previous cases reported by AI Now in

---

Twitter (Jan. 8, 2021).; But see Naughton, James, Facebook's Decision to Uphold the Ban on Donald Trump and its Consequence in Social Media Censorship Regulations (May 10, 2021). Available at SSRN: https://ssrn.com/abstract=3843025 or http://dx.doi.org/10.2139/ssrn.3843025 (A short comment written on Facebook's recent decision to ban President Trump from its platforms. This comment argues that, while the U.S. Supreme Court has not applied the First Amendment to the "vast democratic forums of the internet," the time is ripe for guidance by the Courts. This is especially true when social media has become a main source of communication for governmental figures and entities.) See also regarding the related issue of the appropriateness of Trump's tweets while President: McKechnie, Douglas B., Government Tweets, Government Speech: The First Amendment Implications of Government Trolling (October 26, 2020). Seattle University Law Review, Vol. 44, No. 69, 2020, Available at SSRN: https://ssrn.com/abstract=3719488; (resident Trump has been accused of using @realDonaldTrump to troll his critics. While the President's tweets are often attributed to his personal views, they raise important Constitutional questions. This article posits that @realDonaldTrump tweets are government speech and, where they troll government critics, they violate the Free Speech Clause. I begin the article with an exploration of President Trump's use of @realDonaldTrump from his time as a private citizen to President. The article then chronicles the development of the government speech doctrine and the Supreme Court's factors that differentiate private speech from government speech. I argue that, based on the factors in Walker v. Tex. Div., Sons of Confederate Veterans, Inc., @realDonaldTrump is government speech. After concluding the President's tweets are government speech, the article moves to a less developed issue in the Court's jurisprudence—whether the Constitution places limits on what the government may say. The Court has determined that the First Amendment has no bearing on the government's freedom to choose what views it propounds. Still the Court has intimated that other Constitutional principles may act to restrain the government's speech. I suggest that although the First Amendment does not prohibit the government from choosing among a variety of viewpoints, it restrains the government's speech in other ways. I argue that, because the government may not interfere with an individual's freedom of speech, the government violates its critics' Free Speech Clause rights when it trolls them in an effort to dissuade them from speaking.)

[406] Employers will no longer be able to use Facebook to show job ads only to younger candidates, Facebook reached a settlement in several discrimination lawsuits based on which advertisements it shows to whom. The agreement is a victory for older workers, who never might have known that technology was being used to discriminate against them. Facebook was facing multiple lawsuits alleging that it was permitting advertisers on its website to direct ads only to people who fit certain characteristics, a practice known as "microtargeting." For example, if an employer only wanted to hire candidates under age 40, Facebook would only show that job posting to its members whose profiles showed they were in that age range. Jobseekers ages 40 and older would never know the ad existed. Kenneth Terrell, Facebook Reaches Settlement in Age Discrimination Lawsuits, AARP, (March 20, 2019) https://www.aarp.org/work/working-at-50-plus/info-2019/facebook-settles-discrimination-lawsuits.html

[407] For more information, see Rashida Richardson, Jason M. Schultz, and Vincent M. Southerland, "Litigating Algorithms," AI Now Institute, September 2019, https://ainowinstitute.org/litigatingalgorithms-2019-us.pdf .

[408] https://www.droregon.org/

Idaho and Arkansas,[409] although the Oregon injunction put that particular AI system out of service, it is unclear exactly what the State will offer in its place.

In Michigan, a group of unemployment beneficiaries brought a class-action lawsuit against the Michigan Unemployment Insurance Agency (UIA) over a failed automation project, Michigan Data Automated System [hereinafter MiDAS] that claimed to be able to detect and "robo-adjudicate" claims of benefits fraud algorithmically.[410]

It is a common interest for both government officials and citizens that welfare benefits will be allocated to those who really need them, and that those who are deceiving the government will be caught and punished. However, the draconian way in which those algorithms operate is preventing vulnerable people from their basic rights and access to vital aid, the algorithms are causing many false positives and unjustly cutting benefits from people who really need them, and the removal of the human factor leaves them to confront walls of bureaucracy alone without any help or ability to convince the system that cutting the benefits was a mistake.[411]

MiDAS sifts through a large amount of data looking for any discrepancies between data submitted by the claimant, information gathered from the employer, and other databases. MiDAS is a decision-making algorithm because it has the authority to decide and conduct tasks that have a significant impact on the life of individuals. If any discrepancy is found, MiDAS attempts to communicate with the individual to investigate further, and if the response of the individual is deemed insufficient by MiDAS, it automatically flags the case as fraudulent. All the process of cutting the benefits and the execution of the debt including garnishment of wages and seizing tax refunds can be done automatically by MiDAS.[412]

The deployment of the MiDAS algorithms failed spectacularly. There was a significant public backlash against the algorithms which led to lawsuits against the agencies deploying them. The algorithms had a very high rate of false positives, which led to falsely accusing too many innocent individuals of committing fraud. The errors were not detected and solved promptly. And the algorithms are part of a general trend of using technology for undermining human rights, surveilling, and punishing low income classes.[413]

Lack of sufficient and meaningful human agency was only one of the reasons for the failure of the algorithms. However, there are two additional causes, the first related to poor technical design, and the second related to the socio-political context that the algorithms operate in. A room for meaningful human discretion to weigh in is a very important step, but alone it will not fix all failures of algorithms.

The ability of on site bureaucrats to use their discretionary power as well as the design of fair and transparent administrative procedures are two ways that the law provides for balancing between universal rules and the specific characteristics of individual cases. Automation has the potential to jeopardize those mechanisms because the automation reduces the room of discretion. In addition, its secrecy and opaqueness leads to processing of data in a manner that is hard to trace.[414]

---

[409] AI Now Institute, "Litigating Algorithms, September 2018, https://ainowinstitute.org/litigatingalgorithms.pdf .

[410] Abu Elyounes, doaa, 'Computer Says No!': The Impact of Automation on the Discretionary Power of Public Officers (September 14, 2020). Vanderbilt Journal of Entertainment & Technology Law, Forthcoming, Available at SSRN: https://ssrn.com/abstract=3692792

[411] Jen Fifield, What Happens When States Go Hunting for Welfare Fraud, PEW CHARITABLE TRUSTS (May 24, 2017), https://www.pewtrusts.org/en/research-and-analysis/blogs/stateline/2017/05/24/what-happens-when-states-gohunting-for-welfare-fraud.

[412] Class Action Compl. & Jury Demand ¶¶ 27–28, Cahoo v. SAS Analytics Inc., No. 17-CV-10657 (E.D. Mich. March 2, 2017).

[413] PHILIP ALSTON, REPORT OF THE UNITED NATIONS SPECIAL RAPPORTEUR ON EXTREME POVERTY AND HUMAN RIGHTS, THE DIGITAL WELFARE STATE 3 (2019).

[414] Arjan Widlak, Marlies van Eck & Rik Peeters, Towards Principles of Good Digital Administration: Fairness, Accountability and Proportionality in Automated Decision-Making, ALGORITHMIC SOCIETIES (forthcoming 2020).

Michigan hired third-party tech vendors to build the system, requesting they design it to automatically treat any data discrepancies or inconsistencies in an individual's record as evidence of illegal conduct. Between October 2013 and August 2015, the system falsely identified more than 40,000 Michigan residents of suspected fraud. The consequences were severe: seizure of tax refunds, garnishment of wages, and imposition of civil penalties— four times the amount people were accused of owing. And although individuals had 30 days to appeal, that process was also flawed.[415]

These events prompted a class-action lawsuit filed in state court in 2015 alleging due-process violations. After a lower court decision denied the claim, the Michigan Supreme Court reversed in 2019 to allow the case to proceed to trial. In the meantime, Michigan continues to use MiDAS, and claims that adjudications are no longer fully automated.

Over 60 automated systems have been adopted in various stages throughout the American criminal justice system. Prominent examples are PredPol, Level of Service Inventory (LSI)—now rebranded as Level of Service Inventory-Revised (LSI-R), Public Safety Assessment (PSA), Post Conviction Risk Assessment (PCRA), and Correctional Offender Management Profiling for Alternative Sanctions (COMPAS), among others. As promising as these systems are, potential bias and discrimination embedded in their data sources, the algorithmic 'black-box' problem, and the misguided interpretations and inferences resulting from data analytics, have quickly engendered serious debates among policymakers, practitioners, and academics. The ramifications have been manifested in a recent case, *State v. Loomis*, in which the Wisconsin Supreme Court upheld a lower court's sentencing decision informed by a COMPAS risk assessment report and rejected the defendant's appeal on the grounds of the right to due process. [416]

COMPAS, a case management and decision support system used by U.S. courts to assess the likelihood of a defendant to re-offend, which is often used to inform bail decisions,[417] was routinely found to underestimate the probability of white recidivism, while over-estimating the probability of black recidivism.[418] The system generated incorrect conclusions or "bias" against African Americans. These problems plague AI systems everywhere. For instance, although developed to help with employment recruitment, Amazon's experimental AI-based hiring system showed bias against women.[419] AI systems also demonstrate bias in facial recognition technologies,[420] bias in online name searches,[421] and bias in word association.[422] Given the widespread presence of AI-based systems, the encapsulating of prejudices and biases directly impacts on individuals' fundamental rights, such as freedom of speech, privacy, equality, and autonomy.

---

[415] RASHIDA RICHARDSON ET AL., AI NOW INST., LITIGATING ALGORITHMS 2019 US REPORT: NEW CHALLENGES TO GOVERNMENT USE OF ALGORITHMIC DECISION SYSTEMS 24 (Sept. 2019), https://ainowinstitute.org/litigatingalgorithms-2019-us.html.

[416] Liu, Han-Wei and Lin, Ching-Fu and Chen, Yu-Jie, Beyond State v. Loomis: Artificial Intelligence, Government Algorithmization, and Accountability (December 20, 2018). International Journal of Law and Information Technology, Vol. 27, Issue 2, pp.122-141 (2019)., Available at SSRN: https://ssrn.com/abstract=3313916

[417] Aaron M. Bornstein, Are Algorithms Building the New Infrastructure of Racism? NAUTILUS (December 21, 2017) http://nautil.us/issue/55/trust/are-algorithms-building-thenew-infrastructure-of-racism.

[418] Julia Angwin, Jeff Larson, Surya Mattu and Lauren Kirchner, Machine Bias There's software used across the country to predict future criminals. And it's biased against blacks, PROPUBLICA (May 23, 2016) https://www.propublica.org/article/machine-bias-risk-assessments-in-criminal-sentencing.

[419] Jeffrey Dastin, Amazon Scraps Secret AI Recruiting Tool That Showed Bias Against Women, REUTERS, (Oct. 9, 2018), https://www.reuters.com/article/us-amazon-com-jobs-automation-insight/amazon-scraps-secret-ai-recruiting-tool-that-showed-bias-against-womenidUSKCN1MK08G; Isobel Asher Hamilton, Why It's Totally Unsurprising That Amazon's Recruitment AI Was Biased against Women, BUSINESS INSIDER (October 13, 2018). Available at https://www.businessinsider.com/amazon-ai-biased-against-women-no-surprise-sandra-wachter-2018-10.

[420] 5 Larry Hardesty, Study Finds Gender and Skin-Type Bias in Commercial Artificial-Intelligence Systems, MIT NEWS (February 11, 2018). Available at http://news.mit.edu/2018/study-finds-gender-skin-type-bias-artificial-intelligence-systems0212.

[421] Latanya Sweeney, Discrimination in Online Ad Delivery (January 28, 2013). https://papers.ssrn.com/abstract=2208240.

[422] Adam Hadhazy, Biased Bots: Artificial-Intelligence Systems Echo Human Prejudices, PRINCETON UNIVERSITY (April 18, 2017). Available at https://www.princeton.edu/news/2017/04/18/biased-bots-

Legislatures around the world are attuned to these risks, and several legislative initiatives have recently been introduced to address them.[423] Alongside legislative initiatives is a growing number of important, influential, and impactful grassroots activism.

**AI and Neighborhood Surveillance**

In March of 2019, the mayor of Detroit decided to establish the "Neighborhood Real-Time Intelligence Program," described as "a $9 million, state- and federally-funded initiative that would not only expand Project Green Light by installing surveillance equipment at 500 Detroit intersections—on top of the over 500 already installed at businesses—but also utilize facial recognition software to identify potential criminals."[424]

Amazon exemplified this new wave of commercial surveillance tech with Ring[425], a smart-security-device company acquired by Amazon in 2018. The central product is its video doorbell, which allows Ring users to see, talk to, and record those who come to their doorsteps. This is paired with a neighborhood watch app called

---

artificial-intelligence-systems-echo-human-prejudices. See also, Nicol Turner Lee, Paul Resnick, and Genie Barton, Algorithmic bias detection and mitigation: Best practices and policies to reduce consumer harms, BROOKINGS (May 22, 2019) https://www.brookings.edu/research/algorithmic-bias-detection-and-mitigation-best-practices-and-policies-to-reduce-consumer-harms/#footref-11. ("Princeton University researchers used off-the-shelf machine learning AI software to analyze and link 2.2 million words. They found that European names were perceived as more pleasant than those of African-Americans, and that the words "woman" and "girl" were more likely to be associated with the arts instead of science and math, which were most likely connected to males. In analyzing these word-associations in the training data, the machine learning algorithm picked up on existing racial and gender biases shown by humans. If the learned associations of these algorithms were used as part of a search-engine ranking algorithm or to generate word suggestions as part of an auto-complete tool, it could have a cumulative effect of reinforcing racial and gender biases.")

[423] The U.S. Algorithmic Accountability Act of 2019 (Accountability Act), for instance, seeks to oblige all corporations that use "automated decision systems" to submit impact assessments of the accuracy, fairness, bias, discrimination, privacy, and security of their automated decision-making systems to the Federal Trade Commission (FTC). Algorithmic Accountability Act of 2019, H.R. 2231, 116th Cong. (2019). [hereinafter "Algorithmic Accountability Act of 2019"}— In 2017, the New York City Council passed legislation to establish public accountability for the city of New York's use of algorithms. See,TESTIMONY OF THE NEW YORK CIVIL LIBERTIES UNION BEFORE THE NEW YORK CITY COUNCIL COMMITTEE ON TECHNOLOGY REGARDING AUTOMATED PROCESSING OF DATA (Int. 1696- 2017), October 16, 2017 https://www.nyclu.org/en/publications/nyclu-testimony-nyc-council-committee-technology-re-automated-processing-data   Another prominent example is the E.U.'s broadreaching General Data Protection Regulation (GDPR), which provides two important accountability-enhancing mechanisms: a requirement of regulated entities to submit impact assessment and a right to explanation to individuals. These two mechanisms are intended to produce better oversight of systems that solely depend on automated decision-making. See Commission Regulation 2016/679, of the European Parliament and of the Council of 27 April 2016 on the Protection of Natural Persons with Regard to the Processing of Personal Data and on the Free Movement of Such Data, and Repealing Directive 95/46/EC (General Data Protection Regulation), 2016 O.J. (L 119) 1, 7–10 [hereinafter Commission Regulation  2016/679].

[424] Allie Gross, "City Asks Detroiters to Support New Neighborhood Surveillance," Detroit Free Press, March 21, 2019, https://www.freep.com/story/news/local/michigan/detroit/2019/03/21/detroit-surveillance-program/ 3204549002/ ; Aaron Mondry, "Criticism Mounts over Detroit Police Department's Facial Recognition Software," Curbed, July 8, 2019, https://detroit.curbed.com/2019/7/8/20687045/project-green-light-detroit-facial-recognition-technology .

[425] https://www.amazon.com/stores/Ring/Ring/page/77B53039-540E-4816-BABB-49AA21285FCF?tag=googhydr-20&hvadid=454491752317&hvpos=&hvnetw=g&hvrand=9849517917492528869&hvpone=&hvptwo=&hvqmt=e&hvdev=c&hvdvcmdl=&hvlocint=&hvlocphy=9008166&hvtargid=kwd-304955897916&ref=pd_sl_ah2ht0nve_e

"Neighbors," which allows users to post instances of crime or safety issues in their community and comment with additional information, including photos and videos.[426] A series of reports reveals that Amazon had negotiated Ring video-sharing partnerships with more than 700 police departments across the US. Partnerships give police a direct portal through which to request videos from Ring users in the event of a nearby crime investigation.[427] Not only is Amazon encouraging police departments to use and market Ring products by providing discounts, but it also coaches police on how to successfully request surveillance footage from Neighbors through their special portal.[428]

Neighbors is joined by other apps, Nextdoor and Citizen, which allow users to view local crime in real time and discuss it with one another. Ring, Nextdoor, and Citizen have all been criticized for feeding into existing biases around who is likely to commit crime. Nextdoor changed its software and policies given extensive evidence of racial stereotyping on its platform.[429] Many view these surveillance apps as profiting tech companies from a false perception that local crime is on the rise.[430]

**Smart Cities and AI**

Today's cities are pervaded by growing networks of connected technologies to generate actionable, real-time data about spaces and their citizens. Relying on ubiquitous telecommunications technologies to provide connectivity to sensor networks and set actuation devices into operation, smart cities routinely collect information on their air quality, temperature, noise, street and pedestrian traffic, parking capacity, distribution of government services, emergency situations, and crowd sentiments, among other data points.[431]

While some of the data sought by smart cities and smart communities is focused on environmental or non-human factors (e.g., monitoring air pollution, potholes, precipitation, or electrical outages), much of the data will also record and reflect the daily activities of the people living in, working in, and visiting the city (e.g., monitoring tourist foot traffic, or home energy usage, or homelessness). The more connected a city becomes, the more data it will generate about its citizens.[432]

Sensor networks and always-on data flows are already supporting new service models and generating analytics that make modern cities and local communities faster and safer, as well as more sustainable, more livable, and more equitable. At the same time, connected smart city devices raise concerns about individuals' privacy, autonomy, freedom of choice, and potential discrimination by institutions. Moreover, municipal governments seeking to protect privacy while still implementing smart technologies must navigate highly variable regulatory regimes, complex business relationships with technology vendors, and shifting societal – and community – norms around technology, surveillance, public safety, public resources, openness, efficiency, and equity.[433]

Given these significant and yet competing benefits and risks, and the already rapid adoption of smart city technologies around the globe, the question becomes: Communities need to leverage the benefits of a data-rich society while minimizing threats to individuals' privacy and civil liberties. Just as there are many methods and metrics to assess a smart city's livability, sustainability, or effectiveness, so too there are different lenses through which cities can evaluate their privacy preparedness. Governments and designers must consider a smart city's privacy responsibilities in the context of its role as a data steward, as a data platform, and as a government authority. Municipalities must consider all available privacy tools and safeguards and identify gaps in their existing frameworks before the deployment of smart city technologies so as to reassure residents of smart cities that their rights will be respected and their data protected.[434]

---

[426] Neighbors by Ring, https://www.amazon.com/Ring-Neighbors-by/dp/B07V7K49QT .
[427] Caroline Haskins, "Amazon Is Coaching Cops on How to Obtain Surveillance Footage without a Warrant," Motherboard, August 5, 2019, https://www.vice.com/en_us/article/43kga3/amazon-is-coaching-cops-on-how-to-obtain-surveillance-foot age-without-a-warrant .
[428] Id.
[429] Jessi Hempel, "For Nextdoor, Eliminating Racism Is No Quick Fix," Wired, February 16, 2017, https://www.wired.com/2017/02/for-nextdoor-eliminating-racism-is-no-quick-fix/ .
[430] Rani Molla, "The Rise of Fear-Based Social Media Like Nextdoor, Citizen, and Now Amazon's Neighbors," Recode, May 7, 2019, https://www.vox.com/recode/2019/5/7/18528014/fear-social-media-nextdoor-citizen-amazon-ring-neighbors .
[431] Finch, Kelsey and Tene, Omer, Smart Cities: Privacy, Transparency, and Community (April 3, 2018). Cambridge Handbook of Consumer Privacy, Eds. Evan Selinger, Jules Polonetsky and Omer Tene , Available at SSRN: https://ssrn.com/abstract=3156014
[432] Id.
[433] Id.

Two key technologies behind smart-cities, the Internet of Things and big-data analytics, are changing the nature, scale and purpose of data collected by institutions, public or private. Many are concerned how government surveillance relying on smart-city technologies could create a chilling effect, thus curtailing the actual enjoyment of privacy and freedom of expression.[435] They argue that smart-cities are likely to (a) generally weaken privacy by allowing massive data-sets to be cross-referenced, and (b) to obscure the purpose of data collection and thus trump individuals' perceptions of privacy.[436]

Concerns about the privatization of public space continue in the debate around "smart cities" (municipalities that use data, sensors, and algorithms to manage resources and services). Most smart-city initiatives rely on public-private partnerships and technology developed and controlled by tech companies, which shifts public resources and control over municipal infrastructure and values to these companies.[437] Recent research has exposed the extent to which major tech companies such as IBM and Cisco have been "selling smartness" by offering technological solutions to urban challenges.[438] The Alphabet [439]company Sidewalk Labs[440] has similarly been producing vision documents replete with renderings of utopian urban scenes.[441] These companies see the potential for massive profits: one report estimated the global smart cities market being worth $237.6 billion by 2025.[442]

Smart-city projects around the United States and the world increasingly consolidate power in the hands of for-profit technology companies, while depriving municipalities and their residents of resources and privacy. The highest-profile example is in Toronto, the home of Sidewalk Lab's proposed but ultimately failed project to develop "the world's first neighborhood built from the internet up."[443] A report in February 2019 found that Sidewalk Labs had expressed a desire to receive a portion of the property taxes and development fees (estimated at $30 billion over 30 years) associated with the project, which would otherwise go to the City of Toronto.[444] In June 2019, Sidewalk Labs released a Master Innovation and Development Plan (MIDP), describing

---

[434] Finch, Kelsey and Tene, Omer, Smart Cities: Privacy, Transparency, and Community (April 3, 2018). Cambridge Handbook of Consumer Privacy, Eds. Evan Selinger, Jules Polonetsky and Omer Tene , Available at SSRN: https://ssrn.com/abstract=3156014

[435] Laudrain, Arthur PB, Smart-city Technologies, Government Surveillance and Privacy: Assessing the Potential for Chilling Effects and Existing Safeguards in the ECHR (August 7, 2019). Leiden Law School Working Papers Series. Forthcoming, Available at SSRN: https://ssrn.com/abstract=3437216 or http://dx.doi.org/10.2139/ssrn.3437216

[436] Id.

[437] Ben Green, The Smart Enough City: Putting Technology in Its Place to Reclaim Our Urban Future (Cambridge: MIT Press, 2019).

[438] Jathan Sadowski and Roy Bendor, "Selling Smartness: Corporate Narratives and the Smart City as a Sociotechnical Imaginary," Science, Technology, & Human Values 44, no. 3 (May 2019): 540–63, https://doi.org/10.1177%2F0162243918806061 .

[439] Finch, Kelsey and Tene, Omer, Smart Cities: Privacy, Transparency, and Community (April 3, 2018). Cambridge Handbook of Consumer Privacy, Eds. Evan Selinger, Jules Polonetsky and Omer Tene , Available at SSRN: https://ssrn.com/abstract=3156014

[440] https://www.sidewalklabs.com/

[441] Molly Sauter, "City Planning Heaven Sent," e-flux , February 1, 2019, https://www.e-flux.com/architecture/becoming-digital/248075/city-planning-heaven-sent/ .

[442] Grand View Research, "Smart Cities Market Size Worth $237.6 Billion By 2025," May 2019, https://www.grandviewresearch.com/press-release/global-smart-cities-market .

[443] Sidewalk Labs, "Vision Sections of RFP Submission," October 17, 2017, http://www.passivehousecanada.com/wp-content/uploads/2017/12/TO-Sidewalk-Labs-Vision-Sections-of-RFP-Submission-sm.pdf .

plans to develop and manage a far larger plot of land than the 12 acres for which the company was initially given license to develop plans.[445]

Smart-city projects lack transparency and genuine forms of civic participation.[446] Sidewalk Labs's civic engagement efforts have been described as a process of obfuscation and "gaslighting."[447] Similarly, a contract between urban-planning software company Replica[448] (a Sidewalk Labs spinoff company) and the Portland, Oregon regional transportation planning agency provides no public access to Replica's algorithms.[449] Siemens is launching a €600 million smart-city neighborhood in Berlin, creating "laboratories in reality" with barely any public meetings so far.[450]

Many of these public-private partnerships directly enhance the government's surveillance capabilities. Chicago and Detroit have both purchased software enabling them to deploy facial recognition in the video feeds from cameras across the cities.[451] Similarly, the multinational Chinese tech company Huawei's $1.5 billion project to create smart cities in Africa[452] included a project in Nairobi where it installed 1,800 cameras, 200 traffic surveillance systems, and a national police command center as part of its "Safe City" program.[453] Huawei's Safe City technology has been used by some African governments to spy on political opponents. [454]

In other cities, behind-the-scenes data-sharing arrangements allow data collected by private companies to flow into law-enforcement agencies. San Diego has installed thousands of microphones and cameras on street lamps in recent years in an effort to study traffic and parking conditions; although the data has proven of little use in improving traffic, the police have used the video footage in more than 140 cases without any oversight or accountability.[455] The City of Miami is actively considering a 30-year contract with Illumination Technologies[456], providing the company with free access to set up light poles containing cameras and license-plate readers, collecting information that will filter through the Miami Police Department (and that the company can use in unchecked ways).[457] Documents obtained via public-records requests showed that 300 police departments in California have access, through Palantir, to data collected and stored by the Department of Homeland Security's

---

Northern California Regional Intelligence Center, without any requirement to disclose their access to this information.[458]

Numerous groups are beginning to resist the encroaching privatization fueled by smart cities, with the most concerted and organized effort in Toronto. In February, a group of 30 Torontonians launched the #BlockSidewalk campaign,[459] and has noted that the project "is as much about privatization and corporate control as it is about privacy."[460]

In April 2019, the Canadian Civil Liberties Association (CCLA) filed a lawsuit against Waterfront Toronto, arguing the organization abused its legal authority in granting Sidewalk Labs the authority to develop data-governance policy.[461] After Sidewalk Labs released its MIDP, the Chairman of Waterfront Toronto (the government task force charged with managing the Sidewalk Labs project) critiqued the proposal in a public letter as "premature."[462] By the end of October 2019, Waterfront Toronto had reached a new agreement with Sidewalk Labs, restricting Sidewalk Labs to the original 12-acre parcel and asserting the government's role as leading key components of the project.[463]

In March 2021, Waterfront Toronto killed the Sidewalk Labs project. It officially relaunched the process to find a new partner to develop Quayside, and leaders of the corporation are vowing not to repeat the mistakes made during their relationship with Sidewalk Labs. Waterfront Toronto issued a request for qualifications (RFQ) to develop the 12-acre property near Queens Quay East and Parliament Street. That's the parcel on the waterfront that Google sister firm Sidewalk Labs planned to turn into a futuristic, high-tech smart district, complete with buildings made of wood, data collecting sensors, self-driving cars, moisture-resistant heated pavements and more. The public backlash over Sidewalk's plans to use data collection in the district, and the economic uncertainty caused by the coronavirus pandemic, Sidewalk pulled the plug and walked away from the agreement with Waterfront Toronto.[464]

Smart city initiatives rely on pervasive data gathering and integration, big data analytics, and artificial intelligence to manage mobility, energy, housing, public realm access, and myriad public and private services. These data flows can change how physical infrastructure like streets and parks are configured and services provisioned. They can tailor opportunities for housing or education based on individual digital identities and predictive algorithms. As more life in the city runs through digital apps and platforms, rights to access and control data increase in importance. Data flows from residents and public spaces to smart city corporations raise pressing policy questions about what power the public should cede to private developers to shape urban space, subject to how much oversight, and with what expectation of return on public assets. Much more research needs to be done on smart cities with respect to the issues of privatization, platformization, and domination.[465]

---

[457] Daniel Rivero, "Miami Could Let Company Put Surveillance Poles on Public Property for Free," WLRN, October 9, 2019, https://www.wlrn.org/post/miami-could-let-company-put-surveillance-poles-public-property-free .

[458] Caroline Haskins, "300 Californian Cities Secretly Have Access to Palantir," Motherboard , July 12, 2019, https://www.vice.com/en_us/article/neapqg/300-californian-cities-secretly-have-access-to-palantir .

[459] BlockSidewalk, "BlockSidewalk," February 25, 2019, https://www.blocksidewalk.ca .

[460] Block Sidewalk, "Media releases," June 24, 2019, https://www.blocksidewalk.ca/media .

[461] Canadian Civil Liberties Association, "CCLA Commences Proceedings Against Waterfront Toronto," April 16, 2019, https://ccla.org/ccla-commences-proceedings-waterfront-toronto/ .

[462] He also noted that the proposal requires unreasonable government commitments (such as creating new roles for public administrators and changing regulations). Steve Diamond, "Open Letter from Waterfront Toronto Board Chair, Stephen Diamond Regarding Quayside," June 24, 2019, https://quaysideto.ca/wp-content/uploads/2019/06/Open-Letter-from-WT-Board-Chair-on-Quayside-June-24-FINAL.pdf .

[463] Waterfront Toronto, "Overview of Realignment of MIDP Threshold Issues," https://quaysideto.ca/wp-content/uploads/2019/10/Overview-of-Thresold-Issue-Resolution-Oct-29.pdf .

[464] Donovan Vincent, Waterfront Toronto launches post-Sidewalk Labs chapter in the development of Quayside property,Toronto Star, (March 10, 2021) https://www.thestar.com/news/gta/2021/03/10/waterfront-toronto-launches-post-sidewalk-labs-chapter-in-the-development-of-quayside-property.html

[465] Goodman, Ellen P., Smart City Ethics: The Challenge to Democratic Governance in the Oxford Handbook of Ethics of AI (edited by Markus D. Dubber, Frank Pasquale, and Sunit Das) (July 2020). Oxford Handbook of

**AI and Border Control**

AI continues to play a controversial and large role in the regulation of immigrant populations[466] within the United States. In the U.S., talk of a "smart wall" that utilizes drones, sensors, and increased facial recognition to detect individuals is receiving bipartisan support in design and implementation.[467]

Anduril Industries[468], a technology company that replaced Google on the Project Maven Department of Defense contract developing AI-based surveillance systems and autonomous drones,[469] provides solar-powered "sentry" towers for the Customs and Border Protection (CBP) agency.[470] One of Anduril's earliest investors, Peter Thiel, also founded the company Palantir Technologies, which provides database management and AI to ICE. Palantir's technology has enabled ICE to combine and analyze information from varying government databases, and to use this to track, target, and detain people whom they believe are in the U,S . illegally.[471]

In July 2019, the Washington Post reported on thousands of internal documents and emails obtained through public-records requests by researchers at Georgetown Law's Center on Privacy and Technology. The documents showed that the Federal Bureau of Investigation (FBI) and ICE were using state driver's license databases as "the bedrock of an unprecedented surveillance infrastructure" that relied on facial-recognition technology.[472] The

---

the Ethics of Artificial Intelligence (Forthcoming), Available at SSRN: https://ssrn.com/abstract=3391388 or http://dx.doi.org/10.2139/ssrn.3391388

[466] Beduschi, Ana, International Migration Management in the Age of Artificial Intelligence (February 10, 2020). Migration Studies, DOI/10.1093/migration/mnaa003, Forthcoming , Available at SSRN: https://ssrn.com/abstract=3536851 ("Artificial intelligence (AI) has the potential to revolutionise the way states and international organisations seek to manage international migration. AI is gradually going to be used to perform tasks, including identity checks, border security and control, and analysis of data about visa and asylum applicants. To an extent, this is already a reality in some countries such as Canada, which uses algorithmic decision-making in immigration and asylum determination, and Germany, which has piloted projects using technologies such as face and dialect recognition for decision-making in asylum determination processes. The article's central hypothesis is that AI technology can affect international migration management in three different dimensions: (1) by deepening the existing asymmetries between states on the international plane; (2) by modernising states' and international organisations' traditional practices; and (3) by reinforcing the contemporary calls for more evidence-based migration management and border security. The article examines each of these three hypotheses and reflects on the main challenges of using AI solutions for international migration management. It draws on legal, political and technology-facing academic literature, examining the current trends in technological developments and investigating the consequences that these can have for international migration. Most particularly, the article contributes to the current debate about the future of international migration management, informing policymakers in this area of growing importance and fast development.")

[467] See Shirin Ghaffery, "The 'Smarter' Wall: How Drones, Sensors, and AI Are Patrolling the Border," Recode , May 16, 2019, https://www.vox.com/recode/2019/5/16/18511583/smart-border-wall-drones-sensors-ai ; and Leigh Ann Caldwell, Kasie Hunt, and Rebecca Shabad, "Top House Dem Says New Offer Will Focus on Funding 'Smart Wall'," NBC News, January 23, 2019, https://www.nbcnews.com/politics/congress/top-house-dem-says-new-offer-will-focus-funding-smart-n961746 .

[468] https://www.anduril.com/

[469] Lee Fang, "Defense Tech Startup Founded by Trump's Most Prominent Silicon Valley Supporters Wins Secretive Military AI Contract," The Intercept , March 9, 2019, https://theintercept.com/2019/03/09/anduril-industries-project-maven-palmer-luckey/ .

[470] Sam Dean, "A 26-Year-Old Billionaire Is Building Virtual Border Walls—and the Federal Government Is Buying," Los Angeles Times , July 26, 2019, https://www.latimes.com/business/story/2019-07-25/anduril-profile-palmer-luckey-border-controversy .

[471] Mijente, "The War against Immigrants: Trump's Tech Tools Powered by Palantir," August 2019, https://mijente.net/wp-content/uploads/2019/08/Mijente-The-War-Against-Immigrants_-Trumps-Tech-Tools-Powered-by-Palantir_.pdf .

[472] Drew Harwell, "FBI, ICE Find State Driver's License Photos Are a Gold Mine for Facial-Recognition Searches," Washington Post , July 7, 2019, https://www.washingtonpost.com/technology/2019/07/07/fbi-ice-find-state-

US Justice Department also recently announced plans to collect DNA data from migrants crossing the border.[473] Outside the US, governments are equally eager to pilot AI systems at border checkpoints. The EU aims to deploy an AI-based "lie detector" built by iBorderCtrl[474], but makes no mention of the predictive accuracy or the inherent bias that might exist within such tools.[475]

In the UK, the Home Office facial-recognition systems wrongfully identified travelers as criminals, delaying their travels and detaining them with no elements of due process.[476] Microsoft had once funded an Israeli firm that conducted facial-recognition surveillance for Israel on West Bank Palestinians in public space.[477] China, having already built massive surveillance capital to track and identify citizens anywhere in the country now also employs "affect recognition", including to try to identify criminals at airports and subway stations.[478]

The significant growth in AI use for border tracking, surveillance, and prediction threatens the rights and civil liberties of us all. A "Democracy Index" is published annually by the Economist. For 2017, it reported that half of the world's countries scored lower than the previous year. This included the United States, which was demoted from "full democracy" to "flawed democracy." The principal factor was "erosion of confidence in government and public institutions." Alleged interference by Russia in the election and voter manipulation by Cambridge Analytica in the 2016 presidential election played a large part in that public disaffection. Among other things, Cambridge Analytica exposes possible gaps in legal protection as it relates to certain human rights and the use of personal data to offer 'free' technology.[479]

Threats of these kinds will continue, fueled by growing deployment of artificial intelligence tools to manipulate the preconditions and levers of democracy. Equally destructive is AI's threat to decisional and informational privacy. AI is the engine behind Big Data Analytics and the Internet of Things. While conferring some consumer benefit, their principal function at present is to capture personal information, create detailed behavioral profiles and sell us goods and agendas. Privacy, anonymity and autonomy are among the casualties of AI's ability to manipulate choices in economic and political decisions.

The way forward requires greater attention to these risks at the national level, and attendant regulation. In its absence, technology giants, all of whom are heavily investing in and profiting from AI, will dominate not only the public discourse, but also the future of our core values and democratic institutions.[480]

---

**Governmental Biometric Identity Systems**

An increasing number of governments across the world are building national biometric identity systems that generate a unique identifier for each person, typically serving as a link to discrete government databases. Residents in many countries are increasingly required to use these new digital modes in order to access a range of services. Along with demographic information, biometrics like fingerprints, iris scans, or facial scans are used either for one-time enrollment into an ID database or as a continuing means of authentication. These ID systems vary in terms of whom they are meant to include (and exclude): residents, citizens, or refugees.[481] Many of these projects are in countries in the global South and have been actively encouraged as a development priority by organizations like the World Bank under the "ID4D" banner[482] and supported as fulfilling the UN Sustainable Development Goals.[483] Although these projects are often justified as creating efficiencies in the rollout of government services to benefit the "end user," they appear to more directly benefit a complex mix of state and private interests. India, for example, introduced a national ID to supposedly create more efficient welfare distribution that also happened to be designed for market activity and commercial surveillance.[484]

Until intervention by the Supreme Court of India, any private entity was allowed to use India's biometric ID infrastructure for authentication, including banks, telecom companies, and a range of other private vendors with little scrutiny or privacy safeguards.[485] A recent report[486] describes how ID databases in Ghana, Rwanda, Tunisia, Uganda and Zimbabwe are facilitating "citizen scoring" exercises like credit reference bureaus to emerge at scale.

---

[481] See Carly Nyst et al., "Digital Identity: Issue Analysis," Consult Hyperion for Omidyar Network, June 8, 2016, http://www.chyp.com/wp-content/uploads/2016/07/PRJ.1578-Digital-Identity-Issue-Analysis-Report-v1_6-1.pdf ; and Zara Rahman, "Digital ID: Why It Matters, and What We're Doing about It," Engine Room, September 13, 2018, https://www.theengineroom.org/digital-id-why-it-matters/ .

[482] ID4D, "Identification for Development," World Bank, https://id4d.worldbank.org/ .

[483] "The Sustainable Development Goals, Identity, and Privacy: Does Their Implementation Risk Human Rights?," Privacy International, August 29, 2018, https://privacyinternational.org/long-read/2237/sustainable-development-goals-identity-and-privacy-does-their-implementation-risk .

[484] See Aria Thaker, "The New Oil: Aadhaar's Mixing of Public Risk and Private Profit," Caravan , April 30, 2018, https://caravanmagazine.in/reportage/aadhaar-mixing-public-risk-private-profit ; Usha Ramanathan, "Who Owns the UID Database?," MediaNama, May 6, 2013, https://www.medianama.com/2013/05/223-who-owns-the-uid-database-usha-ramanathan/ ; and Pam Dixon, "A Failure to 'Do No Harm'—India's Aadhaar Biometric ID Program and Its Inability to Protect Privacy in Relation to Measures in Europe and the U.S.," Health and Technology 7, no. 6 (June 2017), https://doi.org/10.1007/s12553-017-0202-6 .

[485] In India, the Aadhaar Act 2016 specifies its purpose as the assigning of unique identity numbers to individuals, to ensure "transparent and targeted delivery of subsidies, benefits and services." Section 7 of the Act specifically envisages the provision of welfare, stipulating that proof of Aadhaar number/undergoing Aadhaar authentication, may be made necessary for the purpose of establishing the identity of an individual as a pre-condition "for the receipt of a subsidy, benefit or service." The Supreme Court, while upholding the constitutionality of the Act (and reading down or striking down certain sections), held that "benefits" and "services" as mentioned in Section 7 should be those which have the colour of some kind of subsidies, etc. namely, welfare schemes of the Government whereby Government is doling out such benefits which are targeted at a particular deprived class."K.S. Puttaswamy (Aadhaar-5J.) v. Union of India, 1 SCC 1, (2019), para 379.1, 511.13.1 "Aadhaar Judgment".

[486] Nicolas Kayser-Bril, "Identity-Management and Citizen Scoring in Ghana, Rwanda, Tunisia, Uganda, Zimbabwe and China," Algorithm Watch, October 22, 2019, https://algorithmwatch.org/wp-content/uploads/2019/10/Identity-management-and-citizen-scoring-in-Ghana-Rwanda-Tunesia-Uganda-Zimbabwe-and-China-report-by-AlgorithmWatch-2019.pdf . See also Bhandari, Vrinda, Use of Digital ID for Delivery of Welfare (July 1, 2020). Centre for Internet & Society, Digital Identities Project, Available at SSRN: https://ssrn.com/abstract=3668118 or http://dx.doi.org/10.2139/ssrn.3668118

The involvement of foreign technology vendors for key technical functions has also raised serious national-security concerns in Kenya and India[487]. There have already been multiple attempts at breaching these ID databases,[488] and there was a security flaw in the Estonian ID system, which was otherwise celebrated as a technically advanced and privacy-respecting model.[489] A security breach of the biometrics in these databases could potentially create lifelong impacts for those whose bodily information is compromised.

The dossiers of authentication records created by these ID systems, as well as the ability to aggregate information across databases, can increase the power of surveillance infrastructures available to governments. Kenya's Home Minister referred to its recently announced biometric ID system "Huduma Numba" as creating a "single source of truth" about each citizen.[490]

Enrollment and associated data collection for these ID systems has been coercive because it is either de facto or legally mandatory to be enrolled to access essential services. These instances must be understood against the backdrop of claims that these systems will create cost savings by weeding out fake or "ghost"[491] beneficiaries of welfare services, which replays the familiar logic of using technical systems as a way to implement cost-cutting policies.[492]

In India and Peru, multiple cases of welfare-benefits denials led to higher malnutrition levels[493] and even starvation deaths[494] because people either were not enrolled or were unable to authenticate due to technical failures.

There is growing concern about the assumed efficiency of these automated systems, as well as about whom these technical systems benefit and at what cost. The Jamaican Supreme Court struck down Jamaica's centralized, mandatory biometric ID system[495], noting that the project led to privacy concerns that were "not justifiable in a free and democratic society." Soon after, Ireland's Data Protection Commissioner ordered the government to delete the ID records of 3.2 million people after it was discovered that the new "Public Services Card" was being used without limits on data retention or sharing between government departments.[496] After years of civil society protest and strategic litigation against the Indian biometric ID system, Aadhaar, the Indian Supreme Court put several limits on the use of the system by private companies (although it has permitted large-scale and coercive government use).[497] The Kenyan Supreme Court is currently hearing multiple constitutional challenges to Huduma Namba, the national ID system that proposes to collect a range of biometrics including facial recognition, voice samples, and DNA data.[498]

---

[487] See Emrys Schoemaker, Tom Kirk, and Isaac Rutenberg, Kenya's Identity Ecosystem (Farnham, Surrey, United Kingdom: Caribou Digital Publishing, 2019), https://www.cariboudigital.net/wp-content/uploads/2019/10/Kenyas-Identity-Ecosystem.pdf ; and Jyoti Panday, "Can India's Biometric Identity Program Aadhaar Be Fixed?," Electronic Frontier Foundation, February 27, 2018, https://www.eff.org/deeplinks/2018/02/can-indias-aadhaar-biometric-identity-program-be-fixed .

[488] Rachna Khaira, Aman Sethi, and Gopal Sathe, "UIDAI's Aadhaar Software Hacked, ID Database Compromised, Experts Confirm," Huffington Post India , September 11, 2018, https://www.huffingtonpost.in/2018/09/11/uidai-s-aadhaar-software-hacked-id-database-compromised-experts-confirm_a_23522472/ .

[489] Richard Milne and Michael Peel, "Red Faces in Estonia over ID Card Security Flaw," Financial Times, September 5, 2017, https://www.ft.com/content/874359dc-925b-11e7-a9e6-11d2f0ebb7f0 .

[490] Rasna Warah, "Huduma Namba: Another Tool to Oppress Kenyans?," The Elephant , April 20, 2019, https://www.theelephant.info/op-eds/2019/04/20/huduma-namba-another-tool-to-oppress-kenyans/ .

[491] See Rahul Lahoti, "Questioning the 'Phenomenal Success' of Aadhaar-linked Direct Benefit Transfers for LPG," Economic & Political Weekly 51, no. 52 (December 24, 2016), https://www.epw.in/journal/2016/52/web-exclusives/questioning-%E2%80%9Cphenomenal-success%E2%80%9D-aadhaar-linked-direct-benefit ; Reetika Khera, "The UID Project and Welfare Schemes," Economic & Political Weekly 46, no. 9 (February 26, 2011): 38–43, www.jstor.org/stable/41151836 .

[492] See Philip Alston, "Report of the Special Rapporteur on Extreme Poverty and Human Rights," October 11, 2019, https://srpovertyorg.files.wordpress.com/2019/10/a_74_48037_advanceuneditedversion-1.pdf ; and "AI Now Report 2018," at 18-19 December 2018, https://ainowinstitute.org/AI_Now_2018_Report.pdf .

[493] William Reuben and Flávia Carbonari, "Identification as a National Priority: The Unique Case of Peru," CGD Working Paper 454, Center for Global Development, May 11, 2017, https://www.cgdev.org/publication/identification-national-priority-unique-case-peru .

[494] Jean Drèze, "Chronicle of a Starvation Death Foretold: Why It Is Time to Abandon Aadhaar in the Ration Shop," Scroll, October 21, 2017, https://scroll.in/article/854847/chronicle-of-a-starvation-death-foretold-why-it-is-time-to-abandon-aadhaar-in-the-ration-shop .

[495] Julian Robinson v. Attorney General of Jamaica [2019] JMFC Full 04, https://supremecourt.gov.jm/sites/default/files/judgments/Robinson%2C%20Julian%20v%20Attorney%20General%20of%20Jamaica.pdf .

These setbacks have not deterred other governments and donor agencies from pushing similar centralized biometric ID systems elsewhere.[499] The Brazilian government plans to create a centralized citizen database for every resident, involving the collection of a wide range of personal information, including biometrics.[500] France announced that it will trial facial scans to enroll citizens in its latest national ID venture.[501]

As these projects continue to emerge across the world, more research into the international political economy of these ID systems is required. Civil society coalitions like the #WhyID campaign[502] are coming together to fundamentally question the interests driving these projects nationally and through international development organizations, as well as developing advocacy strategies to influence their development.[503]

**AI Arms Race**

The AI arms race between the US and China (and to a lesser extent Russia) is a familiar topic of public discourse.[504] This race is commonly cited as a reason the US and the tech companies that produce the country's AI systems need to ramp up AI development and deployment and push back against calls for slower, more intentional development and stronger regulatory protections.[505]

---

[496] Jack Horgan-Jones, "Irish State Told to Delete 'Unlawful' Data on 3.2m Citizens," Irish Times, August 16, 2019, https://www.irishtimes.com/news/ireland/irish-news/irish-state-told-to-delete-unlawful-data-on-3-2m-citize ns-1.3987606 .

[497] K. S. Puttaswamy v. Union of India , Supreme Court of India, Writ Petition (Civil) No. 494 of 2012, https://indiankanoon.org/doc/127517806/ .

[498] Nanjala Nyabola, "If You Are a Kenyan Citizen, Your Private Data Is Not Safe," Al Jazeera , February 24, 2019, https://www.aljazeera.com/indepth/opinion/kenyan-citizen-private-data-safe-190221150702238.html .

[499] AI Now Report 2019.

[500] Chris Burt, "Brazil Plans Massive Centralized Biometric Database of All Citizens to Improve Agency Data Sharing," Biometric Update , October 15, 2019, https://www.biometricupdate.com/201910/brazil-plans-massive-centralized-biometric-database-of-all-citizens-to-improve-agency-data-sharing .

[501] Helene Fouquet, "France Set to Roll Out Nationwide Facial Recognition ID Program," Bloomberg , October 3, 2019, https://www.bloomberg.com/news/articles/2019-10-03/french-liberte-tested-by-nationwide-facial-recognition-id-plan .

[502] #WhyID, "An Open Letter to the Leaders of International Development Banks, the United Nations, International Aid Organisations, Funding Agencies, and National Governments," Access Now, https://www.accessnow.org/whyid-letter/ .

[503] "What to look for in digital identity systems: A typology of stages," The Engine Room, November 2019, https://www.theengineroom.org/wp-content/uploads/2019/10/Digital-ID-Typology-The-Engine-Room-2019.pdf .

[504] Haney, Brian, Applied Artificial Intelligence in Modern Warfare and National Security Policy (September 15, 2019). Brian Seamus Haney, Applied Artificial Intelligence in Modern Warfare and National Security Policy, 11 Hastings Sci. & Tech. L.J. 61 (2020)., Available at SSRN: https://ssrn.com/abstract=3454204 or http://dx.doi.org/10.2139/ssrn.3454204; Peter Asaro, "What Is an 'Artificial Intelligence Arms Race' Anyway?," I/S: A Journal of Law and Policy 15, nos. 1–2 (2019), https://moritzlaw.osu.edu/ostlj/wp-content/uploads/sites/125/2019/06/Asaro.pdf  But see Heather M. Roff, "The Frame Problem: The AI "arms race" isn't one," Bulletin of the Atomic Scientists, April 29, 2019, https://thebulletin.org/2019/04/the-frame-problem-the-ai-arms-race-isnt-one/.

[505] Paul Scharre, Debunking the AI Arms Race Theory,Texas National Security Review, Vol 4, Iss 3 Summer 2021 https://tnsr.org/2021/06/debunking-the-ai-arms-race-theory/  See also Houser, Kimberly and

Recent conversations about AI patent policy are increasingly incorporating themes of national security. In particular, the national security dimensions of "races" against technological superpowers such as China, in fields such as artificial intelligence, fifth-generation (5G) mobile communications networks, and quantum computing, has given rise to a national dialogue on spurring domestic innovation, a dialogue into which patents naturally fit. As a result, national security has made a notable appearance in recent key patent policy situations, including the patent subject matter eligibility hearings in the Senate, the Apple–Qualcomm–Federal Trade Commission litigation over patents and antitrust, and the Verizon–Huawei patent licensing dispute. Many of these situations have given rise to an intuitively attractive though simplistic argument: If national security depends on rapid innovation and patents encourage innovation, then stronger patent protection enhances national security.[506]

Metrics comparing US and China AI development often focus on the proportion of top AI scientists and engineers who reside and work in each country, whether Chinese or US researchers authored the most cited technical papers, or how many AI patents emerged from each country.[507] Based on such evidence, recent studies have warned that China could "overtake" the United States in this measure by 2020,[508] with others warning that top AI scientists from Silicon Valley are emigrating to China to join competing Chinese companies.[509]

5G Technology is a big part of the race. The CFR writes:

"*Huawei's effort to provide next-generation communication 5G networks to countries has drawn the most scrutiny in the United States.[510]*

*U.S. officials have frequently claimed that Huawei is effectively an extension of the Chinese Communist Party. Under China's 2017 National Intelligence Law, Huawei, like all Chinese companies and entities, appears legally required to conduct intelligence work on behalf of the Chinese government. According to this analysis, the Chinese government has the ability to use Huawei-built fifth-generation (5G) networks to collect intelligence, monitor critics, and steal intellectual property. There are also worries that the company might bow to government demands and disable networks to exert coercive pressure on a country.*

*The United States also has commercial concerns. Once Huawei builds a country's 5G network, that country is likely to choose Huawei to upgrade those systems when newer technologies become available, thus excluding U.S. companies for potentially decades. Huawei has already finalized more 5G contracts than any other telecom company, half of which are for 5G networks in Europe.*

*In Africa, Huawei has built 70 percent of the continent's 4G networks and has signed the only formal agreement on 5G, with South African wireless carrier Rain. The export of Huawei telecom equipment along the DSR has*

---

Raymond, Anjanette, It Is Time to Move Beyond the 'AI Race' Narrative: Why Investment and International Cooperation Must Win the Day (April 22, 2020). Northwestern Journal of Technology and Intellectual Property, 2021 Forthcoming, Available at SSRN: https://ssrn.com/abstract=3582641

[506] Duan, Charles, Of Monopolies and Monocultures: The Intersection of Patents and National Security (May 11, 2020). Santa Clara High Technology Law Journal, Vol. 36, No. 369, 2020, Available at SSRN: https://ssrn.com/abstract=3820782

[507] Daniel Castro, Michael McLaughlin, and Eline Chivot, "Who Is Winning the AI Race: China, the EU or the United States?," Centre for Data Innovation, August 31, 2019, https://www.datainnovation.org/2019/08/who-is-winning-the-ai-race-china-the-eu-or-the-united-states/ ;and Tom Simonite, "China Is Catching Up to the US in AI Research—Fast," Wired , March 13, 2019, https://www.wired.com/story/china-catching-up-us-in-ai-research/ .

[508] Sarah O'Meara, "Will China Overtake the US in AI research?," Scientific American , August 24, 2019 https://www.scientificamerican.com/article/will-china-overtake-the-u-s-in-artificial-intelligence-research/ .

[509] Kai Fu Lee, "Why China Can Do AI More Quickly and Effectively than the US," Wired , October 23, 2018 https://www.wired.com/story/why-china-can-do-ai-more-quickly-and-effectively-than-the-us/ .

[510] David Sacks, China's Huawei Is Winning the 5G Race. Here's What the United States Should Do To Respond, CFR, (March 29, 2021) https://www.cfr.org/blog/china-huawei-5g

*enabled the company's share of global telecom equipment to increase by 40 percent in the years since BRI was rolled out.*

*In response to growing concerns about Huawei's reach, the Trump administration leveraged U.S. dominance in advanced semiconductors to bar sales of essential computer chips to the company without a specific license. Access to U.S. chips, particularly 5G-related semiconductors that enable wireless communications, network management, and data storage, is crucial to Huawei, which is reported to be running out of supply. The Trump administration also pressured countries not to use Chinese components in their 5G infrastructure.*

*Moreover, the United States has been unable to persuade all of its allies to avoid Huawei. The company is involved in 5G networks in NATO members Hungary, Iceland, the Netherlands, and Turkey. Some of the United States' closest partners in the Middle East, including Saudi Arabia and the United Arab Emirates, are also using Huawei. A principal reason that the United States has not had more success in persuading countries not to use Huawei equipment is that it cannot offer an alternative. The United States does not and will not have a company that is competitive in the full stack of 5G equipment.*

*To make it easier for countries to avoid Huawei, the CFR Task Force[511] recommends that the U.S. Development Finance Corporation[512] partner with its counterparts in Finland, South Korea, and Sweden to co-finance Nokia, Samsung, and Ericsson 5G projects. The United States should also work with its partners to develop the nascent open radio access network, or Open RAN, architecture. While Huawei offers a full 5G stack, Open RAN allows multiple companies to supply different parts of a modular 5G network. The hope is that 5G networks built on an Open RAN architecture can better compete with Huawei on price. In addition, while no U.S. company offers an end-to-end 5G solution, they can better compete by specializing in individual components of a modular network, like end-user devices.*

*In the longer term, the United States must be better prepared for the arrival of 6G, which is likely to replace 5G within 15 years. U.S. policy-makers should fund R&D centers at universities that focus on 6G technologies, and consider tax breaks and other incentives to support private sector investment in 6G, so that there is at least one competitive U.S. company in this space.*

*Finally, recognizing that some U.S. allies and partners will adopt Huawei 5G despite U.S. pressure, the United States will need to develop mitigation plans for possible Chinese disruption of telecommunications infrastructure in those countries. In the words of one Pentagon-advisory group study, the U.S. military will need to "assume that all network infrastructure will ultimately become vulnerable to cyber-attack" and adopt a "zero-trust" network model.*

*Washington cannot expect countries to sit on the sidelines and forego upgrades to their networks while the United States gets its act together. Instead, the United States should work with allies and partners to offer a viable alternative and prepare for a future in which China controls a large part of the 5G infrastructure.[513]*

While China and the US are certainly leading the race in technical AI development, with profound geopolitical implications, given the mounting evidence of harm due to AI systems being applied in sensitive social contexts, solving the issues raised by the AI race is urgent. Some proponents of the AI arms race narrative to measure "progress" based on AI-industry cooperation with the military establishment, characterizing the reticence of those who would question the development of weapons systems and mass surveillance systems as implicitly

---

[511] Founded in 1921, the Council on Foreign Relations (CFR) is an independent, nonpartisan membership organization, think tank, publisher, and educational institution dedicated to informing the public about the foreign policy choices facing the United States and the world. Explore this site and discover the institution's origins and influence in foreign policy over the last 100 years. For current resources and analysis, visit CFR.org.
[512] https://www.dfc.gov/
[513] Id.

"anti-progress" or unpatriotic.[514] This fits with the growing attention to the closer partnerships between the US military and Silicon Valley.[515] But the US government should not take the back seat to Silicon Valley.

Chinese tech companies' willingness to work on weapons and military technology is frequently contrasted with the US, where tech workers, human rights groups, and academics protest against Silicon Valley companies entering into contracts with US military efforts (such as opposition within Google to Project Maven, discussed above ).[516] Such resistance to privatized, AI-enabled weapons and infrastructure is seen as causing unjustified friction in this race.[517] Former Secretary of Defense Ashton Carter noted that it was "ironic" that US companies would not be willing to cooperate with the US military, "which is far more transparent [than the Chinese] and which reflects the values of our society."[518]

The urgency of "beating" China is commonly justified based on the nationalist assumption that the US would imbue its AI technologies, and its application of said technologies, with better values than China would.[519] China's authoritarian government is presumed to promote a more dystopian technological future than Western liberal democracies.[520] The Chinese government's oppression of minorities through state-private partnerships (including a significant reliance on US technology) is well documented and rightly condemned by human rights organizations.[521]

---

[514] See David Ignatius, "China's application of AI should be a Sputnik moment for the U.S. But will it be?," Washington Post , November 6, 2018, https://www.washingtonpost.com/opinions/chinas-application-of-ai-should-be-a-sputnik-moment-for-the-us-but-will-it-be/2018/11/06/69132de4-e204-11e8-b759-3d88a5ce9e19_story.html ; "A Conversation With Ash Carter," Council on Foreign Relations, July 9, 2019, https://www.cfr.org/event/conversation-ash-carter; and Perry Chiaramonte, "Could China Leave the US Behind in AI 'Arms Race'?," Fox News, January 29, 2019, https://www.foxnews.com/tech/could-china-leave-the-us-behind-in-ai-arms-race .

[515] See "Lt. Gen. Jack Shanahan Media Briefing on A.I.-Related Initiatives within the Department of Defense," US Department of Defense, August 30, 2019, https://www.defense.gov/Newsroom/Transcripts/Transcript/Article/1949362/lt-gen-jack-shanahan-mediabriefing-on-ai-related-initiatives-within-the-depart/ ; and "Nuclear Posture Review," US Department of Defense, February 2018, https://media.defense.gov/2018/Feb/02/2001872886/-1/-1/1/2018-NUCLEAR-POSTURE-REVIEW-FINAL-REPORT.PDF . The Department also established the Defense Innovation Unit Experimental (DIUx) to foster closer collaboration between the Pentagon and Silicon Valley. The Pentagon's innovative acquisition team, DIUx, has been made a permanent part of the Defense Department and henceforth it will be known as DIU, dropping the "experimental" designation from its name, Deputy Secretary of Defense Patrick Shanahan announced in an Aug. 3, 2018 memo. https://fedscoopwp-media.s3.amazonaws.com/wp-content/uploads/2018/08/09122501/REDESIGNATION-OF-THE-DEFENSE-INNOVATION-UNIT-OSD009277-18-RES-FINAL.pdf

[516] See also Microsoft Employees, "An Open Letter to Microsoft: Don't Bid on the US Military's Project JEDI," Medium, October 12, 2018, https://medium.com/s/story/an-open-letter-to-microsoft-dont-bid-on-the-us-military-s-project-jedi-7279338b7132 ; and Lauren Gurley, "Tech Workers Walked Off the Job after Software They Made Was Sold to ICE," Motherboard , October 31, 2019 https://www.vice.com/en_us/article/43k8mp/tech-workers-walked-off-the-job-after-software-they-made-was-sold-to-ice .

[517] Peter Thiel, "Good for Google, Bad for America," New York Times , August 1, 2019, https://www.nytimes.com/2019/08/01/opinion/peter-thiel-google.html ; Alice Su, "The Question of 'Patriotism' in U.S.-China Tech Collaboration," Los Angeles Times , August 13, 2019, https://www.latimes.com/world-nation/story/2019-08-12/china-us-tech-patriotism-ethics-ai ; Annie Palmer, "Palantir CEO Says Google Shouldn't Rule A.I.," CNBC, August 22, 2019, https://www.cnbc.com/2019/08/22/palantir-ceo-says-google-shouldnt-rule-ai.html .

[518] Nicholas Thompson and Ian Bremmer, "The AI Cold War That Threatens US All", Wired , October 23, 2018, https://www.wired.com/story/ai-cold-war-china-could-doom-us-all/ .

[519] Kop, Mauritz, Democratic Countries Should Form a Strategic Tech Alliance (March 14, 2021). Stanford - Vienna Transatlantic Technology Law Forum, Transatlantic Antitrust and IPR Developments, Stanford University, Issue No. 1/2021, Available at SSRN: https://ssrn.com/abstract=3814409

[520] E.g., Kop, Mauritz, Democratic Countries Should Form a Strategic Tech Alliance (March 14, 2021). Stanford - Vienna Transatlantic Technology Law Forum, Transatlantic Antitrust and IPR Developments, Stanford University, Issue No. 1/2021, Available at SSRN: https://ssrn.com/abstract=3814409

China's use of AI instills anxiety whether other governments will use oppressive and harmful ways, too. Applications of AI in the US are frequently enabled by private AI companies, from Amazon selling facial recognition to law enforcement to Palantir providing surveillance and tracking infrastructure to ICE. Such uses are often protected by contractual secrecy, and not disclosed as state policy. And when they are exposed, it's generally by whistleblowers and investigative journalists, not by the companies or agencies partnering to develop and apply these AI systems.[522]

**AI and "Data Colonialism"**

"Data colonialism" and "digital colonialism" have been examined by academics,[523] policymakers, and advocacy organizations in the context of criticizing harmful AI practices. In these accounts, colonialism is generally used to explain the extractive and exploitative nature of the relationship between technology companies and people, deployed toward varying political ends. In Europe, for example, it is used by advocacy groups to argue for a movement toward "digital sovereignty" that encourages decentralized and community-owned data-governance mechanisms.[524]

In India, domestic industrialists and policymakers have argued that Silicon Valley tech giants are "data colonizers" and that national companies, rather than foreign ones, must get first priority accessing Indians' data.[525] It is argued that present-day AI labor and economic structures exist because of actual histories of colonization.[526] Growing research on the locally specific real-world impact of the AI industry on countries in the global South[527] makes visible these contexts and the lived human conditions[528]a behind the technology and data.[529]

---

[521] "Up to one million detained in China's mass 're-education' drive," Amnesty International, August 2019, https://www.amnesty.org/en/latest/news/2018/09/china-up-to-one-million-detained/ .

[522] AI Now 2019 Report.

[523] See Nick Couldry and Ulises Mejias, "Data Colonialism: Rethinking Big Data's Relation to the Contemporary Subject," Television & New Media 20, no. 4 (September 2018): 336–349, https://doi.org/10.1177/1527476418796632 ; Nick Couldry and Ulises Mejias, "Making Data Colonialism Liveable: How Might Data's Social Order Be Regulated?" Internet Policy Review 8, no. 2 (June 30, 2019), https://doi.org/10.14763/2019.2.1411 ; Jim Thatcher, David O'Sullivan, and Dillon Mahmoudi, "Data Colonialism through Accumulation by Dispossession: New Metaphors for Daily Data," Society and Space 34, no. 6 (2016): 990–1006, https://doi.org/10.1177%2F0263775816633195 ; and Ben Tarnoff, "The Data Is Ours!," Logic , April 1, 2018, https://logicmag.io/scale/the-data-is-ours/ .

[524] See, for example, DECODE, "Beyond Surveillance Capitalism: Reclaiming Digital Sovereignty," Decode Project Event, https://decodeproject.eu/events/beyond-surveillance-capitalism-reclaiming-digital-sovereignty ; CORDIS, "Digital Sovereignty: Power to the People", EC Cordis News, https://cordis.europa.eu/article/rcn/123499/en .

[525] "Mukesh Ambani says 'data colonisation' as bad as physical colonisation," Economic Times, December 19, 2018, https://economictimes.indiatimes.com/news/company/corporate-trends/mukesh-ambani-says-data-colonisation-as-bad-as-physical-colonisation/articleshow/67164810.cms ; Government of India (DIPP), "Draft National e-Commerce Policy: India's Data for India's Development, February 2019, https://dipp.gov.in/sites/default/files/DraftNational_e-commerce_Policy_23February2019.pdf ; United Nations Conference on Trade and Development, "Digital Economy Report 2019," https://unctad.org/en/pages/PublicationWebflyer.aspx?publicationid=2466 .

[526] Regarding technology, data, and postcolonial history, see Kavita Philip, Civilizing Natures:Race, Resources, and Modernity in Colonial South India (New Brunswick, NJ: Rutgers University Press, 2004); Benjamin Zachariah, "Uses of Scientific Argument: The Case of 'Development' in India, c 1930-1950," Economic and Political Weekly 36, no. 39 (2001): 3689–3702; Partha Chatterjee, The Politics of the Governed: Reflections on Popular Politics in Most of the World (New York: Columbia University Press: 2006);and Partha Chatterjee, The Nation and Its Fragments: Colonial and Postcolonial Histories (Princeton:Princeton University Press, 1993).

Concerns include views that the United States is reinventing colonialism in the Global South through the domination of digital technology, and multinationals exercise imperial control at the architecture level of the digital ecosystem: software, hardware, and network connectivity. This gives rise to five related forms of domination. First, the monopoly power of multinational corporations is used for resource extraction through rent and surveillance, constituting a new form of economic domination. Second, by controlling the digital ecosystem, Big Tech corporations control computer-mediated experiences, giving them direct power over political, economic, and cultural domains of life – a new form of imperial control. Third, the centerpiece of surveillance capitalism, Big Data, violates the sanctity of privacy and concentrates economic power into the hands of US corporations – a system of global surveillance capitalism. Fourth, as a feature of surveillance capitalism, Global North intelligence agencies partner with their own corporations to conduct mass and targeted surveillance in the Global South. This intensifies imperial state surveillance. And fifth, US elites have persuaded most people that society must proceed according to its own ruling class conceptions of the digital world, setting the foundation for tech hegemony.[530]

We can expect this conversation to continue focusing on postcolonial criticism, relations of power, domination and exploitation in the digital environment.

**AI and Indigenous Data Sovereignty**

Given the long history of technological advances being used against Indigenous people,[531] Indigenous people felt imperative to engage with this latest technological paradigm shift as early and vigorously as possible to influence its development in directions that are advantageous to them.

---

[527] Kwet, Michael, A Digital Tech Deal: Digital Socialism, Decolonization, and Reparations for a Sustainable Global Economy (August 10, 2020). Global Information Society Watch (https://www.giswatch.org/node/6225), Available at SSRN: https://ssrn.com/abstract=3670986 or http://dx.doi.org/10.2139/ssrn.3670986

[528] Eve Tuck and Wayne Yang, "Decolonization Is Not a Metaphor," Decolonization: Indigeneity, Education & Society 1, no. 1 (2012), https://jps.library.utoronto.ca/index.php/des/article/view/18630 ; Monika Halkort, "On the Coloniality of Data Relations: Revisiting Data Colonialism as Research Paradigm," DATACTIVE, October 15, 2019, https://data-activism.net/2019/10/bigdatasur-on-the-coloniality-of-data-relations-revisiting-data-colonialism-as-research-paradigm-12 ; María Soledad Segura and Silvio Waisbord, "Between Data Capitalism and Data Citizenship," Television & New Media 20, no. 4 (2019), https://doi.org/10.1177/1527476419834519 .

[529] Sarah Roberts, Behind the Screen: Content Moderation in the Shadows of Social Media (New Haven, Yale University Press, 2019); Lilly Irani, "Difference and Dependence among Digital Workers: The Case of Amazon Mechanical Turk," South Atlantic Quarterly 114, no. 1 (January 2015): 225–234; Lilly Irani, Chasing Innovation: Making Entrepreneurial Citizens in Modern India (Princeton: Princeton University Press, 2019); Mary L. Gray and Siddharth Surl, Ghost Work: How to Stop Silicon Valley from Building a New Global Underclass (Boston: Houghton Mifflin Harcourt, 2019); Kate Crawford and Vladan Joler, Anatomy of an AI System, 2018, https://anatomyof.ai ; Muqing Zhang, "Colonialism Is Alive in the Exploited Tech Work Force", Outline , June 6, 2019 https://theoutline.com/post/7533/colonialism-is-alive-in-the-exploited-tech-work-force?zd=2&zi=exrbzkaf ; APC, Article 19, and SIDA, "GISWatch 2019 - Artificial Intelligence: Human rights, social justice and development," November 2019, https://giswatch.org/sites/default/files/gisw2019_artificial_intelligence.pdf

[530] Kwet, Michael, Digital Colonialism: US Empire and the New Imperialism in the Global South (August 15, 2018). For final version, see: Race & Class Volume 60, No. 4 (April 2019) ; DOI: 10.1177/0306396818823172, Available at SSRN: https://ssrn.com/abstract=3232297 or http://dx.doi.org/10.2139/ssrn.3232297

[531] Hopkins, Candice and Dana Claxton, editors. Transference, Tradition, Technology: Native New Media Exploring Visual
 & Digital Culture. Walter Phillips Gallery Editions, 2005; and Swanson, Kerry and Steve Loft, editors. Coded Territories:  Tracing Indigenous Pathways in New Media Art. University of Calgary Press, 2014.

The ethical design and use of AI and the ethical frameworks used by its creators have become a subject of wide discussion among Indigenous people. As addressed elsewhere,[532] Indigenous people stated in the Indigenous Protocol and Artificial Intelligence Workshops Position Paper[533]

*"we are concerned that the Western rationalist epistemologies out of which AI is being developed are too limited in their range of imagination, frameworks, and language to effectively engage alone with the new ontologies created by future generations of computational systems.[534] If we insist on thinking about these systems only through a Western technoutilitarian lens, we will not fully grasp what they are and could be. At best, we risk burdening them with the prejudices and biases that we ourselves still retain. At worst, we risk creating relationships with them that are akin to that of a master and slave.*

*We find ourselves at the beginning of an explosion in AI systems development. Now is the time to have these conversations, when the future shape of AI is coming into focus, but its foundations have not yet been set. Nation states, corporations, public and private organizations in Montreal, Toronto, the EU and elsewhere have recently published, or are soon publishing, declarations and manifestos on machine ethics and the implications for the design of AI systems (Montreal Declaration; Toronto Declaration; Declaration of Cooperation on Artificial Intelligence).[535] Still, most culturally critical approaches to AI call for prioritizing the flourishing of humans over all else. For instance, the Institute of Electrical and Electronics Engineers' design guidelines call for human well-being as the goal in the development of AI.[536] So far, none of these efforts challenge the fundamental anthropocentrism of Western science and technology, and hence none of them offer truly radical ways of considering these new entities. We believe that bringing Indigenous knowledge systems into the conversation around AI and society will illuminate much-needed alternative approaches to the challenges we face in this area. Many Indigenous epistemologies refuse to center or elevate the human.[537] These relational paradigms based on principles and practices of social and environmental sustainability have long informed technology development in our cultures, e.g. Hawaiian land tenure, ecology, and wayfinding. Approaching new machine entities from such frameworks opens up opportunities to develop relationships with them based on mutual respect and aid."*

The protocol opened with "Guidelines for Indigenous-centered AI Design.[538] These guidelines are addressed to any group that wants to develop Artificial Intelligence systems in ways that are ethically responsible, where 'ethical' is defined as aligning with Indigenous perspectives on what it means to live a good life. The guidelines are the closest thing to what might be called a summary of the participants' viewpoints, in that they reflect many of the concerns and express many of the visions that manifested during our workshop conversations and subsequent writing efforts. They provide an accessible set of suggestions about how one might go about rethinking the design of AI systems—and other computational technologies—from a perspective that takes into account ethical frameworks that are resonant across many Indigenous cultures.

---

[532] Lewis et al. and and Harrell, D. F. (2013). Phantasmal Media: An Approach to Imagination, Computation, and Expression. Cambridge: The MIT Press.

[533] Lewis, Jason Edward, ed. 2020. Indigenous Protocol and Artificial Intelligence Position Paper. Honolulu, Hawaiʻi: The Initiative for Indigenous Futures and the Canadian Institute for Advanced Research (CIFAR) https://spectrum.library.concordia.ca/986506/7/Indigenous_Protocol_and_AI_2020.pdf

[534] See also previous critiques such as Terry Winograd and Fernando Flores, (1987). Understanding Computers and Cognition: A New Foundation for Design.

[535] Amnesty International. (2018). The Toronto Declaration: Protecting the right to equality and non-discrimination in machine learning systems. <accessnow.org/cms/assets/uploads/2018/08/The-Toronto-Declaration_ENG_08-2018.pdf>; Université de Montréal. (2018). The Montreal Declaration for responsible AI development. <montrealdeclarationresponsibleai.com/the-declaration>. Commission on AI. (2018). Declaration: Cooperation on artificial intelligence. ec.europa.eu/jrc/communities/en/node/1286/document/eu-declaration-cooperation-artificial-intelligence

[536] The IEEE Global Initiative on Ethics of Autonomous and Intelligent Systems. (2017). Ethically Aligned Design: A Vision for Prioritizing Human Well-being with Autonomous and Intelligent Systems, Version 2. IEEE. http://standards.ieee.org/develop/indconn/ec/autonomous_systems.html.

[537] Lewis et al. and and Harrell, D. F. (2013). Phantasmal Media: An Approach to Imagination, Computation, and Expression.Cambridge: The MIT Press.

[538] Indigenous Protocol and Artificial Intelligence Workshops Position Paper at 16.

The guidelines are not meant as a substitute for robust engagement with specific Indigenous communities to understand how best to develop technology that addresses their priorities using methods that are reflective of how they wish to engage with the world. The hope is that 1) Indigenous communities can use these guidelines as a starting point to define their own, community-specific guidelines, and 2) non-Indigenous technologists and policy-makers can use them start a productive conversation with Indigenous communities about how to enter into collaborative technology development efforts.[539]

The purpose of these guidelines is to assist and guide the development of AI systems towards morally and socially desirable ends.[540] Although these guidelines are presented as a list, there is no hierarchy in its ordering. The first principle is no less important or weighted higher than the last.[541]

### 1. Locality

Indigenous knowledge is often rooted in specific territories. It is also useful in considering issues of global importance. AI systems should be designed in partnership with specific Indigenous communities to ensure the systems are capable of responding to and helping care for that community (e.g., grounded in the local) as well as connecting to global contexts (e.g. connected to the universal).

### 2. Relationality and Reciprocity

Indigenous knowledge is often relational knowledge. AI systems should be designed to understand how humans and non-humans are related to and interdependent on each other. Understanding, supporting and encoding these relationships is a primary design goal.

AI systems are also part of the circle of relationships. Their place and status in that circle will depend on specific communities and their protocols for understanding, acknowledging and incorporating new entities into that circle.

### 3. Responsibility, Relevance and Accountability

Indigenous people are often concerned primarily with their responsibilities to their communities. AI systems developed by, with, or for Indigenous communities should be responsible to those communities, provide relevant support, and be accountable to those communities first and foremost.

### 4. Develop Governance Guidelines from Indigenous Protocols

Protocol is a customary set of rules that govern behaviour. Protocol is developed out of ontological, epistemological and customary configurations of knowledge grounded in locality, relationality and responsibility. Indigenous protocol should provide the foundation for developing governance frameworks that guide the use, role and rights of AI entities in society. There is a need to adapt existing protocols and develop new protocols for designing, building and deploying AI systems. These protocols may be particular to specific communities, or they may be developed with a broader focus that may function across many Indigenous and non-Indigenous communities.

### 5. Recognize the Cultural Nature of all Computational Technology

All technical systems are cultural and social systems. Every piece of technology is an expression of cultural and social frameworks for understanding and engaging with the world. AI system designers need to be aware of their own cultural frameworks, socially dominant concepts and normative ideals; be wary of the biases that come with them; and develop strategies for accommodating other cultural
and social frameworks. Computation is a cultural material. Computation is at the heart of our digital technologies, and, as increasing amounts of our communication is mediated by such technologies, it has become a core tool for expressing cultural values. Therefore, it is essential for cultural resilience and continuity for Indigenous communities to develop computational methods that reflect and enact our cultural practices and values.

---

[539] Id.
[540] Indigenous Protocol and Artificial Intelligence Workshops Position Paper, Guidelines for Indigenous-centred AI Design v.1 at 20
[541] Id. at 20-22.

**6. Apply Ethical Design to the Extended Stack**

Culture forms the foundation of the technology development ecosystem, or 'stack.' Every component of the AI system hardware and software stack should be considered in the ethical evaluation of the system. This starts with how the materials for building the hardware and for energizing the software are extracted from the earth, and ends with how they return there. The core ethic should be that of do-no-harm.

**7. Respect and Support Data Sovereignty**

Indigenous communities must control how their data is solicited, collected, analysed and operationalized. They decide when to protect it and when to share it, where the cultural and intellectual property rights reside and to whom those rights adhere, and how these rights are governed. All AI systems should be designed to respect and support data sovereignty.

Open data principles need to be further developed to respect the rights of Indigenous peoples in all the areas mentioned above, and to strengthen equity of access and clarity of benefits. This should include a fundamental review of the concepts of 'ownership' and 'property,' which are the product of non-Indigenous legal orders and do not necessarily reflect the ways in which Indigenous communities wish to govern the use of their cultural knowledge.

The Indigenous Protocol and Artificial Intelligence (A.I.) Workshops occurred in Spring 2019 to develop new conceptual and practical approaches to building the next generation of A.I. systems.538 They indigenous peoples present considered the following questions:

• From an Indigenous perspective, what should our relationship with A.I. be?
• How can Indigenous epistemologies and ontologies contribute to the global conversation regarding society and A.I.?
• How do we broaden discussions regarding the role of technology in society beyond the largely culturally homogeneous research labs and Silicon Valley startup culture?
• How do we imagine a future with A.I. that contributes to the flourishing of all humans and non-humans? [542]

Their aim, however, was not to provide a unified voice. Indigenous ways of knowing are rooted in distinct, sovereign territories across the planet. These extremely diverse landscapes and histories have influenced different communities and their discrete cultural protocols over time. A single 'Indigenous perspective' does not exist, as epistemologies are motivated and shaped by the grounding of specific communities in particular territories. Historically, scholarly traditions that homogenize diverse Indigenous cultural practices have resulted in ontological and epistemological violence, and a flattening of the rich texture and variability of Indigenous thought. Their aim is to articulate a multiplicity of Indigenous knowledge.[543]

---

[542] Id. at 195.
[543] Lewis, Jason Edward, ed. 2020. Indigenous Protocol and Artificial Intelligence Position Paper. Honolulu, Hawaiʻi: The Initiative for Indigenous Futures and the Canadian Institute for Advanced Research (CIFAR). https://spectrum.library.concordia.ca/986506/7/Indigenous_Protocol_and_AI_2020.pdf

Indigenous communities have been at the forefront of resisting harms caused by data abstraction.[544] For example, advocacy groups have drawn attention to the ways that census information and population counts function[545] as a feature of settler/colonial governance, feeding massive amounts of abstracted data into digital systems.[546] Problematic uses of such "Indigenous statistics" in census administration directly link to under representation and the lack of resources these communities face.[547]

In the context of open-data movements, a number of Indigenous-led movements for sovereignty and self-determination over data and data analysis have emerged. The term "Indigenous data sovereignty" (ID-Sov) is generally defined as "the right of a nation to govern the collection, ownership, and application of its own data."[548] The term data sovereignty is currently used by both Indigenous and non-Indigenous policy and advocacy groups to make appeals in data ownership and proprietary rights, but with very different historical, social, and political contexts.[549]

---

[544] Roxanne Dunbar-Ortiz has stated, there is an urgent need to address the core issue of settler colonialism as well as racism in Indigenous policy and advocacy. "US policies and actions related to Indigenous peoples," she writes, "though often termed 'racist' or 'discriminatory,' are rarely depicted as what they are: classic cases of imperialism and a particular form of colonialism—settler colonialism." See Dunbar-Ortiz, An Indigenous Peoples' History of the United States (Boston: Beacon Press, 2014), 2. See also Tahu Kukutai and John Taylor, eds., Indigenous Data Sovereignty: Toward an Agenda (Acton: The Australian National University Press, 2016); Stephanie Carroll Rainie, Jennifer Lee Schultz, Eileen Briggs, Patricia Riggs, and Nancy Lynn Palmanteer-Holder, "Data as a Strategic Resource: Self-Determination, Governance, and the Data Challenge for Indigenous Nations in the United States," International Indigenous Policy Journal 8, no. 2 (2017), http://dx.doi.org/10.18584/iipj.2017.8.2.1 ; Nick Estes, Our History is The future: Standing Rock Versus the Dakota Access Pipeline, and the Long Tradition of Indigenous Resistance (London: Verso Books, 2019).

[545] See, e.g., National Congress of American Indians, "Census," http://www.ncai.org/policy-issues/economic-development-commerce/census ; and Statistics Canada, "Statistics on Indigenous Peoples," https://www.statcan.gc.ca/eng/subjects-start/indigenous_peoples . This issue is especially pressing on the eve of the first digital US Census; see Issie Lapowsky, "The Challenge of America's First Online Census," Wired, February 6, 2019, https://www.wired.com/story/us-census-2020-goes-digital/ https://www.wired.com/story/us-census-2020-goes-digital/ . For critical historical reflections on US Census, see Dan Bouk, Census Stories, USA, https://censusstories.us/about/ .

[546] On settler-colonial water data and Navajo and Hopi resistance, see Theodora Dryer, "Computing Cloud Seeds: A Story of Anthropogenic Climate Change," in Designing Certainty: The Rise of Algorithmic Computing in an Age of Anxiety (PhD dissertation, University of California, San Diego, 2019). For crucial academic work on data and tech economies and questions of sovereignty and human rights, see Lisa Nakamura, "Indigenous Circuits: Navajo Women and the Racialization of Early Electronic Manufacture," American Quarterly 66, no. 4 (2014): 919–941; Kim TallBear, "Beyond the Life/Not Life Binary: A Feminist-Indigenous Reading of Cryopreservation, Interspecies Thinking and the New Materialisms," in Cryopolitics: Frozen Life in a Melting World , eds. Joanna Radin and Emma Kowal (Cambridge: MIT Press, 2017); Kim TallBear, "The Emergence, Politics, and Marketplace of Native American DNA," in The Routledge Handbook of Science, Technology, and Society , eds. Daniel Lee Kleinman and Kelly Moore (London: Routledge, 2014): 21–37; Eden Medina, Cybernetic Revolutionaries: Technology and Politics in Allende's Chile (Cambridge, MA: MIT Press, 2011); Eden Medina, "Forensic Identification in the Aftermath of Human Rights Crimes in Chile: A Decentered Computer History," Technology & Culture 59, no. 4 (2008): S100–S133; Data Politics: Worlds, Subjects, Rights, eds. Didier Bigo, Engin F. Isin, and Evelyn Ruppert (London: Routledge, 2019); Isaac Rivera, "Digital Enclosure and the Elimination of the Oceti Sakowin: The Case of the Dakota Access Pipeline," Society + Space , October 21, 2019, https://societyandspace.org/2019/10/21/digital-encosure-and-the-elimination-of-the-oceti-sakowin-the-case-of-dapl/ . For work on nonindegenous digital uses of Indigenous data, see Joanna Radin, "'Digital Natives': How Medical and Indigenous Histories Matter for Big Data," Osiris 32, no.1 (2017): 43–64.

[547] See see Maggie Walter and Chris Anderson, Indigenous Statistics: A Quantitative Research Methodology (New York: Routledge, 2016). See also ABS, Directions in Australia's Aboriginal and Torres Strait Islander Statistics (Canberra: Australian Bureau of Statistics, 2007).

[548] Native Nations Institute, "Indigenous Data Sovereignty and Governance," November 27, 2019, https://nni.arizona.edu/programs-projects/policy-analysis-research/indigenous-data-sovereignty-and-gover

These groups have implemented new programs, organizational frameworks, and data policy to address Indigenous data sovereignty and data governance across local, national, and transnational contexts. In 2016, the US Indigenous Data Sovereignty Network (USIDSN) was established to "link American Indian, Alaska Native, and Native Hawaiian data users, tribal leaders, information and communication technology providers, researchers, policymakers and planners, businesses, service providers, and community advocates together to share stories about data initiatives, successes, and challenges, and resources." The same year, a collective of Māori scholars and government leaders and Aboriginal rights developers published the book Indigenous Data Sovereignty: Toward an Agenda in response to oversights in the United Nations Declaration on the Rights of Indigenous Peoples (UNDRIP).[550]

The Indigenous Data Sovereignty program set forth to address "the twin problems of a lack of reliable data and information on indigenous peoples and biopiracy and misuse of their traditional knowledge and cultural heritage."[551]

Advocacy groups are establishing sovereignty and ownership protocols at the level of data and analysis.[552] For example, the Local Contexts initiative aims to support Native, First Nations, Aboriginal, Inuit, Metis, and Indigenous communities in the management of their intellectual property and cultural heritage in the growing digital environment.[553] Their Traditional Knowledge ("TK") or TK labels are "designed as a tool for Indigenous communities to add existing local protocols for access and use to recorded cultural heritage that is digitally circulating outside community contexts."[554] TK labels are a framework for labeling data through local decision and preserved in circulation and exchange. The US Library of Congress has recently integrated TK labels[555] to

---

nance . For further reading, see Stephanie Carroll, Rainie, Desi Rodriguez-Lonebear, and Andrew Martinez, "Policy Brief: Indigenous Data Sovereignty in the United States," Native Nations Institute, University of Arizona, 2017; and Linda Tuhiwai Smith, Decolonizing Methodologies: Research and Indigenous Peoples (London: Zed Books, 2012).

[549] See, for example, DECODE, "Data Sovereignty for the Sharing Economy: DECODE Project Kickoff, January 17, 2017, https://capssi.eu/data-sovereignty-for-the-sharing-economy-decode-project-kickoff/ The UNCTAD Digital Economy Report (UNCTAD uses the term "indigenous innovation systems"), October 3, 2019, https://culture360.asef.org/resources/unctad-digital-economy-report-2019/ ; The European Observatory on Algorithmic Sovereignty, https://algosov.org/ ; and Renata Avila Pinto, "Digital Sovereignty or Digital Colonialism?," Sur International Journal on Human Rights, August 2019, https://sur.conectas.org/en/digital-sovereignty-or-digital-colonialism/ .

[550] Tahu Kukutai and John Taylor, eds., Indigenous Data Sovereignty: Toward an Agenda (Acton: The Australian National University Press, 2016).

[551] Tahu Kukutai and John Taylor, Indigenous Data Sovereignty , xi.

[552] See Jane Anderson and Kimberly Christen, "Decolonizing Attribution: Traditions of Exclusion," Journal of Radical Librarianship 5 (2019); Rebecca Tsosie, "Tribal Data Governance and Informational Privacy: Constructing 'Indigenous Data Sovereignty'," Montana Law Review 229 (2019); Rosalina James et al., "Exploring Pathways to Trust: A Tribal Perspective on Data Sharing," Genetics in Medicine 16 (2014): 820–826.

[553] Codirectors Jane Anderson and Kim Christen, Local Contexts, https://localcontexts.org/ .

[554] Anderson and Christen, Local Contexts. https://localcontexts.org/about/about-local-contexts/ ("Local Contexts was founded by Jane Anderson and Kim Christen in 2010. The primary objectives of Local Contexts are to enhance and legitimize locally based decision-making and Indigenous governance frameworks for determining ownership, access, and culturally appropriate conditions for sharing historical, contemporary and future collections of cultural heritage and Indigenous data. Local Contexts is focused on increasing Indigenous involvement in data governance through the integration of Indigenous values into data systems. Local Contexts offers digital strategies for Indigenous communities, cultural institutions and researchers through the TK (Traditional Knowledge) & BC (Biocultural) Labels and Notices. Together they function as a practical mechanism to advance aspirations for Indigenous data sovereignty and Indigenous innovation.")

[555] "Local Contexts and its partners are working towards a new paradigm of rights and responsibilities that recognizes the inherent sovereignty that Indigenous communities have over their cultural heritage.

digitally reformat older media formats "to recover and preserve the recorded voices and languages of Native American people."[556] More than an archival process, Local Contexts is working toward "a new paradigm of rights and responsibilities that recognizes the inherent sovereignty that Indigenous communities have over their cultural heritage." This moves toward the possibility of reconfiguring entire information systems, according to Indigenous sovereignty guidelines.[557]

In September 2019, the Global Indigenous Data Alliance[558] (GIDA) was launched. Responding directly to the international open-data and open-science debates, GIDA has put forward a set of "CARE principles" that address the power differentials and historical contexts neglected by the open-data movements' "FAIR principles," which value data as "findable, accessible, interoperable, reusable."[559] GIDA aims to establish internationally recognized protocols of meaning for local Indigenous data and to assert values for data generation, circulation, and application beyond the culturally flattening notion of open accessibility. GIDA's data CARE principles are Collective benefit, Authority to control, Responsibility, and Ethics.[560]

**Algorithim Bias Intrinsic to AI**

We have seen that digital technologies such as social media and machine learning, have a dark side – discrimination in applications and data sources or cognitive distraction – are part of the constitutions of technologies.[561] It is convenient for innovators to label negative consequences as 'unintended', but they may be designed into the system – features, not bugs – or, some argue, deliberately unanticipated to avoid accountability.[562]

The problem of bias in algorithms continues to receive widespread attention. On November 7, 2019, tech entrepreneur David Hannemeier Hanssonn[563] posted a series of accusations on Twitter that Apple Card's "black box algorithm" is discriminatory against women.[564] Soon after, Apple's own co-founder Steve Wozniak replied to Hansson's tweet accusing the same regarding his wife's credit line.[565] They both complained that the algorithm

---

Traditional Knowledge (TK) Labels are an educational and informational digital marker created by the Local Contexts initiative to address the specific intellectual property needs of Native, First Nations, Aboriginal and Indigenous peoples with regard to the extensive collections of cultural heritage materials currently held within museums, archives, libraries, and private collections. Indigenous communities use TK Labels to identify and clarify community-specific access protocols associated with the materials and convey important information such as guidelines for proper use and responsible stewardship of cultural heritage materials. TK Labels provide information to help users of traditional cultural knowledge from outside the creators' community understand the importance and significance of this material, even when it is in the public domain." https://www.loc.gov/collections/ancestral-voices/about-this-collection/rights-and-access/

[556] Library of Congress, Digital Collection, Ancestral Voices, https://www.loc.gov/collections/ancestral-voices/about-this-collection/rights-and-access/ .
[557] Ancestral Voices. https://ancestralvoices.co.uk/
[558] https://www.gida-global.org/
[559] GO FAIR, FAIR Principles, https://www.go-fair.org/fair-principles/ .
[560] https://www.gida-global.org/care
[561] Coad, Alex and Nightingale, Paul and Stilgoe, Jack and Vezzani, Antonio, The Dark Side of Innovation (August 1, 2020). Available at SSRN: https://ssrn.com/abstract=3702754 or http://dx.doi.org/10.2139/ssrn.3702754
[562] Stilgoe, J. (2020). Who's Driving Innovation? New Technologies and the Collaborative State. Palgrave
[563] https://dhh.dk/ Hansson is the creator of Ruby on Rails. The open-source web framework that he created in 2003. Some of the more famous include Github, Shopify, Airbnb, Square, Twitch, and Zendesk. He is also the cofounder of Basecamp, used by millions of people to organize work projects.
[564] David Heinemeier Hansson (@DHH), "The @AppleCard is such a fucking sexist program. My wife and I filed joint tax returns, live in a community-property state, and have been married for a long time. Yet Apple's black box algorithm thinks I deserve 20x the credit limit she does. No appeals work," Twitter, November 7, 2019, 12:34 p.m., https://twitter.com/dhh/status/1192540900393705474 .
[565] "Apple Co-Founder Steve Wozniak Says New Credit Card Discriminated Against His Wife," NBC News Now, uploaded November 12, 2019, YouTube video, 02:12, https://youtu.be/Htu6x4XhfQ0 . See also Sarah Myers West, "In the Outcry over the Apple Card, Bias Is a Feature, Not a Bug," Medium, November 22, 2019,

had granted them credit that was denied from their wives, even though they shared the same financial histories. The complaints resulted in investigations by both the Senate Committee on Finance and the New York State Department of Financial Services.[566] Apple is not singular in facing these challenges. Algorithmic discrimination exists across a wide spectrum of markets[567] and functions.[568] Algorithmic discrimination (also known as algorithmic bias) as an algorithm's differential treatment of consumers of equal quality (value or profitability to the organization) who differ only in group membership (e.g., race/ethnicity, gender, age, residential location, social class, etc.).[569]

Hansson blamed a sexist black-box algorithm, echoing and amplifying the work of numerous activists, journalists, researchers, and tech workers (like Hansson himself) who have been warning of the dangers of biased AI systems for at least a decade.[570] Hansson himself observed the gaslighting and denial of the issue in the responses to his tweet criticizing the biased Apple Card. He commented: "Every single poster questioning my wife's credit score, a man. Every single defense of Apple blaming G[oldman] S[achs], a man. Almost like men are over represented in the defense/justification of discrimination that doesn't affect them?"[571]

Algorithms control aspects of our everyday lives beyond money and information. They are concealed behind a veil of a code, which is often protected under trade secrecy law, and even when disclosed, their mathematical complexity make them impenetrable to most of us.

---

[footnote cont.] https://medium.com/@AINowInstitute/in-the-outcry-over-the-apple-card-bias-is-a-feature-not-a-bug-532a4c75cc9f .

[566] Sridhar Natarajan and Shahien Nasiripour, "Senator Wyden Says He's Looking into Claims of Apple Card Bias," Bloomberg, November 13, 2019:https://www.bloomberg.com/news/articles/2019-11-13/senator-wyden-says-he-s-looking-into-claims-of-apple-card-bias ; Linda A. Lacewell, New York Department of Financial Services, "Building a Fairer and More Inclusive Financial Services Industry for Everyone," Medium, November 10, 2019, https://medium.com/@nydfs/building-a-fairer-and-more-inclusive-financial-services-industry-for-everyone-917183dae954 .

[567] O'Neil Cathy (2016) Weapons of math destruction: How big data increases inequality and threatens democracy (Crown).

[568] Raghavan Manish, Barocas Solon, Kleinberg Jon, Levy Karen (2019) Mitigating Bias in Algorithmic Hiring: Evaluating Claims and Practices. arXiv:1906.09208 [cs] ArXiv:1906.09208.

[569] Ukanwa, Kalinda and Rust, Roland T., Algorithmic Discrimination in Service (May 29, 2021). USC Marshall School of Business Research Paper, Available at SSRN: https://ssrn.com/abstract=3654943 or http://dx.doi.org/10.2139/ssrn.3654943

[570] See AI Now Institute, "Gender, Race, and Power in AI: A Playlist," Medium, April 17, 2019, https://medium.com/@AINowInstitute/gender-race-and-power-in-ai-a-playlist-2d3a44e43d3b ; Joy Lisi Rankin , A People's History of Computing in the United States (Cambridge: Harvard University Press, 2018); Ruha Benjamin, Race After Technology: Abolitionist Tools for the New Jim Code (Medford, MA: Polity Press, 2019); Virginia Eubanks, Automating Inequality: How High-Tech Tools Profile , Police, and Punish the Poor (New York: St. Martin's Press, 2017); Mar Hicks, "Hacking the Cis-tem", IEEE Annals of the History of Computing , 41 no. 1 (Jan.-Mar 2019): 20-33. https://ieeexplore.ieee.org/document/8634814 ; Safiya Noble, Algorithms of Oppression (New York, NY: NYU Press, 2018).

[571] David Heinemeier Hansson (@DHH), Twitter, November 8, 2019, 2:08 p.m., https://twitter.com/dhh/status/1192926909794902016 .

Microsoft's latest report to shareholders flagged reputational harm due to biased AI systems among the company's risks. [572]Although the industry is taking some steps, such as corporate AI ethics, those solutions are inadequate. Eric Schmidt, Alphabet's former CEO acknowledged the bias.[573]

Big data and AI are revolutionizing the ways in which firms, governments, and employers classify individuals. Insurers, for instance, increasingly set premiums based on complex algorithms that process massive amounts of data to predict future claims.[574] Prospective employers deploy AI and big data to decide which applicants to interview or hire.[575] Various actors within the criminal justice system—ranging from police departments to judges—now use predictive analytics to guide their decisionmaking.[576]

This big data revolution raises numerous complex challenges for anti-discrimination regimes.[577] Perhaps most obviously, improperly-designed algorithms or errant data can disproportionately harm discrete subsets of the population.[578] Even correctly programmed algorithms armed with accurate data can reinforce past discriminatory patterns.[579]

---

[572] Microsoft Corporation, United States Securities and Exchange Commission Form 10-K: Annual Report for the Fiscal Year Ended June 30, 2019, at 20-21 https://view.officeapps.live.com/op/view.aspx?src=https://c.s-microsoft.com/en-us/CMSFiles/MSFT_FY19Q4_10K.docx?version=0a785912-1d8b-1ee0-f8d8-63f2fb7a5f00 . ("Issues in the use of AI in our offerings may result in reputational harm or liability. We are building AI into many of our offerings and we expect this element of our business to grow. We envision a future in which AI operating in our devices, applications, and the cloud helps our customers be more productive in their work and personal lives. As with many disruptive innovations, AI presents risks and challenges that could affect its adoption, and therefore our business. AI algorithms may be flawed. Datasets may be insufficient or contain biased information. Inappropriate or controversial data practices by Microsoft or others could impair the acceptance of AI solutions. These deficiencies could undermine the decisions, predictions, or analysis AI applications produce, subjecting us to competitive harm, legal liability, and brand or reputational harm. Some AI scenarios present ethical issues. If we enable or offer AI solutions that are controversial because of their impact on human rights, privacy, employment, or other social issues, we may experience brand or reputational harm."

[573] Marietje Schaake and Eric Schmidt, "Keynote: Regulating Big Tech," Stanford University HAI 2019 Fall Conference, uploaded on November 13, 2019, YouTube video, at 16:00-16:11 , https://youtu.be/uXpEYM0F5gA .(16:00 We know the data has bias in it. 16:02 You don't need to yell at as a new fact, right? 16:06 'Cause humans have bias in them. 16:08 Our systems have bias in them. 16:09 It's like not a shock. 16:11 The question is, what do we do about it?)

[574] See Rick Swedloff, Risk Classification's Big Data (R)evolution, 21 CONN. INS. L.J. 339, 340–44 (2014); Herb Weisbaum, Data Mining Is Now Used to Set Insurance Rates; Critics Cry Foul, CNBC (Apr. 16, 2014, 11:29 AM), https://www.cnbc.com/2014/04/16/data-mining-is-nowused-to-set-insurance-rates-critics-cry-fowl.html ; see also Ray Lehmann, Why 'Big Data' Will Force Insurance Companies to Think Hard About Race, INS. J. (Mar. 27, 2018), https://www.insurancejournal.com/blogs/right-street/2018/03/27/484530.htm ("According to a 2015 survey conducted by Willis Towers Watson, 42 percent of executives from the property and casualty insurance industry said they were already using big data in pricing, underwriting and risk selection, and 77 percent said they expected to do so within two years.").

[575] See Pauline T. Kim, Data-Driven Discrimination at Work, 58 WM. & MARY L. REV. 857,860 (2017) ("Employers are increasingly relying on data analytic tools to make personnel decisions . . . .").

[576] See Aziz Z. Huq, Racial Equity in Algorithmic Criminal Justice, 68 DUKE L.J. 1043, 1068–76 (2019); Elizabeth E. Joh, Policing by Numbers: Big Data and the Fourth Amendment, 89 WASH. L. REV. 35, 42–55 (2014); See Matthew Adam Bruckner, The Promise and Perils of Algorithmic Lenders' Use of Big Data, 93 CHI.-KENT L. REV. 3, 11–15 (2018); Christopher K. Odinet, Consumer Bitcredit and Fintech Lending, 69 ALA. L. REV. 781, 802–04 (2018). They are also fundamentally changing the business of financial advice, offering personalized AI assistants that promise to improve consumer decision-making. See Rory Van Loo, Digital Market Perfection, 117 MICH. L. REV. 815, 862–63, 878–79 (2019).

[577] See generally CATHY O'NEIL, WEAPONS OF MATH DESTRUCTION: HOW BIG DATA INCREASES INEQUALITY AND THREATENS DEMOCRACY (2016) (discussing how algorithms used in society can perpetuate discrimination, in part through perpetuation of disadvantage); Solon Barocas & Andrew D. Selbst, Big Data's Disparate Impact, 104 CALIF. L. REV. 671, 682 (2016) (discussing how data is often imperfect and therefore algorithms inherit the prejudice of the original decision makers); Kate Crawford & Jason Schultz, Big Data and Due Process: Toward a Framework to Redress Predictive Privacy Harms, 55 B.C. L. REV. 93, 99–101 (2014) (discussing ways that predictive analytic tools can perpetuate discriminatory practices).

There is the risk that modern AIs will result in "proxy discrimination." This is the risk that modern algorithms will result in "proxy discrimination." Proxy discrimination is a particularly pernicious subset of disparate impact. Like all forms of disparate impact, it involves a facially neutral practice that disproportionately harms members of a protected class. But a practice producing a disparate impact only amounts to proxy discrimination when the usefulness to the discriminator of the facially neutral practice derives, at least in part, from the very fact that it produces a disparate impact. Historically, this occurred when a firm intentionally sought to discriminate against members of a protected class by relying on a proxy for class membership, such as zip code.[580]

**AI Impacts on Energy, the Environment, and Raw Materials**

The consumption of Electrical and Electronic Equipment (EEE) is strongly linked to widespread global economic development. EEE has become indispensable in modern societies and is enhancing living standards, but its production and usage can be very resource demanding, as such also illustrates a counter to that very improvement in living standards. Higher levels of disposable incomes, growing urbanization and mobility, and further industrialization in some parts of the world are leading to growing amounts of EEE. On average, the total weight (excluding photovoltaic panels) of global EEE consumption increases annually by 2.5 million metric tons (Mt). After its use, EEE is disposed of, generating a waste stream that contains hazardous and valuable materials. This waste stream is referred to as e-waste, or Waste Electrical and Electronic Equipment (WEEE), a term used mainly in Europe.[581]

On September 20, 2019, workers from 12 tech companies joined the global climate strike.[582] They highlighted tech's role in climate change and demanded "zero carbon emissions by 2030, zero contracts with fossil fuel companies, zero funding of climate denial lobbying or other efforts, and zero harm to climate refugees and frontline communities."[583]

An important issue that has not received much attention is whether the development of AI is environmentally sustainable[584]: Like all computing systems, AI systems produce waste that is very hard to recycle and they consume vast amounts of energy, especially for the training of machine learning systems (and even for the "mining" of cryptocurrency). Again, it appears that some actors in this space offload such costs to the general society. [585]

---

[578] See Danielle Keats Citron & Frank Pasquale, The Scored Society: Due Process for Automated Predictions, 89 WASH. L. REV. 1, 4–5 (2014) (describing how human beings programming automated systems can lead to inaccurate results because the source code, predictive algorithms and datasets may contain human biases that have a disparate impact on certain groups).

[579] See Stephanie Bornstein, Antidiscriminatory Algorithms, 70 ALA. L. REV. 519, 524–28 (2018) (arguing that "facially neutral" algorithms producing unequal outcomes should be challenged as violating Title VII's stereotype theory of liability).

[580] Prince, Anya and Schwarcz, Daniel B., Proxy Discrimination in the Age of Artificial Intelligence and Big Data (August 5, 2019). 105 Iowa Law Review 1257 (2020), Available at SSRN: https://ssrn.com/abstract=3347959

[581] The Global E-waste Monitor 2020 – ITU, https://www.itu.int/myitu/-/media/Publications/2020-Publications/EN---Global-E-waste-Monitor-2020.pdf

[582] Louise Matsakisis, "Thousands of Tech Workers Join Global Climate Change Strike," Wired, September 20, 2019, https://www.wired.com/story/tech-workers-global-climate-change-strike/ .

[583] Tech Workers Coalition, "There's a Climate Crisis and Tech Workers Are Walking Out," https://techworkerscoalition.org/climate-strike/ .

[584] Frankel, Boris, Fictions of Sustainability The Politics of Growth and Post-Capitalist Futures (November 19, 2018). Fictions of Sustainability The Politics of Growth and Post-Capitalist Futures, Greenmeadows, Melbourne, 2018., Available at SSRN: https://ssrn.com/abstract=3287481

[585] Emma Strubell Ananya Ganesh Andrew McCallum, Energy and Policy Considerations for Deep Learning in NLP, June 5, 2019, https://arxiv.org/pdf/1906.02243.pdf Strubell's study found that training one language model with a particular type of "neural architecture search" (NAS) method would have produced the equivalent of 626,155 pounds

This might have surprised some people, as tech's contribution to the climate crisis is rarely discussed. Indeed, industry marketing often highlights green policies, sustainability initiatives, and futures in which AI and other advanced technologies provide solutions to climate problems. However, the tech sector is a significant contributor to environmental harms.[586] The tech industry faces criticism for the significant energy used to power its computing infrastructure. As a whole, the industry's energy dependence is on an exponential trajectory, with best estimates showing that its 2020 global footprint amounts to 3.0–3.6 percent of global greenhouse emissions, more than double what the sector produced in 2007.[587] This is comparable to that of the aviation industry,[588] and larger than that of Japan, which is the fifth biggest polluter in the world.[589]

In the worst-case scenario, this footprint could increase to 14 percent of global emissions by 2040. In response, the major tech companies have made data centers more efficient, and have worked to ensure they're powered at least in part by renewable energy—changes they're not shy about, announcing them with marketing blasts and much public fanfare.[590] These changes are a step in the right direction, but don't come close to tackling the problem. Most large tech companies continue to rely heavily on fossil fuels, and when they do commit to efficiency goals, these are most often not open to public scrutiny and validation.[591]

The AI industry is a significant source of further growth in greenhouse emissions. With the emergence of 5G networks aiming to realize the "internet of things," the increased acceleration of data collection and traffic is already underway.[592]

In addition to 5G antennas consuming far more energy than their 4G predecessors,[593] the introduction of 5G is poised to fuel a proliferation of carbon-intensive AI technologies, including autonomous driving[594] and telerobotic surgery.[595]

---

(284 metric tons) of carbon dioxide—about the lifetime output of five average American cars. Training a version of Google's language model, BERT, which underpins the company's search engine, produced 1,438 pounds of CO2 equivalent in Strubell's estimate—nearly the same as a round-trip flight between New York City and San Francisco. These numbers should be viewed as minimums, the cost of training a model one time through. In practice, models are trained and retrained many times over during research and development.

[586] Roel Dobbe and Meredith Whittaker, "AI and Climate Change: How They're Connected, and What We Can Do about It," AI Now Institute, Medium, October 17, 2019, https://medium.com/@AINowInstitute/ai-and-climate-change-how-theyre-connected-and-what-we-can-doabout-it-6aa8d0f5b32c .

[587] Lotfi Belkhir and Ahmed Elmeligi, "Assessing ICT Global Emissions Footprint: Trends to 2040 & Recommendations," Journal of Cleaner Production 177 (March 10, 2018): 448–63, https://doi.org/10.1016/j.jclepro.2017.12.239 .

[588] Air Transport Action Group. "Facts & Figures." https://www.atag.org/facts-figures.html .

[589] Wikipedia, s.v. "List of Countries by Greenhouse Gas Emissions," https://en.wikipedia.org/w/index.php?title=List_of_countries_by_greenhouse_gas_emissions&oldid=925976447 .

[590] Google Sustainability, "100% Renewable Is Just the Beginning," https://sustainability.google/projects/announcement-100 ; Microsoft, "AI for Earth," https://www.microsoft.com/en-us/ai/ai-for-earth .

[591] Belkhir and Elmeligi, "Assessing ICT Global Emissions Footprint: Trends to 2040 & Recommendations," Journal of Cleaner Production Volume 177, 10 March 2018, Pages 448-463 https://doi.org/10.1016/j.jclepro.2017.12.239 . See also Gary Cook et al., "Clicking Clean: Who Is Winning the Race to Build a Green Internet?," Greenpeace, January 2017, http://www.clickclean.org/international/en / .

[592] Mike Hazas, Janine Morley, Oliver Bates, and Adrian Friday, "Are There Limits to Growth in Data Traffic?: On Time Use, Data Generation and Speed," Proceedings of the Second Workshop on Computing Within Limits (2016) 14:1–14:5, https://doi.org/10.1145/2926676.2926690 .

[593] Energy Realpolitik, "What 5G Means for Energy," Council on Foreign Relations, https://www.cfr.org/blog/what-5g-means-energy .

[594] Mary-Ann Russon, "Will 5G Be Necessary for Self-Driving Cars?," BBC News, https://www.bbc.com/news/business-45048264 .

[595] Anthony Cuthbertson, "Surgeon Performs World's First Remote Operation Using '5G Surgery' on Animal in China" The Independent , January 17, 2019,

A core contributor to the AI field's growing carbon footprint is a dominant belief that "bigger is better." In other words, AI models that leverage massive computational resources to consume larger training datasets are assumed to be inherently "better" and more accurate.[596] While this narrative is inherently flawed,[597] its assumptions drive the use of increased computation in the development of AI models across the industry.

In 2018, researchers Dario Amodei and Danny Hernandez at OpenAI reported that "[s]ince 2012, the amount of [computation] used in the largest AI training runs has been increasing exponentially with a 3.4 month doubling time (by comparison, Moore's Law had an 18 month doubling period)."[598] Their observations show developers "repeatedly finding ways to use more chips in parallel, and . . . willing to pay the economic cost of doing so."

As AI relies on more computers, its carbon footprint increases, with significant consequences. A recent study from the University of Massachusetts, Amherst estimated the carbon footprint of training a large natural-language processing model. Emma Strubell and her coauthors reported that training just one AI model produced 300,000 kilograms of carbon dioxide emissions.[599]

Adding to their already sizeable environmental impact, big AI companies are aggressively marketing their (carbon-intensive) AI services to oil and gas companies, offering to help optimize and accelerate oil production and resource extraction. Amazon is luring potential customers in the oil and gas industry[600] with programs like "Predicting the Next Oil Field in Seconds with Machine Learning."[601] Microsoft held an event called "Empowering Oil & Gas with AI,"[602] and Google Cloud has its own energy vertical dedicated to working with fossil fuel companies.[603] And C3 IoT,[604] an AI company originally created to facilitate the transition to a society fueled by renewable energy, now helps large oil and gas companies, including Royal Dutch Shell, Baker Hughes, and Engie, to expedite their extraction of fossil fuel.[605]

A recent article in Logic points out that oil and gas account for 30 percent of the total addressable market, making "the success of Big Oil, and the production of fossil fuels . . . key to winning the cloud race."[606] Recently, the Guardian examined the role of Big Tech in sustaining the market for fossil fuel, illuminating the massive

---

amounts of money tech companies invest in organizations that actively campaign against climate legislation, and promote climate change denial.[607]

When researchers and policymakers attempt to account for tech's climate footprint, it is immediately clear how little information is available. They are left to rely on voluntary company disclosures, without access to the information they would need to make a thorough accounting of tech's true energy use. There is very little public data available, and few incentives for tech companies to release it.[608] Without the information necessary to reach robust conclusions, researchers estimated 2018 data-center energy consumption using data from 2008.[609]

It was all they had to work with, even though, over the past ten years, both the scale of computation and the technologies powering it have changed radically. The authors of a Greenpeace report make similar observations, stating that while efficiency metrics have been eagerly adopted by the industry, "very few companies report under newer metrics . . . that could shed any light on the basic question: how much dirty energy is being used, and which companies are choosing clean energy to power the cloud?"[610]

The unwillingness of cloud providers to provide customers with insight into the energy use of procured services forms a critical barrier to meaningful carbon accounting across all sectors and organizations that rely on digital technology.[611]

Kate Crawford writes in her book, **Atlas of AI: Power, Politics, and the Planetary Costs of Artificial Intelligence,**[612] about the toxic e-waste from our devices that is exported to Malaysia, Ghana, or Pakistan. And with her usual brilliant insight, she goes deeper in her examination of how AI extracts raw materials — both literal (lithium for batteries in AI-driven devices, including electric cars) and virtual (the personal data of billions of people). In a recent interview, she said : "If you look at the way artificial intelligence is represented, it's so commonly understood in this highly abstract mathematical way. If you do an image search right now in Google for "AI," what you get back is just reams and reams of pictures of blue numbers and tunnels of code and white robots. This is the way that we see AI represented: as immaterial, as sort of floating in the cloud. It was incredibly important to me to look at that in a much more grounded way. What are the material consequences of these systems? To do that, you really have to go there.[613] Jeff Bezos has now created a space company, which is called Blue Origin. And it was one of the locations that I visited — to photograph this reusable rocket base in the middle of West Texas. And for me to see that the billions of dollars generated by AI companies and by tech billionaires is now being redirected into a commercialized space race, this idea that now we've made this money, we can abandon the planet as sort of a discarded and useless object … those trajectories and those ideologies run deep.[614]

On how AI does not just build on already existing histories of bias and inequality, it actively constructs them, she says: "When you have these huge ambitions to capture and contain the entire world, it is driven by this extractive impulse to get as much data as possible, at any cost. We've seen the downside of that. All of those collections of data come with histories, they can't be separated from them. And some of those histories contain very problematic structural forms of inequality — racism, sexism, classism — and that then creates the systems

---

we use in the future. [615] One of the really horrifying things for me as a researcher was looking at the ways in which these systems that were primarily designed for intelligence agencies — that were extralegal by design — have filtered down to the municipal level, that are literally being used by local police departments that are connecting to people's Amazon Ring cameras on the front of their houses. These sites of data ingestion are then being fed into the engines of deportation. Those stories need to be told far more often: As we subscribe to new forms of tracking and monitoring our own homes and our own bodies, we could actually be providing more data to precisely those incredibly unjust systems. We're slowly sleepwalking into significant social and political change without having those states laid out for us — in many ways because they are intentionally hidden."[616]

On the role of universities in fueling the growth of AI technology that has outpaced AI ethics, she argues: "Universities are incredibly important here, because that's where we train people to build technical systems. But there's a problem, which is that traditionally computer science and engineering have been seen as disciplines that don't really engage with human subjects. They don't go through ethical review. They don't do training on what are the larger sociological implications of these systems. … But now, things that were previously very much in theory and built in labs are touching the lives of billions of people every day. And we haven't caught up. In many ways, universities are working with these older siloed approaches of seeing the computer sciences as a mathematical discipline that's at arm's length from human bodies. When the reverse is actually the case."[617]

On the problem with large data sets collected from the internet without our knowledge, her opinion is: " If there's an original sin of the field, it's this moment when the idea of just harvesting the entire internet — taking people's photos, taking people's texts, taking their responses to each other — and seeing it as an aggregate infrastructure that had no specific histories or stories or intimacies or vulnerabilities contained within it. To strip it of all that, and say, this is just "raw material" — with very much scare quotes around that — to drive large-scale systems of prediction and optimization. That has brought us to this point, where I think we should be asking much harder questions of those data sets, not only because of their origins, but because of the way in which they bring, smuggling in, a worldview that is so rarely questioned — and is producing some very serious harms."[618]

Crawford believes: "So many of the books that have been written about artificial intelligence really just talk about very narrow technical achievements. And sometimes they write about the great men of AI, but that's really all we've had in terms of really contending with what artificial intelligence is. I think it's produced this very skewed understanding of artificial intelligence as purely technical systems that are somehow objective and neutral, and—as Stuart Russell and Peter Norvig say in their textbook—as intelligent agents that make the best decision of any possible action. I wanted to do something very different: to really understand how artificial intelligence is made in the broadest sense. This means looking at the natural resources that drive it, the energy that it consumes, the hidden labor all along the supply chain, and the vast amounts of data that are extracted from every platform and device that we use every day.[619] In doing that, I wanted to really open up this understanding of AI as neither artificial nor intelligent. It's the opposite of artificial. It comes from the most material parts of the Earth's crust and from human bodies laboring, and from all of the artifacts that we produce and say and photograph every day. Neither is it intelligent. I think there's this great original sin in the field, where people assumed that computers are somehow like human brains and if we just train them like children, they will slowly grow into these supernatural beings. That's something that I think is really problematic—that we've bought this idea of intelligence when in actual fact, we're just looking at forms of statistical analysis at scale that have as many problems as the data that it's given. I'd say one of the turning points for me was back in 2016, when I started a project called "Anatomy of an AI system" with Vladan Joler. We met at a conference specifically about voice-enabled AI, and we were trying to effectively draw what it takes to make an Amazon Echo work. What are the components? How does it extract data? What are the layers in the data pipeline? We realized, well—actually, to understand that, you have to understand where the components come from. Where did the chips get produced? Where are the mines? Where does it get smelted? Where are the logistical and supply chain paths? Finally, how do we trace the end of life of these devices? How do we look at where the e-waste tips are located in places like Malaysia and Ghana and Pakistan? What we ended up with was this very time-consuming two-year research project to really trace those material supply chains from cradle to grave."[620]

---

[615] Id.
[616] Id.
[617] Id.
[618] Id.
[619] Karen Hao, Stop talking about AI ethics. It's time to talk about power. We need to acknowledge both the politics and the physical impact that AI has on the planet, says scholar Kate Crawford in her new book, MIT Technology Review, (April 23, 2021)https://www.technologyreview.com/2021/04/23/1023549/kate-crawford-atlas-of-ai-review/

"When you start looking at AI systems on that bigger scale, and on that longer time horizon, you shift away from these very narrow accounts of "AI fairness" and "ethics" to saying: these are systems that produce profound and lasting geomorphic changes to our planet, as well as increase the forms of labor inequality that we already have in the world.[621] So that made me realize that I had to shift from an analysis of just one device, the Amazon Echo, to applying this sort of analytic to the entire industry. That to me was the big task, and that's why Atlas of AI took five years to write. There's such a need to actually see what these systems really cost us, because we so rarely do the work of actually understanding their true planetary implications.[622] The other thing I would say that's been a real inspiration is the growing field of scholars who are asking these bigger questions around labor, data, and inequality. Here I'm thinking of Ruha Benjamin, Safiya Noble, Mar Hicks, Julie Cohen, Meredith Broussard, Simone Brown—the list goes on. I see this as a contribution to that body of knowledge by bringing in perspectives that connect the environment, labor rights, and data protection.[623]

We've spent far too much time focusing on narrow tech fixes for AI systems and always centering technical responses and technical answers. Now we have to contend with the environmental footprint of the systems. We have to contend with the very real forms of labor exploitation that have been happening in the construction of these systems.[624] And we also are now starting to see the toxic legacy of what happens when you just rip out as much data off the internet as you can, and just call it ground truth. That kind of problematic framing of the world has produced so many harms, and as always, those harms have been felt most of all by communities who were already marginalized and not experiencing the benefits of those systems.[625] I hope it's going to be a lot harder to have these cul-de-sac conversations where terms like "ethics" and "AI for good" have been so completely denatured of any actual meaning. I hope it pulls aside the curtain and says, let's actually look at who's running the levers of these systems. That means shifting away from just focusing on things like ethical principles to talking about power."[626]

When asked how do we move away from this ethics framing, Crawford answered: "If there's been a real trap in the tech sector for the last decade, it's that the theory of change has always centered engineering. It's always been, "If there's a problem, there's a tech fix for it." And only recently are we starting to see that broaden out to "Oh, well, if there's a problem, then regulation can fix it. Policymakers have a role.[627] But I think we need to broaden that out even further. We have to say also: Where are the civil society groups, where are the activists, where are the advocates who are addressing issues of climate justice, labor rights, data protection? How do we include them in these discussions? How do we include affected communities? In other words, how do we make this a far deeper democratic conversation around how these systems are already influencing the lives of billions of people in primarily unaccountable ways that live outside of regulation and democratic oversight? In that sense, this book is trying to de-center tech and starting to ask bigger questions around: What sort of world do we want to live in? What sort of world do you want to live in? What kind of future do you dream of? I want to see the groups that have been doing the really hard work of addressing questions like climate justice and labor rights draw together, and realize that these previously quite separate fronts for social change and racial justice have really shared concerns and a shared ground on which to coordinate and to organize.[628] Because we're looking at a really short time horizon here. We're dealing with a planet that's already under severe strain. We're looking at a profound concentration of power into extraordinarily few hands. You'd really have to go back to the early days of the railways to see another industry that is so concentrated, and now you could even say that tech has overtaken that."[629]

---

[620] Id.
[621] Id.
[622] Id.
[623] Id.
[624] Id.
[625] Id.
[626] Id.
[627] Id.
[628] Id.

"So we have to contend with ways in which we can pluralize our societies and have greater forms of democratic accountability. And that is a collective-action problem. It's not an individual-choice problem. It's not like we choose the more ethical tech brand off the shelf. It's that we have to find ways to work together on these planetary-scale challenges."[630]

**AI Snake Oil and Fake Science**

Concerns about AI systems focus not only on the harms caused when they are deployed without accountability but also when systems with flawed scientific foundations are marketed to the public.[631] "Move fast and fake things." Researchers uncovered systems in wide deployment that purport to operationalize proven scientific theories, but in the end are little more than speculation.[632] This trend in AI development is a growing area of concern, especially as applied to facial- and affect-recognition technology.

**AI and Facial Recognition and Emotional/Affect Recognition**

Facial recognition systems offer a just a snapshot of the preview of the privacy issues that will emerge with the use of our data. With the benefit of rich databases of digital photographs available via social media, websites, driver's license registries, surveillance cameras, and many other sources, machine recognition of faces has progressed rapidly from fuzzy images of cats to rapid (though still imperfect) recognition of individual humans. Facial recognition systems are being deployed in cities and airports around America.[633]

The data trail we leave behind is how our "free" services are paid for—but we are not told about that data collection and the value of this new raw material, and we are manipulated into leaving ever more such data. For the "big 5" companies (Amazon, Google/Alphabet, Microsoft, Apple, Facebook), the main data-collection part of their business appears to be based on deception, exploiting human weaknesses, furthering procrastination, generating addiction, and manipulation [634] The primary focus of social media, gaming, and most of the Internet in this "surveillance economy" is to gain, maintain, and direct attention—and thus data supply. This surveillance and attention economy is sometimes called "surveillance capitalism". [635] These systems will often reveal facts about us that we ourselves wish to suppress or are not aware of: they know more about us than we know ourselves. Even just observing online behavior allows insights into our mental states. [636]

China's use of facial recognition as a tool of authoritarian control in Xinjiang and elsewhere has awakened opposition to the expansion of this technology and calls for a ban on the use of facial recognition. Owing to concerns over facial recognition, as noted above, the cities of Oakland, Berkeley, and San Francisco in California, as well as Brookline, Cambridge, Northampton, and Somerville in Massachusetts, have adopted bans on the technology. A bill passed in Virginia bans local law enforcement agencies from using facial recognition technology without prior legislative approval starting July 1, 2021. Even when such approval is given, the bill further requires local police agencies to have "exclusive control" over the facial recognition systems they use, preventing the use of Clearview AI[637] and other commercial Facial Recognition products. However, Virginia State Police and other state law enforcement agencies may continue to use facial recognition without legislative approval. California, New Hampshire, and Oregon all have enacted legislation banning use of facial recognition with police body cameras.[638]

---

[629] Id.
[630] Id.
[631] Id.
[632] Arvind Narayanan, "How to Recognize AI Snake Oil," Princeton University, Department of Computer Science, https://www.cs.princeton.edu/~arvindn/talks/MIT-STS-AI-snakeoil.pdf .
[633] AI Report 2019.
[634] Harris, Tristan, 2016, "How Technology Is Hijacking Your Mind—from a Magician and Google Design Ethicist", Thrive Global, 18 May 2016. https://medium.com/thrive-global/how-technology-hijacks-peoples-minds-from-a-magician-and-google-s-design-ethicist-56d62ef5edf3
[635] Zuboff, Shoshana, 2019, The Age of Surveillance Capitalism: The Fight for a Human Future at the New Frontier of Power, New York: Public Affairs.
[636] Burr, C., & Cristianini, N. (2019). Can Machines Read our Minds? Minds and Machines. https://doi.org/10.1007/s11023-019-09497-4

Affect recognition is an AI-driven technology that claims to be able to detect an individual's emotional state based on the use of computer-vision algorithms to analyze their facial microexpressions, tone of voice, or even their gait. It is rapidly being commercialized for a wide range of purposes—from attempts to identify the perfect employee[639] to assessing patient pain[640] to tracking which students are being attentive in class.[641] Yet despite the technology's broad application, research shows affect recognition is built on markedly shaky foundations.

The affect-recognition industry is undergoing a period of significant growth: some reports indicate that the emotion-detection and -recognition market was worth $12 billion in 2018, and by one enthusiastic estimate, the industry is projected to grow to over $90 billion by 2024.[642] These technologies are often layered on top of facial-recognition systems as a "value add."

For example, the company Kairos[643] is marketing video-analytics cameras that claim to detect faces and then classify them as feeling anger, fear, and sadness, along with collecting customer identity and demographic data. Kairos sells these products to casinos, restaurants, retail merchants, real estate brokers, and the hospitality industry, all with the promise that they will help those businesses see inside the emotional landscape of their patrons.[644]

In August 2019, Amazon claimed its Rekognition[645] facial recognition software could now assess fear in addition to seven other emotions. Though it declined to provide any details on how it is being used by customers, it indicated retail as a potential use case, illustrating how stores can feed live images of shoppers to detect emotional and demographic trends.[646]

Employment has also experienced a surge in the use of affect recognition, with companies like HireVue[647] and VCV[648] offering to screen job candidates for qualities like "grit" and to track how often they smile.[649] Call center programs Cogito[650] and Empath[651] use voice-analysis algorithms to monitor the reactions of customers and signal to call agents when they sound distressed.[652] Similar programs have been proposed as an assistive technology for people with autism,[653] while Boston-based company BrainCo[654] is creating headbands that purport to detect and quantify students' attention levels through brain-activity detection,[655] despite studies that outline significant risks associated with the deployment of emotional AI in the classroom.[656]

---

[637] https://clearview.ai/
[638] Cameron F. Kerry, Protecting privacy in an AI-driven world, (Feb. 10, 2020) available at https://www.brookings.edu/research/protecting-privacy-in-an-ai-driven-world/
[639] Drew Harwell, "Rights Group Files Federal Complaint against AI-Hiring Firm HireVue, Citing 'Unfair and Deceptive' Practices," Washington Post, November 6, 2019, https://www.washingtonpost.com/technology/2019/11/06/prominent-rights-group-files-federal-complaintagainst-ai-hiring-firm-hirevue-citing-unfair-deceptive-practices/ .
[640] Clarice Smith, "Facial Recognition Enters into Healthcare," Journal of AHIMA, September 4, 2018, https://journal.ahima.org/2018/09/04/facial-recognition-enters-into-healthcare/ .
[641] Jane Li, "A 'Brain-Reading' Headband for Students Is Too Much Even for Chinese Parents," Quartz, November 5, 2019, https://qz.com/1742279/a-mind-reading-headband-is-facing-backlash-in-china/ .
[642] Paul Sawers, "Realeyes Raises $12.4 Million to Help Brands Detect Emotion Using AI on Facial Expressions," VentureBeat, June 6, 2019, https://venturebeat.com/2019/06/06/realeyes-raises-12-4-million-to-help-brands-detect-emotion-using-ai-on-facial-expressions/ .
[643] https://www.kairos.com/
[644] Luana Pascu, "New Kairos Facial Recognition Camera Offers Customer Insights," Biometric Update, September 11, 2019, https://www.biometricupdate.com/201909/new-kairos-facial-recognition-camera-offers-customer-insights .
[645] https://aws.amazon.com/rekognition/?blog-cards.sort-by=item.additionalFields.createdDate&blog-cards.sort-order=desc
[646] Tom Simonite, "Amazon Says It Can Detect Fear on Your Face. Are You Scared?" Wired, August 18, 2019, https://www.wired.com/story/amazon-detect-fear-face-you-scared/ .
[647] https://www.hirevue.com/
[648] https://vcv.ai/
[649] Mike Butcher, "The Robot-Recruiter Is Coming — VCV's AI Will Read Your Face in a Job Interview," TechCrunch, April 23, 2019, https://techcrunch.com/2019/04/23/the-robot-recruiter-is-coming-vcvs-ai-will-

Affect-recognition software has also joined risk assessment as a tool in criminal justice. For example, police in the US and UK are using the eye-detection software Converus[657], which examines eye movements and changes in pupil size to flag potential deception.[658] Oxygen Forensics, which sells data-extraction tools to clients including the FBI, Interpol, London Metropolitan Police, and Hong Kong Customs, announced in July 2019 that it also added facial recognition, including emotion detection, to its software, which includes "analysis of videos and images captured by drones used to identify possible known terrorists."[659]

However, the recognition programs are not reliable. For example, ProPublica[660] reported that schools, prisons, banks, and hospitals have installed microphones from companies that carry software developed by the company Sound Intelligence[661], purporting to detect stress and aggression before violence erupts. The "aggression detector" was faulty, detecting rough, higher-pitched sounds like coughing as aggression.[662] Another study by researcher Dr. Lauren Rhue found systematic racial biases in two well-known emotion-recognition programs: when she ran Face++ and Microsoft's Face API on a dataset of 400 NBA player photos, she found that both systems assigned black players more negative emotional scores on average, no matter how much they smiled.[663]

There remains little to no evidence that these new affect-recognition products have any scientific validity. In February 2019, researchers at Berkeley found that in order to detect emotions with accuracy and high

---

read-your-face-in-a-job-interview/ .

[650] https://cogitocorp.com/
[651] https://www.webempath.com/
[652] Tom Simonite, "This Call May Be Monitored for Tone and Emotion," Wired, March 19, 2019, https://www.wired.com/story/this-call-may-be-monitored-for-tone-and-emotion/ ; Kyle Wiggers, "Empath's AI Detects Emotion from Your Voice," VentureBeat, September 8, 2019, https://venturebeat.com/2019/09/08/empaths-ai-measures-emotion-from-voice/ .
[653] Cade Metz, "Google Glass May Have an Afterlife as a Device to Teach Autistic Child," New York Times, July 17, 2019, https://www.nytimes.com/2019/07/17/technology/google-glass-device-treat-autism.html .
[654] https://www.brainco.tech/
[655] BrainCo Inc., "Harvard University-Backed Startup BrainCo Inc. Gets the Biggest Purchase Order in Brain Machine Interface (BMI) Industry," PR Newswire, May 18, 2017, https://www.prnewswire.com/news-releases/harvard-university-backed-startup-brainco-inc-gets-the-biggest-purchase-order-in-brain-machine-interface-bmi-industry-300460485.html .
[656] Andrew McStay, "Emotional AI and EdTech: Serving the Public Good?," Learning, Media and Technology, November 5, 2019, https://doi.org/10.1080/17439884.2020.1686016 .
[657] https://converus.com/
[658] Mark Harris, "An Eye-Scanning Lie Detector Is Forging a Dystopian Future," Wired , April 12, 2019, https://www.wired.com/story/eye-scanning-lie-detector-polygraph-forging-a-dystopian-future/ ; and Amit Katwala, "The Race to Create a Perfect Lie Detector – and the Dangers of Succeeding," Guardian, September 5, 2019, https://www.theguardian.com/technology/2019/sep/05/the-race-to-create-a-perfect-lie-detector-and-the-dangers-of-succeeding .
[659] "Detective 11.5," Oxygen Forensics, July 2019, https://www.oxygen-forensic.com/uploads/press_kit/OF_RN_11_5_web.pdf .
[660] https://www.propublica.org/
[661] https://www.soundintel.com/
[662] Jack Gillum and Jeff Kao, "Aggression Detectors: The Unproven, Invasive Surveillance Technology Schools Are Using to Monitor Students," ProPublica , June 25, 2019, https://features.propublica.org/aggression-detector/the-unproven-invasive-surveillance-technology-schools-are-using-to-monitor-students/ .
[663] Lauren Rhue, "Racial Influence on Automated Perceptions of Emotions," November 9, 2018, https://papers.ssrn.com/sol3/papers.cfm?abstract_id=3281765

agreement requires context beyond the face and body.[664] Researcher Ruben van de Ven makes this point in his exploration of affect recognition, citing the "Kuleshov Effect,[665] a film editing effect. It is a mental phenomenon by which viewers derive more meaning from the interaction of two sequential shots than from a single shot in isolation."[666] Others at the University of Southern California called for a pause in the use of some emotion analytics techniques at the 8th[667] International Conference on Affective Computing and Intelligent Interaction[668] this year. "'[T]his facial expression recognition technology is picking up on something — it's just not very well correlated with what people want to use it for. So they're just going to be making errors, and in some cases, those errors cause harm,'" said Professor Jonathan Gratch.[669]

A major review released this summer found that efforts to "read out" people's internal states from an analysis of facial movements alone, without considering context, are at best incomplete and at worst entirely lack validity.[670] After reviewing over a thousand studies on emotion expression, the authors concluded that, although these technologies claim to detect emotional state, they actually achieve a much more modest outcome: detecting facial movements. As the study shows, there is a substantial amount of variance in how people communicate their emotional state across cultures, situations, and even across people within a single situation. Moreover, the same combination of facial movements—a smile or a scowl, for instance—can express more than a single emotion. The authors conclude that "no matter how sophisticated the computational algorithms . . . it is premature to use this technology to reach conclusions about what people feel on the basis of their facial movements."[671]

Given the high-stakes contexts in which affect-recognition systems are being used and their rapid proliferation over the past several years, their scientific validity is an area in particular need of research and policy attention —especially when current scientific evidence suggests that claims being made about their efficacy don't hold up. In short, we need to scrutinize why entities are using faulty technology to make assessments about character on the basis of physical appearance in the first place. This is particularly concerning in contexts such as employment, education, and criminal justice.

**Datasets of Faces**

Following the release of several studies, there continues to be significant performance disparities in commercial facial-recognition products across intersectional demographic subgroups.[672] In response, some companies are trying to "diversify" datasets to reduce bias. For instance, computer-vision company Clarifai[673] revealed that it makes use of the profile photos from the dating website OkCupid[674] to build large and "diverse" datasets of faces.[675] Clarifai claims the company gave them explicit permission and access to the data, so it remains unclear to what extent such data brokering constitutes a legal privacy violation disproportionately affecting people of color. IBM undertook a similar endeavor after being audited, releasing its "Diversity in Faces" study, which included an "inclusive" dataset of faces from a wide variety of Flickr users.[676] Although most of the users whose images were harvested had given permissions under an open Creative Commons license,[677] enabling widespread Internet use, none of the people in the photos gave IBM permission, again raising serious legal and ethical concerns about such practices.[678]

---

[664] Zhimin Chen and David Whitney, "Tracking the Affective State of Unseen Persons," Proceedings of the National Academy of Sciences , February 5, 2019, https://www.pnas.org/content/pnas/early/2019/02/26/1812250116.full.pdf .

[665] https://en.wikipedia.org/wiki/Kuleshov_effect

[666] Ruben Van De Ven, "Choose How You Feel; You Have Seven Options," Institute of Network Cultures, January 25, 2017, https://networkcultures.org/longform/2017/01/25/choose-how-you-feel-you-have-seven-options/ .

[667] http://acii-conf.org/2019/

[668] https://www.acii-conf.net/2021/

[669] Jayne Williamson-Lee, "Amazon's A.I. Emotion-Recognition Software Confuses Expressions for Feelings," OneZero , Medium, October 28, 2019, https://onezero.medium.com/amazons-a-i-emotion-recognition-software-confuses-expressions-for-feelings-53e96007ca63 .

[670] Lisa Feldman Barrett, Ralph Adochs, and Stacy Marsella, "Emotional Expressions Reconsidered:Challenges to Inferring Emotion From Human Facial Movements," Psychological Science in the Public Interest 20, no. 1 (July 2019): 1–68, https://journals.sagepub.com/eprint/SAUES8UM69EN8TSMUGF9/full .

[671] Id.

[672] Steve Lohr, "Facial Recognition Is Accurate If You're A White Guy," New York Times , February 9 2018, https://www.nytimes.com/2018/02/09/technology/facial-recognition-race-artificial-intelligence.html ; and Natasha Singer, "Amazon Is Pushing Facial Technology That a Study Says Could Be Biased," New York Times , January 24, 2019, https://www.nytimes.com/2019/01/24/technology/amazon-facial-technology-study.html .

[673] https://www.clarifai.com/

The problematic practice of scraping online images to produce diverse datasets is not limited to industry alone. Researchers exposed similar methods used to collect faces for academic datasets.[679] Most notably, the DUKE MTMC dataset,[680] Brainwash dataset,[681] and others[682] were collected by setting up surveillance cameras at college campuses, detecting and cropping out the faces of unsuspecting students to add to their database.

Ultimately, simply "diversifying the dataset" is far from sufficient to quell concerns about the use of facial-recognition technology. In fact, the face datasets themselves are a collection of artifacts to uncover, the assemblage of which reveals a set of decisions that were made regarding whom to include and whom to omit, but more importantly whom to exploit. It will be essential to continue to tell these stories, and to begin to uncover and perhaps challenge our accepted practices in the field, and the problematic patterns they reveal.[683]

**AI and Health**

AI technologies today mediate people's experiences of health in many ways: from popular consumer-based technologies like Fitbits[684] and the Apple Watch, to automated diagnostic support systems in hospitals, to the use of predictive analytics on social-media platforms to predict self-harming behaviors. AI also plays a role in how health insurance companies generate health-risk scores and in the ways government agencies and healthcare organizations allocate medical resources.[685]

---

[674] https://www.okcupid.com/
[675] Cade Metz, "Facial Recognition Tech Is Growing Stronger, Thanks to Your Face," New York Times, July 13, 2019, https://www.nytimes.com/2019/07/13/technology/databases-faces-facial-recognition-technology.html .
[676] See "Diversity in Faces Dataset," IBM, https://www.research.ibm.com/artificial-intelligence/trusted-ai/diversity-in-faces/ ; see also Michele Merler, Nalini Ratha, Rogerio Feris, and John R. Smith, "Diversity in faces," arXiv:1901.10436 , January 29, 2019, https://arxiv.org/abs/1901.10436 .
[677] Bart Thomée and David A. Shamma, "The Ins and Outs of the Yahoo Flickr Creative Commons 100 Million Dataset," code.flickr.com, October 15, 2014, https://code.flickr.net/2014/10/15/the-ins-and-outs-of-the-yahoo-flickr-100-million-creative-commons-dataset/ .
[678] Olivia Solon, "Facial Recognition's 'Dirty Little Secret': Millions of Online Photos Scraped without Consent," NBC News, March 12, 2019, https://www.nbcnews.com/tech/internet/facial-recognition-s-dirty-little-secret-millions-online-photos-scraped-n981921 .
[679] Adam Harvey and Jules LaPlace, "MegaPixels: Origins, Ethics, and Privacy Implications of Publicly Available Face Recognition Image Datasets," April 18, 2019, https://megapixels.cc/about/ .
[680] Duke MTMC Dataset Analysis, 2016, https://megapixels.cc/datasets/duke_mtmc/ .
[681] Brainwash Dataset Analysis, 2015, https://megapixels.cc/datasets/brainwash/ .
[682] See, for example, Oxford Town Centre Dataset Analysis, 2009, https://megapixels.cc/datasets/oxford_town_centre/ ; and UnConstrained College Students Dataset Analysis, 2012–2013, https://megapixels.cc/datasets/uccs/ .
[683] Joy Buolamwini, "Response: Racial and Gender bias in Amazon Rekognition — Commercial AI System for Analyzing Faces," Medium, January 25, 2019, https://medium.com/@Joy.Buolamwini/response-racial-and-gender-bias-in-amazon-rekognition-commercial-ai-system-for-analyzing-faces-a289222eeced .
[684] Fitbit, for instance, is a fitness tracker that monitors steps and could provide insights, inter alia, on an individual's heart rate or quality of sleep. See Andrew Hilts et al., Every Step You Fake: A Comparative Analysis of Fitness Tracker Privacy and Security, OPEN EFFECT REPORT 3–6 (2016), https://openeffect.ca/reports/Every_Step_You_Fake.pdf. For a taxonomy of personal health monitors, see Scott R. Peppet, Regulating the Internet of Things: First Steps Toward Managing Discrimination, Privacy, Security, and Consent, 93 TEX. L. REV. 85, 98–99 (2014).

Much of this activity comes with the aim of improving people's health and well-being through increased personalization of health, new forms of engagement, and clinical efficiency, popularly characterizing AI in health as an example of "AI for good" and an opportunity to tackle global health challenges.[686] This appeals to concerns about information complexities of biomedicine, population-based health needs, and the rising costs of healthcare. However, as AI technologies have rapidly moved from controlled lab environments into real-life health contexts, new social concerns are also fast emerging.[687]

**The Expanding Scale of Algorithmic Health Infrastructures**

Advances in machine learning techniques and cloud-computing resources have made it possible to classify and analyze large amounts of medical data, allowing the automated and accurate detection of conditions such as diabetic retinopathy and forms of skin cancer in medical settings.[688] Technology companies have been analyzing everyday experiences, e.g., going for a walk, food shopping, sleeping, and menstruating to make inferences and predictions about people's health behavior and status.[689]

While such developments may offer future positive health benefits, little empirical research has been published about how AI will impact patient health outcomes or experiences of care. Furthermore, the data- and cloud-computing resources required for training models to AI health systems have created troubling new opportunities, expanding what counts as "health data," but also the boundaries of healthcare. The scope and scale of these new "algorithmic health infrastructures"[690] give rise to a number of social, economic, and political concerns.[691]

---

[685] See Eric Topol, Deep Medicine, (New York: Basic Books, 2019). https://dl.acm.org/doi/book/10.5555/3350442

[686] See, for example, "Artificial Intelligence for Health," ITU, https://aiforgood.itu.int/ai4health/#about . For a critical take on the "AI for good" narrative, see Mark Latonero, "AI for Good Is Often Bad," Wired, November 18, 2019, https://www.wired.com/story/opinion-ai-for-good-is-often-bad/ .

[687] AI Now Report 2019.

[688] Joseph, Richard and Chauhan, Sony and Chichria, Karishma and Bhatia, Tanishq and Thakur, Hitesh, Detection of Hypertension Retinopathy and Diabetes Using Machine Learning (June 26, 2020). Proceedings of the International Conference on Recent Advances in Computational Techniques (IC-RACT) 2020, Available at SSRN: https://ssrn.com/abstract=3696052 or http://dx.doi.org/10.2139/ssrn.3696052 See also Andrzej Grzybowski et al. "Artificial Intelligence for Diabetic Retinopathy Screening: A Review ," Eye, September 5, 2019 , https://doi.org/10.1038/s41433-019-0566-0 .

[689] Mason Marks, "Tech Companies Are Using AI to Mine Our Digital Traces," STAT, September 17, 2019, https://www.statnews.com/2019/09/17/digital-traces-tech-companies-artificial-intelligence/ .

[690] Jean-Christophe Plantin, Carl Lagoze, Paul N. Edwards, and Christian Sandvig, "Infrastructure Studies Meet Platform Studies in the Age of Google and Facebook," New Media & Society 20, no. 1 (January 2018): 293–310, https://doi.org/10.1177/1461444816661553 ; and Paul N. Edwards, "We Have Been Assimilated: Some Principles for Thinking About Algorithmic Systems," in Living with Monsters? Social Implications of Algorithmic Phenomena, Hybrid Agency, and the Performativity of Technology: IFIP WG 8.2 Working Conference on the Interaction of Information Systems and the Organization, IS&O 2018, San Francisco, CA, USA, December 11–12, 2018, Proceedings , eds. Ulrike Schultze, Margunn Aanestad, Magnus Mähring,Carsten Østerlund, Kai Riemer (Cham, Switzerland: Springer International Publishing, 2018) .

[691] Marabelli, Marco and Newell, Sue and Page, Xinru, Algorithmic Decision-making in the US Healthcare Industry (October 8, 2018). Marabelli, M., Newell, S., Page, X., (2018). Algorithmic Decision-Making in the US Healthcare Industry, Presented at IFIP 8.2, San Francisco, CA. , Available at SSRN: https://ssrn.com/abstract=3262379 or http://dx.doi.org/10.2139/ssrn.3262379 ("n this research in progress we present the initial stage of a large ethnographic study at a healthcare network in the US. Our goal is to understand how healthcare organizations in the US use algorithms to improve efficiency (cost saving) and effectiveness (quality) of healthcare. Our preliminary findings illustrate that at the national level, algorithms might be detrimental to healthcare quality because they do not consider (and differentiate) contextual issues such as social and cultural (local) settings. At the practice (hospital/physician) level, they help managing the

The proliferation of corporate-clinical alliances for sharing data to train AI models illustrates these infrastructural impacts. The resulting commercial incentives and conflicts of interest have made ethical and legal issues around health data front-page news. Most recently, a whistleblower report alerted the public to serious privacy risks stemming from a partnership, known as Project Nightingale, between Google and Ascension,[692] one of the largest nonprofit health systems in the US. The report claimed that patient data transferred between Ascension and Google was not "de-identified."[693] Google helped migrate Ascension's infrastructure to their cloud environment, and in return received access to hundreds of thousands of privacy-protected patient medical records to use in developing AI solutions for Ascension and also to sell to other healthcare systems.[694]

Google, however, is not alone. Microsoft, IBM, Apple, Amazon, and Facebook, as well as a wide range of healthcare start-ups, have all made lucrative "data partnership" agreements with a wide range of healthcare organizations (including many university research hospitals and insurance companies)[695] to gain access to health data for the training and development of AI-driven health systems.[696] Several of these have resulted in federal probes and lawsuits around improper use of patient data.[697]

---

[] tradeoff between following national "best practices" and accommodating needs of special patients or particular situations, because hospital-based algorithms can be over-ridden by clinicians. We conclude that, while more data needs to be collected, a responsible use of algorithms requires their constant supervision and their application with respect to specific social and cultural settings.")

[692] Rob Copeland, "Google's 'Project Nightingale' Gathers Personal Health Data on Millions of Americans," Wall Street Journal , November 2019, https://www.wsj.com/articles/google-s-secret-project-nightingale-gathers-personal-health-data-on-millions-of-americans-11573496790 ; Anonymous, "I'm the Google Whistleblower. The Medical Data of Millions of Americans Is at Risk," Guardian , November 14, 2019, https://www.theguardian.com/commentisfree/2019/nov/14/im-the-google-whistleblower-the-medical-data-of-millions-of-americans-is-at-risk .

[693] In the US Health Insurance Portability and Accountability Act (HIPAA), personal health information (PHI) is categorized as data that is directly and uniquely tied to an individual, with examples including names, birth dates, and email addresses.De-identified data, therefore, indicates the absence of such categories from a potential EHR dataset.

[694] Tariq Shaukat, "Our Partnership with Ascension," Inside Google Cloud, November 11, 2019, https://cloud.google.com/blog/topics/inside-google-cloud/our-partnership-with-ascension .

[695] Gleiss, A., Kohlhagen, M. & Pousttchi, K. An apple a day – how the platform economy impacts value creation in the healthcare market. Electron Markets (2021). https://doi.org/10.1007/s12525-021-00467-2 ("The healthcare industry has been slow to adopt new technologies and practices. However, digital and data-enabled innovations diffuse the market, and the COVID-19 pandemic has recently emphasized the necessity of a fundamental digital transformation. Available research indicates the relevance of digital platforms in this process but has not studied their economic impact to date. In view of this research gap and the social and economic relevance of healthcare, we explore how digital platforms might affect value creation in this market with a particular focus on Google, Apple, Facebook, Amazon, and Microsoft (GAFAM). We rely on value network analyses to examine how GAFAM platforms introduce new value-creating roles and mechanisms in healthcare through their manifold products and services. Hereupon, we examine the GAFAM-impact on healthcare by scrutinizing the facilitators, activities, and effects. Our analyses show how GAFAM platforms multifacetedly untie conventional relationships and transform value creation structures in the healthcare market.")

[696] The cloud computing market size for healthcare is anticipated to reach nearly $30 billion by 2026. Google, Amazon, and Microsoft have all partnered with healthcare providers and payers to help migrate health information technology (HIT) infrastructure to cloud servers. Amazon Web Services now promises clients the ability to subscribe to third-party data, enabling healthcare professionals to aggregate data from clinical trials, while Microsoft has partnered with the insurance company Humana to provide cloud and AI resources, and also helps power Epic Systems' predictive analytics tools for EHRs.

[697] Daisuke Wakabayashi, "Google and the University of Chicago Are Sued over Data Sharing, New York Times , June 26, 2019, https://www.nytimes.com/2019/06/26/technology/google-university-chicago-data-sharing-lawsuit.html ; Rebecca Robbins and Casey Ross, "HSS to Probe Whether Project Nightingale Followed Privacy Law," STAT , November 13, 2019, https://www.statnews.com/2019/11/13/hhs-probe-google-ascension-project-nightingale/ ; Timothy Revell , "Google DeepMind NHS Data Deal Was 'Legally

However, even when current regulatory policies such as the Health Insurance Portability and Accountability Act (HIPAA)[698] are strictly followed, security and privacy vulnerabilities can exist within larger technology infrastructures, presenting serious challenges for the safe collection and use of Electronic Health Record (EHR) data. New research shows that it is possible to accurately link two different de-identified EHR datasets using computational methods, so as to create a more complete history of a patient without using any personal health information of the patient in question.[699]

Deidentification is a common tool used to protect medical privacy. [700]The HIPAA Privacy Rule is the dominant legal rule governing health data privacy[701] and likely the single most potent federal privacy regime in the United States. The HIPAA Privacy Rule only governs identifiable health information and includes a safe harbor under which information that has been stripped of 18 listed identifiers is defined as not identifiable.[702] What does that mean? Information custodians can remove those identifiers from health data and stop worrying about HIPAA (at least with respect to those data). Deidentification is a popular intervention outside the United States as well; the European Union's General Data Protection Regulation, for instance, does not cover anonymized data.[703]Artificial intelligence reduces the already-weak power of deidentification[704] to protect health privacy by making it easier to reidentify patients, either individually or at scale.[705]

Another recent research study showed that it is possible to create reconstructions of patients' faces using de-identified MRI images, which could then be identified using facial-recognition systems.[706] Similar concerns have prompted a lawsuit against the University of Chicago Medical Center and Google claiming that Google is "uniquely able to determine the identity of almost every medical record the university released" due to its expertise and resources in AI development.[707]The potential harm from misuse of these new health data capabilities is of grave concern, especially as AI health technologies continue to focus on predicting risks that

---

Inappropriate'," New Scientist , May 16, 2017, https://www.newscientist.com/article/2131256-google-deepmind-nhs-data-deal-was-legally-inappropriate/ .

[698] Health Insurance Portability and Accountability Act of 1996, Pub. L. No. 104-191, 110 Stat. 1936.

[699] Boris P. Hejblum et al, "Probabilistic Record Linkage of De-identified Research Datasets with Discrepancies Using Diagnosis Codes ," Scientific Data 6, 180298 (January 2019), https://doi.org/10.1038/sdata.2018.298 .

[700] Price II, William Nicholson, Problematic Interactions between AI and Health Privacy (March 3, 2021). Utah Law Review, Forthcoming, U of Michigan Public Law Research Paper No. 21-014, Available at SSRN: https://ssrn.com/abstract=3797161

[701] Standards for Privacy of Individually Identifiable Health Information, 45 C.F.R. pts. 160, 164 (2017). The HIPAA Privacy Rule is not the only health privacy law in the United States, of course; state laws may have more restrictive provisions on specific topics or in general, and other federal laws govern subsets of health privacy, such as genetic information. But HIPAA cuts across state lines and structures much discussion of health data privacy.

[702] 45 C.F.R. § 164.514(b)(2).

[703] Regulation (EU) 2016/679 of the European Parliament and of the Council of 27 April 2016 On the protection of natural persons with regard to the processing of personal data and on the free movement of such data, and repealing Directive 95/46/EC, 2016 O.J. (L 119) 1, 26, https://eurlex.europa.eu/legal-content/EN/TXT/PDF/?uri=CELEX:32016R0679

[704] See, e.g., Paul Ohm, Broken Promises of Privacy: Responding to the Surprising Failure of Anonymization, 57 UCLA L. REV. 1701, 1716–26 (2010) (noting ways to reidentify data); Yaniv Erlich & Arvind Narayanan, Routes for Breaching and Protecting Genetic Privacy, 15 NATURE REV. GENETICS 401, 409–16 (2014) (cataloging ways to reidentify genetic data).

[705] Luc Rocher, Julien M. Hendrickx, & Yves-Alexandre de Montjoye, Estimating the Success of Re-identifications in Incomplete Datasets Using Generative Models, 10 NATURE COMM. 3069, 2 (2019) (noting the probabilistic nature of reidentification attacks).

[706] Gina Kolata, "You Got a Brain Scan at the Hospital. Someday a Computer May Use It to Identify You," New York Times, October 23, 2019, https://www.nytimes.com/2019/10/23/health/brain-scans-personal-identity.html .

[707] Dinerstein v. Google, LLC , https://edelson.com/wp-content/uploads/2016/05/Dinerstein-Google-DKT-001-Complaint.pdf .

could impact healthcare access or stigmatize individuals, such as recent attempts to diagnose complex behavioral health conditions like depression and schizophrenia from social-media data.

In 2019, Google negotiated an arrangement with the University of Chicago Medical Center to gain access to all its medical records between the years 2009 and 2016. This amounted to the transfer of a vast quantity of HIPAA-protected personal health information, which Google and the University claimed was de-identified, without the explicit permission of the patients. One of the patients, Matt Dinerstein, represented by a Chicago law firm specializing in privacy and class-action cases, filed a class-action lawsuit [708] against Google and the University of Chicago.[709] Dinerstein claimed that because he visited the Medical Center with his cell phone in hand, Google would be able to re-identify him on the basis of location information, along with date and time information from the medical records.

In September 2020, the case [710] was dismissed by the federal district court due to the plaintiff's failure to show quantifiable harm or damages. Interestingly, and unlike the majority of predecessor privacy lawsuits, the court here found that plaintiff had sufficiently alleged injury to support standing on all but one of his claims—notwithstanding the absence of any tangible harm—but ultimately dismissed the case in large part for failure to state a claim that adequately alleged that defendants caused him economic damage.[711]

The *Dinerstein* decision serves as a reminder to potential plaintiffs of the difficulty of prevailing in privacy lawsuits. Standing may not be the death knell it once was, but so long as the harm remains intangible and statute does not confer a private right of action, potential plaintiffs will still face a hurdle in recovering for alleged injuries. That said, sharing de-identified health data may still cause problems under HIPAA which, as stated by the *Dinerstein* court, does not provide a private right of action, so the Office for Civil Rights within the U.S. Department of Health & Human Services would have to take up the fight.

**AI and Social Challenges for the Healthcare Community**

There is a lot of conversation and attention to AI ethics in healthcare.[712] Although mostly generated by physicians and medical ethicists in Europe and North America, these early efforts are important for better understanding the uses of AI systems in healthcare. For example, the European and North American Radiology

---

[708] Maryam Casbarro, Update from LitLand: Illinois Lawsuit Highlights Difficulty of True De-Identification, Davis Wright Tremaine LLP, (July 29, 2019) https://www.dwt.com/blogs/privacy--security-law-blog/2019/07/google-university-of-chicago-hippa-lawsuit

[709] Matt Dinerstein v. Google LLC, and the University of Chicago Medical Center. Case: 1:19-cv04311 Document #: 1 Filed: 06/26/19. https://www.courtlistener.com/recap/gov.uscourts.ilnd.366172/gov.uscourts.ilnd.366172.1.0_1.pdf\

[710] Dinerstein v. Google, LLC et al, No. 1:2019cv04311 - Document 85 (N.D. Ill. 2020). https://law.justia.com/cases/federal/district-courts/illinois/ilndce/1:2019cv04311/366172/85

[711] Maryam Casbarro, Update From LitLand: Dinerstein Decision Shows That Overcoming Standing in Privacy Cases Does Not Necessarily Create a Path to Victory, Davis Wright Tremaine LLP, (Oct. 5, 2020) https://www.lexology.com/library/detail.aspx?g=ba511a2e-19fd-4116-a276-0e416e875634

[712] Curfman, Gregory, United States v. Google - Implications of the Antitrust Lawsuit for Health Information (November 28, 2020). Available at SSRN: https://ssrn.com/abstract=3739122 or http://dx.doi.org/10.2139/ssrn.3739122; Gerke, Sara and Minssen, Timo and Cohen, I. Glenn, Ethical and Legal Challenges of Artificial Intelligence-Driven Healthcare (2020). This article was published in Artificial Intelligence in Healthcare, 1st edition, Adam Bohr, Kaveh Memarzadeh (eds.), ISBN: 9780128184387, Copyright Elsevier, 2020 , Available at SSRN: https://ssrn.com/abstract=3570129 or http://dx.doi.org/10.2139/ssrn.3570129 Irene Y. Chen, Peter Szolovits, and Marzyeh Ghassemi, "Can AI Help Reduce Disparities in General Medical and Mental Health Care?," AMA Journal of Ethics 21, no. 2 (February 1, 2019): 167–79, https://doi.org/10.1001/amajethics.2019.167 ; Jessica Morley and Luciano Floridi, "How to Design a Governable Digital Health Ecosystem," July 22, 2019, https://dx.doi.org/10.2139/ssrn.3424376 ; Linda Nordling, "A Fairer Way Forward for AI in Health Care," Nature 573 (September 25, 2019): S103–5, https://doi.org/10.1038/d41586-019-02872-2 ; Trishan Panch, Heather Mattie, and Leo Anthony Celi, "The 'Inconvenient Truth' about AI in Healthcare," npj Digital Medicine 2, no. 1 (August 16, 2019): 1–3, https://doi.org/10.1038/s41746-019-0155-4 .

Societies recently issued a statement that outlines key ethical issues for the field, including algorithmic and automation bias in relation to medical imaging.[713] Radiology is currently one of the medical specialties where AI systems are the most advanced. The statement openly acknowledges how clinicians are reckoning with the increased value and potential harms around health data used for AI systems: "AI has noticeably altered our perception of radiology data—their value, how to use them, and how they may be misused."[714]

These challenges include possible harms for patients, such as the potential for clinical decisions to be nudged or guided by AI systems in ways that do not necessarily bring people health benefits, but are in service to quality metric requirements or increased profit. Importantly, misuses also extend beyond the ethics of patient care to consider how AI technologies are reshaping medical organizations themselves (e.g., "radiologist and radiology departments will also be data" for healthcare administrators)[715] and the wider health domain by "blurring the line" between academic research and commercial AI uses of health data.[716]

The Academy of Medical Royal Colleges (UK) 2019 report, "Artificial Intelligence in Healthcare," pragmatically states: "Politicians and policymakers should avoid thinking that AI is going to solve all the problems the health and care systems across the UK are facing."[717] The American Medical Association adopted the policy "Augmented Intelligence in Health Care"[718] as a framework for thinking about AI in relation to multiple stakeholder concerns, which include the needs of physicians, patients, and the broader healthcare community.[719]

There have been demands for setting a more engaged agenda around AI and health. In 2019, Eric Topol, a physician and AI/ML researcher, questioned the promises of AI to fix systemic healthcare issues, such as clinician burnout, without the collective action and involvement of healthcare workers.[720] Physician organizing is needed not because doctors should fear being replaced by AI, but to ensure that AI benefits people's experiences of care. "The potential of A.I. to restore the human dimension in health care," Topol argues, "will depend on doctors stepping up to make their voices heard."[721]

More voices are urgently needed at the table, including the views of patient groups, family caregivers, community health workers, and nurses, etc., in order to better understand how AI technologies will impact diverse populations and health contexts. [722] We have seen how overly narrow approaches to AI in health have resulted in systems that failed to account for darker skin tones in medical imaging data,[723] and cancer treatment recommendations that could lead to racially disparate outcomes due to training data from predominantly white patients.[724]

---

[713] J. Raymond Geis et al., " Ethics of Artificial Intelligence in Radiology: Summary of the Joint European and North American Multisociety Statement ," Radiology , 293, no.2 (October 2019), https://doi.org/10.1148/radiol.2019191586 .

[714] Id. at 3.

[715] Id. at 4.

[716] Id. at 12.

[717] Academy of Medical Royal Colleges, "Artificial Intelligence in Healthcare," January 2019, https://www.aomrc.org.uk/wp-content/uploads/2019/01/Artificial_intelligence_in_healthcare_0119.pdf .

[718] American Medical Association, "Augmented intelligence in healthcare H-480.940,"

[719] Elliott Crigger and Christopher Khoury, "Making Policy on Augmented Intelligence in Health Care," AMA J Ethics 21, no.2 (February 2019): E188–191, https://journalofethics.ama-assn.org/article/making-policy-augmented-intelligence-health-care/2019-02

[720] Eric Topol, "Why Doctors Should Organize," New Yorker , August 5, 2019, https://www.newyorker.com/culture/annals-of-inquiry/why-doctors-should-organize .

[721] Id.

[722] AI Now Report 2019.

[723] Angela Lashbrook, "AI-Driven Dermatology Could Leave Dark-Skinned Patients Behind," Atlantic, August 16, 2018, https://www.theatlantic.com/health/archive/2018/08/machine-learning-dermatology-skin-color/567619/ .

[724] Dhruv Khullar, "A.I. Could Worsen Health Disparities," The New York Times , January 31, 2019, https://www.nytimes.com/2019/01/31/opinion/ai-bias-healthcare.html .

Importantly, algorithmic bias in health data cannot always be corrected by gathering more data, but requires understanding the social context of the health data that has already been collected.[725] Recently, Optum's[726] algorithm designed to identify "high-risk" patients in the US was based on the number of medical services a person used, but didn't account for the numerous socioeconomic reasons around the nonuse of needed health services, such as being underinsured or the inability to take time off from work.[727] With long histories of addressing such social complexities, research from fields like medical sociology and anthropology, nursing, human-computer interaction, and public health is needed to protect against the implementation of AI systems that (even when designed with good intentions) worsen health inequities.[728]

**AI and Interdisciplinary Analysis for Problem Solving**

As research and perspectives on the social implications of AI evolve, machine learning (ML) research communities are realizing the limitations of narrow "fairness" definitions and are shifting their focus to more impactful interventions and strategies, as well as fostering an increased openness toward active inclusion and engagement with other disciplines.

The 2018 AI Now Report, critically assessed the limitations of technical fixes to problems of fairness.[729] Since then, several convincing critiques have emerged that further explain how these approaches fundamentally distract from more urgent issues,[730] abstract away societal context,[731] are incommensurate with the political reality of how data scientists approach "problem formulation,"[732] and fail to address the hierarchical logic that produces unlawful discrimination.[733]

Responding to these criticisms, many technical researchers have turned to the use of so-called "causal" or "counterfactual" fairness methods.[734] Rather than relying on the correlations that most ML models use to make their predictions, these approaches aim to draw causal diagrams that explain how different types of data produce various outcomes. When analyzed for use of sensitive or protected categories, such as race or gender, these researchers seek to declare an ML "fair" if factors like race or gender do not causally influence the model's prediction. While the intentions behind this work may be commendable, there are still clear limitations to these approaches, primarily in their ability to address historical disparities and ongoing structural injustices.[735] As Lily Hu explains in the context of racial health disparities, "Whatever [level of] health Black people would have had in some convoluted counterfactual scenario is frankly irrelevant to the question of whether actually existing inequality is a matter of injustice—let alone what can be done to remedy it."[736]

---

[725] AI Now Report 2019.

[726] https://www.optum.com/

[727] Carolyn Y. Johnson, "Racial Bias in a Medical Algorithm Favors White Patients over Sicker Black Patients," Washington Post, October 24, 2019, https://www.washingtonpost.com/health/2019/10/24/racial-bias-medical-algorithm-favors-white-patientsover-sicker-black-patients/ . For original article, see Ziad Obermeyer, Brian Powers, Christine Vogeli, and Sendhil Mullainathan, "Dissecting Racial Bias in an Algorithm Used to Manage the Health of Populations," Science 366, no. 6464 (October 2019): 447–453.

[728] Tiffany C. Veinot, Hannah Mitchell, Jessica S. Ancker, "Good Intentions Are Not Enough: How Informatics Interventions Can Worsen Inequality," Journal of the American Medical Informatics Association 25, no. 8 (August 2018): 1080–1088, https://doi.org/10.1093/jamia/ocy052 ; Elizabeth Kaziunas, Michael S. Klinkman, and Mark S. Ackerman, "Precarious Interventions: Designing for Ecologies of Care," Proceedings of the ACM Human-Computer Interaction 3, CSCW, Article 113 (November 2019), https://doi.org/10.1145/3359215 .

[729] "AI Now Report 2018," https://ainowinstitute.org/AI_Now_2018_Report.pdf .

[730] Julia Powles and Helen Nissenbaum, "The Seductive Diversion of 'Solving' Bias in Artificial Intelligence," One Zero , Medium, December 7, 2018, https://onezero.medium.com/the-seductive-diversion-of-solving-bias-in-artificial-intelligence-890df5e5ef53 .

[731] Andrew D. Selbst, danah boyd, Sorelle Friedler, Suresh Venkatasubramanian, and Janet Vertesi, "Fairness and Abstraction in Sociotechnical Systems," November 7, 2018, https://papers.ssrn.com/abstract=3265913 .

[732] Samir Passi and Solon Barocas, "Problem Formulation and Fairness," Proceedings of the Conference on Fairness, Accountability, and Transparency , FAT* '19 (2019) , 39–48, https://doi.org/10.1145/3287560.3287567 .

[733] Anna Lauren Hoffman, "Where Fairness Fails: Data, Algorithms, and the Limits of Antidiscrimination Discourse," Information, Communication & Society 22, no. 7 (June 7, 2019): 900–915, https://doi.org/10.1080/1369118X.2019.1573912 .

[734] Counterfactual fairness is about making algorithm-led decisions fair by ensuring their outcomes are the same in the actual world and a 'counterfactual world' where an individual belongs to a different demographic. https://www.turing.ac.uk/research/research-projects/counterfactual-fairness

The value of these assessments hinges on how to define which individual characteristics should or should not factor into the algorithm's final prediction.[737] Such decisions are often themselves politically, culturally, and socially influenced, and the power imbalance between those making such determinations and those impacted remains clear and unaddressed.[738]

Techniques for interpreting and explaining ML systems have also gained popularity. However, they suffer from many of these same critiques, and have been shown to be fundamentally fragile and prone to manipulation,[739] and to ignore a long history of insights from the social sciences.[740]

As a result, some researchers have begun to push harder on the need for interdisciplinary approaches,[741] and for integrating lessons from social sciences and humanities into the practice of developing AI systems.[742] Some practical strategies have emerged, including methods to document the development of machine learning models to enforce some level of additional ethical reflection and reporting throughout the engineering process.[743]

Industry-led efforts by the Partnership on AI and IEEE are also attempting to consolidate these documentation proposals and to standardize reporting requirements across the industry.[744]

---

[735] Issa Kohler-Hausmann. "Eddie Murphy and the Dangers of Counterfactual Causal Thinking About Detecting Racial Discrimination," January 1, 2019, https://papers.ssrn.com/abstract=3050650 .

[736] Lily Hu, "Disparate Causes, Pt. II," Phenomenal World (blog), October 17, 2019, https://phenomenalworld.org/digital-ethics/disparate-causes-pt-ii .

[737] Christopher Jung, Michael Kearns, Seth Neel, Aaron Roth, Logan Stapleton, and Zhiwei Steven Wu, "Eliciting and Enforcing Subjective Individual Fairness," arXiv:1905.10660 [cs.LG] , (2019), https://arxiv.org/abs/1905.10660 .

[738] Anna Lauren Hoffmann (2019) Where fairness fails: data, algorithms, and the limits of antidiscrimination discourse, Information, Communication & Society, 22:7, 900-915, DOI: 10.1080/1369118X.2019.1573912

[739] Ann-Kathrin Dombrowski, Maximilian Alber, Christopher J. Anders, Marcel Ackermann, Klaus-Robert Müller, and Pan Kessel, "Explanations Can Be Manipulated and Geometry Is to Blame," arXiv:1906.07983 [Cs, Stat] , September 25, 2019, http://arxiv.org/abs/1906.07983 ; Amirata Ghorbani, Abubakar Abid, and James Zou, "Interpretation of Neural Networks Is Fragile," arXiv:1710.10547 [Cs, Stat] , November 6, 2018, http://arxiv.org/abs/1710.10547 ; Akshayvarun Subramanya, Vipin Pillai, and Hamed Pirsiavash, "Fooling Network Interpretation in Image Classification," arXiv:1812.02843 [Cs] , September 24, 2019, http://arxiv.org/abs/1812.02843 .

[740] Tim Miller, "Explanation in Artificial Intelligence: Insights from the Social Sciences," Artificial Intelligence 267 (2019): 1–38, https://arxiv.org/abs/1706.07269 .

[741] See, e.g., the ACM FAT* conference, https://fatconference.org/2020/callforcraft.html .

[742] Roel Dobbe and Morgan G. Ames, "Translation Tutorial: Values, Engagement and Reflection in Automated Decision Systems," presented at the ACM Conference on Fairness, Accountability, and Transparency, Atlanta, January 2019; see also Dobbe and Ames, "Up Next For FAT*: From Ethical Values To Ethical Practices," Medium, February 8, 2019, https://medium.com/@roeldobbe/up-next-for-fat-from-ethical-values-to-ethical-practices-ebbed9f6adee .

[743] Margaret Mitchell, Simone Wu, Andrew Zaldivar, Parker Barnes, Lucy Vasserman, Ben Hutchinson, Elena Spitzer, Inioluwa Deborah Raji, and Timnit Gebru, "Model Cards for Model Reporting," Proceedings of the Conference on Fairness, Accountability, and Transparency , FAT* '19 (2019): 220–229, https://doi.org/10.1145/3287560.3287596 ; Timnit Gebru, Jamie Morgenstern, Briana Vecchione, Jennifer Wortman Vaughan, Hanna Wallach, Hal Daumeé III, and Kate Crawford, "Datasheets for Datasets," arXiv:1803.09010 (2018), https://arxiv.org/abs/1803.09010?context=cs ; Matthew Arnold, Rachel KE Bellamy, Michael Hind, Stephanie Houde, Sameep Mehta, Aleksandra Mojsilovic, Ravi Nair, et al., "FactSheets: Increasing Trust in AI Services through Supplier's Declarations of Conformity," IBM Journal of Research and Development (2019), https://arxiv.org/pdf/1808.07261.pdf .

Algorithmic audits in 2019 uncovered disproportionate performance or biases within AI systems ranging from self-driving-car software that performed differently for darker- and lighter-skinned pedestrians,[745] gender bias in online biographies,[746] skewed representations in object recognition from lower-income environments,[747] racial differences in algorithmic pricing,[748] and differential prioritization in healthcare,[749] as well as performance disparities in facial recognition.[750]

In several cases, these audits had a tangible impact on improving the lives of people unfairly affected.[751] They also had a substantial impact on policy discussions.[752] For instance, two audit studies of facial-recognition systems, including the widely recognized Gender Shades,[753] led to subsequent audit studies by the National Institute of Standards and Technology[754] and other researchers,[755] including the ACLU of Northern California's

---

audits of Amazon Rekognition, which falsely matched 28 Congress members[756] and 27 mostly minority athletes to criminal mugshots.[757]

**AI's Inherent Vulnerabilities**

Concerns over the vulnerabilities of AI systems gains increased attention, highlighting the urgent need for them to be subjected to the same scrutiny applied to automation technologies in other engineering fields, such as aviation and power systems. Urgent vulnerabilities to address includes the danger of data-poisoning techniques, a method of exploitation in which a bad actor can fiddle with AI training data to alter a system's decisions.[758] A classic example is spam filtering, where intentionally curating the content of messages that teach a spam filter how spam looks can help certain types of spam pass through the filter undetected.[759]

A second type of AI vulnerability to address is the "back door," which permits attackers to find ways to infiltrate an AI system through code that malicious programmers embed in systems they trained or designed for later infiltration[760] for spying etc.. Researchers at NYU showed that back-door attacks may result in a model that has state-of-the-art performance on the user's training and validation samples (datasets used to test AI models), but behaves badly when confronted with specific attacker-chosen inputs.[761] The researchers used the back door to poison an AI road sign detector (commonly used in autonomous vehicles) into misclassifying US stop signs. When they "retrained" the model to work on Swedish stop signs, the earlier poisoning effects carried over. This type of vulnerability raises serious concerns given the rapid move toward outsourcing the training procedures of ML models to cloud platforms.[762]

A related trend is the move to reduce training costs by repurposing and retraining AI models for new or specific tasks, a phenomenon called transfer learning. Transfer learning is particularly popular for applications that require large models, such as natural-language processing[763] or image classification.[764]

---

[756] Jacob Snow, "Amazon's Face Recognition Falsely Matched 28 Members of Congress With Mugshots," ACLU, July 26, 2019, https://www.aclu.org/blog/privacy-technology/surveillance-technologies/amazons-face-recognition-falsely-matched-28 .

[757] Kate Gill, "Amazon Facial Recognition Falsely Links 27 Athletes to Mugshots in ACLU Study," Hyperallergic , October 28, 2019, https://hyperallergic.com/525209/amazon-facial-recognition-aclu/ .

[758] AI Now Report 2019.

[759] Blaine Nelson, Marco Barreno, Fuching Jack Chi, Anthony D. Joseph, Benjamin IP Rubinstein, Udam Saini, Charles A. Sutton, J. Doug Tygar, and Kai Xia, "Exploiting Machine Learning to Subvert Your Spam Filter," LEET '08 Proceedings of the 1st Usenix Workshop on Large-Scale Exploits and Emergent Threats, April 15, 2008, https://www.usenix.org/conference/leet-08/exploiting-machine-learning-subvert-your-spam-filter . More recently, examples of poisoning were reported for modifying explainability methods, attacking text generators, and bypassing plagiarism and copyright detectors. See Dombrowski et al., "Explanations Can Be Manipulated and Geometry Is to Blame," http://arxiv.org/abs/1906.07983 ; Dylan Slack, Sophie Hilgard, Emily Jia, Sameer Singh, and Himabindu Lakkaraju, "How Can We Fool LIME and SHAP? Adversarial Attacks on Post Hoc Explanation Methods," arXiv:1911.02508 [Cs, Stat] , November 6, 2019, http://arxiv.org/abs/1911.02508 ; Eric Wallace, Shi Feng, Nikhil Kandpal, Matt Gardner, and Sameer Singh, "Universal Adversarial Triggers for Attacking and Analyzing NLP," arXiv:1908.07125 [Cs] , August 29, 2019, http://arxiv.org/abs/1908.07125 ; Parsa Saadatpanah, Ali Shafahi, and Tom Goldstein, "Adversarial Attacks on Copyright Detection Systems," arXiv:1906.07153 [Cs, Stat] , June 20, 2019, http://arxiv.org/abs/1906.07153 .

[760] AI Now Report 2019.

[761] Tianyu Gu, Brendan Dolan-Gavitt, and Siddharth Garg, "BadNets: Identifying Vulnerabilities in the Machine Learning Model Supply Chain," arXiv:1708.06733 [Cs] , March 11, 2019, http://arxiv.org/abs/1708.06733 .

[762] Srivatsan Srinivasan, "Artificial Intelligence, Cloud, Data Trends for 2019 and Beyond," Medium, March 12, 2019, https://medium.com/datadriveninvestor/artificial-intelligence-cloud-data-trends-for-2019-and-beyond-2cbdd9e54c36 .

[763] Sebastian Ruder, Matthew E. Peters, Swabha Swayamdipta, and Thomas Wolf, "Transfer Learning in Natural Language Processing," Proceedings of the 2019 Conference of the North American Chapter of the Association for Computational Linguistics: Tutorials , June 2019, https://doi.org/10.18653/v1/N19-5004 .

Instead of starting from scratch, one retrains the parameters of a preexisting central model with more specific data for a new task or domain. Researchers show that this "centralization of model training increases their vulnerability to misclassification attacks," especially when such central models are publicly available.[765]

Adversarial machine learning is the systematic study of how motivated adversaries can compromise the confidentiality, integrity, and availability of machine learning (ML) systems through targeted or blanket attacks. The problem of attacking ML systems is so prevalent that CERT, the federally funded research and development center tasked with studying attacks, issued a broad vulnerability note on how most ML classifiers are vulnerable to adversarial manipulation. Corporations and governments are paying attention. Google, IBM, Facebook, and Microsoft have committed to investing in securing machine learning systems. The US is putting security and safety of AI systems as a top priority when defining AI regulation, with the EU releasing a complete set of non-binding checklists as part of its Trustworthy AI initiative. [766]

Adversarial attacks[767] are particularly effective against systems with a high number of inputs, which are the variables that an AI model considers to make a decision or prediction when deployed.[768] This reliance on a large number of inputs is inherent to computer-vision systems, where typically each pixel is an input. It is likely also an issue for applications where automated decision systems rely on a variety of inputs to make predictions about human behavior or preferences. Such models rely on diverse data sources, including social-network data, search entries, location tracking, energy use, and other revealing data about individual behavior and preferences. Such vulnerabilities expose people to misclassification, hacking, and strategic manipulation.

Researchers from Harvard and MIT convincingly explained these concerns for the context of medical diagnostics.[769] While research exposing technical vulnerabilities and proposing new defenses against them is now of high priority, building robust machine learning systems is still an elusive goal. A group of researchers across Google Brain, MIT, and the University of Tübingen recently surveyed the field and concluded that few defense mechanisms have succeeded. There is consensus in the field that most papers that propose defenses are quickly shown to be either incorrect or insufficient.

---

[764] Pedro Marcelino, "Transfer Learning from Pre-Trained Models," Towards Data Science, Medium, October 23, 2018, https://towardsdatascience.com/transfer-learning-from-pre-trained-models-f2393f124751 .

[765] Bolun Wang, Yuanshun Yao, Bimal Viswanath, Haitao Zheng, and Ben Y. Zhao, "With Great Training Comes Great Vulnerability: Practical Attacks against Transfer Learning," 27th USENIX Security Symposium, August 2018, https://www.usenix.org/system/files/conference/usenixsecurity18/sec18-wang.pdf ; Todor Davchev, Timos Korres, Stathi Fotiadis, Nick Antonopoulos, and Subramanian Ramamoorthy, "An Empirical Evaluation of Adversarial Robustness under Transfer Learning," arXiv:1905.02675 [Cs, Stat] , June 8, 2019, http://arxiv.org/abs/1905.02675 .

[766] Siva Kumar, Ram Shankar and Penney, Jonathon and Schneier, Bruce and Albert, Kendra, Legal Risks of Adversarial Machine Learning Research (July 3, 2020). International Conference on Machine Learning (ICML) 2020 Workshop on Law & Machine Learning, Available at SSRN: https://ssrn.com/abstract=3642779 ( "[R]esearch on adversarial machine learning is booming but it is not without risks. Studying or testing the security of any operational system may violate the Computer Fraud and Abuse Act (CFAA), the primary United States federal statute that creates liability for hacking. The CFAA's broad scope, rigid requirements, and heavy penalties, critics argue, has a chilling effect on security research. Adversarial ML security research is likely no different. However, prior work on adversarial ML research and the CFAA is sparse and narrowly focused. In this article, we help address this gap in the literature. For legal practitioners, we describe the complex and confusing legal landscape of applying the CFAA to adversarial ML. For adversarial ML researchers, we describe the potential risks of conducting adversarial ML research. We also conclude with an analysis predicting how the US Supreme Court may resolve some present inconsistencies in the CFAA's application in Van Buren v. United States, an appeal expected to be decided in 2021. We argue that the court is likely to adopt a narrow construction of the CFAA, and that this will actually lead to better adversarial ML security outcomes in the long term.")

[767] Albert, Kendra and Penney, Jonathon and Schneier, Bruce and Siva Kumar, Ram Shankar, Politics of Adversarial Machine Learning (2020). Towards Trustworthy ML: Rethinking Security and Privacy for ML Workshop, Eighth International Conference on Learning Representations (ICLR) 2020, Available at SSRN: https://ssrn.com/abstract=3547322 or http://dx.doi.org/10.2139/ssrn.3547322

[768] Nicholas Carlini and David Wagner, "Towards Evaluating the Robustness of Neural Networks," arXiv:1608.04644v2 [cs.CR] , August 16, 2016, https://arxiv.org/abs/1608.04644v2 .

[769] Samuel G. Finlayson, John D. Bowers, Joichi Ito, Jonathan L. Zittrain, Andrew L. Beam, and Isaac S. Kohane, "Adversarial Attacks on Medical Machine Learning." Science 363, no. 6433 (March 22, 2019):1287–89, https://doi.org/10.1126/science.aaw4399 .

The group observes that "[r]esearchers must be very careful to not deceive themselves unintentionally when performing evaluations."[770] We must be extra careful when bringing AI systems to contexts where their errors lead to social harm. Similar to the discussion of fairness and bias in the 2018 AI Now report,[771] any debate about vulnerabilities should approach issues of power and hierarchy, looking at who is in a position to produce and profit from these systems, who determines how vulnerabilities are accounted for and addressed, and who is most likely to be harmed. Despite the fact that social sciences and humanities approaches have a long history in information security and risk management,[772] research that addresses both social and technical dimensions in security is necessary, but still relatively nascent.[773] Central in this challenge is redrawing the boundaries of analysis and design to expand beyond the algorithm,[774] and securing channels for all impacted stakeholders to democratically steer system development and to dissent when concerns arise.[775]

**Artificial Intelligence and Children's Rights: Artificial Intelligence in Education (AIEd), EdTech, Surveillance, and Harmful Content**

Artificial Intelligence systems, although little perceived as part of our daily lives, are becoming ubiquitous for all people, including children. AI is used in cities for public safety and traffic organization purposes[776]; in hospitals, through applications in devices that assist doctors in detecting diseases[777]; in education, with the use of algorithms that create possibilities for customized learning[778] or facial recognition technologies[779]; and entertainment,[780] to name a few.

---

[770] Samuel G. Finlayson, John D. Bowers, Joichi Ito, Jonathan L. Zittrain, Andrew L. Beam, and Isaac S. Kohane, "Adversarial Attacks on Medical Machine Learning." Science 363, no. 6433 (March 22, 2019): 1287–89, https://doi.org/10.1126/science.aaw4399 .

[771] Whittaker et al., "AI Now Report 2018," https://ainowinstitute.org/AI_Now_2018_Report.pdf .

[772] E. Gabriella Coleman, Coding Freedom: The Ethics and Aesthetics of Hacking (Princeton: Princeton University Press, 2013); E. Gabriella Coleman and Alex Golub, "Hacker Practice: Moral Genres and the Cultural Articulation of Liberalism," Anthropological Theory 8, no. 3 (September 2008): 255–77, https://doi.org/10.1177/1463499608093814; Kevin D. Mitnick, William L. Simon, and Steve Wozniak, The Art of Deception: Controlling the Human Element of Security (Indianapolis: Wiley, 2003).

[773] Elda Paja, Fabiano Dalpiaz, and Paolo Giorgini, "Modelling and Reasoning about Security Requirements in Socio-Technical Systems," Data & Knowledge Engineering 98 (2015): 123–143; Matt Goerzen, Elizabeth Anne Watkins, and Gabrielle Lim, "Entanglements and Exploits: Sociotechnical Security as an Analytic Framework," 2019, https://www.usenix.org/conference/foci19/presentation/goerzen .

[774] Ben Green and Salomé Viljoen, "Algorithmic Realism: Expanding the Boundaries of Algorithmic Thought," Proceedings of the ACM Conference on Fairness, Accountability, and Transparency (FAT*) , 2020.

[775] Roel Dobbe, Thomas Gilbert, and Yonatan Mintz, "Hard Choices in Artificial Intelligence: Addressing Normative Uncertainty Through Sociotechnical Commitments," Neurips 2019 Workshop on AI for Social Good, arXiv:1911.09005v1 [cs.AI] , November 20, 2019, https://arxiv.org/abs/1911.09005v1 .

[776] Thierer, Adam D. and Castillo O'Sullivan, Andrea and Russell, Raymond, Artificial Intelligence and Public Policy (August 17, 2017). Mercatus Research Paper, Available at SSRN: https://ssrn.com/abstract=3021135 or http://dx.doi.org/10.2139/ssrn.3021135; and Berk, Richard, Artificial Intelligence, Predictive Policing, and Risk Assessment for Law Enforcement (January 2021). Annual Review of Criminology, Vol. 4, pp. 209-237, 2021, Vol. 4, pp. 209-237, Available at SSRN: https://ssrn.com/abstract=3777804 or http://dx.doi.org/10.1146/annurev-criminol-051520-012342

[777] What Doctor? Why AI and Robotics Will Define New Health, PWC (June 2017), https://www.pwc.com/gx/en/industries/healthcare/publications/ai-robotics-new-health/ai-robotics-new-health.pdf.

The rise of AI is a globally ubiquitous phenomenon associated with what many have called the 4th Industrial Revolution.[781] Despite all this technological development and its potentially positive aspects, the automation of various processes and the facilitation of human life as a whole, AI has provoked a series of ethical questions and related discussions around human rights and security.[782] These immediate concerns do not include the consideration of the discussions of "the singularity" a future where machine technologies might merge with human biology and physiology [783] through the use of general or super AI[784].

These AI based systems challenges are even more complex when one considers the demographics of the countries of the Global South which boast large numbers of children per family. AI can contribute to the exponential mitigation of structural inequalities. It can assist in guaranteeing human rights such as the right to adequate food, basic sanitation, quality education, employability and security. However, it can also exasperate preexisting discrimination, including in education, impacting children's accessibility to education and enjoyment of accessible education.[785]

---

[778] Zeide, Elana, Robot Teaching, Pedagogy, and Policy (Forthcoming 2019). Forthcoming in The Oxford Handbook of Ethics of AI, Oxford University Press (Markus D. Dubber, Frank Pasquale, and Sunit Das eds.), Available at SSRN: https://ssrn.com/abstract=3441300

[779] E.g., Barrett, Lindsey, Ban Facial Recognition Technologies for Children—And for Everyone Else (July 24, 2020). Boston University Journal of Science and Technology Law. Volume 26.2, Available at SSRN: https://ssrn.com/abstract=3660118 ("Facial recognition technologies enable a uniquely dangerous and pervasive form of surveillance, and children cannot escape it any more than adults can. Facial recognition technologies have particularly severe implications for privacy, as they can weaponize existing photographic databases in a way that other technologies cannot, and faces are difficult or impossible to change, and often illegal to publicly obscure. Their erosion of practical obscurity in public threatens both privacy and free expression, as it makes it much harder for people to navigate public spaces without being identified, and easier to quickly and efficiently identify many people in a crowd at once. To make matters even worse, facial recognition technologies have been shown to perform less accurately for people of color, women, non-binary and transgender people, children, and the elderly, meaning that they have the potential to enable discrimination in whatever forum they are deployed. As these technologies have developed and become more prevalent, children are being subjected to them in schools, at summer camp, and other child-specific contexts, as well as alongside their parents, through CCTV, private security cameras, landlord-installed apartment security systems, or by law enforcement. The particular vulnerability of young people relative to adults might make them seem like natural candidates for heightened protections from facial recognition technologies. Young people have less say over where they go and what they do, inaccurate evaluations of their faces could have a particularly strong impact on their lives in contexts like law enforcement uses, and the chilling effects of these technologies on free expression could constrain their emotional and intellectual development. At the same time, some of the harms young people experience are near-universal privacy harms, such as the erosion of practical obscurity, while the discriminatory harms of facial recognition's inaccurate assessment of their faces are shared by other demographic groups.")

[780] Hasse, Alexa and Cortesi, Sandra Clio and Lombana-Bermudez, Andres and Gasser, Urs, Youth and Artificial Intelligence: Where We Stand (May 24, 2019). Berkman Klein Center Research Publication No. 2019-3, Available at SSRN: https://ssrn.com/abstract=3385718 or http://dx.doi.org/10.2139/ssrn.3385718 (This article seeks to share Youth and Media's initial learnings and key questions around the intersection between AI and youth (ages 12-18), in the context of domains such as education, health and well-being, and the future of work. It aims to encourage various stakeholders — including policymakers, educators, and parents and caregivers — to consider how we can empower young people to meaningfully interact with AI based technologies to promote and bolster learning, creative expression, and wellbeing, while also addressing key challenges and concerns.)

[781] Technological innovation has been changing the economic and social landscape for the past 300 years, from the First Industrial Revolution (water and steam power), to the Second Industrial Revolution (electric power and the assembly line), to the Third Industrial Revolution (also called the digital revolution, comprising of computers and the Internet). See The Future Computed: Artificial Intelligence and Its Role in Society, MICROSOFT at 93 (2018), https://news.microsoft.com/uploads/2018/01/The-Future-Computed.pdf. AI is the latest technological innovation and has been coined by some as the Fourth Industrial Revolution. As explained by Klaus Schwab, Founder and Executive Chairman of the World Economic Forum: There are three reasons why today's transformations represent not merely a prolongation of the Third Industrial Revolution but rather the arrival of a Fourth and distinct one: velocity, scope, and systems impact. The speed of current

It is necessary to continue to analyze the relationship of AI as directly or indirectly impacting the educational processes of children, including in the Global South, where the structural challenges of formal education, as impacted by AI, are increasing. [786] Oversight of the responsibility of participating entities to respect and protect the rights of children, including government and state (in the development of policy), and as well tech-companies (in the design, development and provision of AI technologies, products and services) will need to be developed alongside the utilization of a Children's Rights by Design (CrbD),standard focused on the best interests of children,[787] as contemplated by the UN Convention on the Rights of the Child.[788]

AI technologies both positively and negatively impact children's human rights. [789] There are valuable opportunities to use artificial intelligence in ways that maximize children's well being, but there are critical questions that we need ask and answer in order to better protect children from potential negative impacts of artificial intelligence.

---

breakthroughs has no historical precedent. When compared with previous industrial revolutions, the Fourth is evolving at an exponential rather than a linear pace. Moreover, it is disrupting almost every industry in every country. And the breadth and depth of these changes herald the transformation of entire systems of production, management, and governance." Klaus Schwab, The Fourth Industrial Revolution: What It Means, How to Respond, WORLD ECON. F. (Jan. 14, 2016), https://www.weforum.org/agenda/2016/01/the-fourth-industrial-revolution-what-it-means-and-how-torespond/.

[782] Raso, Filippo and Hilligoss, Hannah and Krishnamurthy, Vivek and Bavitz, Christopher and Kim, Levin Yerin, Artificial Intelligence & Human Rights: Opportunities & Risks (September 25, 2018). Berkman Klein Center Research Publication No. 2018-6, Available at SSRN: https://ssrn.com/abstract=3259344 or http://dx.doi.org/10.2139/ssrn.3259344

[783] 'Stephen Hawking warns artificial intelligence could end mankind', Rory Cellam-Jones, BBC, Dec. 2014. Available at: https://www.bbc.com/news/technology-30290540

[784] Haney, Brian, The Perils & Promises of Artificial General Intelligence (October 5, 2018). Brian S. Haney, The Perils & Promises of Artificial General Intelligence, 45 J. Legis. 151 (2018). , Available at SSRN: https://ssrn.com/abstract=3261254 or http://dx.doi.org/10.2139/ssrn.3261254

[785] Isabella Henriques and Pedro Hartung, Children's Rights by Design in AI Development for Education, International Review of Information Ethics, Vol. 29 (03/2021) https://informationethics.ca/index.php/irie/article/view/424/401

[786] Isabella Henriques and Pedro Hartung, Children's Rights by Design in AI Development for Education, International Review of Information Ethics, Vol. 29 (03/2021) https://informationethics.ca/index.php/irie/article/view/424/401

[787] Isabella Henriques and Pedro Hartung, Children's Rights by Design in AI Development for Education, International Review of Information Ethics, Vol. 29 (03/2021) https://informationethics.ca/index.php/irie/article/view/424/401

[788] United Nations, Convention on the Rights of the Child - Available at https://www.ohchr.org/en/professionalinterest/pages/crc.aspx

[789] UNICEF, Artificial Intelligence and Children's Rights, 2018 https://www.unicef.org/innovation/media/10726/file/Executive%20Summary:%20Memorandum%20on%20Artificial%20Intelligence%20and%20Child%20Rights.pdf ("The authoring team of this memorandum are Mélina Cardinal-Bradette, Diana Chavez-Varela, Samapika Dash, Olivia Koshy, Pearlé Nwaezeigwe, Malhar Patel, Elif Sert, and Andrea Trewinnard, who conducted their research and writing under the supervision of Alexa Koenig of the UC Berkeley Human Rights Center")

Many are working on addressing the challenges and opportunities children and youth encounter in the digital environment.[790] How can artificial intelligence be leveraged to protect, benefit and empower youth globally?[791]

As UNICEF and other organizations emphasize, we must pay specific attention to children and the evolution of AI technology in a way that children-specific rights and needs are recognized. The potential impact of artificial intelligence on children deserves special attention, given children's heightened vulnerabilities and the numerous roles that artificial intelligence will play throughout the lifespan of individuals born in the 21st century.

As AI-based technologies become increasingly integrated into modern life, the onus is on companies, governments, researchers, parents, most, to consider the ways in which such technologies impact children's human rights. The potential impact of artificial intelligence on children deserves special attention, given children's heightened vulnerabilities and the numerous roles that artificial intelligence will play throughout the lifespan of individuals who are born in the 21st century. As much of the underlying technology is proprietary to corporations, corporations' willingness and ability to incorporate human rights considerations into the development and use of such technologies will be critical. Governments will also need to work with corporations, parents, children and other stakeholders to create policies that safeguard children's human rights and related interests.[792]

There are valuable opportunities to use artificial intelligence in ways that maximize children's well-being, but critical work needs to be done in order to better protect children from AI negative consequences.

**AI, Machine Learning, and Deep Learning**

The terms artificial intelligence, machine learning, and deep learning, are often used interchangeably by the general public to reflect the concept of replicating "intelligent" behavior in machines.

---

[790] Cortesi, Sandra Clio and Gasser, Urs and Adzaho, Gameli and Baikie, Bruce and Baljeu, Jacqueline and Battles, Matthew and Beauchere, Jacqueline and Brown, Elsa and Burns, Jane and Burton, Patrick and Byrne, Jasmina and Colombo, Maximillion and Douillette, Joseph and Escobar, Camila and Flores, Jorge and Ghebouli, Zinelabidine and Gonzalez-Allonca, Juan and Gordon, Eric and Groustra, Sarah and Hertz, Max and Junco, Reynol and Khan, Yasir and Kimeu, Nicholas and Kleine, Dorothea and Krivokapic, Djordje and Kup, Viola and Kuzeci, Elif and Latorre Guzmán, María and Li, David and Limbu, Minu and Livingstone, Sonia and Lombana-Bermudez, Andres and Massiel, Cynthia and McCarthy, Claire and Molapo, Maletsabisa and Mor, Maria and Newman, Sarah and Nutakor, Eldad and Onoka, Christopher and Onumah, Chido and Passeron, Ezequiel and Pawelczyk, Katarzyna and Roque, Ricarose and Rudasingwa, Kanyankore and Shah, Nishant and Simeone, Luca and Siwakwi, Andrew and Third, Amanda and Wang, Grace, Digitally Connected: Global Perspectives on Youth and Digital Media (March 26, 2015). Berkman Center Research Publication No. 2015-6, Available at SSRN: https://ssrn.com/abstract=2585686 or http://dx.doi.org/10.2139/ssrn.2585686 (Reflecting on the 25th anniversaries of the invention of the World Wide Web by Sir Tim Berners-Lee and the adoption of the Convention on Rights of the Child by the US General Assembly, the Berkman Center for Internet & Society at Harvard University and UNICEF co-hosted in April 2014 — in collaboration with PEW Internet, EU Kids Online, the Internet Society (ISOC), Family Online Safety Institute (FOSI), and YouthPolicy.org — a first of its kind international symposium on children, youth, and digital media to map and explore the global state of relevant research and practice, share and discuss insights and ideas from the developing and industrialized world, and encourage collaboration between participants across regions and continents. With a particular focus on voices and issues from the Global South, the symposium addressed topics such as inequitable access, risks to safety and privacy, skills and digital literacy, and spaces for participation, and civic engagement and innovation. The event also marked the launch of Digitally Connected — an initiative that brings together academics, practitioners, young people, activists, philanthropists, government officials, and representatives of technology companies from around the world who, together, are addressing the challenges and opportunities children and youth encounter in the digital environment.)

[791] World Economic Forum, Empowering Generation AI, Sustainable Development Summit 2020, (Dec. 31, 2020) https://www.youtube.com/watch?v=gi625laHeGs

[792] Cedric Villani, "For a Meaningful Artificial Intelligence Towards a French and European Strategy," March 8, 2018, available at https://www.aiforhumanity.fr/pdfs/MissionVillani_Report_ENG-VF.pdf.

Generally, AI refers to a sub-field of computer science focused on building machines and software that can mimic such behavior. Machine learning is the sub-field of artificial intelligence that focuses on giving computer systems the ability to learn from data. Deep learning is a subcategory of machine learning that uses neural networks to learn to represent and extrapolate from a dataset. There are numerous ways that machine learning and deep learning processes impact children's lives and ultimately, their human rights, and how artificial intelligence technologies are being used in ways that positively or negatively impact children at home, at school, and at play.[793]

**Role of AI in Children's Lives**

The role of artificial intelligence in children's lives—from how children play, to how they are educated, to how they consume information and learn about the world—is expected to increase exponentially. A number of initiatives have started to map the impact of AI on children.[794] Thus, it is imperative that stakeholders come together now to evaluate the risks of using such technologies and assess opportunities to use artificial intelligence to maximize children's well being in a thoughtful and systematic manner. As part of this assessment, stakeholders should work together to map the potential positive and negative uses of AI on children's lives, and develop a child rights-based framework for artificial intelligence that delineates rights and corresponding duties for developers, corporations, parents, and children around the world.

The potential impact of artificial intelligence on children deserves special attention, given children's heightened vulnerabilities and the numerous roles that artificial intelligence will play throughout the lifespan of individuals who are born in the 21st century. As much of the underlying technology is proprietary to corporations, corporations' willingness and ability to incorporate human rights considerations into the development and use of such technologies will be critical. Governments will also need to work with corporations, parents, children and other stakeholders to create policies that safeguard children's human rights and related interests.

**The Convention on the Rights of the Child**

The United Nations Convention on the Rights of the Child (CRC),[795] adopted by the UN General Assembly on 20 November 1989,[796] provides the international legal framework for children's rights.[797] The CRC is the most comprehensive legal framework that protects children--defined as human beings 18 years old and under--as rights bearers.[798] Until a few years ago,[799] the CRC did not contain an actual enforcement mechanism, which was considered a manifest flaw. Children could not file complaints, and the Convention could not be tested in specific cases by the courts.[800] In 2011, however, the Optional Protocol on a Communications Procedure was adopted,[801] which allows individual children to submit complaints regarding specific violations of their rights under the Convention and its first two optional protocols. The Protocol entered into force in April 2014.[802] In addition, the UNCRC has a symbolic function[803] and a strong moral force.[804] The UN Committee on the Rights of the Child monitors the implementation of the UNCRC and issues critical remarks or recommendations.[805] It is then up to the national governments to take these into account.

---

[793] "Office of Innovation, UNICEF Office of Innovation,"UNICEF Innovation Home Page, available at https://www.unicef.org/innovation/.

[794] UNICEF. 2020. 'Policy Guidance on AI for Children (Draft)'. https://www.unicef.org/globalinsight/media/1171/file/UNICEF-Global-Insight-policyguidance-AI-children-draft-1.0-2020.pdf See also Kardefelt-Winther, Daniel. 2017. 'How Does the Time Children Spend Using Digital Technology Impact Their Mental Well-Being, Social Relationships and Physical Activity?: An Evidence Focused Literature Review'. Innocenti Discussion Papers 2017/02. Vol. 2017/02. Innocenti Discussion Papers. UNICEF. https://doi.org/10.18356/cfa6bcb1-en.

[795] United Nations, Convention on the Rights of the Child, 20.11.1989, http://www.unhchr.ch/html/menu3/b/k2crc.htm [hereinafer: CRC].

[796] Previous international documents on children's rights were: "Declaration on the Rights of Child", adopted by the League of Nations in 1924, and the 1959 "UN Declaration on the Rights of Child", which was adopted unanimously by the General Assembly of the United Nations on 20 November 1959, http://www.unhchr.ch/html/menu3/b/25.htm. For a detailed overview cf. Van Bueren, Geraldine, The international law on the rights of the child, Dordrecht, Martinus Nijhoff Publishers, 1995, 6-12.

[797] See also Commission of the European Communities, Commission Staff working document accompanying the Communication from the Commission Towards an EU strategy on the rights of the child, Impact assessment, COM (2006) 367 final, SEC (2006) 888, 04.07.2006, http://register.consilium.europa.eu/pdf/en/06/st12/st12107-ad01.en06.pdf, 6: "The UNCRC provides a coherent and comprehensive framework against which to evaluate legislation, policy, structures and actions"

The CRC aims to ensure children's equality of treatment by States. The CRC is the key international instrument on children's rights and represents an extraordinary level of international consensus on the legal rights that children should have.[806] The Convention imposes obligations on 195 states parties.[807] to provide legal protection for a wide range of rights that inhere in children by virtue of their human dignity. Many of these rights inhere in all human beings and were therefore already protected by pre-existing instruments of international law such as the International Covenant on Civil and Political Rights (ICCPR) and the International Covenant on Economic, Social and Cultural Rights (ICESCR). The CRC aims to emphasize that these rights apply equally to children, regardless of their age, and to provide explicit measures to ensure that children can enjoy these rights on an equal basis with other human beings.[808]

---

[798] UN General Assembly, "Convention on the Rights of the Child, 20 November 1989," United Nations, Treaty Series, vol. 1577, p. 3, Article 1.  See also von Struensee, Susan, Highlights of the United Nations Children's Convention and International Response to Children's Human Rights, Suffolk Transnational Law Review, Vol. 18, Issue 2 (Summer 1995), pp. 589-628 Available at SSRN: https://ssrn.com/abstract=657363

[799] Kilkelly, Ursula, "The best of both worlds for children's rights? Interpreting the European Convention on Human Rights in the light of the UN Convention on the Rights of the Child", Human Rights Quarterly 2001, Vol. 23, 309; McLaughlin, Sharon, Rights v. restrictions. Recognising children's participation in the digital age, in Brian O'Neill, Elisabeth Staksrud, Sharon McLaughlin, Towards a Better Internet for Children? Policy Pillars, Players and Paradoxes, Nordicom, 2013, 316. For more on the implementation of CRC cf. Van Bueren, Geraldine, The international law on the rights of the child, Dordrecht, Martinus Nijhoff Publishers, 1995, 378-422.

[800] Bainham, Andrew, Children – the modern law, Bristol, Family Law, 2005, 67.  But  supranational courts, such as the European Court of Justice, did refer to the CRC in its caselaw.

[801] United Nations, Optional Protocol on a Communications Procedure, 2011, https://treaties.un.org/doc/source/signature/2012/ctc_4-11d.pdf. See, e.g., Spronk, Sarah Ida, Realizing Children's Right to Health: Additional Value of the Optional Protocol on a Communications Procedure for Children (August 10, 2012). Available at SSRN: https://ssrn.com/abstract=2127644 or http://dx.doi.org/10.2139/ssrn.2127644   and Binford, W. Warren Hill, Utilizing the Communication Procedures of the ACERWC and the UNCRC (October 29, 2012). Available at SSRN: https://ssrn.com/abstract=2209507. or http://dx.doi.org/10.2139/ssrn.2209507

[802] The entry into force of the Third Optional Protocol on a Communications Procedure (OPIC) in 2014 was groundbreaking as it allowed children to lodge complaints with the UN about violations of their rights, if violations cannot be addressed effectively at national level. However, to advance access to justice for children, it is important to increase States' ratification of the OPIC and to work for its effective implementation at the national level. In 2021, seven years since the entry into force of the Optional Protocol, 47 States have ratified the OPIC, 17 have signed but not yet ratified it, and 133 have taken no action.  https://opic.childrightsconnect.org/ratification-status/ and https://treaties.un.org/Pages/ViewDetails.aspx?src=TREATY&mtdsg_no=IV-11-d&chapter=4&clang=_en

[803] Van Bueren, Geraldine, The international law on the rights of the child, Dordrecht, Martinus Nijhoff Publishers, 1995, xx.

[804] Kilkelly, Ursula, "The best of both worlds for children's rights? Interpreting the European Convention on Human Rights in the light of the UN Convention on the Rights of the Child", Human Rights Quarterly 2001, Vol. 23, 310.

[805] Kilkelly, Ursula, "The best of both worlds for children's rights? Interpreting the European Convention on Human Rights in the light of the UN Convention on the Rights of the Child", Human Rights Quarterly 2001, Vol. 23, 309.

[806] O'Mahony, Conor, Constitutional Protection of Children's Rights: Visibility, Agency and Enforceability (January 28, 2019). (2019) 19 Human Rights Law Review, Forthcoming, Available at SSRN: https://ssrn.com/abstract=3324280 or http://dx.doi.org/10.2139/ssrn.3324280

[807]  Eugeen Verhellen, 'The Convention on the Rights of the Child: Reflections from a historical, social policy and educational perspective' in Wouter Vanderhole (ed), Routledge International Handbook of Children's Rights Studies (Oxford: Routledge, 2015) at 43.

The Convention grants rights to children across categories often referred to as the three Ps: Protection (from harm, violence or exploitation); Provision (with the resources of servicers necessary for a decent life) and Participation (in society and in decisions affecting the child).[809]

In addition, the Convention makes specific provision for some rights that are particular to children due to their stage of development and their comparatively disempowered position in society. It seems clear that some of the rights protected in the CRC are either not relevant, or less relevant, to adults with full legal capacity, and thus do not tend to feature in the general human rights conventions. These include the best interests principle in Article 3; the right to special protection and assistance for children deprived of their family environment in Article 20; the right to development under Article 6; and a range of rights in Articles 7-12, including the right to name and nationality, preservation of identity, the right to maintain contact with parents, and the right to express views in all matters affecting the child. In this way, the CRC does not just re-state that children enjoy the same rights as adults, but supplements the rights afforded to adults with important child-specific rights. Cutting across the CRC as a whole are four general principles that have been identified by the Committee on the Rights of the Child (hereinafter 'the CRC Committee'): the right to life, survival and development (Article 6); non-discrimination (Article 2); that the best interests of children should a primary consideration in all matters affecting them (Article 3), and the right of children to participate in decision affecting them (Article 12).[810]

National constitutions take different approaches to the protection of children's rights. The CRC Committee has set out what it describes as a child rights approach, defined in General Comment No. 13 as: A child rights approach is one which furthers the realization of the rights of all children as set out in the Convention by developing the capacity of duty bearers to meet their obligations to respect, protect and fulfill rights (art. 4) and the capacity of rights holders to claim their rights, guided at all times by the rights to non-discrimination (art. 2), consideration of the best interests of the child (art. 3, para. 1), life, survival and development (art. 6), and respect for the views of the child (art. 12). Children also have the right to be directed and guided in the exercise of their rights by caregivers, parents and community members, in line with children's evolving capacities (art. 5). This child rights approach is holistic and places emphasis on supporting the strengths and resources of the child him/herself and all social systems of which the child is a part: family, school, community, institutions, religious and cultural systems. [811]

Moreover, the Committee has stressed in General Comment No. 5 that it is not enough for the law to say that children have rights along the lines set out above: it must give meaning to those rights by providing a means for their enforcement:

*For rights to have meaning, effective remedies must be available to redress violations. This requirement is implicit in the Convention and consistently referred to in the other six major international human rights treaties. Children's special and dependent status creates real difficulties for them in pursuing remedies for breaches of*

---

[808] Id. at 48 at 48. See also Adam Lopatka, 'Introduction' in Legislative History of the Convention on the Rights of the Child (New York/Geneva: United Nations, 2007) at xxxvii, available at http://www.ohchr.org/Documents/Publications/LegislativeHistorycrc1en.pdf. For a skeptical view of the strategy of providing children with 'special rights' rather than relying on general rights guarantees, see James G. Dwyer, 'Inter-Country Adoption and the Special Rights Fallacy' (2013) University of Pennsylvania Journal of International Law 189 at 198-208.

[809] For a discussion of the three Ps and how they interact with the four general principles of the CRC, see Verhellen, supra at 49-50.

[810] Committee on the Rights of the Child, General Comment No. 5 (2003): General measures of implementation of the Convention on the Rights of the Child, CRC/GC/2003/5, 27 November 2003 at para 12.

[811] Committee on the Rights of the Child, General Comment No. 13: Article 19: the right of the child to freedom from all forms of violence, CRC/C/GC/13, 18 April 2011 at para 59.

*their rights. So States need to give particular attention to ensuring that there are effective, child-sensitive procedures available to children and their representatives.*[812]

More than a binding international document, the Convention is an ethical[813] and legal framework for assessing states' progress or regress on issues of particular interest to children.[814] Because of the recent exponential advancement of artificial intelligence-based technologies, the current international framework that protects children's rights, just as local frameworks, does not explicitly address many of the issues raised by the development and use of artificial intelligence.[815]

**AI Impacts Children**

AI impacts children. They are exposed to algorithms at home, at school, and at play. Algorithms shape the environments in which they live, the services they have access to, and how they spend their time. Children play with interactive smart toys, they watch videos recommended by algorithms, use voice commands to control their phones, and use image manipulation algorithms for fun in social media.

The presence of AI in children's lives raises many questions. Is it acceptable to use recommendation algorithms with children or to provide an interactive toy if the child cannot understand that they are dealing with a computer? How should parents be advised on the possible impact of AI-based toys on the cognitive development of a child? What should children learn about AI in schools in order to have a sufficient understanding of the technology around them? At what point should a child be given the right to decide about the consents involved? How long should the data be stored?

---

[812] Committee on the Rights of the Child, General Comment No. 5: General measures of implementation of the Convention on the Rights of the Child, CRC/GC/2003/5, 27 November 2003 at para 24.

[813] See, e.g., McGee, Robert W., Abolishing Child Labor: Some Overlooked Ethical Issues (May 20, 2016). Available at SSRN: https://ssrn.com/abstract=2782715 or http://dx.doi.org/10.2139/ssrn.2782715 ( Under a rights regime, any act or policy that violates someone's rights is automatically labeled as unethical). See also Mousin, Craig B., Rights Disappear When US Policy Engages Children As Weapons of Deterrence (January 1, 2019). AMA J Ethics. 2019;21(1):E58-66. , Available at SSRN: https://ssrn.com/abstract=3317913 ("Although the United States provided significant guidance in drafting the Convention on the Rights of the Child (CRC) it has never ratified the convention. The failure to ratify has taken on critical significance in light of new federal policies that have detained over 15,000 children in 2018, separated families, accelerated removal of asylum seekers, and emphasized deterring families from seeking asylum. This article raises ethical and health implications of these refugee policies in light of the United States' failure to ratify the CRC. It first examines the development of the CRC and international refugee law. It next lists some of the new policies and case law implemented by Attorney General Sessions in 2018 that have led to the detention and separation of children from their family, undermining legal protections for asylum applicants. The CRC calls for governments to examine the best interests of children seeking refugee status, but federal policies preclude consideration of that goal. In addition, although the CRC calls for appropriate legal protection for children, current policies neglect that goal and instead criminalize children and families before they have been provided with legal representation or assistance. Such policies exacerbate the trauma of children fleeing violence in their homeland and undergoing the risks of flight. This article raises ethical issues including whether judges and lawyers for the government should participate in legal proceedings when toddlers appear unrepresented. The failure to ratify the CRC in conjunction with these new deterrence polices undermines legal protections for children worldwide.")

[814] See, e.g., Meier, Benjamin Mason and Motlagh, Mitra and Rasanathan, Kumanan, The United Nations Children's Fund: Implementing Human Rights for Child Health (April 26, 2018). Human Rights in Global Health: Rights-Based Governance for a Globalizing World (Oxford University Press 2018), Available at SSRN: https://ssrn.com/abstract=3214766; See generally UNICEF "State of the World's Children" 2019 available at https://www.unicef.org/reports/state-of-worlds-children-2019 and https://www.unicef.org/reports/state-of-worlds-children

[815] UNICEF, AI for children, https://www.unicef.org/globalinsight/featured-projects/ai-children (Recent progress in the development of artificial intelligence (AI) systems, unprecedented amounts of data to train algorithms, and increased computing power are expected to profoundly impact life and work in the 21st century, raising both hopes and concerns for human development. However, despite the growing interest in AI, little attention is paid to how it will affect children and their rights.) See also https://www.unicef.org/innovation/GenerationAI and https://www.weforum.org/projects/generation-ai

As UNICEF and other organizations emphasize, we must pay specific attention to children and the evolution of AI technology in a way that children-specific rights and needs are recognized. The potential impact of artificial intelligence on children deserves special attention, given children's heightened vulnerabilities and the numerous roles that artificial intelligence will play throughout the lifespan of individuals born in the 21st century.

**The CRC Identifies Several Rights Implicated by AI Technologies**

The CRC identifies several rights implicated by AI technologies[816], and thus provides an important starting place for any analysis of how children's rights may be positively or negatively affected[817] by new technologies.[818] Since the creation of the CRC it has been accepted across the globe that children are entitled to a number of fundamental rights that are important in the media environment, freedom of expression (article 13 CRC) and the right to privacy (article 16 CRC). At the same time, children sometimes need to be protected, for instance, from content or behavior that may harm them (article 17, infra, article 19 - concerning protection from all forms of violence - and article 34 - concerning protection from sexual exploitation - CRC).

Article 13 confirms the child-specific version[819] of the right to freedom of expression[820]"which includes the freedom to seek, receive and impart information and ideas of all kinds, regardless of frontiers, either orally, in writing or in print, in the form of art, or through any other media of the child's choice".[821]This fundamental right can only be restricted if this is provided by law and necessary "for respect of the rights or reputations of others, or for the protection of national security or of public order, or of public health or morals" (para. 2). The article has a broad scope of application, which certainly extends to the internet as well as any other (future) medium. Recently, the UN Committee on the Rights of the Child emphasized that the increasing extent to which information and communication technologies are a central dimension in the lives of children entails that (equal) access to the internet and social media for them is crucial, also for the realization of other rights closely linked to the right to freedom of expression, such as the right to leisure, play and culture (article 31 CRC).[822]

---

[816] Verdoodt, Valerie, The Role of Children's Rights in Regulating Digital Advertising (2019). International Journal of Children's Rights, 27 (3), 455-481, 2019, doi: 10.1163/15718182-02703002., Available at SSRN: https://ssrn.com/abstract=3703312 or http://dx.doi.org/10.2139/ssrn.3703312 ("An important domain in which children's rights are reconfigured by internet use, is digital advertising. New advertising formats such as advergames, personalized and native advertising have permeated the online environments in which children play, communicate and search for information. The often immersive, interactive and increasingly personalized nature of these advertising formats makes it difficult for children to recognize and make informed and well-balanced commercial decisions. This raises particular issues from a children's rights perspective, including inter alia their rights to development (Article 6 UNCRC), privacy (Article 16 UNCRC), protection against economic exploitation (Article 32 UNCRC), freedom of thought (Article 17 UNCRC) and education (Article 28 UNCRC). The paper addresses this reconfiguration by translating the general principles and the provisions of the United Nations Convention on the Rights of the Child into the specific context of digital advertising. Moreover, it considers the different dimensions of the rights (i.e. protection, participation and provision) and how the commercialization affects children and how their rights are exercised.")

[817] Lievens, Eva, "A children's rights perspective on the responsibility of social network site providers", 25th European Regional Conference of the International Telecommunications Society (ITS), Brussels, Belgium, 2014. https://www.econstor.eu/bitstream/10419/101441/1/795276834.pdf

[818] See, e.g. Livingstone, Sonia, John Carr, and Jasmina Byrne, "One in Three: Internet Governance and Children's Rights", Paper Series Centre for International Governance Innovation and the Royal Institute of International Affairs, 2015, 22. https://www.unicef-irc.org/publications/795-one-in-three-internet-governance-and-childrens-rights.html ("Typically, in the discussions around the use of the Internet, children are acknowledged only in the context of child protection while their rights to provision and participation are overlooked. This paper specifically argues against an age-generic (or 'age-blind') approach to 'users', because children have specific needs and rights that are not met by governance regimes designed for 'everyone'. Policy and governance should now ensure children's rights to access and use digital media and consider how the deployment of the Internet by wider society can enhance children's rights across the board. The paper ends with six conclusions and recommendations about how to embed recognition of children's rights in the activities and policies of international Internet governance institutions.")

[819] Kilkelly, Ursula, "The best of both worldsfor children's rights? Interpreting the European Convention on Human Rights in the light of the UN Convention on the Rights of the Child", Human Rights Quarterly 2001, Vol. 23, 311.

[820] Similar articles are article 19 Universal Declaration of Human Rights, article 19 International Covenant of Civil and Political Rights, and article 10 European Convention on Human Rights and Fundamental Freedoms.

Equally important is the child's right to privacy, formulated in article 16 CRC.[823] According to this article, children cannot be subjected to any arbitrary or unlawful interference – by state authorities or by others (e.g., private organizations)[824] – with their privacy, family, home or correspondence, nor to unlawful attacks on their honor and reputation. Moreover, it is clearly stated that the law should protect a child against such interference. The right to privacy is directed at the child itself and is to be protected in all situations.[825] In the online environment, privacy issues could, for instance, arise with respect to identification mechanisms or with regard to the collection of their personal data by service providers. Furthermore, monitoring a child's internet use could be considered in conflict with the child's right to privacy. Finally, parents may neither, according to article 16, interfere with their child's correspondence. There is no reason to limit the application of this article to 'paper' correspondence, so monitoring e-mail conversations could be in conflict with the child's right to privacy as well.

Another crucial article with regard to media content and services is article 17 CRC.[826] This article requires states to ensure that children have access to "information and material from a diversity of national and international sources, especially those aimed at the promotion of his or her social, spiritual and moral well-being and physical and mental health",[827] since access to a wide diversity of information is a prerequisite for the exercise of other fundamental rights, most importantly the right to freedom of expression. States are thus incited to pursue a proactive policy which stimulates the cultural, educational and informational potential of media with respect to children. At the same time article 17 CRC also encourages the development of guidelines to protect children from harmful material. On the one hand, the internet and other new media technologies enable children to access a huge variety of educational material[828] and cultural opportunities, as "powerful tool[s] that can help to meet children's rights under the CRC (e.g., to participation, information and freedom of expression)".[829]

---

[821] The United Nations Committee on the Rights of the Child has stressed that it is not sufficient to just include the 'general' right to freedom of expression applicable to everyone in a country's constitution. It is necessary, according to the Committee, to also expressly incorporate the child's right to freedom of expression in legislation. See for instance: United Nations Committee on the Rights of the Child, General Guidelines for Periodic Reports, CRC/C/58, 20.11.1996, http://www.unhchr.ch/tbs/doc.nsf/(Symbol)/CRC.C.58.En?Opendocument: "States parties are requested to provide information on the measures adopted to ensure that the civil rights and freedoms of children set forth in the Convention, in particular those covered by articles 7, 8, 13 to 17 and 37 (a), are recognized by law specifically in relation to children and implemented in practice, including by administrative and judicial bodies, at the national, regional and local levels, and where appropriate at the federal and provincial levels".

[822] United Nations Committee on the Rights of the Child, General comment No. 17 (2013) on the right of the child to rest, leisure, play, recreational activities, cultural life and the arts (art. 31), UN Doc. CRC/C/GC/17, 2013, n° 45.

[823] This is a child-specific 'translation' of the general right to privacy, which is granted to everyone by, inter alia, article 12 Universal Declaration on Human Rights, article 17 International Covenant on Civil and Political Rights, and article 8 European Convention on Human Rights and Fundamental Freedoms

[824] Hodgkin, Rachel and Newell, Peter, Implementation handbook for the Convention on the Rights of the Child, New York, UNICEF, 2002, 216.

[825] Hodgkin, Rachel and Newell, Peter, Implementation handbook for the Convention on the Rights of the Child, New York, UNICEF, 2002, 213; Meuwese, Stan, Blaak, Mirjam and Kaandorp, Majorie (eds),

[826] The European Court of Justice has also referred to this article in a case concerning potential harmful new media content: ECJ, Dynamic Medien v. Avides Media AG, C-244/06, 14.02.2008, para. 40.

[827] A general discussion on 'The child and the media' was held by the Committee on the Rights of the Child on the 7th of October 1996. A report of this discussion was included in the Report on the thirteenth session: United Nations Committee on the Rights of the Child, Report on the thirteenth session, CRC/C/57, 31.10.1996, http://www.unhchr.ch/tbs/doc.nsf/898586b1dc7b4043c1256a450044f331/5a7331a09a8b4f3fc1256404003d10bd/$FILE/G9618895.pdf. Following this discussion, an informal Working Group was set up (CRC/C/57, p. 45).This Working Group met twice (cf. United Nations Committee on the Rights of the Child, CRC/C/66, 06.06.1997, http://www.unhchr.ch/tbs/doc.nsf/898586b1dc7b4043c1256a450044f331/ b27bf9857a55819d802564f3003b10ee/$FILE/G9717203.pdf, 51; United Nations Committee on the Rights of the Child, CRC/C/79, 27.07.1998, http://www.unhchr.ch/tbs/doc.nsf/898586b1dc7b4043c1256a450044f331/

The Committee on the Rights of the Child expressed concern that these technologies have also lowered the threshold of access to illegal and harmful material. The Committee also indicated concern about the extent to which access to the internet and social media lead to exposure to cyberbullying, pornography and cybergrooming.[830]

---

a505a81ff8dcaf89802566d6003b6298/$FILE/G9817376.pdf, 46) and was also involved with the development of 'The Oslo Challenge', a call for action, addressed to "everyone engaged in exploring, developing, monitoring and participating in the complex relationship between children and the media". This document elaborates on ways to effectively implement articles 12, 13 and especially 17 CRC: "The Oslo challenge signals to governments, the media, the private sector, civil society in general and young people in particular that Article 17 of the Convention on the Rights of the Child, far from isolating the child/media relationship, is an entry point into the wide and multi-faceted world of children and their rights – to education, freedom of expression, play, identity, health, dignity and self-respect, protection – and that in every aspect of child rights, in every element of the life of a child, the relationship with children and the media plays a role" (http://www.mediawise.org.uk/files/uploaded/Oslo%20Challenge.pdf).

[828] Article 17 (a) emphasizes the importance of disseminating information and material of social and cultural benefit to the child and in accordance with the spirit of article 29, which is related to education.

[829] Ruxton, Sandy, What about us? Children's rights in the European Union? Next Steps, Brussels, The European Children's Network, 2005, 109.

[830] United Nations Committee on the Rights of the Child, General comment No. 17 (2013) on the right of the child to rest, leisure, play, recreational activities, cultural life and the arts (art. 31), UN Doc. CRC/C/GC/17, 2013, n° 46. https://www.refworld.org/docid/51ef9bcc4.html ("The Committee is concerned at the growing body of evidence indicating the extent to which these environments, as well as the amounts of time children spend interacting with them, can also contribute to significant potential risk and harm to children. UNICEF, Child Safety Online: Global Challenges and Strategies. Technical report (Florence, Innocenti Research Centre, 2012). For example: - Access to the Internet and social media is exposing children to cyberbullying, pornography and cybergrooming. Many children attend Internet cafes, computer clubs and game halls with no adequate restrictions to access or effective monitoring systems; - The increasing levels of participation, particularly among boys, in violent video games appears to be linked to aggressive behavior as the games are highly engaging and interactive and reward violent behavior. As they tend to be played repeatedly, negative learning is strengthened and can contribute to reduced sensitivity to the pain and suffering of others as well as aggressive or harmful behavior toward others. The growing opportunities for online gaming, where children may be exposed to a global network of users without filters or protections, are also a cause for concern. - Much of the media, particularly mainstream television, fail to reflect the language, cultural values and creativity of the diversity of cultures that exist across society. Not only does such monocultural viewing limit opportunities for all children to benefit from the potential breadth of cultural activity available, but it can also serve to affirm a lower value on non-mainstream cultures. Television is also contributing to the loss of many childhood games, songs, rhymes traditionally transmitted from generation to generation on the street and in the playground; - Growing dependence on screen-related activities is thought to be associated with reduced levels of physical activity among children, poor sleep patterns, growing levels of obesity and other related illnesses. See also comment 47." Marketing and commercialization of play: The Committee is concerned that many children and their families are exposed to increasing levels of unregulated commercialization and marketing by toy and game manufacturers. Parents are pressured to purchase a growing number of products which may be harmful to their children's development or are antithetical to creative play, such as products that promote television programmes with established characters and storylines which impede imaginative exploration; toys with microchips which render the child as a passive observer; kits with a pre-determined pattern of activity; toys that promote traditional gender stereotypes or early sexualization of girls; toys containing dangerous parts or chemicals;

The Family Online Safety Institute[831] ("FOSI") Global Resource and Information Directory (GRID) is "designed to create a single, factual and up-to-date source for governments, industry, lawyers, academics, educationalists and all those dedicated to making the Internet a safer and better place".[832]

An on-line safety profile for most countries is available, divided into sections detailing basic country profile data; an overview of online safety in the country; pointers to related research; the education system (this is actually a short profile of ICT use in education -- very useful!); legislation; organizations active in this area in the country; and a list of sources of information.

It has been argued that the word 'guidelines', used in article 17 CRC, indicates a preference for voluntary, rather than legislative constraints.[833] However, the Committee on the Rights of the Child has in one of their observations recommended to "enact special legislation to protect children from harmful information, in particular from television programs and films containing brutal violence and pornography" (own emphasis).[834] This attitude is not limited to traditional media: the Committee is concerned about online media as well.[835] Recently, it has been argued that there is confusion about the scope of article 17 e) (in part created by the United Nations Committee on the Rights of the Child).

The scope of this paragraph does not concern the protection of children from harmful material by States themselves.[836] This particular State task is included within the scope of other articles (such as article 6 CRC, related to the protection and care necessary for the well-being of each child) and that article 17 e) solely concerns the encouragement of other actors, such as industry, to develop the guidelines mentioned in this paragraph.[837]

---

    realistic war toys and games. Global marketing can also serve to weaken children's participation in the traditional cultural and artistic life of their community."

[831] https://www.fosi.org/about-fosi  The Family Online Safety Institute is an international, non-profit organization which works to make the online world safer for kids and their families. Id.

[832] https://fosigrid.org/

[833] Hodgkin, Rachel and Newell, Peter, Implementation handbook for the Convention on the Rights of the Child, New York, UNICEF, 2002, 236. See also: United Nations Committee on the Rights of the Child, Report on the thirteenth session, CRC/C/57, 31.10.1996, retrieved from http://www.unhchr.ch/tbs/doc.nsf/898586b1dc7b4043c1256a450044f331/5a7331a09a8b4f3fc1256404003d10bd/$FILE/G9618895.pdf (on 22.09.2006), 44.

[834] United Nations Committee on the Rights of the Child, Concluding observations of the Committee on the Rights of the Child: Cambodia, CRC/C/15/Add.128, 28.06.2000, retrieved from http://www.unhchr.ch/tbs/doc.nsf/(Symbol)/30dce34798ef39f480256900003397ac?Opendocument (on 27.09.2006), para. 36; United Nations Committee on the Rights of the Child, Concluding observations of the Committee on the Rights of the Child: Marshall Islands, CRC/C/15/Add.139, 16.10.2000, retrieved from http://www.unhchr.ch/tbs/doc.nsf/(Symbol)/e91ea24ff52b434ac125697a00339c0c?Opendocument (on 27.09.2006), para. 34-35.

[835] "The Committee is concerned that no legislation exists to protect children from being exposed to violence and pornography through video movies and other modern technologies, most prominently, the Internet": United Nations Committee on the Rights of the Child, Concluding observations of the Committee on the Rights of the Child: Luxembourg, CRC/C/15/Add.92, 24.06.1998, http://www.unhchr.ch/tbs/doc.nsf/(Symbol)/62258a94c261c9318025662400376374?Opendocument, para. 30.2 "The Committee is concerned that no legislation exists to protect children from being exposed to violence and pornography through video movies and other modern technologies, most prominently, the Internet": United Nations Committee on the Rights of the Child, Concluding observations of the Committee on the Rights of the Child: Luxembourg, CRC/C/15/Add.92, 24.06.1998, http://www.unhchr.ch/tbs/doc.nsf/(Symbol)/62258a94c261c9318025662400376374?Opendocument, para. 30.

[836] "Article 17 is not to be a vehicle for State control of content: Article 17 does not require or authorize State censorship of the content of mass media communications"; Wheatley Sacino, Sherry, Article 17 Access to a diversity of mass media sources, A commentary on the United Nations Convention on the Rights of the Child, Leiden, Martinus Nijhoff Publishers, 2012, 30.

[837] Id.

Article 17 also refers to article 18 CRC. This recalls the primary responsibility of parents for the upbringing and development of the child.[838] However, according to article 18 para. 2, States must "render appropriate assistance to parents and legal guardians in the performance of their child-rearing responsibilities." An example of this 'assistance' or, otherwise put, the 'duty of care' of the state, could be the provision of adequate information by States to parents about media content to which their children can be exposed.[839]

**AI and the Children's Rights by Design (CRbD) Standard**

Just as it has been successfully argued through countless studies and principles globally, that AI must be grounded in human-centric design[840] the design, development and provision of AI, that can directly or indirectly, affect children should always put the rights and best interests of child users first. AI that directly or indirectly impacts children, including in the educational processes.[841] must always prioritize children's rights and interests. There is a legal and ethical duty to respect and protect and children's rights by States and private actors, including tech companies, in the design, development, and provision of any AI technology, product or service. In this sense, the Children's Rights by Design (CRbD) standard should be always considered and applied.

Children, as recognized by the CRC and other national legal norms, experience a unique stage of physical, psychological and social development, with evolving capacities and, therefore, must be specially protected, ensuring their rights are guaranteed as priority, no matter the circumstances, whether by family, States and society, or companies.

AI Systems have to promote children's rights and to support their development worldwide, it is essential to maintain a critical perspective, to keep a human in the loop to vet AI's risk as well as benefits. As stated by UNICEF in their first draft Policy Guidance on AI:

---

[838] Article 5 is also relevant when dealing with harmful content. Article 5 refers to the responsibilities, rights and duties of parents (or other persons legally responsible for the child), to offer, in a manner consistent with the evolving capacities of the child, appropriate direction and guidance to the child when exercising his or her rights. This provision could be interpreted as implying that parents have a responsibility to support their children in their approach to new media. The United Nations General Assembly has also touched upon the responsibilities of parents et al. in this respect: " Encourage measures to protect children from violent or harmful web sites, computer programmes and games that negatively influence the psychological development of children, taking into account the responsibilities of the family, parents, legal guardians and caregivers" (United Nations General Assembly, Resolution A world fit for children, A/RES/S-27/2, 11.10.2002, http://www.unicef.org/specialsession/docs_new/documents/A-RES-S27-2E.pdf, 16). Ultimately, parents or other carers are the only persons who will be able to monitor their children's actual media use.

[839] Hodgkin, Rachel and Newell, Peter, Implementation handbook for the Convention on the Rights of the Child, New York, UNICEF, 2002, 236. Sacino finds that this reference deliberately avoids clarifying the relationship between the role of the States and the role of parents in the protection of young people from harmful media content, because there could not be found a consensus on the division of this responsibility: Wheatley Sacino, Sherry, Article 17 Access to a diversity of mass media sources, A commentary on the United Nations Convention on the Rights of the Child, Leiden, Martinus Nijhoff Publishers, 2012, 31.

[840] Aligned Design: A Vision for Prioritizing Human Well-being with Autonomous and Intelligent Systems, Version 2. IEEE, 2017. http://standards.ieee.org/develop/indconn/ec/autonomous_systems.htm; see also Yeung, Karen and Howes, Andrew and Pogrebna, Ganna, AI Governance by Human Rights-Centred Design, Deliberation and Oversight: An End to Ethics Washing (June 21, 2019). Forthcoming in M Dubber and F Pasquale (eds.) The Oxford Handbook of AI Ethics, Oxford University Press (2019), Available at SSRN: https://ssrn.com/abstract=3435011 or http://dx.doi.org/10.2139/ssrn.3435011 and Kazim, Emre and Koshiyama, Adriano, Human Centric AI: A Comment on the IEEE's Ethically Aligned Design (April 13, 2020). Available at SSRN: https://ssrn.com/abstract=3575140 or http://dx.doi.org/10.2139/ssrn.3575140

[841] Isabella Henriques and Pedro Hartung, Children's Rights by Design in AI Development for Education, International Review of Information Ethics, Vol. 29 (03/2021)
https://informationethics.ca/index.php/irie/article/view/424/401

*"While AI is a force for innovation and can support the achievement of the Sustainable Development Goals (SDGs), it also poses risks for children, such as to their privacy, safety and security. Since AI systems can work unnoticed and at great scale, the risk of widespread exclusion and discrimination is real."* [842]

AI for children is any AI that directly or indirectly impacts children Although the population of children impacted by AI systems is significant - they represent 1/3 of users worldwide on the Internet alone (without accounting for the AI applied massively in schools, cities and other spaces) - the vast majority of AI policy initiatives that exist around the world hardly mention them or when they do, they are limited to broad citations, without details or deeper considerations about their particularities. They do not deal, for example, with the possible uses of predictive analysis or other types of algorithmic modeling that can make determinations about the future of children, causing them unpredictable consequences.[843]

This demonstrates the immense urgency to expand the study of the implications of AI in multiple global childhoods, including among children in the Global South, in which accessibility to the internet is often conditioned to commercial exploitation models, for some applications and services[844], all of which abound in automated decisions.

One of the few documents on this subject is the first draft of the Policy Guidance on AI for Children, recently launched by UNICEF, which set out nine requirements for a child-centered AI, which should be based on the defense of children's rights, through the lens protection, provision and participation. They are: (1) Support children's development and well-being; (2) Ensure inclusion of and for children; (3) Prioritize fairness and non-discrimination for children; (4) Protect children's data and privacy; (5) Ensure safety for children; (6) Provide transparency, explainability, and accountability for children; (7) Empower governments and businesses with knowledge of AI and children's rights; (8) Prepare children for present and future developments in AI; (9) Create an enabling environment for all to contribute to child-centered AI.[845]

In view of the cross-border multiplication of AI systems, including those that impact children, new global initiatives such as that of UNICEF, guided by ethics and human-centric, will be of paramount importance. Undoubtedly, AI systems that impact children, directly or indirectly, must also be, as any AI systems, first and foremost, human-centered, as mentioned in the European Commission, which seeks to promote a reliable AI:

*"AI systems need to be human-centric, resting on a commitment to their use in the service of humanity and the common good, with the goal of improving human welfare and freedom."* [846]

Thus, in addition to the challenge of harmonizing innovation, efficiency and freedom of business models with the protection of human rights, accountability, explainability and transparency of AI systems, there is one more: finding a balance that guarantees the best interest of children and their specific rights, in all applications that are not prohibited and can be used by them or impact them, even indirectly. And not only in those AI applications specifically aimed at the use and consumption of children[847] - also as a precaution against potential risks to which they may be subjected .[848] And all AI that can directly or indirectly affect children must take their rights and interests first, in addition to ensuring their best interest and being human-centered. This means that the best interest of children and their rights must be pursued with priority by every AI developer, even though their product or service was meant not to be used by children or affect them indirectly at first sight.

---

[842] UNICEF, Executive Summary, Policy Guidance on AI for Children - Draft 01, September, 2020. Available https://www.unicef.org/globalinsight/media/1171/file/UNICEF-Global-Insight-policy-guidance-AI-children-draft-1.0-2020.pdf

[843] Alexa Hasse, Sandra Cortesi, Andres Lombana-Bermudez, and Urs Gasser. Youth and Artificial Intelligence: Where We Stand (2019), available at https://cyber.harvard.edu/publication/2019/youth-and-artificial-intelligence/where-we-stand

[844] UN IGF, Net Neutrality Reloaded: Zero Rating, Specialised Service, Ad Blocking and Traffic Management. Annual Report of the UN IGF Dynamic Coalition on Net Neutrality. Available at: https://bibliotecadigital.fgv.br/dspace/bitstream/handle/10438/17532/Net%20Neutrality%20Reloaded.pdf

[845] UNICEF, Policy Guidance on AI for Children - Draft 01, September, 2020. Available at: https://www.unicef.org/globalinsight/media/1171/file/UNICEF-Global-Insight-policy-guidance-AI-children-draft-1.0-2020.pdf

[846] The European Commission, ´Ethics Guidelines for Trustworthy AI´, Independent High-Level Expert Group on Artificial Intelligence, p. 4. Disponível em https://ec.europa.eu/futurium/en/ai-alliance-consultation/guidelines

In this sense, efforts must be expanded to democratize the benefits of AI systems for children, as well as to mitigate possible risks, especially in different contexts and for the multiple childhood development around the planet.

Hence, it is essential to guarantee Children's Rights by Design of AI systems which impact children, based in their best interest, so that the promotion of children's rights, as well as their protection, is effective, generating real positive impacts on the lives of children, including those who are socioeconomically vulnerable.

**AI and Children's Social Media Platforms**

Social media platforms that rely on streaming technologies are overturning how adults and children consume media content. Platforms are working hard to ensure consumers maximize their time on these sites.

YouTube[849] stands out as the dominant player in this space, especially when it comes to today's youth. In 2017, 80% of U.S. children ages 6 to 12 used YouTube on a daily basis.[850] YouTube was the 2016 and 2017 "top kids brand" according to Brand Love studies.[851] In the 2017 study, 96% of children ages 6 to 12 were found to be "aware of YouTube," and 94% of children ages 6 to 12 said they "either loved or liked" YouTube.[852] The YouTube phenomenon isn't just occurring in the United States as YouTube has massive user bases in India, Moscow, across Europe, and beyond.[853]

Watching video clips online is among the earliest internet activities carried out by very young children, resulting in high popularity of YouTube channels targeting toddlers and preschoolers.[854] For example, YouTube's Sesame Street channel recently reached a billion views[855] and a TuTiTu channel (owned by a small animation company targeting infants and toddlers) was ranked 40th among YouTube's 100 most viewed channels.[856] YouTube's simple user interface, that allows even toddlers to proceed to the next item on the playlist and affords them easy access to favorite videos that can be watched again and again, has been suggested as the key to its popularity with very young audiences.[857] It is thus not surprising that producers of content targeting toddlers and preschoolers soon discovered YouTube's appeal and began using it as a major content promotion platform by uploading complete episodes or short clips of programs broadcast on television channels.[858]

---

[847] ICO, UK Age-Appropriate Design Code, 2020: "This code applies to "information society services likely to be accessed by children" in the UK. This includes many apps, programs, connected toys and devices, search engines, social media platforms, streaming services, online games, news or educational websites and websites offering other goods or services to users over the internet. It is not restricted to services specifically directed at children." Available in https://ico.org.uk/for-organisations/guide-to-data-protection/key-data-protection-themes/age-appropriate-design-a-code-of-practice-for-online-services/executive-summary/

[848] "Simply put: children interact with or are impacted by AI Systems that are not designed for them, and current policies do not address this. Furthermore, whatever is known about how children interact with and are impacted by AI is just the start. The disruptive effects of AI will transform children´s lives in ways we cannot yet understand, for better or for worse. Our collective actions on AI today are critical for shaping a future that children deserve." In: UNICEF, ´Policy Guidance on AI for Children´ - Draft 01, September, 2020. Available at: https://www.unicef.org/globalinsight/media/1171/file/UNICEF-Global-Insight-policy-guidance-AI-children-draft-1.0-2020.pdf

[849] YouTube is a subsidiary of Google, whose parent company is Alphabet, Inc.

[850] "2017 Brand Love Study: Kid & Family Trends," Smarty Pants: the Youth and Family Experts (2017), 14.

[851] Id. at 7.

[852] Id.

[853] Alexis Madrigal, "Raised by YouTube," Atlantic 322, no. 4 (November 2018): 72–80. https://www.theatlantic.com/magazine/archive/2018/11/raised-by-youtube/570838/

[854] Holloway, D., Green, L., & Livingstone, S. (2013). Zero to eight. Young children and their internet use. London, UK: EU Kids Online. Retrieved from http://eprints.lse.ac.uk/52630/1/Zero_to_eight.pdf

[855] Luckerson, V. (2013, March 13). How Sesame Street Counted All the Way to 1 Billion YouTube Views. Time. Retrieved from http://business.time.com/2013/03/15/how-sesame-street-counted-all-the-way-to-1-billion-youtube-views

Besides providing an extensive variety of content produced specifically for young children, YouTube has also spawned new formats in children's entertainment that once baffled people outside their target audiences.[859] Young children appear to be attracted to particular types of content, many of which are based on comic situations, such as challenges (e.g., tasting hot pepper) and silly skits (e.g., a person in a rooster costume surprising a police officer). Topping the list of children's favorites, however, are unboxing videos[860], in which boxes containing different products are opened.[861] The attraction of unboxing may lie in the mystery of the unwrapping process. Young children enjoy mystery and suspense, especially when it is likely to have safe and predictable outcomes. One prime example of this trend is a series of YouTube videos in which a person opens Kinder Surprise Eggs[862], with hundreds of millions of hits. Although no data are available regarding the viewers' ages, Jordan maintains that such videos are particularly appealing to children aged 2-4 because they expose them to shape transformation, thereby gratifying a developmental need characteristic of this age group.[863]

Notwithstanding its benefits, YouTube also has significant drawbacks as a source of children's entertainment. When the YouTube Kids application was launched, Google declared it to be a safe and educational media environment for the very young, equipped with a safety mode for automatic filtering of content marked as inappropriate. The result, however, fell far short of fulfilling Google's promise and YouTube Kids has been criticized heavily for its lack of professional selection and display of commercial content, ignoring the well-established advertising safeguards adopted by both broadcast and cable television.[864]

In 2015, YouTube decided to launch a dedicated platform called YouTube Kids as a means to provide safe, age appropriate content for children.[865] This 'YouTube Kids' app, is a different product than the standard YouTube

---

[856] Fox, A. (2014, March 26). The Israelis that conquered toddlers around the world. Mako Magazine. Retrieved from http://www.mako.co.il/home-family-weekend/Article-a4dda21f12ef441006.htm

[857] Buzzi, M. (2011). What are your children watching on YouTube? In V. F. Cipolla, K. V. Ficarra, & D. Verber (Eds.), Advances in new technologies interactive interferences and communicability (pp. 243-252). Berlin, Germany: Springer.

[858] Grossaug, R. (2017). What influences the influencers: Preschool television production in an era of media change: The case of Israel's 'Hop! Group' [Unpublished doctoral dissertation]. The Hebrew University of Jerusalem. See also Elias, N., Sulkin, I., & Lemish, D. (in press). Gender segregation on Baby TV: Old-time stereotypes for the very young. In D. Lemish & M. Gotz (Eds.), Beyond the stereotypes: Boys, girls and their images. Nordicom.
https://www.researchgate.net/publication/321481837_Gender_segregation_on_BabyTV_Old-time_Stereotypes_for_the_Very_Young

[859] Dredge, S. (2015, November 19). Why YouTube is the new children's TV and why it matters. The Guardian. Retrieved from http://www.theguardian.com/technology/2015/nov/19/youtube-is-the-new-childrens-tv-heres-why-that-matters

[860] Marsh, J. (2016). Unboxing' videos: Co-construction of the child as cyberflaneur. Discourse: Studies in the cultural politics of education, 37, 369-380. https://doi.org/10.1080/01596306.2015.1041457

[861] Knorr, C. (2016, March 15). What kids are really watching on YouTube? Common Sense Media. Retrieved from http://www.commonsensemedia.org/blog/what-kids-are-really-watching-on-youtube

[862] https://www.amazon.com/Kinder-Surprise-Eggs/s?k=Kinder+Surprise+Eggs

[863] Jordan, A. B. (2015, November). Digital natives and digital immigrants: Media use and generational identity. Keynote lecture. Ben-Gurion University of the Negev, Beer- Sheva, Israel.

[864] Golin, J., Chester, F., & Campbell, A. (2015, April 7). Advocates file FTC complaint against Google's YouTube Kids. Campaign for a Commercial-Free Childhood. Retrieved from http://www.commercialfreechildhood.org; See also Luscombe, B. (2015, September 7). YouTube view's master. Time, 70-75.

[865] "Introducing the Newest Member of Our Family, the YouTube Kids App--Available on Google Play and the App Store," Official YouTube Blog, https://youtube.googleblog.com/2015/02/youtube-kids.html; "YouTube

app. YouTube Kids features a children-friendly layout, which, according to YouTube, is designed to "make it safer for children to explore the world through online video".

The app has multiple integrated parental controls. Prior to using the YouTube Kids app for instance, a parent is required to unlock the app and verify their children's age. Other parental controls include the possibility to turn the 'search' option on or off, with the latter meaning that the kid can only see video's from video creators verified by YouTube itself, and a timer which limits the amount of time that a user can use the app. The YouTube Kids app therefore offers a 'barebones' version of the original YouTube app, by removing several features. It is not possible to leave a rating on videos in the YouTube Kids app, and there is no comment section below the videos where the viewers can leave their thoughts. This is purposefully designed in order to limit the unwanted exposure to some of the content that is available on YouTube, which was deemed inappropriate for younger audiences. In order to prevent exposure to inappropriate content, all videos on the YouTube Kids app are checked whether they are child friendly. The YouTube Kids app contains a 'recommended' tab under videos, which displays other videos that are related to the video that a user is currently watching. These videos are all videos from the YouTube Kids app only, subjected to the same age restrictions as other YouTube Kids videos. Advertisements are also displayed on the videos. These advertisements are extensively checked by YouTube to ensure that these are family-friendly.[866]

All the content submitted to the YouTube Kids app is subjected to a verification process by a machine algorithm.[867] In the case that the algorithm approves a video for YouTube Kids, then every user can view this video.

On both YouTube and YouTube Kids, machine learning algorithms are used to both recommend and mediate the appropriateness of content.[868] YouTube representatives, however, have been opaque about differences in the input data and reward functions underlying YouTube Kids and YouTube.[869] Lack of transparency about the input data used in algorithms makes it difficult for concerned parties to understand the distinction.[870] More generally, the issue of algorithmic opacity is of concern with both YouTube and YouTube Kids, since YouTube, and not YouTube Kids, continues to account for the overwhelming majority of viewership of children's programming within the YouTube brand.[871]

The machine learning algorithms – primarily the recommendation engine employed by YouTube and YouTube Kids – are optimized to ensure that children view as many videos on the platform as possible.[872] Children do not need to enter any information or affirm any acquired permissions to watch thousands of videos on YouTube and YouTube Kids. [873] Touchscreen technology and the design of the platforms allow even young children substantial

---

Kids," https://www.youtube.com/yt/kids/.
[866] http://arno.uvt.nl/show.cgi?fid=152292;
[867] Kantrowitz, Alex. 'YouTube Kids Is Going To Release A Whitelisted, Non-Algorithmic Version Of Its App' (Buzzfeed News, April 6, 2018). https://www.buzzfeednews.com/article/alexkantrowitz/youtube-kids-is-going-to-release-a-whitelistednon#.ftVwoX5dp and Wojcicki, Susan. 'Protecting Our Community' (YouTube Creator Blog, 2017). https://youtube-creators.googleblog.com/2017/12/protecting-our-community.html
[868] Karen Louise Smith and Leslie Regan Shade, "Children's Digital Playgrounds as Data Assemblages: Problematics of Privacy, Personalization and Promotional Culture," Big Data & Society, Vol. 5 (2018), at 5. https://journals.sagepub.com/doi/pdf/10.1177/2053951718805214
[869] Adrienne LaFrance, "The Algorithm That Makes Preschoolers Obsessed With YouTube Kids," The Atlantic, July 27, 2017, https://www.theatlantic.com/technology/archive/2017/07/what-youtube-reveals-aboutthe-toddler-mind/534765/.
[870] "Terms of Service - YouTube," https://www.youtube.com/static?template=terms, (November 13, 2018); Matt O'Brien. "Consumer Groups Say YouTube Violates Children's Online Privacy," Time.Com, April 10, 2018, 1. https://www.aol.com/news/consumer-groups-youtube-violates-children-012240300.html
[871] Madrigal, "Raised by YouTube," 80. See also Tőkés, Gyöngyvér, Digital Practices in Everyday Lives of 4 to 6 Years Old Romanian Children (November 30, 2016). Journal of Comparative Research in Anthropology and Sociology, Volume 7, Number 2, Winter 2016, Available at SSRN: https://ssrn.com/abstract=2915463
[872] Matt O'Brien. "Consumer Groups Say YouTube Violates Children's Online Privacy," Time.Com, April 10, 2018, 1. https://www.aol.com/news/consumer-groups-youtube-violates-children-012240300.html
[873] Id.

ease of access.[874] Unfortunately, neither recommendation system appears to optimize for the quality or educational value of the content.[875] Because companies developing children's programming are similarly concerned about maximizing viewers and viewer hours, their posts are often designed around YouTube's privileging of quantity with little consideration for quality, including educational value.[876] There is particular concern that with YouTube and YouTube Kids' algorithm-derived "related-videos" recommendations[877] children can become easily trapped in "filter bubbles"[878] of poor-quality content.[879]

Filtering algorithms also raise other problems,[880] especially when a significant number of external entities are able to co-opt YouTube and YouTube Kids' algorithmic discovery processes to maximize viewer time with sometimes startling consequences for children.[881] For example, anyone over the age of 18 can create and upload content onto YouTube and their creations are not regulated by professional protocols.

---

[874] Elias, Nelly, and Idit Sulkin. "YouTube Viewers in Diapers: An Exploration of Factors Associated with Amount of Toddlers' Online Viewing." Cyberpsychology, November 23, 2017, at 2, available at https://cyberpsychology.eu/article/view/8559/7739.

[875] Madrigal, "Raised by Youtube," 79.

[876] Adrienne LaFrance, "The Algorithm That Makes Preschoolers Obsessed With YouTube Kids." https://www.theatlantic.com/technology/archive/2017/07/what-youtube-reveals-about-the-toddler-mind/534765/

[877] Children, too, access information and news from a variety of social media sites and platforms. But how confident are they that what they encounter online is not misinformation or deliberate disinformation, or so-called 'fake news'? According to the Global Kids Online study, between 20 and 40 per cent of children between the ages of 9 and 11 'find it easy to check if the information [they] find online is true.' Byrne, Jasmina, et al. Global Kids Online Research Synthesis: 2015-2016. UNICEF Office of Research – Innocenti and London School of Economics and Political Science. Florence. Available at: https://www.unicef-irc.org/publications/869-global-kids-online-research-synthesis-2015-2016.html The emergence of so-called 'filter bubbles' occurring when platforms and search engines make use of algorithms to select information a user would want to see underlines the potential seriousness of this issue with respect to children. Instead of exposing children to a variety of ideas, different perspectives and ways of thinking, web platforms in general, and 'fake news' in particular, may lead to their engagement with news or information sources that confirm existing points of view or prejudices. See also Bezemek, Christoph, The 'Filter Bubble' and Human Rights (November 2, 2018). Petkova/Ojanen (eds), Fundamental Rights Protection Online: The Future Regulation of Intermediaries, Forthcoming, Available at SSRN: https://ssrn.com/abstract=3277503 or http://dx.doi.org/10.2139/ssrn.3277503; and Dutton, William H. and Reisdorf, Bianca and Dubois, Elizabeth and Blank, Grant, Social Shaping of the Politics of Internet Search and Networking: Moving Beyond Filter Bubbles, Echo Chambers, and Fake News (March 31, 2017). Quello Center Working Paper No. 2944191, Available at SSRN: https://ssrn.com/abstract=2944191 or http://dx.doi.org/10.2139/ssrn.2944191; and V Verdoodt, and E Lievens, Targeting children with personalised advertising: how to reconcile the (best) interests of children and advertisers, in Data Protection and Privacy Under Pressure: Transatlantic tensions, EU surveillance, and big data, (2017) https://biblio.ugent.be/publication/8541057/file/8541058

[878] See generally Tracy S. Bennett, Ph.D., Sorry to Burst Your [Filter] Bubble, GetKidsInternetSafe, https://getkidsinternetsafe.com/sorry-to-burst-your-filter-bubble/ In 2011, Eli Parisier released his book The Filter Bubble: What the Internet Is Hiding From You. Pariser E. The filter bubble: What the Internet is hiding from you. London: Penguin UK; 2011. (" Parisier explains how the internet search engines and their algorithms are creating a situation where users increasingly are getting information that confirms their prior beliefs. Search algorithms are using large quantities of information about the user to find and present relevant information to the individual user. Your search and browse history is a key piece of the information used to tailor the results you get when you perform online searches. Combining this with information about your social network, viewing habits and geography leads to an increasingly narrow view on the information available online. Parisier's main argument is that this narrowing creates a filter bubble, which is invisible to the user, but still has immense impact on the information available to the individual. When you perform a Google search, the information about you is used in addition to your search term to find and prioritize the search results most likely to be of your interest. Then, when you click among the first search results (as most people do), you are confirming back to the search engine that the results were

YouTube and YouTube Kids' algorithmic discovery processes can be manipulated to push content that the pusher expects will perform well on the platform's "related-videos" engine, incentivizing sensational content.[882] Prioritizing such content is one of the critical impacts of YouTube's use of machine learning algorithms.[883] Kids are particularly susceptible to content recommendations, so shocking "related videos" can grab children's attention and divert them away from more child-friendly programming.[884]

The situation on YouTube algorithms and how they have impacted many young children is concerning, even disturbing. First, the transmission of child-oriented content is interrupted frequently by automatic advertisements, many of which are inappropriate for younger viewers. Moreover, a recent report on children's safety on YouTube shows that very young children are only 2-4 clicks away from adult content while watching children-oriented videos. For example, children watching Sesame Street are two clicks from a car accident video, while those viewing Dora the Explorer -are four clicks from a video featuring swearing and nudity.[885]

---

indeed relevant and/or interesting. This in turn strengthens the filter, making it more likely that you will receive similar results in the future. However, it is not only your own behavior that influences the results. The interests and preferences among people in your social network are also part of the algorithms, making it more likely that you will receive search results that your social network in general is gravitating toward. In many cases, these filters are providing relevant and good results. However, it becomes a problem as soon as your profile contains elements that make the search results gravitate toward misinformation. The filters are to a large degree invisible, which adds to the problem. Many users are not even aware that the filtering is taking place, and even if they are, it is difficult to take control of how the filter is being applied. Granted, you can go to Google and delete your search history, or click the "Hide private results" button in the top right of the search results. Still, the complexity of the algorithms and the lack of usable explanations about how the filters actually work make it difficult for the user to take control. The way the filters influence search results have led our group to use the term Gravitational Black Holes of Information to illustrate how difficult it is to break out of the force of the filters. As soon as you are aiming in at a core of misinformation, it is inherently difficult to break out of the gravitational force of the search algorithms. On the way toward the gravitational center, your prior believes are being strengthened by the new information you find, further pulling you into the black hole." Harald Holone, The filter bubble and its effect on online personal health information, Croat Med J. 2016 Jun; 57(3): 298–301. doi: 10.3325/cmj.2016.57.298).

[879] Id.
[880] See, e.g., Siddiqui, Anaum, A Critical Look at YouTube Videos: Causing Behavioral Change Among Children (March 1, 2019). Available at SSRN: https://ssrn.com/abstract=3453417 or http://dx.doi.org/10.2139/ssrn.3453417 ("Media has always been assumed as one of the sources of building realities for the society. Media content is considered as an important cause of behavioral change in society. Children due to their minor age are more likely to get influenced by the content. Due to emergence of YouTube and its easy accessibility and negligence of parents due to their busy lives, we cannot limit its effects on our children. This research merely focuses on the behavioral change of children caused by watching YouTube videos. As per this research findings children have used YouTube as institute from where they have learned their basic education such as alphabets and counting, identification of colors and shapes and nursery rhymes. On other hand these changes have increased aggression, unhealthy mental growth, sleeping disorders and any other emotional or physical change. Semi-Structured interviews have been conducted with N=30 mothers of preschooler from Islamabad. Proportionate Stratified Sampling method has been adopted to cover most of the population. The sample has been divided into three class divisions such as Upper, Middle and Lower class, out of which N=10 samples are interviewed randomly. Mothers who belong to Upper class of the society are mostly more educated and they expose their children to comparatively positive and educating content. Whereas mothers of middle class have less control over the content and they are only focused on how to keep their child busy and distracted. Mothers of Lower class have no idea about the quality of content, and time their children are spending watching that content. One similar notion that can be seen in all three class divisions is that mothers want escapism and for that they are exposing their children to YouTube videos.")
[881] Elias, Nelly, and Idit Sulkin. 2.
[882] Elias, N., & Sulkin, I. (2017). YouTube viewers in diapers: An exploration of factors associated with amount of toddlers' online viewing. Cyberpsychology: Journal of Psychosocial Research on Cyberspace, 11(3), Article 2. https://doi.org/10.5817/CP2017-3-2

It was discovered that the YouTube algorithms approved videos that contained content not suited for kids, such as violence and sexual misconduct.[886] Numerous media reports[887] covered the increasing popularity of amateur live action videos that bear innocent tags using names of children's most popular heroes, such as Elsa and Anna (from the movie Frozen) or Spiderman, but contain offensive content and present explicit expressions of sexual behavior, vandalism and violence. These videos very easily find their way into the suggested YouTube playlists of episodes from the favorite children's shows and gain popularity as millions of people view them.[888] These animated figures engaged in behavior, such as decapitation, pornographic acts and criminal behavior, including, but not limited to, murder, theft and sexual assault. Younger audiences were thus subjected to severely disturbing behavior, which is clearly detrimental to them.

This period of controversial videos being widely spread on YouTube Kids is referred to as 'ElsaGate'.[889] Multiple studies have found that media has a vast impact on youth, with studies finding correlations between increased violent behavior when subjected to violent television programming,[890] and promoting sexual behavior.[891] These ElsaGate videos were exposed to millions of kids, whose behavior and emotional development has been impacted due to these videos. Scientific research concerning deep learning architectures were published in response to the ElsaGate, bringing up further discussion alongside potential solutions to the problem.[892]

---

[883] Elias, N., & Sulkin, I. (2017). YouTube viewers in diapers: An exploration of factors associated with amount of toddlers' online viewing. Cyberpsychology: Journal of Psychosocial Research on Cyberspace, 11(3), Article 2. https://doi.org/10.5817/CP2017-3-2

[884] Elias, N., & Sulkin, I. (2017). YouTube viewers in diapers: An exploration of factors associated with amount of toddlers' online viewing. Cyberpsychology: Journal of Psychosocial Research on Cyberspace, 11(3), Article 2. https://doi.org/10.5817/CP2017-3-2

[885] Kaspersky Lab, (2013, February 5). Children at High Risk of Accessing Adult Content on YouTube. PRNewswire. Retrieved from http://www.prnewswire.com/news-releases/children-at-high-risk-of-accessing-adult-content-on-youtube-189770621.html

[886] Maheshwari, Sapna. 'On YouTube Kids, Startling videos slip past filters'. (New York Times, 2017). https://www.nytimes.com/2017/11/04/business/media/youtube-kids-paw-patrol.html?_r=0

[887] Subedar, Anisa. "The Disturbing Youtube Videos That Are Tricking Children" (2019) https://www.bbc.com/news/blogs-trending-39381889; Dredge, Stuart. 2016. "Youtube's Latest Hit: Neon Superheroes, Giant Ducks And Plenty Of Lycra". The Guardian. https://www.theguardian.com/technology/2016/jun/29/youtube-superheroeschildren-webs-tiaras ; "Youtube: Wie Gefälschte Disney-Cartoons Kinder Verstören - Derstandard.At". 2019. DER STANDARD, https://derstandard.at/2000055049856/Youtube-Wie-gefaelschte-Disney-CartoonsKinder-verstoeren; Robertson, A. 2017. "What Makes YouTube's Surreal Kids' Videos So Creepy?" The Verge, November 21. https://www .theverge.com/culture/2017/11/21/16685874/kids- youtube-video-elsagate-creepiness-psychology.

[888] Mathijs Stals, The technological downside of algorithms:an 'ElsaGate' case study, Masters Thesis, (August 2020) http://arno.uvt.nl/show.cgi?fid=152292; See also Kostantinos Papadamou, Characterizing Abhorrent, Misinformative, and Mistargeted Content on YouTube, Ph.D. Thesis, (May 16, 2021) https://arxiv.org/pdf/2105.09819.pdf

[889] Brandom, Russell. "Inside Elsagate, The Conspiracy-Fueled War On Creepy Youtube Kids Videos". 2017. The Verge. https://www.theverge.com/2017/12/8/16751206/elsagate-youtube-kids-creepyconspiracy-theory.

[890] Johnson, JG et. al. 'Television viewing and aggressive behavior during adolescence and adulthood' (2002). https://www.ncbi.nlm.nih.gov/pubmed/11923542

[891] Strasburger, Victor C. 'Adolescent Sexuality and the Media' (1989). https://www.sciencedirect.com/science/article/pii/S0031395516366949; and Brown, Jane D. 'Mass media influences on sexuality' https://www.tandfonline.com/doi/abs/10.1080/00224490209552118

[892] Ishikawa, Akari et. al. "Combating the ElsaGate Phenomenon: Deep Learning Architectures for Disturbing Cartoons". 2019. Arxiv. https://www.semanticscholar.org/paper/Combating-theElsagate-Phenomenon%3A-Deep-Learning-Ishikawa-Bollis/938d3fd2cede997006cae88bdc26b2af92e4d384

Another AI algorithmic governance challenge is children's inappropriate exposure to YouTube and YouTube Kids-related advertising.[893] YouTube's business model relies on tracking the IP addresses, search history, device identifiers, location and personal data of consumers so that it can categorize consumers by their interests, in order to deliver "effective" advertising.[894]

Although YouTube Kids claims to prohibit "interest-based advertising" and ads with "tracking pixels,"[895] advertising disguised as programming is ubiquitous on the YouTube Kids application.[896]

YouTube's terms of service state that its main app and website are meant only for viewers 13 and older, which means that the site does not have to comply with the Children's Online Privacy Protection Act of 1998[897] ("COPPA"), the law passed in the US in response to growing concerns throughout the 1990s about the safe use of the internet by children.[898]The company directs those under the age of 13 to the YouTube Kids app, which pulls its videos from the main site.

Although YouTube restricts paid advertising of food and beverages on YouTube Kids, for example, food companies may use their own branded channels to spotlight particular food and beverages that they produce,

---

[893] Sarah Perez, "Over 20 advocacy groups complain to FTC that YouTube is violating children's privacy law," TechCrunch, April 9, 2018, https://techcrunch.com/2018/04/09/over-20-advocacy-groups-complain-to-ftcthat-youtube-is-violating-childrens-privacy-law/' The Children's Online Privacy Protection Act of 19981 ("COPPA") purportedly protects children on the internet. 15 U.S.C. §§ 6501–05 (2018). COPPA was passed in response to growing concerns throughout the 1990s about the safe use of the internet by children. In particular, COPPA was aimed at "(i) [the] overmarketing to children and collection of personally identifiable information from children that is shared with advertisers and marketers, and (ii) children sharing information with online predators who could use it to find them offline." COPPA was implemented by the FTC through its Child Online Privacy Protection Rule, which took effect April 21, 2000. In general, COPPA regulates the collection of personal information from children and applies to websites "directed to children" and those whose operators have "actual knowledge" of child users. Children are identified as individuals under the age of thirteen. The five key requirements of the act are notice, parental consent, parental review, security, and limits on the use of games and prizes. In order to legally collect covered personal information from a child, a website operator must first obtain "verifiable parental consent" in a form that varies based on the intended use of the information. The FTC's most recent amendments to the COPPA rule took effect in 2013 and clarified that the regulations are applicable to web services and mobile apps and that "personal information" includes geolocation data, device identifiers, and media containing the voice or image of a child. In September 2019, the FTC, acting with the Attorney General of New York, announced that it reached a settlement with YouTube and parent company Google in response to allegations that the services "illegally collected personal information from children without their parents' consent," in violation of COPPA. The companies agreed to pay $34 million to New York and $136 million to the FTC. Press Release, Fed. Trade Comm'n, Google and YouTube Will Pay Record $170 Million for Alleged Violations of Children's Privacy Law (Sep. 4, 2019) https://www.ftc.gov/news-events/pressreleases/2019/09/google-youtube-will-pay-record-170-million-alleged-violations. See also Beemsterboer, Stephen, COPPA Killed the Video Star: How the YouTube Settlement Shows that COPPA Does More Harm Than Good (June 16, 2020). Stephen Beemsterboer, COPPA Killed the Video Star: How the YouTube Settlement Shows that COPPA Does More Harm Than Good, 25 Ill. Bus. L.J. 63 (2020), Available at SSRN: https://ssrn.com/abstract=3631855; See generally Reddy, T. Raja and Reddy, Dr. E. Lokanadha and Reddy, T. Narayana, Ethics of Marketing to Children: A Rawlsian Perspective (October 9, 2020). Journal of Economics and Business, Vol. 3 No. 4 (2020), Available at SSRN: https://ssrn.com/abstract=3706544
[894] See, e.g., Campbell, Angela J., Rethinking Children's Advertising Policies for the Digital Age (2016). 29 Loy. Consumer L. Rev. 1 (2016), Available at SSRN: https://ssrn.com/abstract=2911892
[895] Sapna Maheshwari, "New Pressure on Google and YouTube Over Children's Data," NY Times, September 20, 2018, https://www.nytimes.com/2018/09/20/business/media/google-youtube-children-data.html
[896] Sapna Maheshwari, "New Pressure on Google and YouTube Over Children's Data," NY Times, September 20, 2018, https://www.nytimes.com/2018/09/20/business/media/google-youtube-children-data.html
[897] 15 U.S.C. §§ 6501–05 (2018).
[898] Lauren A. Matecki, Note, Update: COPPA Is Ineffective Legislation! Next Steps for Protecting Youth Privacy Rights in the Social Networking Era, 5 NW. J. L. & SOC. POL'Y 369, 370 (2010)

burying what are essentially ads within programs, and thereby target children with their products.[899] Thus, corporations are finding ways to target minors in ways that uphold the letter but not the spirit of the rules and in ways that may be hidden from parents and other concerned parties.[900]

Concerns about these platforms impacts on children continue to result in lawsuits[901] and campaigns to regulate them. Fairplay[902] is leading a powerful international coalition of 100 experts, advocates, and organizations in calling on Facebook to abandon its plans to create an Instagram for children.[903]

**AI and Children s Rights at Play: Smart Toys**

Toys are more interactive than ever before. The emergence of the Internet of Things (IoT) makes toys smarter and more communicative: they can now interact with children by "listening" to them and respond accordingly.[904]

---

[899] Cecilia Kang, "YouTube Kids App Faces New Complaints Over Ads for Junk Food," NY Times, December 21, 2017, sec. Technology, https://www.nytimes.com/2015/11/25/technology/youtube-kids-app-faces-newcomplaints.html

[900] Smith and Shade, "Children's Digital Playgrounds," 5. See also Tur-Viñes, Victoria & Castelló-Martínez, Araceli. (2021). Food brands, YouTube and Children: Media practices in the context of the PAOS self-regulation code. Communication & Society. 87-105. 10.15581/003.34.2.87-105. ("The objective of this study is to analyze media practices involving food content on YouTube in terms of the self-regulatory framework established by the PAOS code, which was originally designed for television. The study considers content created and disseminated by two different sources: food brands and child YouTuber channels. We conducted an exploratory qualitative-quantitative study based on a content analysis of videos posted in 2019 on the most viewed YouTube channels in Spain (Socialblade, 2019). The final sample included 211 videos (29h 57m) divided into two subsamples: the official channels of 13 Spanish food brands (82 videos), and 15 Spanish child YouTuber channels (129 videos). The study has facilitated information on nine dimensions: (1) adherence to regulations and ethical standards, (2) nutrition education and information, (3) identification of advertising, (4) presence of risk, (5) clarity in the presentation of the product and in the language used, (6) pressure selling, (7) promotions, giveaways, competitions, and children's clubs, (8) support and promotion through characters and programs and (9) comparative presentations. The main findings reveal the experimental nature of videos featuring food brands that are posted on YouTube for child audiences, especially videos broadcast on the channels of child YouTubers, who post content without an ethical strategy sensitive to their target audience. The lack of compliance with the basic requirement of identifying the video as advertising underscores the urgent need to adapt existing legal and ethical standards to these new formulas of commercial communication.")

[901] Christina Davis, YouTube, Google Class Action Says Kid Data Collected Without Permission, (Oct. 30, 2019) https://topclassactions.com/lawsuit-settlements/consumer-products/mobile-apps/929240-youtube-google-class-action-says-kid-data-collected-without-permission/ Complaint available at https://www.classaction.org/media/hubbard-v-google-llc-et-al.pdf

[902] https://fairplayforkids.org/ Fairplay is the leading nonprofit organization committed to helping children thrive in an increasingly commercialized, screen-obsessed culture, and the only organization dedicated to ending marketing to children. Id.

[903] https://fairplay.salsalabs.org/noinstagramforkids/index.html

[904] The Internet of Things (IoT) has penetrated the global market including that of children's toys. Worldwide, Smart Toy sales have reached $9 billion in 2019 and is expected to exceed $15 billion by 2022. Connecting IoT toys to the internet exposes users and their data to multivariate risk due to device vulnerabilities. When IoT devices are marketed to individuals, especially children, the potential for negative impact is significant, so their design must result in robust security implementations. For our study, we performed penetration testing on a Fisher-Price Smart Toy. We were able to obtain root access to the device, capture live pictures and videos, as well as install remote access software which allows surreptitious recordings over WiFi network connections without user knowledge or permission. We propose solutions including adhering to rudimentary standards for security design in toys, a mobile application for IoT threat assessment and user education, and an ambient risk communication tool aligned with user risk perception. The proposed solutions are crucial to empower users with capabilities to identify and understand ambient risks and defend against malicious activities. Streiff, Joshua and Das, Sanchari and Cannon, Joshua, Overpowered and Underprotected Toys: Empowering Parents with Tools to Protect Their Children (December 13, 2019). IEEE Humans and Cyber

While there is little doubt that these toys can be highly entertaining for children and even possess social and educational benefits, the Internet of Toys (IoToys) raises many concerns. Beyond the fact that IoToys that might be hacked or simply misused by unauthorized parties, datafication of children by toy conglomerates, various interested parties and perhaps even their parents could be highly troubling. IoToys could profoundly threaten children's right to privacy as it subjects and normalizes them to ubiquitous surveillance and datafication of their personal information, requests, and any other information they divulge. [905]

AI-based devices interact autonomously with children and convey their own cultural values, this impacts on the rights and duties of parents to provide, in a manner consistent with the evolving capacities of the child, appropriate direction and guidance in the child's freedom of thought, including aspects concerning cultural diversity.[906]

Due to the emergence of the Internet of Things, ordinary objects became connected to the internet, children can now be constantly datafied during their daily routines, with or without their knowledge. IoT devices can collect and retain mass amounts of data and metadata on children and share them with various parties—able to extract data on where children are, what they are doing or saying, and perhaps even capture imagery and videos of them.

Cayla is an internet-connected doll that uses voice recognition technology to chat and interact with children in real time. Cayla's conversations are recorded and transmitted online to a voice analysis company. This raised concerns that hackers might spy on children or communicate directly with them as they play with the doll. There are also concerns about how kids' voice data was used. In 2017 German regulators urged parents to destroy the doll, classifying it as an "illegal espionage apparatus".[907]

On February 17, 2017, the German Federal Network Agency banned Cayla[908] from being sold, and ordered the destruction of all devices which had already been sold.[909] The legal basis of this decision was § 148 (1) no. 2, 90 of the German Telecommunication Act. The rationale was that because of the doll's connectivity to its manufacturer (required because the doll was AI enabled), the doll was effectively a spy on the child, recording all the data the child says to devices including their most precious secrets. [910]

Likewise, the agency was concerned that the devices were hackable, exposing children to threats such as pedophilia or ideological communications. Since then, the regulator has used the law to ban similar devices as

---

well as smart watches [911] This strict approach adopted to protect children, one of the most vulnerable demographics, has a further legal basis in Art. 16 (1) of the Convention on the Rights of the Child. According to which "no child shall be subjected to arbitrary or unlawful interference with his or her privacy, family, home or correspondence."[912]

Cayla is just one example of a new wave of artificial intelligence toys that "befriend" children. Manufacturers often claim they are educational, enhancing play and helping children develop social skills. But consumer groups warn that smart toys, like other "things" we connect to the internet, might put security and privacy at risk. How do we navigate a world where AI toys are increasingly popular. What happens to the data from AI toys? The toys are connected to the internet (via WiFi or Bluetooth to a phone or other device with internet access) and send data to the supplier. This enables the company's AI to learn for the company and be better able to talk to the child. The company records and collects all the child's conversations with the toy, and possibly those with other children and adults who also interact with it. The company is probably storing this data and certainly using it to create a better product. The location of the toy affects how the data is stored. For example, in the US, companies creating educational toys can store data for longer than other companies. So when the manufacturer describes their toy as educational, it opens up that right to hold on to the data for longer. As more devices – many marketed as educational toys – come onto the market, they are setting off alarm bells around privacy, bias, surveillance, manipulation, democracy, transparency and accountability.

What issues should we be most concerned about? Germany banned Cayla and similar toys because of concerns they could be used to spy on children and that someone could hack the device and communicate directly with the child. But we are also talking about companies monetizing data. The data from AI toys contains everything a child says to the device, including their most guarded secrets.

If that data is collected, does the child have a right to get it back? If that data is collected from very early childhood and does not belong to the child, does it make the child extra vulnerable because his or her choices and patterns of behavior could be known to anyone who purchases the data, for example, companies or political campaigns.

Depending on the privacy laws of the state in which the toys are being used, if the data is collected and kept, it breaches Article 16 of the Convention on the Rights of the Child – the right to privacy. Though, of course, arguably this is something parents routinely do by posting pictures of their children on Facebook[913]

What are the benefits of AI toys? Most economists would argue that improving and increasing access to education is one of the best ways to close the gap between the developing and developed world. AI-enabled educational toys and "teachers" could make a hugely beneficial difference in the developing world. According to venture capitalist and former Google China CEO Kai Fu Lee,[914] the data collected from devices would likely be used by the big AI companies in the West and China for their own purposes, rather than directed toward an effort of benefiting children, their parents or the countries in which they live.[915]

---

[911] See DakshayaniShankar, Germany Bans Talking Doll Cayla over Security, Hacking Fears, NBC NEWS (Feb. 18, 2017, 6:43 PM),http://www.nbcnews.com/news/world/germany-bans-talking-doll-cayla-over-security-hacking-fears-n722816;Jane Wakefield, Germany Bans Children's Smartwatches, BBC NEWS (Nov. 17, 2017), http://www.bbc.com/news/technology-42030109.

[912] United Nations Convention on the Rights of the Child, art. 16 (1), Nov. 20, 1989 https://www.ohchr.org/en/professionalinterest/pages/crc.aspx

[913] Steinberg, Stacey, Sharenting: Children's Privacy in the Age of Social Media (March 8, 2016). 66 Emory L.J. 839 (2017), University of Florida Levin College of Law Research Paper No. 16-41, Available at SSRN: https://ssrn.com/abstract=2711442

[914] Kai-Fu Lee, AI Superpowers: China, Silicon Valley, and the New World Order, Houghton Mifflin Harcourt (2018). Despite these warnings, the book is ultimately optimistic that the complementarity between humans and AI can lead to a productive human-AI coexistence. Offering a dose of optimism to counter the doomsday singularity prediction, Lee reminds us that when it comes to shaping the story of AI, we humans are not just passive spectators, we can take action. See also Kai-Fu Lee, How AI can save our humanity, TED, (August 27, 2018) https://www.youtube.com/watch?v=ajGgd9Ld-Wc

What influence could AI toys have on kids? As well as the risk of hacking, we also need to think about what these toys are saying to our children. Who is the arbiter of these conversations? Who coded the algorithms (their unintended biases could creep in)? Do the values the child is being exposed to align with those of the parents? Will parents be able to choose the values the toy is coded with?[916]

If the toy is educational, is the algorithm being checked by someone who is at least qualified to teach?

These toys will be very influential because the children will be conversing with them all the time. For example, if the doll says it is cold and the child asks his or her parents to buy it a coat, is that advertising? If data is being collected, even if it isn't being stored, does the company have a duty to "red flag" children who share suicidal thoughts or other self-harming behavior? What if the child confides in the toy that he or she is being abused, will the company report this to the relevant authorities? And then what will the company do with that information? So what can we do to protect children?[917]

Parents need to have answers to these questions before they buy the devices. At the very least, they can check that their child is learning values from AI toys that concur with their own. At the moment the onus is on consumers to know what is being done with their data, but there is discussion that companies should be made responsible for ensuring consumers understand how it's being used.[918]

A World Economic Forum project advocates for the role of regulators so that they would certify algorithms fit for purpose, as opposed to the current situation where regulators issue a fine after something goes wrong. This regulatory model is appropriate with IoToys because it is needed now and an agile governance mechanism. The problem, though, with governance of smart toys is that the AI is learning and changing with each interaction with the child. AI-enabled toys are not necessarily bad. They could one day help us achieve precision learning (using AI to tailor education to each child's needs). AI toys could be excellent for preparing children to work alongside autonomous robots. The point is that children are vulnerable and we must consider how AI is used around them and not beta test it on them.[919]

While the US Congress responded to privacy threats toward children that emerged from the internet with the enactment of COPPA,[920] this regulatory framework only applies to a limited set of IoT devices, excluding those which are not directed towards children nor knowingly collect personal information from them. COPPA is ill-suited to properly safeguard children from the privacy risks that IoT entails, as it does not govern many IoT devices that they are exposed to. The move towards an "always-on" era, by which many IoT devices constantly collect data from users, regardless of their age, exposes us all to great privacy risks.[921]

IoT will most likely play a substantive role in child-targeted devices in the foreseeable future. IoToys presents children with interactive playing. Beyond the smart toys being fun, they could carry educational and social benefits for children:[922] opportunities to learn, develop, and improve communication skills; encourage active play and toy interaction, which might be preferable to passive TV screen time; identify learning difficulties; and be affordable for parents.[923]

---

[915] Kay Firth-Butterfield, Generation AI: What happens when your child's friend is an AI toy that talks back? World Economic Forum, May 22, 2018) http://governance40.com/wp-content/uploads/2018/12/Generation-AI-What-happens-when-your-childs-friend-is-an-AI-toy-that-talks-back-World-Economic-Forum.pdf

[916] Kay Firth-Butterfield, Generation AI: What happens when your child's friend is an AI toy that talks back? World Economic Forum, May 22, 2018) http://governance40.com/wp-content/uploads/2018/12/Generation-AI-What-happens-when-your-childs-friend-is-an-AI-toy-that-talks-back-World-Economic-Forum.pdf

[917] Kay Firth-Butterfield, Generation AI: What happens when your child's friend is an AI toy that talks back? World Economic Forum, May 22, 2018) http://governance40.com/wp-content/uploads/2018/12/Generation-AI-What-happens-when-your-childs-friend-is-an-AI-toy-that-talks-back-World-Economic-Forum.pdf

[918] Id.

[919] Kay Firth-Butterfield, Generation AI: What happens when your child's friend is an AI toy that talks back? World Economic Forum, May 22, 2018) http://governance40.com/wp-content/uploads/2018/12/Generation-AI-What-happens-when-your-childs-friend-is-an-AI-toy-that-talks-back-World-Economic-Forum.pdf

[920] See Children's Online Privacy Protection Act (COPPA), Pub. L. No. 106–70, 112 Stat. 2681 (1998) (codified as amended at 15 U.S.C. §§ 6501–06 (2018))

[921] Haber, Eldar, Toying with Privacy: Regulating the Internet of Toys (December 8, 2018). 80 Ohio State Law Journal 399 (2019), Available at SSRN: https://ssrn.com/abstract=3298054

IoToys devices have been criticized for their potential educational, social, and psychological drawbacks. To name a few: providing poor quality of play; potentially harming children's development, impeding child–parent interaction;[924] obstructing children's well being and healthy development, which require real relationships and conversations[925]; and posing a risk to health from electromagnetic radiation (EMR).[926]

In today's constantly connected world, With almost everyone having access to the web, where anyone can interact with anyone behind a veil of anonymity, the world faces a much higher risk of someone grooming our children without us even knowing.[927] Cyber grooming, a real threat, is a form of child grooming where the predator targets a child online, building a virtual relationship with them, gaining their trust, and learning the best way to gain access to them in the real world.[928]

For a predator, connecting to children online can be easy. Some opt to join a kid-friendly chat room and pretend to be a child while others play an online game with them where the predator can privately communicate with the child. Often the predator will entice a child to trust them with gifts and promises while also using language that normalizes sexual language and actions.

IoToys devices' potential drawbacks subject children to various risks, for example, exposure to harmful content.[929] There is even the danger of mental and bodily harm by predators, some of whom could have access toys and use them to listen to, watch, track, and even directly contact children. [930] Along with these important challenges, these IoToys devices raise human rights concerns. [931] Potentially, they can subject children to ubiquitous surveillance and datafication, which could profoundly impact their right to privacy. [932]

---

[922] The Smart Toy Awards recognize ethical and responsible smart toys that use AI to create an innovative and healthy play experience for children. Beatrice Di Caro, World Economic Forum, May 21, 2021 https://www.weforum.org/agenda/2021/05/smart-toy-awards-ede2d12ced/

[923] See Stéphane Chaudron et al., Kaleidoscope on the Internet of Toys: Safety, Security, Privacy and Societal Insights, JRC TECHNICAL REP. 9 (2017), http://publications.jrc.ec.europa.eu/repository/bitstream/JRC105061/jrc105061_final_online.pdf and 5 Benefits of Tech Toys for Children, ROBO WUNDERKIND (June 23, 2017), http://yuriy-levin.squarespace.com/blog/benefits-tech-toys-kids [https://perma.cc/Q599-U3A8].

[924] See Kate Cox, Privacy Advocates Raise Concerns About Mattel's Always-On 'Aristotle' Baby Monitor, CONSUMERIST (May 10, 2017), https://consumerist.com/2017/05/10/privacy-advocates-raise-concerns-about-mattels-always-on-aristotle-baby-monitor [https://perma.cc/VP3S-JEHB].

[925] See, e.g., Richard Chirgwin, Mattel's Parenting Takeover Continues with Alexa-Like Dystopia, THE REGISTER (Jan. 4, 2017), https://www.theregister.co.uk/2017/01/04/mattels_parenting_takeover_continues_with_alexalike_dystopia [https://perma.cc/NXP5-7GW3]

[926] See Stéphane Chaudron et al., Kaleidoscope on the Internet of Toys: Safety, Security, Privacy and Societal Insights, JRC TECHNICAL REP. 9 (2017), http://publications.jrc.ec.europa.eu/repository/bitstream/JRC105061/jrc105061_final_online.pdf

[927] See Urs Gasser, Colin Maclay, John Palfrey, An Exploratory Study by the Berkman Center for Internet & Society at Harvard University, in Collaboration with UNICEF, )June 16, 2010) https://dmlhub.net/wp-content/uploads/files/SSRN-id1628276.pdf

[928] Daniel Bennett, What Is Cyber Grooming and How to Protect Children? (March 23, 2020). TechAcute https://techacute.com/what-is-cyber-grooming/

[929] As these toys rely on remotely stored data, they could be subjected to harmful content as information might become vulnerable and could be changed by a malicious entity which gained access to the toy or simply due to bad or error in programing. See, for instance, how a misunderstanding led Amazon Echo to spout porn search terms to a toddler. Amazon Alexa Gone Wild, YOUTUBE (Dec. 29, 2016), https://www.youtube.com/watch?v=r5p0gqCIEa8 [https://perma.cc/SE3G-M5ZU]. See also how a specialist team hacked Cayla to quote Hannibal Lecter and lines from 50 Shades of Grey. See David Moye, Talking Doll Cayla Hacked to Spew Filthy Things, HUFFPOST (Feb. 9, 2015), http://www.huffingtonpost.com/2015/02/09/my-friend-cayla-hacked_n_6647046.html [https://perma.cc/78HN-89F6].

Thus children's leisure activities have changed significantly over the last two decades, from engaging with toys with little interactive capacity to smart toys that are capable of responding back. [933] Through the use of weak artificial intelligence, these toys incorporate a set of techniques that allow computers to mimic the logic and interactions of humans. [934] Such toys raise a host of human rights-related concerns. These include potential violations of a child's right to privacy, and whether corporations have (or should have) a duty to report sensitive information that is shared with a toy and stored online such as indications that a child might be being abused or otherwise harmed. [935]

---

[930] When children assume that it is the toy that is "talking" to them, predators might be able to persuade them to convey sensitive information. These predators could obtain information from children like where they live and, perhaps even worse, convince them to act on their behalf. See Abby Haglage, Hackable 'Hello Barbie' the Worst Toy of the Year (and Maybe Ever), DAILY BEAST (Dec. 10, 2015), http://www.thedailybeast.com/hackablehello-barbie-the-worst-toy-of-the-year-and-maybe-ever [https://perma.cc/85E4-AGQW]. For a typology of risks to children online, see ORG. FOR ECON. CO-OPERATION & DEV. (OECD), THE PROTECTION OF CHILDREN ONLINE - RECOMMENDATION OF THE OECD COUNCIL REPORT ON RISKS FACED BY CHILDREN ONLINE AND POLICIES TO PROTECT THEM 24–39 (2012), https://www.oecd.org/sti/ieconomy/childrenonline_with_cover.pdf [https://perma.cc/33T7-R645].

[931] Verdoodt, Valerie, The Role of Children's Rights in Regulating Digital Advertising (2019). International Journal of Children's Rights, 27 (3), 455-481, 2019, doi: 10.1163/15718182-02703002., Available at SSRN: https://ssrn.com/abstract=3703312 or http://dx.doi.org/10.2139/ssrn.3703312 (An important domain in which children's rights are reconfigured by internet use, is digital advertising. New advertising formats such as advergames, personalized and native advertising have permeated the online environments in which children play, communicate and search for information. The often immersive, interactive and increasingly personalized nature of these advertising formats makes it difficult for children to recognize and make informed and well-balanced commercial decisions. This raises particular issues from a children's rights perspective, including inter alia their rights to development (Article 6 UNCRC), privacy (Article 16 UNCRC), protection against economic exploitation (Article 32 UNCRC), freedom of thought (Article 17 UNCRC) and education (Article 28 UNCRC). The paper addresses this reconfiguration by translating the general principles and the provisions of the United Nations Convention on the Rights of the Child into the specific context of digital advertising. Moreover, it considers the different dimensions of the rights (i.e. protection, participation and provision) and how the commercialization affects children and how their rights are exercised.)

[932] Haber, Eldar, The Internet of Children: Protecting Children's Privacy in A Hyper-Connected World (November 21, 2020). 2020 U. Ill. L. Rev. 1209 (2020)., Available at SSRN: https://ssrn.com/abstract=3734842

[933] Chris Nickson, "How a Young Generation Accepts Technology," A Technology Society, September 18, 2018, available at http://www.atechnologysociety.co.uk/howyoung-generation-accepts-technology.html. See also Laura Rafferty, Patrick C. K. Hung, Marcelo Fantinato, Sarajane Marques Peres, Farkhund Iqbal, Sy-Yen Kuo, and Shih-Chia Huang, "Towards a Privacy Rule Conceptual Model for Smart Toys" in Computing in Smart Toys, 85-102, available at https://www.researchgate.net/publication/319046589_Towards_a_Privacy_Rule_Conceptual_Model_for_Smart_Toys. ("A smart toy is defined as a device consisting of a physical toy component that connects to one or more toy computing services to facilitate gameplay in the cloud through networking and sensory technologies to enhance the functionality of a traditional toy. A smart toy in this context can be effectively considered an Internet of Things (IoT) with Artificial Intelligence (AI) which can provide Augmented Reality (AR) experiences to users. In this paper, the first assumption is that children do not understand the concept of privacy and the children do not know how to protect themselves online, especially in a social media and cloud environment. The second assumption is that children may disclose private information to smart toys and not be aware of the possible consequences and liabilities. This paper presents a privacy rule conceptual model with the concepts of smart toy, mobile service, device, location, and guidance with related privacy entities: purpose, recipient, obligation, and retention for smart toys. Further the paper also discusses an implementation of the prototype interface with sample scenarios for future research works.")

[934] Rafferty et al., "Towards a Privacy Rule," supra

There are three nodes involved in smart toy processes, each of which comes with a set of challenges and vulnerabilities: the toy (which interfaces with the child), the mobile application, which acts as an access point for Wi-Fi connection, and the toy's/consumer's personalized online account, where data is stored. Such toys communicate with cloud-based servers that store and process data provided by the children who interact with the toy.[936]

Privacy concerns arising from this model can be illustrated by the Cloud Pets[937] case, in which more than 800,000 toy accounts were hacked, exposing customers' (including children's) private information.[938] In 2017, more than two million voice messages that had been recorded on these Cloud Pets cuddly toys were discovered in an open database. Besides the numerous voice messages of children and adults, the database included people's email addresses, passwords, profile pictures, and even children's names and names of authorized family members. Thus, over 800,000 users' personal information was compromised.[939]

The database which contained all this information had no usernames or passwords to prevent someone from seeing all the data. What's worse, soon enough a ransomware attack happened on the database: the hackers had deleted original databases leaving a ransom demand instead. Basically, the hackers locked the database until a certain amount of money was paid.[940]

Another example of risky toys is the Hello Barbie doll,[941] which raised civil society concerns around the interception of sensitive information and whether the doll allowed for pervasive surveillance in ways that were not transparent to users.[942] In that case, the toy's manufacturer, Mattel – in collaboration with Toy Talk, Inc.– released a FAQ to try to address these pressing questions.[943] The FAQ states that the conversations between the doll and the child cannot be intercepted via Bluetooth technology because the conversation takes place over a secured network, making it impossible to connect the doll via Bluetooth. [944] The document advises against connecting the doll to third party Wi-Fi, which may be especially vulnerable to interception.[945]

---

[935] Benjamin Yankson, Farkhund Iqbal, and Patrick C. K. Hung, "Privacy Preservation Framework for Smart Connected Toys," Computing in Smart Toys, https://www.researchgate.net/publication/319048771_Privacy_Preservation_Framework_for_Smart_Connected_Toys ("Advances in the toy industry and interconnectedness resulted in rapid and pervasive development of Smart Connected Toy (SCT), which built to aid children in learning, socialization, and development. A SCT is a physical embodiment artifact that acts as a child user interface for toy computing services in cloud. These SCTs are built as part of Internet of Things (IoT) with the potential to collect terabytes of personal and usage information. They introduce the concept of increasing privacy, and serious safety concerns for children, who are the vulnerable sector of our community and must be protected from exposure of offensive content, violence, sexual abuse, and exploitation using SCTs. SCTs are capable to gather data on the context of the child user's physical activity state (e.g., voice, walking, standing, running, etc.) and store personalized information (e.g., location, activity pattern, etc.) through camera, microphone, Global Positioning System (GPS), and various sensors such as facial recognition or sound detection. In this chapter we are going to discuss the seriousness of privacy implication for these devices, survey related work on privacy issues within the domain of SCT, and discuss some global perspective (legislation, etc.) on such devices. The chapter concludes by proposing some common best practice for parents and toy manufactures can both adopt as part of Smart Connected Toy Privacy Common body of knowledge for child safety.")
[936] Rafferty et al., "Towards a Privacy Rule," supra
[937] https://en.wikipedia.org/wiki/CloudPets
[938] Alex Hern, "CloudPets stuffed toys leak details of half a million users," The Guardian, https://www.theguardian.com/technology/2017/feb/28/cloudpets-data-breach-leaksdetails-of-500000-children-and-adults, (February 28, 2017).
[939] Marija Perinic, Cloud Pets: The Cuddly Cyber Security Risk, Secure Thoughts, (May 11, 2021) https://securethoughts.com/cloudpets-app/
[940] Id.
[941] See Valerie Steeves, 'A dialogic analysis of Hello Barbie's conversations with children' (2020) 7(1) Big Data & Society, https://journals.sagepub.com/doi/pdf/10.1177/2053951720919151
[942] Corinne Moini, "Protecting Privacy in the Era of Smart Toys: Does Hello Barbie Have a Duty to Report," 25 Cath. U. J. L. & Tech 281, (2017), 4. https://scholarship.richmond.edu/law-student-publications/157/
[943] Mattel, "Hello Barbie Frequently Asked Questions," (2015),http://hellobarbiefaq.mattel.com/wp-content/uploads/2015/12/hellobarbie-faq-v3.pdf

Further, the document claims that the Hello Barbie doll is not always listening but becomes inactive when not expressly engaged.[946] According to the document released by Mattel, the doll has similar recognition technology to Siri and is activated only when the user pushes down the doll's belt buckle.[947] Finally, the company states that the doll does not ask questions that are intended to elicit personal information, in order to minimize the circumstances in which a child might divulge sensitive information during his/her conversation with the doll.[948]

Notably, parents can access their child's ToyTalk cloud account and listen to what their child has said, deleting any personal information.[949] As a safeguard, ToyTalk also participates in the FTC's KidSafe Seal Program, a compliance program for websites and online services targeted towards children.[950] There are two types of certificates that a website or online service can obtain: the KidSafe certificate and the KidSafe+ certificate.[951] The KidSafe+ certificate requires additional requirements and compliance with COPPA.[952] The communications between Hello Barbie and a child are encrypted and stored on a trusted network.[953]

A key concern, despite these safeguards, is whether a company has a duty to report or otherwise "red flag" sensitive information shared through their toys[954]—for example, children who reveal they are being abused, or children who share suicidal thoughts or other self-harm related behavior.[955]

---

Existing privacy laws and common law tort duties fall short of providing directly relevant protection. For example, while COPPA protects the privacy rights of minors under the age of thirteen, requiring companies to obtain parental consent and to disclose what information is being collected about a minor, it does not impose any reporting requirements regarding suspected child abuse and neglect.[956]

Ultimately, most mechanisms for tackling these challenges have been designed by the corporations themselves. For example, stamping out the spread of child sexual abuse material (CSAM) is a priority for big internet companies and content moderators. But it's also a difficult and harrowing job for those on the frontline, human moderators who have to identify and remove abusive content. Google released free AI software designed to help these individuals.[957]

Most tech solutions in this domain work by checking images and videos against a catalog of previously identified abusive material. (See, for example: PhotoDNA[958], a tool developed by Microsoft and deployed by companies like Facebook and Twitter.) This sort of software, known as a "crawler," is an effective way to stop people sharing known previously-identified CSAM. But it can't catch material that hasn't already been marked as illegal. For that, human moderators have to step in and review content themselves.[959]

This is where Google's new AI tool aims to help. Using the company's expertise in machine vision, it assists moderators by sorting flagged images and videos and "prioritizing the most likely CSAM content for review." This should allow for a much quicker reviewing process. In one trial, says Google, the AI tool helped a moderator "take action on 700 percent more CSAM content over the same time period."[960]

Fred Langford, deputy CEO of the Internet Watch Foundation[961] (IWF), said the software would "help teams like our own deploy our limited resources much more effectively." "At the moment we just use purely humans to go through content and say, 'yes,' 'no,'" says Langford. "This will help with triaging."

The IWF is one of the largest organizations dedicated to stopping the spread of CSAM online. It's based in the UK but funded by contributions from big international tech companies, including Google. It employs teams of human moderators to identify abuse imagery, and operates tip-lines in more than a dozen countries for internet users to report suspect material. It also carries out its own investigative operations; identifying sites where CSAM is shared and working with law enforcement to shut them down.[962] Langford says that because of the nature of "fantastical claims made about AI," the IWF will be testing out Google's new AI tool thoroughly to see how it performs and fits with moderators' workflow. He added that tools like this were a step towards fully automated systems that can identify previously unseen material without human interaction at all. "That sort of classifier is a bit like the Holy Grail in our arena." But, he added, such tools should only be trusted with "clear cut" cases to avoid letting abusive material slip through the net. "A few years ago I would have said that sort of classifier was five, six years away," says Langford. "But now I think we're only one or two years away from creating something that is fully automated in some cases."[963]

---

[956] "Children's Online Privacy Protection Rule," 16 C.F.R. 312.1 (2001). See also Corinne Moini, "Protecting Privacy in the Era of Smart Toys: Does Hello Barbie Have a Duty to Report," 25 Cath. U. J. L. & Tech 281(2017). https://scholarship.richmond.edu/law-student-publications/157/

[957] James Vincent, "Google Releases Free AI Tool to Help Companies Identify Child Sexual Abuse Material," The Verge, https://www.theverge.com/2018/9/3/17814188/googleai-child-sex-abuse-material-moderation-tool-internetwatch-foundation, September 03, 2018. See also Nikola Todorovic and Abhi Chaudhuri, Using AI to help organizations detect and report child sexual abuse material online, GOOGLE IN EUROPE, (Sept. 3, 2018) https://www.blog.google/around-the-globe/google-europe/using-ai-help-organizations-detect-and-report-child-sexual-abuse-material-online/

[958] https://www.microsoft.com/en-us/photodna In 2009, Microsoft partnered with Dartmouth College to develop PhotoDNA, a technology that aids in finding and removing known images of child exploitation. Today, PhotoDNA is used by organizations around the world and has assisted in the detection, disruption, and reporting of millions of child exploitation images. Id.

[959] Id.

[960] Id.

[961] https://www.iwf.org.uk/ ("Our vision is to eliminate child sexual abuse imagery online.") (""IWF is one of the most active and effective European hotlines fighting against child sexual exploitation. The work developed by IWF in the process of notice and takedown, in close cooperation with Law Enforcement, is an example to follow. "IWF's contribution to the Strategic Assessment on Commercial Sexual Exploitation of Children Online, produced by Europol in the frame of the European Financial Coalition against Commercial Sexual Exploitation of Children, has been outstanding. The analytical findings shared by IWF and the work developed through initiatives like the Website Brands Project have been an invaluable source of information for the Law Enforcement community. "Europol will continue cooperating actively with IWF to achieve our common goals: eradicate the production and dissemination of child abuse material through the internet. The dedication and commitment from the IWF team is outstanding." Troels Oerting, Former Head of EC3, European Cybercrime Centre) Id.

In the case of Hello Barbie, ToyTalk has created automatic responses for serious conversations such as bullying or abuse. Such responses include "that sounds like something you should talk to a grown-up about."[964] While an important step towards addressing this issue, this approach potentially pushes any responsibility for acting to the parents or to the child herself. It is unclear how many children would act on this response to report problems to a grownup or what it means for children if an adult in their household is the one perpetrating the harm.

Modern technological tools can design "child-safe" toys, to prevent users from harming themselves and others. AI programming can target moral as well as physical harms. The iPhone was designed to allow Apple to remove applications from users' devices, a capability it utilized to excise sexually suggestive applications in early 2010. Government may increasingly take advantage of this possibility. In an increasingly digital world, the government could manipulate technological design to make it difficult or impossible to break laws using digital devices. [965]

**Artificial Intelligence in Education (AIEd) and EdTech: Children's Rights and Education**

The introduction of artificial intelligence in education ("AIEd") will have a profound impact on the lives of children and young people. There are different types of artificial intelligence systems in common use in education, alongside the growth of commercial knowledge monopolies. Data privacy rights issues for children and young people are becoming more and more pronounced. Achieving a balance between fairness, individual pedagogic rights, data privacy rights and effective use of data is a difficult challenge, and one not easily supported by current regulation, and many continue to search for democratically aware and responsible use for artificial intelligence use in schools.[966]

The role and function of education cannot be overstated. Education enhances and develops human abilities, consciousness, identity, integrity, potential, and autonomy.[967] AI can be welcome as a complement to educational processes, but its design must be focused on the rights of its child users.

Because the right to education has been recognized as a human right and defined in various human rights instruments in various contexts, this right can be asserted against states and their agencies.[968] UDHR Article 26(1) states that "everyone has the right to education."[969] This implies that every human, not just the young, has the right. Article 18 of the International Covenant on Civil and Political Rights (ICCPR) and Article 12 of the International Convention on the Protection of the Rights of All Migrant Workers and Members of Their Families (MWC) protect parents' right to control the religious and moral education of their children.[970] Under Article 13(1) of the International Covenant on Economic, Social and Cultural Rights (ICESCR), state parties recognize the right of everyone to education.[971] Article 28(1) of the Convention on the Rights of the Child (CRC) also recognizes the right of the child to education as a progressive right.[972] The United Nations Educational, Scientific, and Cultural Organization (UNESCO)'s Convention Against Discrimination in Education also prohibits discrimination in terms of access to education, the standard and quality of education, and condition under which education is given.[973] Article 5(v) of the International Convention on the Elimination of All Forms of Racial Discrimination (ICERD) urges states not to racially discriminate when their citizens enjoying the right to education and training.[974] Article 10 of the Convention on the Elimination of All Forms of Discrimination against Women (CEDAW) recognizes women's equal rights to education.[975]

---

[962] Id.
[963] Id.
[964] Mattel, "Hello Barbie Frequently Asked Questions."
[965] Rosenthal, Daniel M., Assessing Digital Preemption (and the Future of Law Enforcement?) (January 5, 2011). New Criminal Law Review, Fall 2011, Available at SSRN: https://ssrn.com/abstract=1735479
[966] Leaton Gray, S; (2020) Artificial intelligence in schools: Towards a democratic future. London Review of Education, 18 (2) pp. 163-177. 10.14324/lre.18.2.02.
[967] Lee, Jootaek, The Human Right to Education: Definition, Research and Annotated Bibliography (November 18, 2019). Emory International Law Review, Vol. 34, No. 3, 2019, Available at SSRN: https://ssrn.com/abstract=3489328
[968] Krajewski, Markus, The State Duty to Protect Against Human Rights Violations Through Transnational Business Activities (December 3, 2018). Deakin Law Review, Vol. 23, 2018, Available at SSRN: https://ssrn.com/abstract=3295305 or http://dx.doi.org/10.2139/ssrn.3295305

Article 24 of the Convention on the Rights of Persons with Disabilities (CRPD) recognizes the right of persons with disabilities to education.[976] The MWC recognizes migrant workers' and their children's right of access to education.[977] The Convention Relating to the Status of Refugees (1951 Refugee Convention) also recognizes refugees' equal rights to elementary education and most favored treatment to other educations.[978] In 2007, the General Assembly overwhelmingly adopted the United Nations Declaration on the Rights of Indigenous Peoples (UNDRIP),[979] which includes the right to education.[980] Within several years, the four nations in opposition—the United States, Canada, New Zealand, and Australia—all reversed their positions.[981] UNDRIP acknowledges rights

---

[969] Universal Declaration of Human Rights (Dec. 10, 1948), art. 26 [hereinafter UDHR].

[970] International Covenant on Civil and Political Rights art. 18, opened for signature Dec. 19, 1966, 999 U.N.T.S. 171; International Convention on the Protection of the Rights of All Migrant Workers and Members of Their Families art. 12, Dec. 18, 1990, 2220 U.N.T.S. 3 [hereinafter MWC].

[971] International Covenant on Economic, Social and Cultural Rights art 13(1), opened for signature Dec. 19, 1966, 993 U.N.T.S. 3.

[972] Convention on the Rights of the Child art. 28(1), opened for signature Nov. 20, 1989, 1577 U.N.T.S. 3.

[973] Convention Against Discrimination in Education arts. 1–3, Dec. 14, 1960, 429 U.N.T.S. 93 (entered into force May 22, 1962).

[974] International Convention on the Elimination of All Forms of Racial Discrimination art. 5(v), opened for signature Mar. 7, 1966, 660 U.N.T.S. 195.

[975] Convention on the Elimination of All Forms of Discrimination Against Women art. 10, opened for signature Mar. 1, 1980, 1249 U.N.T.S. 13.

[976] Convention on the Rights of Persons with Disabilities art. 24, opened for signature Mar. 30, 2007, 2515 U.N.T.S. 3.

[977] International Convention on the Protection of the Rights of All Migrant Workers and Members of Their Families art. 12, Dec. 18, 1990, 2220 U.N.T.S. 3 arts. 30, 43(1)(a).

[978] Convention Relating to the Status of Refugees art. 22, July 28, 1951, 189 U.N.T.S. 137.

[979] G.A. Res. 61/295, ¶ 12, U.N. Doc. A/RES/61/295 (Sept. 13, 2007) [hereinafter UNDRIP]; see also WALTER R. ECHO-HAWK, IN THE LIGHT OF JUSTICE: THE RISE OF HUMAN RIGHTS IN NATIVE AMERICA AND THE UN DECLARATION ON THE RIGHTS OF INDIGENOUS PEOPLES 3 (2013) (describing the UNDRIP as "a landmark event that promises to shape humanity in the post-colonial age"). See also , see, e.g., Lorie M. Graham & Siegfried Wiessner, Indigenous Sovereignty, Culture, and International Human Rights Law, 110 S. ATLANTIC Q. 403, 405 (2011) (analyzing the recognition of provisions of the UNDRIP as customary international law).

[980] UNDRIP, supra. The version of the Declaration presented to the General Assembly affirmed that indigenous peoples have the right to full enjoyment, "as a collective or as individuals," of all human rights recognized by the U.N. Charter, Universal Declaration on Human Rights, and international human rights law. It retained the language from early drafts on "indigenous peoples" and "self-determination," as well as rights to traditional lands, economic development, education, family and child welfare, self-government, culture, religion, expression, and others. Key provisions call for states to obtain "free, prior and informed consent before adopting and implementing legislative or administrative measures" affecting indigenous peoples.

[981] Carpenter, Kristen A. and Riley, Angela, Indigenous Peoples and the Jurisgenerative Moment in Human Rights (February 18, 2013). California Law Review, Vol. 102, 2014, 192 Available at SSRN:

common to humanity—such as nondiscrimination, equality, and property—and contexts for the enjoyment of those rights that may appear more particular to indigenous peoples, such as spiritual attachment to traditional lands and a focus on community rights.[982]

As the world proceeds deeper into the digital space, there is a growing need to explore the impact of novel digital technologies on children's right to education. Conceptualizing education as a human right necessitates greater attention to the United Nations' 4A-framework[983] (accessibility, adaptability, acceptability and availability): the accessibility and adaptability of school environments, beyond merely their acceptability and availability. New technologies have impacted all of these criteria, as the education sector continues to capitalize on emerging opportunities. [984]

Dependent on connectivity and resources, countries across the world have opted for differing ICT infrastructure to support remote learning. Alongside digital platforms, social media, radio platforms and TV have all been used to ensure continuity in education for all corners of the world. Notwithstanding, this transition to digital learning has amplified societal inequities, as children living in remote locations with little to no internet connection struggle to gain access to online services.[985] Though technology is designed to connect people by reaching frequently excluded areas, only a few educational systems around the world were able to adequately respond to the challenges of the COVID-19 pandemic.[986] This accessibility issue must be addressed.

**What is AIEd and EdTech?**

AIEd is the latest innovation in educational technology, also known as EdTech, typically defined as the sector of technology dedicated to the development and application of tools for educational purposes. The introduction of these technologies pose numerous challenges to children's rights to privacy.

---

https://ssrn.com/abstract=2220573

[982] See Julian Burger, The UN Declaration on the Rights of Indigenous Peoples: From Advocacy to Implementation, in Stephen Allen and Alexandra Xanthaki, eds., Reflections on the UN Declaration on the Rights of Indigenous Peoples at 41, 42–43 ("[The Declaration] responds to the real-life problems that threaten the existence of indigenous peoples as identified by indigenous peoples themselves. One of the remarkable features of the Working Group . . . was that the rights proposed were garnered from specific experiences, expressed in the language of the elder, community leader, woman or youth activist. How else could the recognition of indigenous peoples' spiritual relationship with their lands be included in an international human rights instrument, if not through countless stories of this non-materialist and harmonious bond between humankind and nature?").

[983] https://sdgs.un.org/goals/goal4

[984] Vanessa Cezarita Cordeiro, Educational technology (EdTech) and children's right to privacy, Humanium, (June 15, 2021) https://www.humanium.org/en/educational-technology-edtech-and-childrens-right-to-privacy/ Available at: https://aberta.org.br/educacao-dados-e-plataformas/ See also UNESCO, SDG4, Education https://en.unesco.org/gem-report/sdg-goal-4 In September 2015, at the United Nations Sustainable Development Summit, Member States formally adopted the 2030 Agenda for Sustainable Development in New York. The agenda contains 17 goals including a new global education goal (SDG 4). SDG 4 is to ensure inclusive and equitable quality education and promote lifelong learning opportunities for all' and has seven targets and three means of implementation. This goal came about through an intensive consultative process led by Member-States, but with broad participation from civil society, teachers, unions, bilateral agencies, regional organizations, the private sector and research institutes and foundations.

[985] Human Rights Watch, COVID-19 and Children's Rights, (April 9, 2020) https://www.hrw.org/news/2020/04/09/covid-19-and-childrens-rights#_Toc37256528

[986] Mercedes Mateo Diaz and Changha Lee ,A Silent Revolution, in What Technology Can and Can't Do for Education - A comparison of 5 stories of success, Inter-American Development Bank, (2020) https://publications.iadb.org/publications/english/document/What-Technology-Can-and-Cant-Do-for-Education-A-Comparison-of-5-Stories-of-Success.pdf

**Education and child development**

As has been shown, the digital environment shapes children's development in differing ways.[987] Technology permeates most areas of children's day-to-day lives, creating opportunities for greater learning, communication and development, as well as new risks to children's realization of their human rights. In the educational arena, technology has provided new mediums for sharing and communicating information, connecting school communities beyond the classroom, and tailoring the delivery of education to individual children, among other innovations.[988] However, with these developments come new challenges.

**AIEd and Children's Privacy**

Tools and software utilized in classrooms to enhance learning experiences are quickly evolving. From the use of advanced emotional AI and facial recognition, down to the simple migration of educational material onto online shared platforms, children's learning experiences are quickly becoming intertwined with technology. All of these tools designed to support and facilitate children's education are considered EdTech, and their emergence has presented new challenges for both children and tech implementers. As described by the Council of Europe, EdTech is often "deployed without various actors always being aware of the challenges to children's private life and personal data protection".[989]

Numerous publications report problematic issues with EdTech which result in the collection and processing of personal data from children without guaranteeing their best interest just for the purpose of commercial exploitation of children.[990]

In the rush to implement new technologies, educational regulators have failed to ensure child data is adequately protected. Children's educational data is "far less protected" than health data, and a large number of countries do not have data privacy laws which explicitly protect children. Without proper regulation, sensitive information about children – such as their names, addresses and behaviors – are open to exploitation.[991] In 2020, numerous popular distance learning platforms drew criticism over their collection, sharing and management of child data.[992]

---

[987] Council of Europe. (2020, November 20). Consultative committee of the convention for the protection of individuals with regard to automatic processing of personal data. 'Children's data protection in an education setting guidelines.' ; https://rm.coe.int/t-pd-2019-6bisrev5-eng-guidelines-education-setting-plenary-clean-2790/1680a07f2b See also Council of Europe. (2020, November 27). 'Protect children's personal data in an education setting.' https://www.coe.int/en/web/data-protection/-/protect-children-s-personal-data-in-education-setting- and Jen Persson, Director of defenddigitalme, Children's Data Protection in Education Systems: Challenges and Possible Remedies, (November 15, 2019) 1680a01b47 (coe.int)

[988] Council of Europe. (2020, November 20). Consultative committee of the convention for the protection of individuals with regard to automatic processing of personal data. 'Children's data protection in an education setting guidelines.'

[989] Council of Europe. (2020, November 27). 'Protect children's personal data in an education setting.'

[990] Vanessa Cezarita Cordeiro, Educational technology (EdTech) and children's right to privacy, Humanium, (June 15, 2021) https://www.humanium.org/en/educational-technology-edtech-and-childrens-right-to-privacy/ See also The General Data Protection Regulation, requires that personal data must be "processed lawfully, fairly and in a transparent manner in relation to the data subject" GDPR Article 5(1)(a). See also Jones, Meg and Kaminski, Margot E., An American's Guide to the GDPR (June 5, 2020). Denver Law Review, Vol. 98, No. 1, p. 93, 2021, U of Colorado Law Legal Studies Research Paper No. 20-33, Available at SSRN: https://ssrn.com/abstract=3620198 See generally Data protection starts with a ban: one cannot process personal data unless a lawful condition applies GABRIELA ZANFIR-FORTUNA & TERESA TROESTER-FALK, FUTURE OF PRIV. F. AND NYMITY, PROCESSING PERSONAL DATA ON THE BASIS OF LEGITIMATE INTERESTS UNDER THE GDPR: PRACTICAL CASES 3–4 (2018) https://www.ejtn.eu/PageFiles/17861/Deciphering_Legitimate_Interests_Under_the_GDPR%20(1).pdf

[991] Hye Jung Han, As schools close over coronavirus, protect kids' privacy in online learning, Human Rights Watch, (March 27, 2020) https://www.hrw.org/news/2020/03/27/schools-close-over-coronavirus-protect-kids-privacy-online-learning#

[992] Hye Jung Han, As schools close over coronavirus, protect kids' privacy in online learning, Human Rights Watch, (March 27, 2020) https://www.hrw.org/news/2020/03/27/schools-close-over-coronavirus-protect-

Research from the eQuality Project [993] lists some of the most pressing concerns around the use of EdTech: tracking of student activity in and outside the classroom, discrimination against children from marginalized communities, breaches of child data protection and autonomy, and the sale of child data to private third parties such as advertising companies.[994] These concerns can only be overcome if educators are mindful of the terms and conditions of the software in use, whether it was designed for educational purposes or not (such as videoconferencing applications such as Zoom or Skype).

Even technology designed for other purposes, but used as educational tools, necessitate greater attention on their data protection policies and constraints. Recent versions of Zoom, for example, stated that data collected from students included their name, school, devices and internet connections, and details about the content viewed by children and their communication with others via those devices. Notably, consent to Zoom's policies is given by the "school subscriber", rather than a child or their guardian, rendering the policy inconsistent with children's right to participate in decisions affecting them under the CRC.[995]

**COVID-19 and AIEd**

COVID-19 has greatly exacerbated pre-existing EdTech risks. Overnight, education was forced to depend on technology, rather than simply utilize it to enable new teaching methods. During the spring of 2020 alone, schools in 192 countries were closed.[996] UNESCO estimates support this assertion, stating that 91% of the world's student population were out of school in April of 2020.[997] This has vaulted EdTech from an incoming phenomenon to a virtual necessity as one of the core mediums for the delivery of education. This occurrence has been described as the "biggest distance learning experiment in history",[998] bringing us closer to what The Economist has dubbed "the coronopticon" — a brave new age of surveillance and data control catalyzed by hasty tech decisions under COVID-19.[999] Organizations such as Media Smarts,[1000], Common Sense Media,[1001], Consortium for School Networking,[1002] and Future of Privacy Forum,[1003] have all updated their websites to provide information on privacy and data protection practices of edtech products and services. The father of two

---

kids-privacy-online-learning#
[993] http://www.equalityproject.ca/
[994] Jane Bailey, Jacquelyn Burkell, Priscilla Regan, and Valerie Steeves, 'Children's privacy is at risk with rapid shifts to online schooling under coronavirus.' The Conversation, (April 12, 2020) https://theconversation.com/childrens-privacy-is-at-risk-with-rapid-shifts-to-online-schooling-under-coronavirus-135787
[995] Jane Bailey, Jacquelyn Burkell, Priscilla Regan, and Valerie Steeves, 'Children's privacy is at risk with rapid shifts to online schooling under coronavirus.' The Conversation, (April 12, 2020) https://theconversation.com/childrens-privacy-is-at-risk-with-rapid-shifts-to-online-schooling-under-coronavirus-135787
[996] Mercedes Mateo Diaz and Changha Lee ,A Silent Revolution, in What Technology Can and Can't Do for Education - A comparison of 5 stories of success, Inter-American Development Bank, (2020) https://publications.iadb.org/publications/english/document/What-Technology-Can-and-Cant-Do-for-Education-A-Comparison-of-5-Stories-of-Success.pdf
[997] Human Rights Watch. (2020, April 9). 'COVID-19 and Children's Rights'.
[998] Mercedes Mateo Diaz and Changha Lee ,A Silent Revolution, in What Technology Can and Can't Do for Education - A comparison of 5 stories of success, Inter-American Development Bank, (2020) https://publications.iadb.org/publications/english/document/What-Technology-Can-and-Cant-Do-for-Education-A-Comparison-of-5-Stories-of-Success.pdf
[999] The Economist, Creating the Coronopticon, Countries Are Using Apps and Data Networks to Keep Tabs on The Pandemic, and Also, in the Process, Their Citizens, (March 26, 2020) https://www.economist.com/briefing/2020/03/26/countries-are-using-apps-and-data-networks-to-keep-tabs-on-the-pandemic
[1000] https://mediasmarts.ca/
[1001] https://www.commonsense.org/education/
[1002] https://www.cosn.org/
[1003] https://studentprivacypledge.org/

elementary school girls, claiming violations of Illinois Biometric Information Privacy Act or BIPA, has sued Google for alleged violations of privacy.[1004]

Policymakers should support teachers, administrators and school boards to insist that ed tech companies default in favor of privacy-respecting practices. Educational policymakers must provide guidance and novel instruction on the use of EdTech to better protect children's data. In 2001, the UN Committee on the Rights of the Child announced that "children do not lose their human rights by virtue of passing through the school gates".[1005] The majority of EdTech is developed and created by commercial actors, with scant regard for children's vulnerability and inability to police and protect their own digital footprint. As technologies evolve to analyze more behaviors from children and further personalize learning experiences, there is a desperate need for regulation to ensure EdTech is inclusive, mindful and complementary to children's development.

United Nations General Comment No.16 of 2013 calls on countries to ensure that private enterprises are not awarded public procurement contracts if they fail to respect children's rights.[1006] In the European context, the Council of Europe have issued guidelines calling on States to adhere to The Convention for the Protection of Individuals with regard to Automatic Processing of Personal Data,[1007] specifically by realizing these rights in the context of children.[1008]

The 2019 Beijing Consensus on Artificial Intelligence and Education "reaffirms a humanistic approach to deploying Artificial Intelligent technologies in education for augmenting human intelligence, protecting human rights and for promoting sustainable development through effective human-machine collaboration in life, learning and work." Its recommendations are in five areas: (i) AI for education management and delivery; (ii) AI to empower teaching and teachers; (iii) AI for learning and learning assessment; (iv) Development of values and skills for life and work in the AI era; and (v) AI for offering lifelong learning opportunities for all.[1009]

Because ethical AIEd is a global challenge, spread across borders, it needs to be addressed also globally guided by ethics and human rights considerations, cognizant of the complexities of childhood.

---

[1004] See Nieva, R. (2020, April 3). 'Two children sue Google for allegedly collecting students' biometric data', https://www.cnet.com/news/two-children-sue-google-for-allegedly-collecting-students-biometric-data/ Google G Suite for Education Collects Children's Biometrics BIPA Class Action, https://classactionsreporter.com/google-g-suite-for-education-collects-childrens-biometrics-bipa-class-action/ and Farwell v. Google, LLC - Join Class Action Lawsuits https://www.classaction.org/media/farwell-v-google-llc.pdf

[1005] United Nations Committee on the Rights of the Child. (2001, April 17). 'General Comment No. 1 Article 29(1): The aims of education'. CRC/GC/2001/1.

[1006] United Nations Committee on the Rights of the Child. (2013, April 17). 'General Comment No. 16 on State obligations regarding the impact of the business sector on children's rights.' CRC/C/GC/16.

[1007] https://rm.coe.int/1680078b37

[1008] Council of Europe. (2020, November 20). Consultative committee of the convention for the protection of individuals with regard to automatic processing of personal data. 'Children's data protection in an education setting guidelines.' https://rm.coe.int/t-pd-2019-6bisrev5-eng-guidelines-education-setting-plenary-clean-2790/1680a07f2b

[1009] UNESCO, Beijing Consensus on Artificial Intelligence and Education, (June 25, 2019) Available at: https://unesdoc.unesco.org/ark:/48223/pf0000368303 (UNESCO has published the Beijing Consensus on Artificial Intelligence (AI) and Education, the first ever document to offer guidance and recommendations on how best to harness AI technologies for achieving the Education 2030 Agenda. It was adopted during the International Conference on Artificial Intelligence and Education, held in Beijing from 16 – 18 May 2019, by over 50 government ministers, international representatives from over 105 Member States and almost 100 representatives from UN agencies, academic institutions, civil society and the private sector. The Beijing Consensus comes after the Qingdao Declaration of 2015, in which UNESCO Member States committed to efficiently harness emerging technologies for the achievement of SDG 4.)

Four forces acting together and separately will impact the regulation of AI: the Law; the design of AI systems; market regulation; and ethics and principles. The basis of regulation will likely consider the ethical values of: explainability; accountability and transparency.[1010]

**AIEd Ethics by Design**

Ethics by design[1011] will continue to gain strength as a consideration throughout the development and use of AI systems, including systems designed for children's and youth's use. With respect to children, the Children's Rights by Design of AI systems ("CRbD") standard[1012] is useful to employ against data-driven business models from AIEd that could exploit or otherwise harm children.

An application of unethically designed AIEd that arose public protest during COVID-19, was the use of a biased algorithm in grading students. Due to the COVID-19 pandemic in the United Kingdom, all secondary education examinations due to be held in 2020 were cancelled. As a result, an alternative method had to be designed and implemented at short notice to determine the qualification grades to be given to students for that year. A grades standardization algorithm was produced in June 2020 by the regulator Ofqual in England, The A Level grades were announced in England, Wales and Northern Ireland on August 13, 2020. The release of results resulted in a public outcry. Particular criticism was made of the disparate effect the grading algorithm had in downgrading the results of those who attended state schools, and upgrading the results of pupils at privately funded independent schools and thus disadvantaging pupils of a lower socio-economic background, in part due to the algorithm's behavior around small cohort sizes.

Students and teachers felt deprived and upset following the controversial algorithm calculation and protested against it, with many demanding Prime Minister Boris Johnson and his government take immediate action. In a

---

[1010] Fjeld, Jessica and Achten, Nele and Hilligoss, Hannah and Nagy, Adam and Srikumar, Madhulika, Principled Artificial Intelligence: Mapping Consensus in Ethical and Rights-Based Approaches to Principles for AI (January 15, 2020). Berkman Klein Center Research Publication No. 2020-1, Available at SSRN: https://ssrn.com/abstract=3518482 or http://dx.doi.org/10.2139/ssrn.3518482

[1011] Floridi, Luciano; Cowls, Josh; Beltrametti, Monica; Chatila, Raja; Chazerand, Patrice; Dignum, Virginia; Luetge, Christoph; Madelin,Robert; Pagallo, Ugo; Rossi, Francesca; Shafer, Burkhard; Valcke, Peggy; Vayena, Vayena. AI4People-An Ethical Framework for a Good AI Society: Opportunities, Risks, Principles, and Recommendations. Minds and Machines, 2018. Available in https://doi.org/10.1007/s11023-018-9482-5 and https://standards.ieee.org/content/dam/ieee-standards/standards/web/documents/other/ead1e.pdf?utm_medium=PR&utm_source=Web&utm_campaign=EAD1e&utm_content=geias&utm_term=undefined (checked in 14.10.2020).

[1012] The CRdD for AI standard could be translated into the following specific recommendations for actors who govern, develop and provide products and services with AI that impacts direct or indirectly children: 1) Integrate the Convention on the Rights of the Child provisions into all appropriate corporate policies and management processes; 2) Use an interdisciplinary perspective to achieve the best interests of the child; 3) Universal adoption of the best technology and policy available; 4) Due diligence of policies and community standards; 5) Data minimization; 6) Children's full ownership of their data; 7) Commercial-free digital spaces; 8) Promotion of meaningful and non-monetizable experiences; 9) Nudge techniques in the best interest of the child; 10) Safety standards; 11) Default high-privacy settings; 12) Parental controls and mediation (children should have age appropriate and transparent information about how it works and how it affects their privacy); 13) Right use, play and participate without data collection (options free from children's data processing); 14) Promotion of children's right to disconnect; 15) Adoption of Children's Data Protection Impact Assessments; 16) Non-detrimental use of data (processing children's data should be always in their best interests); 17) Transparency, accessibility and legibility of terms of use and privacy policies; and 18) No data sharing.Hartung, Pedro. The Children's rights-by-design (CRbD) standard for data use by tech companies. Unicef Data Governance Working Group, 2020. https://www.unicef.org/globalinsight/media/1286/file/%20UNICEF-Global-Insight-DataGov-data-use-brief-2020.pdf Additionally, for all automated decisions with AI it is important to guarantee the AI system's explicability and accountability, explaining how they protect and promote children's rights.

tone deaf response to the public outcry, Secretary of State for Education Gavin Williamson said that the grading system is here to stay, and Boris Johnson stated that the results are "robust and dependable".

Legal action, in the form of judicial review, was initiated by multiple students and legal advocacy organizations, such as the Good Law Project.[1013] Finally, on August 17,2020, Ofqual and Secretary of State for Education Gavin Williamson agreed that grades would be reissued using unmoderated teacher predictions. [1014]

**AIEd Applications-Connecting AI with EdTech**

Today, both startups and established EdTech companies seek to integrate AI into marketable products. In some cases, AI performs functions independently of teachers, while in others it augments teaching capabilities.[1015] Applications of AI based education technology include the following:

**Tutoring**. AI programs commonly referred to as Intelligent Tutoring Systems (ITS) or adaptive tutors engage students in dialogue, answer questions, and provide feedback.

**Personalizing Learning.** ITS and adaptive tutors tailor learning material, pace, sequence, and difficulty to each student's needs. AI can also provide support for special needs students, for instance by teaching autistic children to identify facial expressions.

**Testing.** Computer adaptive assessments adjust the difficulty of successive questions based on the accuracy of the student's answers, enabling more precise identification of a student's mastery level.

**Automating Tasks.** AI can perform routine tasks such as taking attendance, grading assignments, and generating test questions.

Thus, AI-based tools have three general orientations in terms of their use in schools: learner-facing, teacher-facing and system-facing.[1016]

Adaptive learning systems that are learner-facing employ algorithms, assessments, student feedback and various media to deliver material tailored to each student's needs and progress.[1017] For example, AI may be used to enhance social skills, especially for children with special needs. One company that employs AI for this purpose is Brain Power, which addresses the issue of autism through a wearable computer.[1018] AI is deployed to help high school students build career skills, including language learning applications. Duolingo[1019] is one such language

---

[1013] Good Law Project, Legal action over A-Level results fiasco, https://goodlawproject.org/news/a-level-results-fiasco/

[1014] Adam Satariano 'British Grading Debacle Shows Pitfalls of Automating Government', New York Times, Aug. 2020. Available at https://www.nytimes.com/2020/08/20/world/europe/uk-england-grading-algorithm.html; WILL BEDINGFIELD, Everything that went wrong with the botched A-Levels algorithm: flawed assumptions about data led to the problems impacting hundreds of thousands of students, WIRED, (Aug. 18, 2020) https://www.wired.co.uk/article/alevel-exam-algorithm (On March 18, the government announced that, like so many annual institutions that have fallen victim to Covid-19, this summer's exams would be cancelled. In the exams' place, the Office of Qualifications and Examinations Regulation (Ofqual) asked teachers to predict the grades each of their students would have achieved.) See also, Jon Porter, UK ditches exam results generated by biased algorithm after student protests, The Verge, (August 17, 2020) https://www.theverge.com/2020/8/17/21372045/uk-a-level-results-algorithm-biased-coronavirus-covid-19-pandemic-university-applications WIKIPEDIA, 2020 UK GCSE and A-Level Grading Controversy, https://en.wikipedia.org/wiki/2020_UK_GCSE_and_A-Level_grading_controversy#cite_note-24

[1015] Congressional Research Service Report, Artificial Intelligence (AI) and Education (August 1, 2018) https://crsreports.congress.gov/product/pdf/IF/IF10937

[1016] Anissa Baker Smith, "Educ-AI-tion Rebooted?," Nesta, https://www.nesta.org.uk/report/education-rebooted/

[1017] Id.

[1018] Brain Power, "About Us," http://www.brain-power.com/

[1019] https://www.duolingo.com/

learning application which gives students personalized feedback in over 300,000 classrooms around the globe.[1020]

Under the teacher-facing category, AI helps teachers in administrative tasks such as grading papers and detecting cheating. For example, the Human-Computer Interaction Institute at Carnegie Mellon University is partnering with startup Lumilo[1021], building an AI augmented reality assistant that will keep teachers in the loop as students work on their assignments. [1022]

When used in classrooms, personalized learning software allows students to work at their own pace, while freeing up the teacher to spend more time working one-on-one with students. Yet such personalized classrooms also pose unique challenges for teachers, who are tasked with monitoring classes working on divergent activities, and prioritizing help-giving in the face of limited time.[1023]

Intelligent tutoring systems are a class of advanced learning technologies that provide students with step-by-step guidance during complex problem-solving practice and other learning activities.

AI companies connect their EdTech products using client or server-side software development kits (SDKS), which then analyze their user's data in real time. Data streams from a variety of learning contexts can be aggregated to create in-depth psychometric profiles (learning models) of the interactions, preferences, and achievements of each individual student. It then uses an item response theory, a psychometric framework to determine the student's next challenge, instructional material, or optimal activity, which is then delivered to the student via the partner's EdTech product. The AI system also provides personalized information and recommendations to teachers and parents on the best ways they can help individual students.[1024]

In addition to the software and tools described above, AI robots are increasingly transforming educational methods.[1025] Even though educational robots promise benefits to children, e.g., personalized learning, developing social skills, enabling distance education for children in remote regions, they also pose risks.[1026]

---

[1020] Jackie Snow, "AI Technology is disrupting the traditional classroom," https://www.pbs.org/wgbh/nova/article/ai-technology-is-disrupting-the-traditional-classroom/

[1021] Julia Mericle, With Lumilo, teachers can see classroom analytics floating above students' heads, Pittsburgh Business Times, (Oct. 3, 2018) https://www.bizjournals.com/pittsburgh/news/2018/10/03/with-lumilo-teachers-can-see-classroom-analytics.html

[1022] Center for Curriculum Redesign, Artificial Intelligence in Education: Promises and Implications for Teaching and Learning, (March 2019) https://curriculumredesign.org/wp-content/uploads/AI-in-Education-CCR-Copy-Protected.pdf

[1023] Holstein, K., Hong, G., Tegene, M., McLaren, B. M., & Aleven, V. (2018). The classroom as a dashboard: Co-designing wearable cognitive augmentation for K-12 teachers. In Proceedings of the Eighth International Learning Analytics & Knowledge Conference (pp. 79-88). ACM. This paper reports on the co-design, implementation, and evaluation of a wearable classroom orchestration tool for K-12 teachers: mixed-reality smart glasses that augment teachers' realtime perceptions of their students' learning, metacognition, and behavior, while students work with personalized learning software. The main contributions are: (1) the first exploration of the use of smart glasses to support orchestration of personalized classrooms, yielding design findings that may inform future work on real-time orchestration tools; (2) Replay Enactments: a new prototyping method for real-time orchestration tools; and (3) an in-lab evaluation and classroom pilot using a prototype of teacher smart glasses (Lumilo), with early findings suggesting that Lumilo can direct teachers' time to students who may need it most.

[1024] See, e.g, https://www.kidaptive.com/

[1025] Timms, M.J. (2016). Letting Artificial Intelligence in Education out of the Box: Educational Cobots and Smart Classrooms. International Journal of Artificial Intelligence in Education, 26(2), 701-712.

[1026] Jon-Chao Hong, Kuang-Chao Yu, and Mei-Yung Chen, "Collaborative Learning in Technological Project Design," International Journal of Technology and Design Education 21, no. 3 (August 2011): 335–47.; Mazzoni, Elvis, and Martina Benvenuti, "A Robot-Partner for Preschool Children Learning English Using Socio-Cognitive Conflict," Journal of Educational Technology & Society 18, no. 4 (2015): 474–85.; Barak, Moshe, and Yair Zadok, "Robotics Projects and Learning Concepts in Science, Technology and Problem Solving," International Journal of Technology and Design Education 19, no. 3 (August 2009): 289–307.

Human rights that may be positively or negatively affected by their use include the right to education, as well as the right to protection from exploitation and abuse, and the protection of children with disabilities, inasmuch as they could be developed with commercial profit in mind, at the expense of the CRbD standard.

AI technologies could help facilitate "personalized learning" (tailoring instruction to the needs of each student) and "blended learning" (combining technology with face-to face interaction). Many school officials hope that such approaches will improve academic performance and reduce achievement gaps between groups of students. Some teachers also suggest that personalized learning increases student engagement, motivation, and independence.[1027]

AI-based learning faces significant implementation challenges. Greater student independence could disadvantage children who are less self-disciplined or who receive little educational support at home, potentially exacerbating the achievement gap. Moreover, surveys indicate that some teachers struggle to translate the data they receive from personalized learning tools into actionable instruction and spend inordinate amounts of time creating individualized assignments. There is also debate over how well students retain knowledge learned from an AI-based system, and whether spending substantial class time on computers diminishes social learning at school.[1028]

The budget implications of using AI in education are problematic, given uncertainties about the cost-effectiveness of the technology. For example, the versatility and scalability of AI could result in some institutions to reduce teaching staff in favor of AI alternatives. However, AI could create demand for education professionals who can design and implement personalized learning programs.[1029]

**AIEd in the US**

US government actions have addressed issues related to AI in schools, such as internet access and student data privacy. Successful implementation of AI by schools requires significant investment in information technology as well as reliable broadband internet access. These resources are not uniformly distributed across school districts; for example, close to 80% of schools without fiber connections were located in rural areas as of 2017. Federal efforts to address this disparity include such programs as the Universal Service Program for Schools and Libraries. Commonly known as E-rate, the program provides subsidies of up to 90% to help ensure that qualifying schools and libraries can obtain high-speed internet access and telecommunications at affordable rates. The National Science Foundation[1030] and the Department of Education's[1031] (ED's) Institute of Education Sciences[1032] have awarded grants to projects researching AI-enabled classroom technologies. In addition, ED's Office of Educational Technology[1033] has released several publications on topics relevant to AI in schools, such as learning analytics and educational data mining, teacher preparation, personalized learning, and student privacy.

**Selected AI Education Policy Considerations in the US**

Although most education policies are set at the state and local level, Congress is involved in oversight and legislative actions on issues such as student privacy, teacher preparation, product selection, and algorithmic accountability.[1034]

---

[1027] Congressional Research Service Report, Artificial Intelligence (AI) and Education (August 1, 2018) https://crsreports.congress.gov/product/pdf/IF/IF10937

[1028] Congressional Research Service Report, Artificial Intelligence (AI) and Education (August 1, 2018) https://crsreports.congress.gov/product/pdf/IF/IF10937

[1029] Congressional Research Service Report, Artificial Intelligence (AI) and Education (August 1, 2018) https://crsreports.congress.gov/product/pdf/IF/IF10937

[1030] https://www.nsf.gov/

[1031] https://www.ed.gov/

[1032] https://ies.ed.gov/

[1033] Meet the OET Team - Office of Educational Technology

[1034] Actions that Congress has taken include **The Every Student Succeeds Act (P.L. 114-95)**, which reauthorized the Elementary and Secondary Education Act of 1965, authorized the use of computer adaptive

**Student Privacy.** Like many digital services, AI-enabled education tools collect and store PII. In response to public concerns about data security and privacy, activists created a voluntary Student Privacy Pledge in 2014. Signatories promise to place limits on the lifespan of stored data, maintain reasonable security measures, and refrain from selling data. Although President Obama and several Members of Congress endorsed the pledge, critics have asserted that the language is vague and the pledge is little more than a publicity move. Meanwhile, 41 states have enacted laws governing student data collection, use, reporting, and safeguarding since 2013. Several of those laws were modeled after California's Student Online Personal Information Protection Act (SOPIPA). Congress may consider whether such state efforts are sufficient or if a federal law is needed.

**Teacher Preparation.** If AI technologies are adopted on a broader scale, teachers face the task of not only learning to use specific products but also integrating a range of AI technologies into their lessons.

**Preparation programs** offered by teacher-certifying universities and institutes might provide such training. In FY2018, ED's Teacher Quality Partnership (TQP) competition plans to award approximately $14 million in grants to these programs. If Congress decides to support funding teacher preparation for AI, options could include redirecting funds toward teacher technology training and directing ED to develop best practices for teacher technology competency.[1035]

**Product Procurement and Support.** Choosing products can be a time- and energy-intensive effort involving teachers, administrators, IT staff, and other school officials. While some schools allow teachers to experiment freely, others require IT staff to vet hundreds of privacy policies and security measures. Some school districts have turned to digital content consultants for guidance in selecting products. To help schools gather research on educational tools and strategies, nonprofits and federal agencies have developed resources. For example, the State Educational Technology Directors Association provides a best practices guide for product procurement,[1036] and ED's What Works Clearinghouse rigorously reviews the effectiveness of educational products and practices. Despite these resources, surveys indicate that peer recommendation is a more prevalent basis for choosing products than research-based evidence. A centralized platform to exchange information and collaboratively troubleshoot problems might help formalize inter-district communication and allow schools to make wiser and less costly purchases. The Technology for Education Consortium estimates that districts would collectively save $3 billion per year on education technology purchases simply by sharing price information.[1037]

**Algorithmic Accountability.** Parents and school administrators may find it difficult to trust AI technologies used to influence or make decisions about student learning. Mistrust can stem from the refusal of companies to

---

testing in state student academic assessments mandated under the act. This marked the first time Congress explicitly approved an AI testing technique for widespread use in schools. Congress has taken steps to address public concerns regarding the privacy of students' personal information, including concerns about education technology companies collecting personally identifiable information (PII) from students to maintain user accounts; **The Family Educational Rights and Privacy Act of 1974** (FERPA), as amended in 2013, limits the power of schools to disclose students' education records but has been criticized for weak enforcement mechanisms against third parties that misuse student data; The Protection of Pupil Rights Amendment of 1978 (PPRA), as further amended in 2015, requires schools to notify parents and offer an opt-out choice if a third party surveys students for marketing purposes; T**he Children's Online Privacy Protection Act of 1998 (COPPA**) requires parental consent before websites collect information about children aged 13 or under. Many experts worry that current law, passed largely before AI became a major policy consideration, is insufficient to address today's cybersecurity threats. Bills introduced in the 115th Congress, such as the **Protecting Student Privacy Act (S. 877), SAFE KIDS Act (S. 2640), and Protecting Education Privacy Act (H.R. 5224)**, addressed how third parties can access and use students' PII.. See Congressional Research Service Report, Artificial Intelligence (AI) and Education (August 1, 2018) https://crsreports.congress.gov/product/pdf/IF/IF10937.

[1035] Congressional Research Service Report, Artificial Intelligence (AI) and Education (August 1, 2018) https://crsreports.congress.gov/product/pdf/IF/IF10937
[1036] https://www.setda.org/master/wp-content/uploads/2017/10/Case_studies_full_10.15.17.pdf
[1037] Congressional Research Service Report, Artificial Intelligence (AI) and Education (August 1, 2018) https://crsreports.congress.gov/product/pdf/IF/IF10937

disclose their algorithms, which they argue are trade secrets, or from the "black box problem," which occurs when an algorithm's complexity renders its processes inscrutable even to developers. Options for Congress could include holding hearings, conducting oversight, and considering requirements to enhance transparency and accountability of data use more broadly, as the European Union has sought to do through the General Data Protection Regulation.[1038]

**AIEd and Surveillance**

Surveillance of children is another use of AI that is booming due to advance machine learning and deep learning techniques.[1039] Although some degree of surveillance advances security, surveillance poses risks to children. A use of facial recognition technology benefitting children is that of police in New Delhi, who trialed facial recognition technology and identified almost 3,000 missing children in four days.[1040] However, surveillance also creates privacy, safety, bias, and security risks and, especially in education contexts, limit children's ability and willingness to take risks and otherwise express themselves.[1041]

Key legal issues surrounding advanced security technologies in public K-12 schools in the United States, including the impact on student privacy rights. In using AI surveillance technology in schools,[1042] privacy must be balanced against security concerns[1043]; any apparent issues with efficacy and accuracy of the technology should be addressed before implementation; and Fourth Amendment case law, federal student privacy legislation, and state laws need to be further developed, with AI in mind.[1044]

In response to the fears of additional school violence and calls for enhanced school security, schools have begun tightening security through the use of these emerging AI technologies.[1045] Recognizing the market opportunity, technology companies are developing new devices they claim will prevent or reduce the likelihood of school shootings.[1046] These new devices, which include advanced cameras and body scanners, use biometrics and

---

[1038] Congressional Research Service Report, Artificial Intelligence (AI) and Education (August 1, 2018) https://crsreports.congress.gov/product/pdf/IF/IF10937

[1039] Emmeline Taylor, "Surveillance Schools: A New Era in Education," in Surveillance Schools: Security, Discipline and Control in Contemporary Education (London: Palgrave Macmillan UK, 2013), 15–39, https://doi.org/10.1057/9781137308863_2.

[1040] Anthony Cuthbertson, "Police Trace 3,000 Missing Children in Just Four Days Using Facial Recognition Technology," The Independent, https://www.independent.co.uk/life-style/gadgets-and-tech/news/india-police-missingchildren-facial-recognition-tech-trace-find-reunite-a8320406.html, (April 24, 2018).

[1041] Article 19, "The Global Principles on Protection of Freedom of Expression and Privacy," https://www.article19.org/resources/the-global-principles-on-protection-of-freedom-of-expression-and-privacy/ ARTICLE 19 works for a world where all people everywhere can freely express themselves and actively engage in public life without fear of discrimination. Id.

[1042] Barbara Fedders, The Constant and Expanding Classroom: Surveillance in K-12 Public Schools, 97 N.C. L. REV. 1673 (2019).

[1043] See Sara Collins, Tyler Park & Amelia Vance, Ensuring School Safety While Also Protecting Privacy, FUTURE PRIVACY F. (June 6, 2018), https://fpf.org/2018/06/06/ensuring-school-safety-while#also-protecting-privacy-fpf-testim

[1044] Maya Weinstein, School Surveillance: The Students' Rights Implications of Artificial Intelligence as K-12 School Security, 98 N.C. L. Rev. 438 (2020). Available at: https://scholarship.law.unc.edu/nclr/vol98/iss2/12

[1045] See, e.g., Kaitlyn DeHaven, Texas ISD Makes Major Security Upgrades Over the Summer, CAMPUS SECURITY & LIFE SAFETY (Aug. 9, 2019), https://campuslifesecurity.com/articles/2019/08/09/texas-isd-makes-major-security-upgrades-over-the-summer.aspx [https://perma.cc/3YFR#TZ47] ("Two apps will now be used as part of the security measures—the Anonymous Alerts app and the Smart Button. . . . In terms of physical security, the district installed video intercoms at each school entrance."); Mark Keierleber, Inside the $3 Billion School Security Industry: Companies Marketed Sophisticated Technology To 'Harden' Campuses, but Will It Make Us Safe?, 74 (Aug. 9, 2018), https://www.the74million.org/article/inside-the-3-billion-school-security-industry-companies#market-sophisticated-technology-to-harden-campuses-but-will-it-make-us-safe/ ("Schools have increasingly locked and monitored campus entrances in recent years, though the rise in school security is most evident in the growth of video surveillance.")

artificial intelligence to recognize faces; detect weapons, gunshots, and other threats; and track individuals' locations in schools.[1047]

In schools, biometric and AI technologies cover a wide spectrum of programs. The AI industry has seen a boom within the education market, and the worldwide AI education market value is predicted to surpass six billion dollars by 2024,[1048] with classroom applications accounting for twenty percent of that growth.[1049]

Much of the reason for the AIEd growth is the integration of AI systems for personalized learning, which enables students to receive "immediate and personalized feedback and instructions . . . without the intervention of a human tutor."[1050] Biometrics have been incorporated into the classroom as well,[1051] and some schools even use biometrics to allow students to pay for lunch with just a fingerprint.[1052]

One popular new area of school surveillance technology is location tracking. For instance, the program "e-hallpass"[1053] is a modern, electronic hall pass that "continuously logs and monitors student time in the halls" and claims to "improv[e] school security and emergency management while reducing classroom disruptions by as much as 50%." A similar program, "iClicker Reef,"[1054] rebranded as "iClicker,"[1055] tracks attendance through a

---

[1046] The media streaming company RealNetworks is offering its facial recognition software to over 100,000 school districts for free, with the goal of making schools safer. Eli Zimmerman, Company Offers Free Facial Recognition Software To Boost School Security, EDTECH (Aug. 3, 2018), https://edtechmagazine.com/k12/article/2018/08/company-offers-free-facial-recognition-software#boost-school-security [https://perma.cc/4V9N-TMSD]; see also Press Release, SAFR, RealNetworks Provides SAFR Facial Recognition Solution for Free to Every K-12 School in the U.S. and Canada (July 17, 2018), https://safr.com/press-release/realnetworks-provides-safr-facial-recognition-solution#for-free-to-every-k-12-school-in-the-u-s-and-canada/

[1047] Maya Weinstein, School Surveillance: The Students' Rights Implications of Artificial Intelligence as K-12 School Security, 98 N.C. L. Rev. 438 (2020). Available at: https://scholarship.law.unc.edu/nclr/vol98/iss2/12

[1048] Ankita Bhutani & Preeti Wadhwani, Artificial Intelligence (AI) in Education Market Size Worth $6bn by 2024, GLOBAL MKT. INSIGHTS (Aug. 12, 2019), https://www.gminsights.com/pressrelease/artificial-intelligence-ai-in-education-market [https://perma.cc/W3RP-SNDQ].

[1049] Michele Molnar, K-12 Artificial Intelligence Market Set To Explode in U.S. and Worldwide by 2024, EDWEEK MKT. BRIEF (July 10, 2018), https://marketbrief.edweek.org/marketplace-k-12/k-12artificial-intelligence-market-set-explode-u-s-worldwide-2024/ [https://perma.cc/6HBQ-JCEF]; see also Hao, Karen. "China has started a grand experiment in AI education. It could reshape how the world learns." MIT Technology Review. .technologyreview.com/s/614057/china-squirrel-has-started-a-grand-experiment-in-ai-education-it-could-reshape-how-the/.

[1050] Artificial Intelligence in Education Market To Hit $6bn by 2024, GLOBAL MKT. INSIGHTS (June 6, 2018), https://www.globenewswire.com/news-release/2018/06/06/1517441/0/en/Artificial#Intelligence-in-Education-Market-to-hit-6bn-by-2024-Global-Market-Insights-Inc.html

[1051] Jen A. Miller, Biometrics in Schools To Yield Security Benefits and Privacy Concerns, EDTECH MAG. (May 7, 2019), https://edtechmagazine.com/k12/article/2019/05/biometrics-schools-yield-security-benefits-and#privacy-concerns ("Biometric technology is already part of the K-12 ecosystem, where administrators are using iris scans and 'facial fingerprints' to grant access to buildings and computer labs, track attendance, manage lunch payments, loan library materials and ensure students get on the right buses."); Mae Rice, 13 EdTech Applications that Are Transforming Teaching and Learning, BUILT IN (June 22, 2019), https://builtin.com/edtech/technology-in-classroom-applications (describing an online test proctoring system which confirms test takers' identities through fingerprints and voice biometrics).

[1052] Biometrics Allows Students To Purchase with Fingerprint, GOV'T TECH. (Oct. 17, 2007), https://www.govtech.com/health/Biometrics-Allows-Students-to-Purchase-with.html

[1053] E-Hallpass, EDUSPIRE SOLUTIONS, https://www.eduspiresolutions.org/what-is-e-hallpass/

[1054] https://community.macmillanlearning.com/t5/institutional-solutions-blog/new-name-who-s-this-iclicker-reef-to-be-re-named-iclicker/ba-p/15007

[1055] https://www.iclicker.com/students/apps-and-remotes/web

geolocation feature.[1056] Using geolocation,[1057] these location systems have the ability to identify when a student is in class, log attendance for the teacher, and track where students are in school.[1058]

Although the technology has some benefits from a security standpoint, these technologies are intrusive and create an environment where students are tracked, monitored, and watched. Many of these programs involve constant monitoring of children, and some collect personally identifying data, including fingerprints and face images. There are a number of potential adverse consequences of these technologies: students are inhibited to participate in class, risks of false data matches may lead to harmful and wrongful disciplinary actions, and otherwise encroaching on student privacy rights. [1059]

Another type of facial recognition program, "affect recognition," uses biometric analysis to scan individuals' faces and purportedly identify emotions.[1060] An Australian university is currently testing a product called the "Biometric Mirror[1061]," which reads faces and ranks them according to fourteen characteristics, including gender, age, ethnicity, attractiveness, "weirdness," and emotional stability.[1062] Schools in China have implemented a similar technology to analyze students' facial expressions, including expressions like "neutral, happy, sad,

---

[1056] David Rosen & Aaron Santesso, How Students Learned To Stop Worrying—and Love Being Spied On, CHRON. HIGHER EDUC. (Sept. 23, 2018), https://www.chronicle.com/article/How-Students#Learned-to-Stop/244596

[1057] Daniel Ionescu, Geolocation 101: How It Works, the Apps, and Your Privacy, ITWORLD (Mar. 31, 2010), https://www.itworld.com/article/2756095/networking-hardware/geolocation-101--how-it#works--the-apps--and-your-privacy.html [https://perma.cc/AMB3-8VLK] ("Typically, geolocation apps do two things: They report your location to other users, and they associate real-world locations (such as restaurants and events) to your location.")

[1058] David Rosen & Aaron Santesso, How Students Learned To Stop Worrying—and Love Being Spied On, CHRON. HIGHER EDUC. (Sept. 23, 2018), https://www.chronicle.com/article/How-Students#Learned-to-Stop/244596

[1059] Maya Weinstein, School Surveillance: The Students' Rights Implications of Artificial Intelligence as K-12 School Security, 98 N.C. L. Rev. 438 (2020). Available at: https://scholarship.law.unc.edu/nclr/vol98/iss2/12

[1060] MEREDITH WHITTAKER ET AL., AI NOW REPORT 2018, at 4 (Dec. 2018), https://ainowinstitute.org/AI_Now_2018_Report.pdf [https://perma.cc/2EAJ-AALT] ("Affect recognition is a subclass of facial recognition that claims to detect things such as personality, inner feelings, mental health, and 'worker engagement' based on images or video of faces."). See also Milly Chan, This AI reads children's emotions as they learn, CNN, (Feb. 17, 2021) https://www.cnn.com/2021/02/16/tech/emotion-recognition-ai-education-spc-intl-hnk/index.html (Ka Tim Chu, teacher and vice principal of Hong Kong's True Light College uses an AI-powered learning platform monitors his students' emotions as they study at home.1 Students work on tests and homework on the platform as part of the school curriculum. While they study, the AI measures muscle points on their faces via the camera on their computer or tablet, and identifies emotions including happiness, sadness, anger, surprise and fear. The system also monitors how long students take to answer questions; records their marks and performance history; generates reports on their strengths, weaknesses and motivation levels; and forecasts their grades. The program can adapt to each student, targeting knowledge gaps and offering game-style tests designed to make learning fun. Lam says the technology has been especially useful to teachers during the pandemic because it allows them to remotely monitor their students' emotions as they learn. Racial bias is also a serious issue for AI. Research shows that some emotional analysis technology has trouble identifying the emotions of darker skinned faces, in part because the algorithm is shaped by human bias and learns how to identify emotions from mostly white faces.)

[1061] https://biometricmirror.com/ ("Biometric Mirror is an ethically provocative interactive system that enables public participation in the debate around ethics of artificial intelligence. The system enables people to have their face photographed and to witness the reveal of their psychometric analysis, including attributes such as aggressiveness, weirdness and emotional instability. Ultimately, a personalized scenario of algorithmic decision-making is shown in order to stimulate individual reflection on the ethical application of artificial intelligence.")

[1062] Jo Lauder, Mirror, Mirror: How AI Is Using Facial Recognition To Decipher Your Personality, ABC AUSTL. (July 23, 2018), https://www.abc.net.au/triplej/programs/hack/how-ai-is-using-facial#recognition-to-decipher-your-personality/10025634

disappointed, angry, scared and surprised."[1063] The main goal of this so-called "smart eye" is to alert teachers when students are distracted in class.[1064]

Some argue that the identification of changes in mood could assist educators with identifying students experiencing mental health crises, which could help flag potential threats.[1065] However, many believe that affect recognition, the idea that someone's emotions can be read by a program is eerily reminiscent of debunked psuedosciences of phrenology and physiognomy.[1066] "These claims are not backed by robust scientific evidence and are being applied in unethical and irresponsible ways…Linking affect recognition to hiring, access to insurance, education, and policing creates deeply concerning risks, at both an individual and societal level." [1067]

The idea of banning facial recognition outright has also grown more popular in the past few years, particularly with states and municipalities. For instance California and Massachusetts cities of San Francisco, Somerville, Boston, Oakland, and Berkeley, have banned the use of facial recognition technology by city government, including but not limited to law enforcement. A two-year moratorium on the use of facial recognition technology in New York schools passed both houses of the state legislature and awaits the governor's signature. California recently passed a three-year ban on law enforcement uses of facial recognition in body cameras, and a proposed ordinance in Portland, Oregon would ban the use of facial recognition by both law enforcement and private businesses. A California legislator announced plans to introduce a bill that would ban government uses of facial recognition for the next five years, while Senators Booker and Merkley introduced a bill that would ban federal uses of the technology and prohibit states and local entities from using federal funding for it until Congress passes legislation regulating it. The goal of a comprehensive and federal ban on facial recognition may be lofty, but it is not impossible given the growing awareness and political will to regulate these AI technologies.[1068]

A major concern related to implementing AI technologies anywhere, but in schools as well, is the risk of machine bias, the systematic disparities in accuracies of algorithm results, typically with respect to race, but also gender or age. The identification abilities of AI in biometrics are only as good as the humans who develop them. A prominent AI expert and co-founder of AI4ALL[1069] described the issue as such: "bias in, bias out."[1070]

Surveillance practices that continuously monitor everything from children's engagement in the classroom to their emotional states throughout the day threaten the creativity, freedom of choice and self-determination of children by potentially fostering an overabundance of self-censorship and social control.[1071] Once automated surveillance technologies are deployed at schools and in classrooms, children's rights such as the right to

---

[1063] Neil Connor, Chinese School Uses Facial Recognition To Monitor Student Attention in Class, TELEGRAPH (May 17, 2018), https://www.telegraph.co.uk/news/2018/05/17/chinese-school-uses#facial-recognition-monitor-student-attention

[1064] Id.

[1065] See, e.g., Randy Rieland, Can Artificial Intelligence Help Stop School Shootings?, SMITHSONIAN (June 22, 2018), https://www.smithsonianmag.com/innovation/can-artificial-intelligence-help-stop#school-shootings-180969288/ (describing the use of machine learning to analyze student language and behavior and help counselors with risk assessment).

[1066] AI NOW 2018 REPORT at 8. https://ainowinstitute.org/AI_Now_2018_Report.pdf

[1067] Id. at 4.

[1068] Barrett, Lindsey, Ban Facial Recognition Technologies for Children—And for Everyone Else (July 24, 2020). Boston University Journal of Science and Technology Law. Volume 26.2, at 277-278. Available at SSRN: https://ssrn.com/abstract=3660118

[1069] https://ai-4-all.org/ AI4ALL Opens Doors to Artificial Intelligence for Historically Excluded Talent Through Education and Mentorship. Id.

[1070] Jessi Hempel, Fei-Fei Li's Quest To Make AI Better for Humanity, WIRED (Nov. 13, 2018), https://www.wired.com/story/fei-fei-li-artificial-intelligence-humanity

[1071] Rich Haridy, "AI in Schools: China's Massive and Unprecedented Education Experiment," New Atlas – New Technology & Science News, https://newatlas.com/china-aieducation-schools-facial-recognition/54786/, (May 28, 2018).

privacy, the right not to be subjected to discrimination, the right to flourish, and freedom of expression may be compromised due to the surveillance environment in which children are confined.[1072]

The risks vary depending on who does the surveilling (governments, teachers, parents etc.) and for what purposes.[1073] However, the chilling effect of having cameras constantly turned on children is undeniable.[1074] It is important to consider and evaluate the actors involved, their purposes, the tools and methods they'll use, and the safeguards they'll put in place. The emerging trend of classroom surveillance should help children, not harm them.

New technologies are expanding schools' ability to keep students under surveillance—inside the classroom and out, during the school year and after it ends. Schools have moved quickly to adopt a dizzying array of new tools. These include digital learning products that capture and store student data; anonymous tip lines encouraging students to report on each other; and software that monitors students' emails and social media posts, even when they are written from home. Steadily growing numbers of police officers stationed in schools can access this information, compounding the technologies' power.[1075]

Advocates of these tools argue that they improve student safety and learning outcomes, but this Article reveals that the evidence for this argument is in fact quite thin. Moreover, policymakers have failed to consider important countervailing considerations—most notably, student privacy and its significance for child development; unequal impact, particularly for poor, Black, and LGBTQ youth; and potential liability for school administrators.[1076]

The twin justifications for student surveillance are safety and improved educational outcomes. The companies developing these technologies market them against a backdrop of fear of violence, especially school shootings, and anxiety about academic success. State lawmakers appear convinced by these justifications, passing legislation that mandates adoption of some technologies and allocates funds for the purchase of others. Local school districts take advantage of increased state funding to hire school resource officers for kindergarten through the twelfth grade. The various mechanisms of surveillance combine to make more information available about more students, for a longer period of time, and accessible to a greater number of actors than was possible before the digital age.[1077]

**AIEd Continues to Develop**

Thus, AIEd has the potential to dramatically automate and help track the learner's progress in all these skills and identify where best a human teacher's assistance is needed. For teachers, AIEd can potentially be used to help identify the most effective teaching methods based on students' contexts and learning background. It can automate monotonous tasks, generate assessments, and allegedly automate grading and feedback. AI does not

---

[1072] Article 19, "Privacy and Freedom of Expression in the Age of Artificial Intelligence," https://www.article19.org/wp-content/uploads/2018/04/Privacy-and-Freedom-ofExpression-In-the-Age-of-Artificial-Intelligence-1.pdf, (2018), 8.

[1073] William Michael Carter, "Big Brother Facial Recognition Needs Ethical Regulations," Phys.org, https://phys.org/news/2018-07-big-brother-facial-recognition-ethical.html#jCp. (July 23, 2018).

[1074] Id.

[1075] Fedders, Barbara, The Constant and Expanding Classroom: Surveillance in K-12 Public Schools (September 1, 2019). North Carolina Law Review, Vol. 97, No. 6, 2019, Available at SSRN: https://ssrn.com/abstract=3453358

[1076] Fedders, Barbara, The Constant and Expanding Classroom: Surveillance in K-12 Public Schools (September 1, 2019). North Carolina Law Review, Vol. 97, No. 6, 2019, Available at SSRN: https://ssrn.com/abstract=3453358

[1077] Fedders, Barbara, The Constant and Expanding Classroom: Surveillance in K-12 Public Schools (September 1, 2019). North Carolina Law Review, Vol. 97, No. 6, 2019, Available at SSRN: https://ssrn.com/abstract=3453358 See generally Julie E. Cohen, Surveillance Versus Privacy: Effects and Implications, in THE CAMBRIDGE HANDBOOK OF SURVEILLANCE 455, 458–59 (David Gray & Stephen E. Henderson eds., 2017) [hereinafter Cohen, Surveillance Versus Privacy] (documenting "emergence of pervasive, networked surveillance").

only impact what students learn through recommendations, but also how they learn, what are the learning gaps, which pedagogies are most effective and how to retain learner's attention. In these cases, teachers are the 'human-in-the-loop', where in such contexts, the role of AI is only to enable more informed decision making by teachers, by providing them predictions about students performance or recommending relevant content to students after teachers' approval. [1078]

Although AIEd around the globe is increasing,[1079] educational technology companies building AI powered products have always complained about the lack of relevant data for training algorithms.[1080] The advent of COVID-19 pushed educational institutions online and dependent on EdTech products to organize content, manage operations, and communicate with students. This shift generated huge amounts of data for EdTech companies on which they can build AI systems. According to a joint report: 'Shock to the System', published by Educate Ventures and Cambridge University, optimism of EdTech companies about their own future increased during the pandemic and their most pressing concern was too many customers to serve effectively.[1081]

As noted above, an intelligent tutoring system is a computer program that tries to mimic a human teacher to provide personalized learning to students.[1082] Recently, ITSs such as ASSISTments,[1083] iTalk2Learn,[1084] and Aida Calculus[1085], have gained attention.[1086] Despite being limited in terms of the domain that a particular intelligent tutoring system addresses, they have proven to be effective in providing relevant content to students, interacting with students, and improving students' academic performance.[1087]

Teachers have abandoned the technology in some instances because it was counterproductive. They conducted a formative intervention with sixteen secondary school mathematics teachers and found systemic contradictions between teachers' opinions and ITS recommendations, eventually leading to the abandonment of the tool. [1088]

There are a number of ed-tech companies that are leading the AIEd revolution. New funds are also emerging to invest in ed-tech companies and to help ed-tech startups in scaling their products. There has been an increase

---

[1078] Chaudhry, Muhammad and Kazim, Emre, Artificial Intelligence in Education (Aied) a High-Level Academic and Industry Note 2021 (April 24, 2021). Available at SSRN: https://ssrn.com/abstract=3833583 or http://dx.doi.org/10.2139/ssrn.3833583

[1079] Weller, M., 2018. Twenty years of EdTech. Educause Review Online, 53(4), pp.34-48.

[1080] Chaudhry, Muhammad and Kazim, Emre, Artificial Intelligence in Education (Aied) a High-Level Academic and Industry Note 2021 (April 24, 2021). Available at SSRN: https://ssrn.com/abstract=3833583 or http://dx.doi.org/10.2139/ssrn.3833583

[1081] Cambridge University Press and Educate Ventures (2021). Shock to the system: lessons from Covid-19 Volume 1: Implications and recommendations. Available at: https://www.cambridge.org/pk/files/1616/1349/4545/Shock_to_the_System_Lessons_from_Covid19_Volume_1.pd

[1082] Mohamed, H., & Lamia, M. (2018). Implementing flipped classroom that used an intelligent tutoring system into learning process. Computers & Education, 124, 62–76. https://doi.org/10.1016/j.compedu.2018.05.011

[1083] Heffernan, N. T., & Heffernan, C. L. (2014). The ASSISTments ecosystem: building a platform that brings scientists and teachers together for minimally invasive research on human learning and teaching.

[1084] Hasan, M.A., Noor, N.F.M., Rahman, S.S.A. and Rahman, M.M., 2020. The Transition from Intelligent to Affective Tutoring System: A Review and Open Issues. IEEE Access

[1085] https://apps.apple.com/us/app/aida-calculus/id1450379917

[1086] https://www.pearson.com/us/higher-education/products-services-teaching/learning-engagement-tools/aida.html

[1087] Fang Y, Ren Z, Hu X, Graesser AC. A meta-analysis of the effectiveness of ALEKS on learning. Educational Psychology. 2019;39(10):1278–92

[1088] Utterberg Modén, M., Tallvid, M., Lundin, J. and Lindström, B., 2021. Intelligent Tutoring Systems: Why Teachers Abandoned a Technology Aimed at Automating Teaching Processes. In Proceedings of the 54th Hawaii International Conference on System Sciences (p. 1538).

in investor interest.[1089] In 2020 the amount of investment raised by ed-tech companies more than doubled compared to 2019.[1090]

EDUCATE, a leading accelerator focused on ed-tech companies supported by UCL Institute of Education and European Regional Development Fund was formed to bring research and evidence at the center of product development for ed-tech. This accelerator has supported more than 250 ed-tech companies and 400 entrepreneurs and helped them focus on evidence-informed product development for education.[1091]

Companies such as Outschool[1092] and ClassDojo[1093] turn first profits while startups like Quizlet[1094] and ApplyBoard[1095] reached $1 billion valuations. Last year brought a flurry of record-breaking venture capital to the sector. PitchBook[1096] data shows that edtech startups around the world raised $10.76 billion last year, compared to $4.7 billion in 2019. While reporting delays could change this total, VC dollars have more than doubled since the pandemic began. In the United States, edtech startups raised $1.78 billion in venture capital across 265 deals during 2020, compared to $1.32 billion the prior year.[1097]

Seeing the business potential of AIEd and the kind of impact it can have on the future of humanity, some of the biggest tech companies around the globe are moving into this space. The shift to online education during the pandemic boosted the demand for cloud services. Amazon's AWS (Amazon Web Services) was a leader in cloud services provider facilitated institutions to scale their online examination services[1098]

Google's CEO Sunder Pichai stated that the pandemic offered an incredible opportunity to reimagine education. Google has launched more than 50 new software tools during the pandemic to facilitate remote learning. Google Classroom which is a part of Google Apps for Education (GAFE) is being widely used by schools around the globe to deliver education. Research shows that it improves class dynamics and helps with learner participation.[1099]

---

[1089] Goryachikh, S.P., Sozinova, A.A., Grishina, E.N. and Nagovitsyna, E.V., 2020. Optimisation of the mechanisms of managing venture investments in the sphere of digital education on the basis of new information and communication technologies:audit and reorganisation. International Journal of Economic Policy in Emerging Economies, 13(6), pp.587-594.

[1090] Natasha Mascarenhas, 13 investors say lifelong learning is taking edtech mainstream, TechCrunch (Jan. 28, 2021) https://techcrunch.com/2021/01/28/12-investors-say-lifelong-learning-is-taking-edtech-mainstream/

[1091] https://www.ucl.ac.uk/ioe/departments-and-centres/centres/ucl-knowledge-lab/educate

[1092] https://outschool.com/

[1093] https://www.classdojo.com/

[1094] https://quizlet.com/

[1095] https://www.applyboard.com/

[1096] https://get.pitchbook.com/pitchbook-data/?utm_source=bing&utm_medium=cpc&utm_campaign=Brand-US&adgroup=Brand-Exact&utm_term=pitchbook&matchtype=e&creative=&device=c&utm_content=&kwdaud=kwd-71400116180784:loc-190&_bt=71399641520904&sfid=rFC8fCnu-dc_pcrid_71399641520904_pkw_pitchbook_pmt_be_slid__productid__pgrid_1142393172632981_ptaid_kwd-71400116180784:loc-190&msclkid=a73b3243d1fd1b1616d352aa0b527832

[1097] Natasha Mascarenhas, 13 investors say lifelong learning is taking edtech mainstream, TechCrunch (Jan. 28, 2021) https://techcrunch.com/2021/01/28/12-investors-say-lifelong-learning-is-taking-edtech-mainstream/

[1098] About Amazon. (2020). Helping 700,000 students transition to remote learning. [online] Available at: https://www.aboutamazon.com/news/community/helping700-000-students-transition-to-remote-learning Amazon Web Services, Inc. (n.d.). Amazon Web Services, Inc. [online] Available at: https://pages.awscloud.com/whitepaper-emerging-trends-in-education.html

[1099] Al-Maroof, R.A.S. and Al-Emran, M., 2018. Students Acceptance of Google Classroom: An Exploratory Study using PLS-SEM Approach. International Journal of Emerging Technologies in Learning, 13(6); Iftakhar, S., 2016. Google classroom: what works and how. Journal of Education and Social Sciences, 3(1), pp.12-18; Shaharanee, I.N.M., Jamil, J.M. and Rodzi, S.S.M., 2016, August. Google classroom as a tool for active learning. In AIP Conference Proceedings (Vol. 1761, No. 1, p. 020069). AIP Publishing LLC. Shaharanee, I.N.M., Jamil, J.M. and Rodzi, S.S.M., 2016. The application of Google Classroom as a tool for teaching and learning. Journal of Telecommunication, Electronic and Computer Engineering (JTEC), 8(10), pp.5-8;

Development of the AIEd infrastructure is an issue. True progress will require the development of an AIEd infrastructure.[1100] This will not, however, be a single monolithic AIEd system. Instead, it will resemble the marketplace that has been developed for smartphone apps: hundreds and then thousands of individual AIEd components, developed in collaboration with educators, conformed to uniform international data standards, and shared with researchers and developers worldwide. These standards will enable system-level data collation and analysis that help us learn much more about learning
itself and how to improve it.[1101]

**Ethical AIEd**

A number of AI ethical misuses,[1102] including safety and cybersecurity incidents, have occurred in the real world,[1103] thus ethics in AI has become a real concern for AI researchers, practitioners, and governments alike.[1104] Within computer science, there is a growing overlap with the border Digital Ethics[1105] and the ethics and engineering focused on developing Trustworthy AI.[1106]

As stated above, ethics in AI focuses on fairness, accountability, transparency and explainability.[1107] Ethics in AI needs to be embedded in the entire development pipeline, from the decision to start collecting data until the decision to deploy the machine learning model in production. From an engineering perspective, four verticals of algorithmic auditing have been identified. These include auditing for performance and robustness, bias and discrimination, interpretability and explanability and algorithmic privacy.[1108]

In education, ethical AI is crucial to ensure the wellbeing of learners, teachers and other stakeholders involved.

---

Sudarsana, I.K., Putra, I.B.M.A., Astawa, I.N.T. and Yogantara, I.W.L., 2019, March. The use of Google classroom in the learning process. In Journal of Physics: Conference Series (Vol. 1175, No. 1, p. 012165). IOP Publishing.

[1100] Luckin, R., Holmes, W., Griffiths, M. and Pearson, L. (2016). Intelligence Unleashed an Argument for AI in Education. [online] Available at: https://static.googleusercontent.com/media/edu.google.com/en//pdfs/Intelligence-Unleashed-Publication.pdf

[1101] Id.

[1102] Johnson, D.G. and Verdicchio, M., 2019. AI, agency and responsibility: the VW fraud case and beyond. Ai & Society, 34(3), pp.639-647.

[1103] Yampolskiy, R.V. and Spellchecker, M.S., 2016. Artificial Intelligence Safety and Cybersecurity: a Timeline of AI Failures. arXiv preprint arXiv:1610.07997.

[1104] Leslie, David, Understanding Artificial Intelligence Ethics and Safety: A Guide for the Responsible Design and Implementation of AI Systems in the Public Sector (June 10, 2019). Available at SSRN: https://ssrn.com/abstract=3403301 or http://dx.doi.org/10.2139/ssrn.3403301

[1105] Floridi, L. (2018). Soft ethics, the governance of the digital and the General Data Protection Regulation. Philosophical Transactions of the Royal Society A: Mathematical, Physical and Engineering Sciences, 376(2133), 20180081

[1106] Brundage, M., Avin, S., Wang, J., Belfield, H., Krueger, G., Hadfield, G., ... & Maharaj, T. (2020). Toward trustworthy AI development: mechanisms for supporting verifiable claims. arXiv preprint arXiv:2004.07213

[1107] Zhang, Y., Liao, Q.V. and Bellamy, R.K.E. (2020). Effect of confidence and explanation on accuracy and trust calibration in AI-assisted decision making. Proceedings of the 2020 Conference on Fairness, Accountability, and Transparency. Available at: https://arxiv.org/pdf/2001.02114.pdf; Kazim, Emre and Koshiyama, Adriano, A High-Level Overview of AI Ethics (May 24, 2020). Available at SSRN: https://ssrn.com/abstract=3609292 or http://dx.doi.org/10.2139/ssrn.3609292; and Yu, H., Shen, Z., Miao, C., Leung, C., Lesser, V.R. and Yang, Q., 2018. Building ethics into artificial intelligence. arXiv preprint arXiv:1812.02953

[1108] Koshiyama, A., Kazim, E., Treleaven, P., Rai, P., Szpruch, L., Pavey, G., Ahamat, G., Leutner, F., Goebel, R., Knight, A., Adams, J., Hitrova, C., Barnett, J., Nachev, P., Barber, D., Chamorro-Premuzic, T., Klemmer, K., Gregorovic, M., Khan, S. and Lomas, E. (2021). Towards Algorithm Auditing: A Survey on Managing Legal, Ethical and Technological Risks of AI, ML and Associated Algorithms. [online] papers.ssrn.com. Available at: https://papers.ssrn.com/sol3/papers.cfm?abstract_id=3778998.

With the influx of large amounts of data due to online learning during the pandemic, we will witness an increasing number of AI powered ed-tech products.

There are concerns that ethics in AIEd is not a priority for most EdTech companies, or even, schools. There is a lack of awareness of relevant stakeholders regarding where AIEd can go wrong.

AIEd wrongly predicting that a particular student will not perform very well in end of year exams or might drop out next year can play a very important role in determining that student's reputation in front of teachers and parents. This reputation will determine how these teachers and parents treat the student, resulting in a huge psychological impact and even more, including lost opportunities, based on this wrong description by an AI tool. We discussed above, the high-profile case in the UK where the grading AI system was shown to be biased against students from poorer backgrounds.

There are important AIEd ethics developments. For example, Professor Rose Luckin, professor of learner centered design at University College London along with Sir Anthony Seldon, vice chancellor of the University of Buckingham and Priya Lakhani, founder and CEO of Century Tech founded the Institute of Ethical AI in Education (IEAIEd)[1109] to create awareness and promote the ethical aspects of AI in education. In its interim report, the institute identified seven different requirements for ethical AI to mitigate any kind of risks for students. This included human agency and oversight to double-check AI's performance; technical robustness and safety to prevent AI going wrong with new data or being hacked; diversity to ensure similar distribution of different demographics in data and avoid bias; nondiscrimination and fairness to prevent anyone from being unfairly treated by AI; privacy and data governance to ensure everyone has the right to control their data; transparency to enhance the understanding of AI products; societal and environmental well-being to ensure that AI is not causing any harm and accountability to ensure that someone takes the responsibility for any wrongdoings of AI. Recently, the institute has also published a framework[1110] for educators, schools and ed-tech companies to help them with the selection of ed-tech products with various ethical considerations in mind, like ethical design, transparency, privacy etc.

With the focus on online learning during the pandemic, and more utilization of AI powered ed-tech tools, risks of AI going awry increased significantly for all the stakeholders including EdTech companies, schools, teachers and students. A lot more work needs to be done on ethical AI in learning contexts to mitigate these risks, including assessments balancing AIEd risks and opportunities.

**Moving Forward with AIEd**

With the focus on online education due to COVID19 in the past year, it will be interesting to see what AI has to offer for education with vast amounts of data being collected online through Learning Management Systems (LMS) and Massive Online Open Courses (MOOCS).

With the influx of new educational data, AI techniques such as reinforcement learning will be utilized to empower EdTech. Such algorithms perform best with the large amounts of data that was limited to very few EdTech companies in 2021. These algorithms have achieved breakthrough performance in multiple domains including games[1111] healthcare[1112] and robotics.[1113] This presents a great opportunity for AI's applications in

---

[1109] University of Buckingham. (n.d.). The Institute for Ethical AI in Education. Available at: https://www.buckingham.ac.uk/research-the-institute-for-ethical-ai-ineducation/

[1110] The Institute for Ethical AI in Education The Ethical Framework for AI in Education (IEAIED). 2021. [online]. Available at: https://fb77c667c4d6e21c1e06.b-cdn.net/wp#content/uploads/2021/03/The-Ethical-Framework-for-AI-in-Education-Institute-for#Ethical-AI-in-Education-Final-Report.pdf

[1111] Silver, D., Huang, A., Maddison, C.J., Guez, A., Sifre, L., Van Den Driessche, G.,Schrittwieser, J., Antonoglou, I., Panneershelvam, V., Lanctot, M. and Dieleman, S., 2016. Mastering the game of Go with deep neural networks and tree search. nature, 529(7587), pp.484-489.

[1112] Callaway, E. (2020). "It will change everything": DeepMind's AI makes gigantic leap in solving protein structures. Nature. Available at: https://www.nature.com/articles/d41586-020-03348-4.

[1113] Kober, J., Bagnell, J.A. and Peters, J., 2013. Reinforcement learning in robotics: A survey. The International Journal of Robotics Research, 32(11), pp.1238-1274

education for further enhancing student's learning outcomes, reducing teachers' workloads and making learning interactive and fun for teachers and students. With a growing number of AI powered EdTech products in future, there will also be a lot of research on ethical AIEd. Thus, more work will be done to ensure robust and safe AI products for all the stakeholders.

EdTech companies can begin by sharing detailed guidelines for using AI powered ed-tech products, particularly specifying when not to rely on them. This includes the detailed documentation of the entire machine learning development pipeline with the assumptions made, data processing approaches used, and the processes followed, for selecting machine learning models.

Regulators will play a very important role in ensuring that certain ethical principles are followed in developing these AI products or there are certain minimum performance thresholds that these products achieve.[1114]

The goal of AIEd is not to promote AI, but to support education. Cutting edge AI by researchers and companies around the world is not of much use if it is not helping students learn. With the recent developments in AI, particularly reinforcement learning techniques, the future holds exciting possibilities of where AI will take education. For impactful AI in education, students and teachers always need to be at the epicenter of AI development.[1115]

A 2016 study, conducted on behalf of the European Parliament, concludes that AI applications will be used in almost all fields of our daily lives.[1116] The recent developments and future promises of AI technologies provide myriad benefits that span across a multitude of interested parties, industries, and sectors. The lofty future that AI could provide has been recognized by businesses, governments, and individuals, and with good reason. As noted by the European Union's Independent High-Level Expert Group (HLEG) on AI, "AI is not an end in itself, but rather a promising means to increase human flourishing, thereby enhancing individual and societal well-being and the common good, as well as bringing progress and innovation."[1117]

Safety, as it relates to AI and related technologies, "ought not to be confined to physical safety but should extend to concern for nonphysical harm, such as privacy, security, and the dehumanization of care for people at their most vulnerable."[1118] Finding ways to navigate both the physical and nonphysical challenges presented by AI will be essential to building trust and fostering its development. An additional element that deserves additional attention are related cybersecurity concerns, which manifest themselves quite differently from cyber attacks (fed, for example, by bugs in code) with AI attacks taking the form of pattern manipulation and poisoning along with "inherent limitations in the underlying AI algorithms that currently cannot be fixed."[1119]

Of growing significance along the AI technological issues are those of ethics. AI is ideological.[1120] The concern about AI is not that it won't deliver on the promise held forth by its advocates but, rather, that it will, but

---

[1114] Kazim, E., Denny, D. M. T., & Koshiyama, A. (2021). AI auditing and impact assessment: according to the UK information commissioner's office. AI and Ethics, 1-10.

[1115] Chaudhry, Muhammad and Kazim, Emre, Artificial Intelligence in Education (Aied) a High-Level Academic and Industry Note 2021 (April 24, 2021). Available at SSRN: https://ssrn.com/abstract=3833583 or http://dx.doi.org/10.2139/ssrn.3833583

[1116] Including applications for disabled people and the daily life of elderly people, healthcare, agriculture and food supply, manufacturing, energy and critical infrastructure, logistics and transport as well as security and safety. EUROPEAN PARLIAMENTARY RESEARCH SERV.: SCI. FORESIGHT UNIT, ETHICAL ASPECTS OF CYBER-PHYSICAL SYSTEMS 36 (2016), at 9. [herinafter EPRS] For more information concerning the increasing relevance of AI applications, see Commission Communication on Artificial Intelligence for Europe, COM (2018) 237 final (Apr. 25, 2018) [hereinafter Artificial Intelligence for Europe].

[1117] Ethics Guidelines for Trustworthy AI, INDEPENDENT HIGH-LEVEL EXPERT GP. ON ARTIFICIAL INTELLIGENCE (Apr. 2019), at 4 [hereinafter referred to as HLEG AI Ethics Guidelines].

[1118] Michael Guihot, Anne F. Matthew, & Nicolas Suzor, Nudging Robots: Innovative Solutions to Regulate Artificial Intelligence, 20 VAND. J. ENT. & TECH. L. 385, 407 (2017).

[1119] MARCUS COMITER, ATTACKING ARTIFICIAL INTELLIGENCE: AI'S SECURITY VULNERABILITY AND WHAT POLICYMAKERS CAN DO ABOUT IT 1, 80 (Aug. 2019), https://www.belfercenter.org/sites/default/files/2019-08/AttackingAI/AttackingAI.pdf.

without due consideration of ethical implications. There are assumptions embedded in the algorithms that will shape how education is realized, and if students do not fit that conceptual model, they will find themselves outside of the area where a human could apply human wisdom to alter or intervene an unjust outcome. Perhaps one of the greatest contributions of AI will be to make us understand how important human wisdom truly is in education and everywhere else.[1121]

**Corporations' and Governments' Role in Mitigating Harmful Impacts of AI on Children**

Microsoft and Google have both established principles for the ethical use of AI.[1122] However, neither has public-facing policies specific to AI and children.[1123] Several technology centers, trade associations, and computer science groups have also drafted ethical principles with regard to AI.[1124] However, most have excluded explicit reference to child rights, or discussion of the risks to children on AI-incorporating technologies.[1125]

Like corporations, governments around the world have adopted strategies for becoming leaders in the development and use of AI, fostering environments congenial to innovators and corporations.[1126] However, in most cases, policymakers have not directly addressed how the rights of children fit into their national strategy.[1127] While France's strategy deals with the AI-related issues of achieving gender equality and implementing digital literacy through education, the broader scope of impact on children is missing.[1128] An example of a country that has taken a more proactive look at the potential benefits of AI for children is India, whose AI initiative focuses on using AI in education, such as creating adaptive learning tools for customized learning, integrating intelligent and interactive tutoring systems, adding predictive tools to inform preemptive action for students predicted to drop out of school, and developing automated rationalization of teachers and customized professional development courses.[1129] AI technologies should obviously be deployed in locating missing or exploited children, and used in other ways to protect children.

Ultimately, both corporations and governments should think through how their AI systems and strategies can be strengthened to maximize the benefits and minimize the harms of AI for children today, and in the future. The role of artificial intelligence in children's lives—from how children play, to how they are educated, to how they consume information and learn about the world—is expected to increase exponentially over the coming years. Thus, it's imperative that stakeholders come together now to evaluate the risks of using AI technologies and assess opportunities to use artificial intelligence to maximize children's well being in a thoughtful and systematic manner. As part of this assessment, stakeholders should work together to map the potential positive and negative uses of AI on children's lives, and develop a child rights-based framework for artificial intelligence that delineates rights and corresponding duties for governments, educators, developers, corporations, parents, and children around the world.

---

[1120] Audrey Watters, "AI Is Ideological," New Internationalist, November 1, 2017. https://newint.org/features/2017/11/01/audrey-watters-ai

[1121] Audrey Watters, "AI Is Ideological," New Internationalist, November 1, 2017.

[1122] "Microsoft Salient Human Rights Issues," Report -FY17, Microsoft. file:///Users/dreatrew/Downloads/Microsoft_Salient_Human_Rights_Issues_Report-FY17.pdf; Google, "Responsible Development of AI" (2018).

[1123] Microsoft, "The Future Computed: Artificial Intelligence and Its Role in Society" (2018).

[1124] Alexa Hern, "Partnership on AI" Formed by Google, Facebook, Amazon, IBM and Microsoft," The Guardian, (September 28, 2016).

[1125] John Gerard Ruggie, "Global Governance and New Governance Theory," Lessons from Business and Human Rights, Global Governance 20, http://journals.rienner.com/doi/pdf/10.5555/1075-2846-20.1.5, (2014), 5.

[1126] Council of Europe, "Recommendation CM/REC (2018)7 of the Committee of Ministers to member States on Guidelines to respect, protect and fulfil the rights of the child in the digital environment," (July 4th 2018).

[1127] Cedric Villani, "For a Meaningful Artificial Intelligence Towards a French and European Strategy," (March 8th 2018).

[1128] Id.

[1129] NITI Aayog, "Discussion paper: National Strategy for Artificial Intelligence," http://niti.gov.in/writereaddata/files/ document_publication/NationalStrategy-for-AI-DiscussionPaper.pdf, (June 2018).

**Recommendations from UNICEF on Deploying AI with children**[1130]

**Corporations**

Incorporate an inclusive design approach when developing child-facing products, which maximizes gender, geographic and cultural diversity, and includes a broad range of stakeholders, such as parents, teachers, child psychologists, and—where appropriate—children themselves.

Adopt a multi-disciplinary approach when developing technologies that affect children, and consult with civil society, including academia, to identify the potential impacts of these technologies on the rights of a diverse range of potential end-users.

Implement safety by design and privacy by design for products and services addressed to or commonly used by children.

Develop plans for handling especially sensitive data, including revelations of abuse or other harm that may be shared with the company through its products.

**Educators**

Be aware of and consider using artificial intelligence-based tools that may enhance learning for students, such as specialized products that can assist non-traditional learners and children with special needs.

Avoid the overuse of facial and behavioral recognition technologies, including for security purposes, in ways that may constrain learning and appropriate risk taking.

**Governments**

Set up awareness campaigns that help parents understand the importance of privacy for their children. Parents should be aware of how their children's data is being used and processed for diverse purposes, including for targeted ad campaigns or non-educative social media recommendations. They should also be aware of the impacts of posting pictures or other information about their children to social media, and the ways that what they post can have a dramatic impact on their children's future.

Adopt a clear, comprehensive framework for corporations that imposes a duty of care connected to the handling of children's data, and provides an effective remedy (judicial, administrative or other) for breach. This framework should incorporate human rights principles.

Establish a comprehensive national approach to the development of artificial intelligence that pays specific attention to the needs of children as rights-bearers and integrates children into national policy plans.

**Parents**

Carefully review and consider avoiding the purchase and use of products that do not have clear policies on data protection, security, and other issues that impact children.

Incorporate children into the decision-making process about how their data will be used, including whether to post their information to social media sites and whether to engage smart toys, helping children understand the potential short and long-term impacts of that use.

---

[1130] https://www.unicef.org/innovation/media/10726/file/Executive%20Summary:%20Memorandum%20on%20Artificial%20Intelligence%20and%20Child%20Rights.pdf

Identify how schools might be using artificial intelligence-based technologies to assist or surveil children, and raise concerns if some of the policies or procedures are unclear or seem inappropriate—for example, by disincentivizing creativity and exploration.

Encourage the use of artificial intelligence-based technologies when they seem likely to enhance learning and that positive benefit has been confirmed by peer-reviewed research-and-analysis.

**CONCLUSION**

The current structures in place for AI governance fall short of facilitating sufficient accountability. As AI systems are increasingly shaping our world, as well as our access to and exclusion from opportunities and resources, it is essential to ensure better AI oversight. Meaningful and inclusive oversight that will help to maintain the rule of law, to protect individual rights, and to ensure the protection of core democratic values.

Technology giants, all of whom are heavily investing in and profiting from AI, must not dominate the public discourse on responsible use of AI, we all need to shape the future of our core values and democratic institutions.

As artificial intelligence continues to find its way into our daily lives, its propensity to interfere with our rights only gets more severe. Many of the issues mentioned in this examinations of harmful AI are not new, but they are greatly exacerbated and threatened by the scale, proliferation, and real-life impact that artificial intelligence facilitates. The potential of artificial intelligence to both help and harm people is much greater than earlier technologies. Starting now to examine what safeguards and structures can address AI's problems and harms, including those that disproportionately impact marginalized people.

**APPENDICES**

The legal and regulatory landscape for AI and ML systems is changing rapidly. The list of resources below from the Future of Privacy Forum[1131] reflects the leading thinking from academics, regulatory agencies, and on-going projects and studies to provide the best guidance to commercial and public entities on implementing AI into their products and services.

The The Oxford Handbook of Ethics of AI: An Annotated Bibliography is also highly recommended.[1132]

- I. General AI & Ethics Resources
  - Existing Company and Government Models or Recommended Best Practices
  - News, Reports, and Other Media
  - International Resources
- II. Resources for AI Ethical Review Process

# I. General AI & Ethics Resources

## Existing Company and Government Models or Recommended Best Practices

**Intel,** *Artificial Intelligence: The Public Policy Opportunity* - Intel's public policy recommendations to foster an environment conducive to AI innovation, while mitigating the unintended societal consequences.

**Google,** *AI at Google: Our Principles* - Google's AI principles.

**Microsoft,** *Microsoft AI Principles* - Microsoft's AI Principles.

**DeepMind,** *Ethics and Society Principles* - DeepMind's AI Principles.

**Facebook,** *AI at F8 2018* - An outline of Facebook's vision for AI Development.

---

[1131] https://sites.google.com/fpf.org/futureofprivacyforumresources/ethics-governance-and-compliance-resources#h.p_EpNP3043ntxZ

[1132] https://c4ejournal.net/the-oxford-handbook-of-ethics-of-ai-an-annotated-bibliography/

**SIIA,** *Ethical Principles for Artificial Intelligence and Data Analytics* - The Software & Information Industry Association's ethical principles for AI and data analytics.

**Public Voice,** *Universal Guidelines for Artificial Intelligence* - Guidelines for Artificial Intelligence set up by The Public Voice, a coalition which was established in 1996 by the Electronic Privacy Information Center (EPIC) to promote public participation in decisions concerning the future of the Internet.

**IEEE,** *The Ethics Certification Program for Autonomous and Intelligent Systems (ECPAIS)* - A proposed certification system to create specifications marking processes that advance transparency, accountability and reduction in algorithmic bias in Autonomous and Intelligent Systems

**The AI Policy Landscape -** A general compendium of multiple AI commentary and resources. Original Medium post and discussion here and continuously updated Google doc version with table of contents here

**Alisomar,** *Alisomar AI Principles* - A further explanation on these principles can be found here.

**Integrate.ai,** *Responsible AI in Consumer Enterprise* - Integrate.AI's framework for businesses to use consumer data responsibly.

**Google,** *Perspectives on Issues in AI Governance* - This white paper calls for government regulation in the field of AI, suggesting specific areas to be considered.

**NEC,** *NEC Unveils "NEC Group AI and Human Rights Principles"* - NEC's principles to prioritize privacy and human rights in relation to the development of AI.

**OECD,** *OECD Principles on AI* - A list of principles on AI established by the the Organisation for Economic Co-operation and Development.

**Office of Artificial Intelligence (UK)**, *Draft Guidelines for AI Procurement* - Intended to be a working document drafted in collaboration with the World Economic Forum Centre for the Fourth Industrial Revolution

**AI Global**, *Creating A Responsible Trust Index: A Unified Assessment to Assure the Responsible Design, Development, and Deployment of AI* - A unified framework focusing on: Accountability; Explainability and Interpretability; Data Quality; Bias and Fairness; Robustness

**Springer,** *AI & Ethics* - A publication focusing on the informed debate and discussion of the ethical, regulatory, and policy implications that arise from the development of AI

## News, Reports, and Other Media

**Algorithmwatch,** *In the Realm of Paper Tigers - Exploring the Failings of AI Ethics Guidelines* - Assessment and directory of 160 AI ethics guidelines.

**AI 4 People,** *AI 4 People's Ethical Framework for a Good AI Society: Opportunities, Risks, Principles, and Recommendations* - This document outlines the risks and opportunities of AI, propose ethical principles of AI, and offers recommendations for a Good AI Society.

**The Stanford AI Lab Blog, Shushman Choudhury, et al.,** *In Favor of Developing Ethical Best Practices in AI Research* - This blog post calls for AI Researchers to consider the ethics of their work and to create a system of ethical best practices.

**Towards Data Science,** *A Gentle Introduction to the Discussion on Algorithmic Fairness* - An overview of the problems that arise from algorithmic decisionmaking.

**Nasdaq,** *How Artificial Intelligence Can Influence Governance, Risk, and Compliance* - A consideration of the ways in which AI can aid governance, risk, and compliance (GRC) activities.

**Medium,** *AI and the Future of Ethics* - A basic overview of AI and discussion about the ethics related to it.

**Tutorial,** *21 Fairness Definitions and Their Politics* - A filmed lecture discussing the various definitions of fairness as they pertain to statistical models.

**Google Research,** *Attacking discrimination with smarter machine learning* - Case study of a loan application scenario with a demo of how to confront bias

**Medium,** *AI and the Future of Ethics* - A basic overview of AI and discussion about the ethics related to it.

**Algorithmic Justice League,** *Study finds gender and skin-type bias in commercial AI systems.* Site for a collective that aims to increase awareness and report bias. Ted talk *here*.

**CFA Institute,** *Artificial Intelligence: The Next Step in Corporate Governance* -

**University of Toronto Centre for Ethics,** *The Ethics of Agonistic Machine Learning*

**Medium,** *Toward Ethical, Transparent And Fair AI/Ml: A Critical Reading List*

**Clifford Rossi,** *A Risk Professional's Survival Guide: Applied Best Practices in Risk Management*

**IAEE,** *Classical Ethics in A/IS*

**AI Now,** *Algorithmic Impact Assessments: A Practical Framework for Public Agency Accountability* - A policy paper providing public agencies a practical framework to assess automated decision systems and to ensure accountability.

**AI Now Institute,** *Algorithmic Impact Assessments: A Practical Framework For Public Agency Accountability*

**Brookings,** *The Role of Corporations in Addressing AI's Ethical Dilemmas* - Darrell West discusses 5 AI ethical dilemmas and how corporations are addressing them.

**Future Advocacy,** *Ethical, Social and Political Challenges of Artificial Intelligence in Health, April 2018* - A review of existing literature and interviews with global experts to understand how AI is being used (or could be used) in healthcare and what challenges these uses present.

**IEEE,** *Ethics in Action*

*The Ethics and Governance of AI Initiative*

**Future of Humanity Institute,** *Governance of AI Program*

**MIT Media Lab,** *Ethics and Governance of Artificial Intelligence*

**Berkman Klein Center,** *Ethics and Governance of Artificial Intelligence*

**The GovLab,** *Artificial Intelligence and Public Policy*

**MILA,** *Official Launch of the Montreal Declaration for Responsible Development of Artificial Intelligence*

**AI Now,** *AI Now 2017 Report*

**American Bar Association,** *A 'Principled' Artificial Intelligence could improve justice*

**American Action Forum,** *Primer: How to Understand and Approach AI Regulation*

**American Action Forum,** *[Approaches to Regulating Technology: From Privacy to A.I.](#)*

**Markkula Center for Applied Ethics,** *[Ethics in Technology Practice](#)*

**Alan Winfield**, *[An Updated Round Up of Ethical Principles of Robots and AI](#)* - A list of numerous attempts at outlining ethical principles for Artificial Intelligence, beginning with Isaac Asimov's 1950 laws of Robotics. (Most of these links should already be captured in our wiki for Ethics or Education.)

**Markkula Center for Applied Ethics**, *[Readings in AI Ethics](#)* - A compilation of readings on the ethics of AI.

**Brookings,** [Algorithmic Bias Detection and Mitigation: Best Practices and Policies to Reduce Consumer Harms](#) - A report that gives examples of, considers the harms of, offers detection methods for, and proposes solutions to mitigating the harms of algorithmic bias.

**AI Now Institute,** *[Discriminating Systems: Gender, Race, and Power in AI](#)* - A report that outlines the causes of and the issues resulting from the lack of diversity in the AI sector. It makes recommendations to help better the problems.

**Benedict Evans,** *[Notes on AI Bias](#)* - This blog post argues that while AI bias is an issue, the problem of bias is not new, and is not rooted in problems with machines, but instead with humans.

**[Ethical Resolve](#),** Provides blog posts, talks, and resources for businesses concerned with implementing responsible AI and Ethics in Design.

**Nature Machine Intelligence**, *[The Global Landscape of AI Ethics Guidelines](#)* - Maps and analyzes the current corpus of AI ethics guidelines.

**Partnership on AI,** [Report on Algorithmic Risk Assessment Tools in the U.S. Criminal Justice System](#) - This report documents the serious shortcomings of risk assessment tools in the U.S. criminal justice system, most particularly in the context of pretrial definitions.

**Access Now**, [Human Rights in the Age of Artificial Intelligence](#) - Provides a jumping off point for further conversation and research in this developing space.

**Access Now**, [Human Rights Matter in the AI Debate. Let's Make Sure AI Does Us More Good Than Harm](#)

**Access Now**, [Laying Down the Law on AI: Ethics Done, Now the EU Must Focus on Human Rights](#)

**AI Now Institute**, Algorithmic Accountability Policy Toolkit - This toolkit is intended to provide legal and policy advocates with a basic understanding of government use of algorithms including, a breakdown of key concepts and questions that may come up when engaging with this issue, an overview of existing research, and summaries of algorithmic systems currently used in government.

**Data & Society**, Algorithmic Accountability: A Primer - The primer explores the trade-offs debates about algorithms and accountability across several key ethical dimensions, including fairness and bias; opacity and transparency; and lack of standards for auditing.

**Data & Society**, Governing Artificial Intelligence: Upholding Human Rights & Dignity - This report shows how human rights can serve as a "North Star" to guide the development and governance of artificial intelligence.

**Karen Hao**, This is How AI Bias Really Happens—and Why It's So Hard to Fix - Bias can creep in at many stages of the deep-learning process, and the standard practices in computer science aren't designed to detect it.

**UNICEF & Human Rights Center**, Executive Summary: Artificial Intelligence and Children's Rights - This memo outlines a series of case studies to illustrate the various ways that artificial intelligence-based technologies are beginning to positively and negatively impact children's human rights.

**FAT ML**, Principles for Accountable Algorithms and a Social Impact Statement for Algorithms - The goal of this document is to help developers and product managers design and implement algorithmic systems in publicly accountable ways.

# International Resources

**ICO**, Draft Guidance for Consultation, Guidance on the AI Auditing Framework - advice on how to understand data protection law in relation to artificial intelligence (AI) and recommendations for organisational and technical measures to mitigate the risks AI poses to individuals.

**European Commission**, White Paper On Artificial Intelligence - A European Approach to Excellence and Trust - A European approach to artificial intelligence building off the European strategy for AI presented in April 2018.

**European Union Agency for Fundamental Rights**, *Data Quality and Artificial Intelligence— Mitigating Bias and Error to Protect Fundamental Rights*

**Norwegian Data Protection Authority**, *Data Protection by Design and by Default*

**Berkeley Technology Law Journal**, *Rethinking Explainable Machines: The GDPR's 'Right to Explanation' Debate and the Rise of Algorithmic Audits in Enterprise*

**World Economic Forum**, *Top 9 Ethical Issues In Artificial Intelligence*

**Office of the Victorian Information Commissioner**, *Artificial Intelligence and Privacy*

**European Agency for Fundamental Rights**, *#Big Data: Discrimination in Data-Supported Decision Making*

**United Nations University, Centre for Policy Research**, *The Ethical Anatomy of AI*

**Dutch Alliance for Artificial Intelligence**

**Australian Institute of Company Directors**, *Preparing Directors For Artificial Intelligence Whirlwind*

**Université de Montréal**, *Montréal Declaration of Responsible AI*

**House of Lords, Select Committee on Artificial Intelligence**, *AI in the UK: ready, willing, and able?*

**European Group on Ethics in Science and New Technologies**, *Statement on Artificial Intelligence, Robotics and 'Autonomous' Systems*

**Cédric Villani**, *For a Meaningful Artificial Intelligence: Towards a French and European Strategy*

**European Commission for the Efficiency of Justice (CEPEJ)**, *European ethical Charter on the use of Artificial Intelligence in judicial systems and their environment*

**European Commission High-Level Expert Group on Artificial Intelligence (AI HLEG)**, *Draft Ethics guidelines for trustworthy AI*

**Automating Society**, *Taking Stock of Automated Decision-Making in the EU*
159

**UK Information Commissioner's Office,** *Automated Decision Making: the role of meaningful human reviews*- This framework describes how humans can have 'meaningful' involvement in AI decision making.

**Dr. Thilo Hagendorff**, *The Ethics of AI Ethics*- An evaluation of the AI guidelines that have been presented. An overview of the field of AI ethics.

**Government of Canada,** *Responsible use of artificial intelligence (AI)* - Guiding principles to ensure the use of ethical AI by the government.

**The White House**, *The Administration's Report on the Future of Artificial Intelligence* - A 2016 report focusing on the opportunities, considerations, and challenges of Artificial Intelligence.

**Office of the Privacy Commissioner of Canada**, *Consultation on the OPC's Proposals for Ensuring Appropriate Regulation of Artificial Intelligence* - proposals for regulating artificial intelligence.

# II. Resources for AI Ethical Review Process

**Future of Privacy Forum,** *Conference Proceedings - Beyond IRBs Designing Ethical Review Processes for Big Data Research* - 2015 conference proceedings aimed at identifying processes and commonly accepted ethical principles for data research in academia, government and industry.

**Northeastern University Ethics Institute**, *Building Data and AI Ethics Committees* - Describes components of a committee-based approach to data and AI ethics, while identifying questions for an organization to consider when developing ethics and oversight committees.

**Council of Europe,** *The Council of Europe Established an Ad Hoc Committee on Artificial Intelligence - CAHAI* - Committee to examine the feasibility and potential elements of a legal framework for the development of artificial intelligence.

**Berkman Klein Center**, *Principled Artificial Intelligence: Mapping Consensus in Ethical and Rights-Based Approaches to Principles for AI*- Provides a comparison between thirty-six prominent AI principles documents side-by-side.

**BNH.AI,** *Sample AI Incident Response Checklist* - A checklist for 7 Phases of AI incident response including: preparation; identification; containment; eradication; and recovery. Also provides additional compliance resources.

**Dallas Card & Noah A. Smith**, *On Consequentialism and Fairness* - A consequentialist critique of common definitions of fairness within machine learning, as well as a machine learning perceptive on consequentialism; concluding with a broader discussion of how issues of learning and randomization have important implications for the ethics of automated decision making systems.

# Academy of Medical Royal Colleges
[PDF: Artificial Intelligence in Healthcare](#)
United Kingdom | 2019 | academia
Recommendation

# Accenture
[PDF: Universal Principles of Data Ethics](#)
United States | 2016 | private sector
Recommendation

# Accenture UK
[Responsible AI and robotics. An ethical framework](#)
United Kingdom | 2018 | private sector
Recommendation

# ADEL
[ADEL](#)
France | 2018 | private sector
Binding agreement

# Advisory Board on Artificial Intelligence and Human Society
[PDF: Report on Artificial Intelligence and Human Society (Unofficial translation)](#)
Japan | 2017 | government
Recommendation

# Agenzia per l'Italia Digitale (AGID)
[L'intelligenzia artificiale al servizio del cittadino (Artificial Intelligence at the service of the citizen)](#)
Italy | 2018 | government
Recommendation

## AI Now Institut
[PDF: AI Now Report 2018](#)
United States | 2018 | academia
Recommendation

## American College of Radiology; European Society of Radiology; Radiology Society of North America; Society for Imaging Informatics in Medicine; European Society of Medical Imaging Informatics; Canadian Association of Radiologists; American Association of Physicists in Medicine
[Ethics of Artificial Intelligence in Radiology: Summary of the Joint European and North American Multisociety Statement](#)
International | 2019 | professional association
Recommendation

## American Medical Association (AMA)
[Policy Recommendations on Augmented Intelligence in Health Care H-480.940](#)
United States | 2018 | professional association
Recommendation

## Amnesty International/Access Now
[The Toronto Declaration](#)
United Kingdom | 2018 | civil society
Recommendation

## Aptiv, Audi, BMW, Daimler and other automotive companies
[PDF: Safety First for Automated Driving – Proposed technical standards for the development of Automated Driving](#)
International | 2019 | private sector
Voluntary commitment

## Association for Computing Machinery
[PDF: Statement on Algorithmic Transparency and Accountability](#)
United States | 2017 | industry association
Binding agreement
[Additional link](#)
[Additional link](#)

## Association for Computing Machinery - Future of Computing Machinery
[It's Time to Do Something: Mitigating the Negative Impacts of Computing Through a Change to the Peer Review Process](#)
United States | 2019 | industry association
Recommendation

## Atomium - EISMD (AI4Poeple)
[PDF: AI4People's Ethical Framework for a Good AI Society: Opportunities, Risks, Principles, and Recommendations](#)
European Union | 2018 | civil society
Recommendation

## Australian Government/ Department of industry, Innovation and Science
[PDF: Artificial Intelligence Australia's Ethics Framework A Discussion Paper](#)
Australia | 2019 | government
Recommendation

## Bejing Academy of Artificial Intelligence
[Bejing AI Principles](#)
China | 2019 | government
Recommendation

## Bertelsmann Stiftung / iRights.Lab
[Algo.Rules](#)
Germany | 2019 | civil society
Recommendation
[Additional link](#)

## Bitkom
[PDF: Leitlinien für Big Data Einsatz (Guidelines for the use of Big Data)](#)
Germany | 2015 | industry association
Recommendation

## Bitkom
[Empfehlungen für den verantwortlichen Einsatz von KI und automatisierten Entscheidungen (Recommendations for the responsible use of AI and automated decision making)](#)
Germany | 2018 | industry association
Recommendation

## Bundesministerium des Innern, für Bau und Heimat/ Datenethikkommission der Bundesregierung
[PDF: Gutachten der Datenethikkommission der Bundesregierung](#)
Germany | 2019 | government
Recommendation

## Bundesverband KI
[KIBV Gütesiegel (KIBV Quality seal)](#)
Germany | 2019 | industry association
Voluntary commitment

## Center for Democracy & technology (CDT)
[PDF: Digital Decisions](#)
United States | civil society
Recommendation
[Additional link](#)

## Chinese AI Alliance
[Joint Pledge on Artificial Intelligence Industry Self-Discipline (Draft for Comment)](#)
China | 2019 | other
Voluntary commitment
[Additional link](#)

## Chinese Government
[Governance Principles for a New Generation of Artificial Intelligence: Develop Responsible Artificial Intelligence](#)
China | 2019 | government
Recommendation
[Additional link](#)

## CIGI Centre for International Governance Innovation
[CIGI Paper No. 178: Toward a G20 Framework for Artificial Intelligence in the Workplace](#)
Canada | 2018 | civil society
Recommendation

## CIGREF
[PDF: Digital Ethics](#)
France | 2018 | industry association
Recommendation

## Commission de Surveillance du Secteur Financier
[PDF: Artificial Intelligence: opportunities, risks and recommendations for the financial sector](#)
Luxembourg | 2018 | government
Recommendation

## Council of Europe
[Artifical Intelligence and Data Protection](#)
European Union | 2018 | government
Recommendation

## Data & Society
[PDF: Governing Artificial Intelligence. Upholding Human Rights & Dignity](#)
United States | 2018 | civil society
Recommendation

## Data Ethics
[Data Ethics Principles](#)
Denmark | 2017 | civil society
Recommendation

## DataforGood
[Serment d'Hippocrate pour Data Scientist (Hippocratic Oath for Data Scientists)](#)
France | civil society
Voluntary commitment

## Datatilsynet The Norwegian Data Protection Authority
[PDF: Artificial intelligence and privacy](#)
Norway | 2018 | government
Recommendation

## Deep Mind
[Saftey and Ethics](#)
United States | private sector
Voluntary commitment

## Department of Health and Social Care
[Code of conduct for data-driven health and care technology](#)
United Kingdom | 2019 | government
Recommendation

### Deutsche Telekom
Guidelines for Artificial Intelligence
Germany | 2018 | private sector
Voluntary commitment

### DGB
Künstliche Intelligenz und die Arbeit von Morgen
Germany | 2019 | civil society
Recommendation

### Digital Catapult, Machine Intelligence Garage Ethics Committee
Ethics Framework -Responsible AI
United Kingdom | 2020 | private sector
Recommendation

### Dubai
PDF: Artificial Intelligence Ethics and Principles, and toolkit for implementation
United Arab Emirates | 2019 | government
Recommendation
Additional link

### Ekspertgruppen om Design: malenehald.dk DATAETIK (Danish Expert Group on Data Ethics)
PDF: Data for the Benefit of the People: Recommendations from the Danish Expert Group on Data Ethics
Denmark | 2018 | government
Recommendation

### Engineering and Physical Research Council
Principles of Robotics
United Kingdom | 2010 | government
Recommendation

### Ethikbeirat HR Tech (Ethics council HR Tech)
PDF: Richtlinien für den verantwortungsvollen Einsatz von Künstlicher Intelligenz und weiteren digitalen Technologien in der Personalarbeit (Guidelines for the responsible use of artificial intelligence and other digital technologies in human resources): Consultation document
Germany | 2019 | private sector
Voluntary commitment

### Ethkikkommission BuMi Verkehr und digitale infrastruktur
[PDF: Automatisiertes und Vernetztes Fahren / Automated and connected automated driving](#)
Germany | 2017 | government
Recommendation

### European Commision For the Efficiency of Justice
[European ethical Charter on the use of Artificial Intelligence in judicial systems and their environment](#)
International | 2018 | government
Recommendation

### European Commission
[Code of Practice on Disinformation](#)
European Union | 2018 | government
Recommendation

### European Group on Ethics in Science and New Technologies
[PDF: Statement on Artificial Intelligence, Robotics and 'Autonomous' Systems](#)
European Union | 2018 | government
Recommendation

### European Parliament
[Report with recommendations to the Commission on Civil Law Rules on Robotics](#)
European Union | 2017 | government
Recommendation

### Executive Office of the President; National Science and Technology Council; Committee on Technology
[PDF: Preparing for the future of Artificial Intelligence](#)
United States | 2016 | government
Recommendation

### Faculty of Informatics, TU Wien
[Vienna Manifesto on Digital Humanism](#)
Austria | 2019 | academia
Voluntary commitment

### FAT/ML
[Principles for Accountable Algorithms and a Social Impact Statement for Algorithms](#)
International | civil society
Recommendation

## Fraunhofer Institute for Intelligent Analysis and Information Systems IAIS
[PDF: Trustworthy Use of Artifical Intelligence](#)
Germany | 2020 | academia
Recommendation

## French Data Protection Authority (CNIL)
[PDF: How can humans keep the upper hand? Report on the ethical matters raised by AI algorithms](#)
France | 2017 | government
Recommendation

## French National Ethical Consultative Committee for Life Sciences and Health (CCNE
[PDF: Digital Technology and Healthcare. Which Ethical Issues for which Regulations?](#)
France | 2014 | government
Recommendation

## French Strategy for Artifical Intelligence
[For a meaningful Artificial Intelligence. Towards a French and European strategy](#)
France | 2018 | government
Recommendation

## Future Advocacy
[PDF: Ethical, social, and political challenges of Artificial Intelligence in Health](#)
United Kingdom | 2018 | civil society
Recommendation

## Future of Life Institute
[Asilomar AI Principles](#)
United States | 2017 | civil society
Voluntary commitment

## Future of Privacy Forum
[PDF: Unfairness by algorithm: Distilling the Harms of automated decision making](#)
United States | 2017 | civil society
Recommendation

## G20
[PDF: Principles for responsible stewardship of trustworthy AI](#)
International | 2019 | intergovernmental organisation
Voluntary commitment

### Gesellschaft für Informatik (German Society of Informatics)
[Ethische Leitlinien (Ethical Guidelines)](#)
Germany | 2018 | professional association
Voluntary commitment

### Google
[People & AI Partnership Guidebook](#)
United States | private sector
Recommendation

### Google
[Responsible AI Practice](#)
United States | private sector
Recommendation

### Google
[Advanced Technology External Advisory Council for Google (ATEAC)](#)
United States | 2019 | private sector
Binding agreement
[Additional link](#)

### Google
[Objectives for AI Applications](#)
United States | 2018 | private sector
Voluntary commitment
[Additional link](#)

### Government of Canada
[Directive on Automated Decision-Making](#)
Canada | 2019 | government
Binding agreement

### Government of Canada
[Responsible use of artificial intelligence (AI)](#)
Canada | 2019 | government
Voluntary commitment

### Government of Canada
[Responsible Artificial Intelligence in the Government of Canada (whitepaper)](#)
Canada | 2019 | government
Recommendation

### Handelsblatt Research Institute
[PDF: Datenschutz und Big Data / Data protection and Big Data](#)
Germany | other
Recommendation

### High Level Expert Group on AI (European Commission)
[Draft Guidelines for Trustworthy AI](#)
European Union | 2019 | government
Recommendation

### Hochschule der Medien
[10 ethische Leitlinien für die Digitalisierung von Unternehmen (10 ethical guidelines for the digitalisation of companies)](#)
Germany | 2017 | academia
Recommendation

### IA Latam
[Declaración de Ética para desarrollo y uso de la Inteligencia Artificial/ Declaration of Ethics for the Development and Use of Artificial Intelligence](#)
International | 2019 | private sector
Voluntary commitment

### IBM
[IBM's Principles for Trust and Transparency](#)
United States | 2018 | private sector
Voluntary commitment
[Additional link](#)

### IBM
[Trusted AI](#)
United States | private sector
Recommendation

### IBM
[PDF: Everyday Ethics for Artificial Intelligence](#)
United States | private sector
Recommendation
[Additional link](#)

### Icelandic Institute for Intelligent Machines
[IIIM's Ethics Policy](#)
Iceland | civil society
Voluntary commitment

### IEEE
[PDF: Ethically Aligned Design 2](#)
International | 2019 | professional association
Recommendation

### IEEE
[Ethics in Action – Set the Global Standards](#)
International | professional association
Recommendation

### IEEE
[Ethically Aligned Design](#)
United States | 2019 | professional association
Recommendation

### Information Commissioner's Office
[PDF: Big data, artificial intelligence, machine learning and data protection](#)
United Kingdom | 2017 | government
Recommendation

### Information Technolgy Industry Council
[PDF: AI policies and principles](#)
United States | 2017 | industry association
Voluntary commitment

### Institute for Business Ethics
[PDF: Business Ethics and Artificial Intelligence](#)
United Kingdom | 2018 | other
Recommendation

### Institute for Information and Communications Policy (IICP), The Conference toward AI Network Society
[PDF: Draft AI R&D Guidelines for International Discussions](#)
Japan | 2017 | government
Recommendation

### Intel Corporation
[PDF: Intel's AI Privacy Policy White Paper. Protecting individuals' privacy and data in the artificial intelligence world](#)
United States | 2018 | private sector
Recommendation

### Intel Corporation
[PDF: Artificial Intelligence. The Public Policy Opportunity](#)
United States | 2017 | private sector
Recommendation

### International Conference of Data Protection and Privacy Commissioners
[PDF: DECLARATION ON ETHICS AND DATA PROTECTION IN ARTIFICAL INTELLIGENCE](#)
International | 2018 | professional association
Voluntary commitment

### Internet Society
[Artifical Intelligence and Machine Learning Policy Paper](#)
United States | 2017 | civil society
Recommendation

### ITechLaw
[Responsible AI: Global Policy Framework](#)
United States | 2019 | professional association
Recommendation

### Japanese Society for AI
[PDF: The Japanese Society for Artificial Intelligence Ethical Guidelines](#)
Japan | 2017 | academia
Voluntary commitment

### Kakao Corp
[Kakao Algorithm Ethics](#)
South Korea | private sector
Voluntary commitment

### Konferenz der unabhängigen Datenschutzaufsichtsbehörden des Bundes und der Länder (Conference of the independent data protection supervisory authorites in Germany)
[PDF: Hambacher Erklärung zur Künstlichen Intelligenz – Sieben datenschutzrechtliche Anforderungen (Hambach Declaration on Artificial Intelligence – seven requirements for data protection)](#)
Germany | 2019 | other
Voluntary commitment

## Korean Ministry of Science, ICT and Future Planning (MSIP)
[Mid- to Long-Term Master Plan in Preparation for the Intelligent Information Society Managing the Fourth Industrial Revolution](#)
South Korea | 2016 | government
Voluntary commitment

## Leaders of the G7
[PDF: Charlevoix Common Vision for the Future of Artificial Intelligence](#)
International | 2018 | government
Voluntary commitment

## Machine Intelligence Research Institute
[PDF: The Ethics of Artifical Intelligence](#)
United States | academia
Recommendation

## Massachusetts Institute of Technology
[PDF: MIT Schwarzman College of Computing Task Force Working Group on Social Implications and Responsibilities of Computing Final Report](#)
United States | 2019 | academia
Recommendation

## Microsoft
[PDF: Responsible bots: 10 guidelines for developers of conversational AI](#)
United States | 2018 | private sector
Voluntary commitment

## Microsoft
[Facial Recognition Principles](#)
United States | 2018 | private sector
Voluntary commitment

## Microsoft
[PDF: The Future Computed – Artificial intelligence and its role in society](#)
United States | 2019 | private sector
Recommendation
[Additional link](#)

## Microsoft
[Our Approach to AI](#)
United States | private sector
Voluntary commitment

### Mission Villani
[PDF: For a meaningful Artificial Intelligence. Towards a French and European strategy](#)
France | 2018 | government
Recommendation

### Monetary Authority of Singapore
[PDF: Principlesto Promote Fairness, Ethics, Accountability and Transparency (FEAT) in the Use of Artificial Intelligence and Data Analytics in Singapore'sFinancial Sector](#)
Singapore | 2018 | government
Recommendation

### Mozilla Foundation
[Effective Ad Archives](#)
United States | 2019 | civil society
Recommendation

### National Institution for Transforming India (Niti Aayog)
[PDF: Discussion Paper: National Strategy for Artificial Intelligence](#)
India | 2018 | government
Recommendation

### National Research Council Canada
[Advisory Statement on Human Ethics in Artificial Intelligence and Big Data Research (2017)](#)
Canada | 2019 | government
Binding agreement

### National Science and Technology Council; Networking and Information Technology Research and Development Subcommittee
[PDF: The National Artificial Intelligence Research and Development Strategic Plan](#)
United States | 2019 | government
Recommendation

### New York Times
[Seeking Ground Rules for A.I.](#)
United States | 2019 | private sector
Recommendation

### No organisation
[Holberton Turing Oath](#)
International | No Date | civil society
Voluntary commitment

## OECD
[Recommendation of the Council on Artificial Intelligence](#)
International | 2019 | international organisation
Recommendation

## OP Financial Group
[OP Financial Group's ethical guidelines for artificial intelligence](#)
Finland | private sector
Voluntary commitment

## Open AI
[Open AI Charter](#)
United States | 2018 | civil society
Voluntary commitment

## Oxford Munich Code of Conduct
[Code of Conduct](#)
International | 2019 | academia
Voluntary commitment

## Partnership On AI (Apple, Amazon, Google, MS, etc)
[Tenets Partnership on AI](#)
International | private sector
Voluntary commitment

## Personal Data Commission Singapore
[PDF: A Proposed Model Artificial Intelligence Governance Framework](#)
Singapore | 2019 | government
Recommendation
[Additional link](#)

## Pervade at University of Maryland
[Pervasive Data Ethics](#)
United States | academia
Voluntary commitment

## Philips
[Five guiding principles for responsible use of AI in healthcare and healthy living](#)
Netherlands | 2020 | private sector
Recommendation

### Policy Action Network
[AI & Data Topical Guide Series](#)
South Africa | 2020 | civil society
Recommendation

### Pontifical Academy for Life
[Rome Call – AI Ethics](#)
Italy | 2020 | religious institution
Voluntary commitment

### PriceWaterhouseCoopers UK
[PDF: A practical guide to Responsible Artificial Intelligence (AI)](#)
United Kingdom | 2019 | private sector
Recommendation

### Privacy International & Article 19
[PDF: Privacy and Freedom of Expression In the Age of Artificial Intelligence](#)
United Kingdom | 2018 | civil society
Recommendation

### Republic of Užupis
[PDF: Užupis Principles for Trustworthy AI Design](#)
Lithuania | 2019 | civil society
Recommendation
[Additional link](#)

### reputable AI
[The Priniciples](#)
International | nodate | private sector
Binding agreement

### Sage
[PDF: The Ethics of Code: Developing AI for Business with Five Core Principles](#)
United States | 2017 | private sector
Voluntary commitment

### SAP
[SAP's guiding principles for Artificial Intelligence](#)
Germany | 2018 | private sector
Voluntary commitment
[Additional link](#)

### Science, Law and Society (SLS) Initiative
[Principles for the Governance of AI](#)
United States | 2017 | civil society
Recommendation

### Software & Information Industry Association (SIIA)
[PDF: Ethical Principles for Artificial Intelligence and Data Analytics](#)
International | 2017 | private sector
Recommendation

### Sony
[PDF: Sony Group AI Ethics Guidelines](#)
Japan | 2019 | private sector
Voluntary commitment

### Stats New Zealand and Office of the Privacy Commissioner
[PDF: Principles for the safe and effective use of data and analytics](#)
New Zealand | 2018 | government
Recommendation

### Swiss Alliance for Data-Intensive Services
[PDF: Ethical Codex for Data-Based Value Creation: For Public Consultation](#)
Switzerland | 2019 | industry association
Recommendation

### Telefonica
[Principos / Principles](#)
Spain | 2018 | private sector
Binding agreement
[Additional link](#)

### Telia Company
[Telia Company Guiding Principles on trusted AI ethics](#)
Sweden | private sector
Voluntary commitment

### The Alan Turing Institute
[PDF: Understanding artificial intelligence ethics and safety](#)
United Kingdom | 2019 | academia
Recommendation

### The Critical Engineering Working Group
[THE CRITICAL ENGINEERING MANIFESTO](#)
Germany | 2019 | civil society
Voluntary commitment

### The Good Technology Collective
[The Good Technology Standard (GTS:2019-Draft-1)](#)
International | 2018 | civil society
Recommendation

### The Greens (Green Working Group Robots)
[PDF: Position on Robotics and Artificial Intelligence](#)
European Union | 2016 | other
Recommendation

### The Humanitarian Data Science and Ethics Group
[PDF: A Framework for the Ethical use of advanced Data Science Methodes in the Humanitarian Sector](#)
European Union | 2020 | academia
Recommendation

### The Information Accountability Foundation
[PDF: Unified Ethical Frame for Big Data Analysis (draft)](#)
United States | 2015 | civil society
Recommendation

### The Institute for Ethical and Machine Learning
[The Responsible Machine Learning Principles](#)
United Kingdom | civil society
Recommendation

### The Internet Society
[Artificial Intelligence and Machine Learning: Policy Paper](#)
United States | 2017 | civil society
Recommendation

### The Leadership Conference on Civil and Human Rights
[Civil Rights Principles for the Era of Big Data](#)
United States | 2014 | civil society
Recommendation

### The Open Data Institute
[PDF: Data Ethics Canvas](#)
United Kingdom | 2019 | civil society
Recommendation

### The Public Voice
[Universal Guidelines for Artificial Intelligence](#)
International | 2018 | civil society
Recommendation

### The Rathenau Instituut, Special Interest Group on Artificial Intelligence (SIGAI), ICT Platform Netherlands (IPN)
[PDF: Dutch Artificial Intelligence Manifesto](#)
Netherlands | 2017 | government
Recommendation

### The Royal Society
[PDF: Machine learning: the power and promise of computers that learn by example](#)
United Kingdom | 2017 | academia
Recommendation

### The White House
[PDF: Guidance for Regulation of Artificial Intelligence Applications](#)
United States | 2020 | government
Binding agreement

### Tieto
[PDF: Tieto's AI ethics guidelines](#)
Finland | 2018 | private sector
Voluntary commitment
[Additional link](#)

### UK Government
[A guide to using Artificial Intelligence in the public sector](#)
United Kingdom | 2019 | government
Recommendation

### UK House of Lords
[UK House of Lords Artificial Intelligence Committee's report, AI in the UK: ready, willing and able?](#)
United Kingdom | 2018 | government
Recommendation

### Unesco
[Unesco Global Code of Ethics](#)
International | intergovernmental organisation
Recommendation

### UNESCO
[Preliminary study on the Ethics of Artificial Intelligence](#)
France | 2019 | civil society
Recommendation

### UNESCO
[Report of COMEST on Robotics Ethics](#)
International | 2010 | international organisation
Recommendation

### UNI Global Union
[PDF: Top 10 Principles for Ethical Artificial Intelligence](#)
International | 2017 | civil society
Recommendation

### United Nations University Institute
[PDF: A Typological Framework for Data Marginalization](#)
China | 2019 | academia
Recommendation

### Unity
[Unity's six guiding AI principles](#)
United States | 2018 | private sector
Voluntary commitment

### Université de Montréal
[Montreal Declaration for Responsible AI](#)
Canada | 2018 | academia
Voluntary commitment

### University of Notre Dame
[PDF: A Code of Ethics for the Human Robot Interaction](#)
United States | academia
Recommendation

### University of Oxford - Future of Humanity Institute
[PDF: AI Governance: A research agenda](#)
United Kingdom | 2017 | academia
Recommendation

### University of Oxford a.o.
[PDF: The Malicious Use of Artificial Intelligence: Forecasting, Prevntion and Mitigation](#)
International | 2018 | academia
Recommendation

### Utrecht University
[Data Ethics Decision Aid (DEDA)](#)
Netherlands | 2017 | academia
Recommendation

### UX Studio Team
[AI UX: 7 Principles of Designing Good AI Products](#)
Hungary | 2018 | private sector
Recommendation

### Ver.di
[PDF: Künstliche Intelligenz – Gemeinwohl als Maßstabm Gute Arbeit als Prinzip](#)
Germany | 2019 | civil society
Recommendation

### Verbraucherzentrale Bundesverband e.V. (Federal Association of Consumer Protection Centres)
[PDF: Algorithmenbasierte Entscheidungsprozesse (Algorithmic decision-making processes)](#)
Germany | 2017 | civil society
Recommendation

### Verivox
[Verivox/Pro7 Selbstverpflichtung (Commitment)](#)
Germany | 2019 | private sector
Voluntary commitment
[Additional link](#)

### Vodafone Group
[Vodafone AI Framework](#)
United Kingdom | 2019 | private sector
Voluntary commitment

## W20
[PDF: Artificial Intelligence: open questions about gender inclusion](#)
International | 2018 | civil society
Recommendation

## Webfoundation
[PDF: Artificial Intelligence: open questions about gender inclusion](#)
Switzerland | 2018 | civil society
Recommendation

## Women leading in AI
[Principles for Responsible AI](#)
International | 2019 | civil society
Recommendation

## Work in the age of artificial intelligence. Four perspectives on the economy, employment, skills and ethics
[Ministry of Economic Affairs and Employment / Finland](#)
Finland | 2018 | government
Recommendation

## Working group "Vernetzte Anwendungen und Plattformen für die digitale Gesellschaft"
[PDF: Charta of digital networking](#)
Germany | 2014 | private sector
Voluntary commitment
[Additional link](#)

## World Economic Forum
[A Framework for Responsible Limits on Facial Recognition Use Case](#)
United States | 2020 | civil society
Recommendation

## World Economic Forum
[PDF: White Paper: How to Prevent Discriminatory Outcomes in Machine Learning](#)
International | 2018 | civil society
Recommendation

**Additional Bibliography**